%% file: thesis.tex
\documentclass[12pt,PhD]{thesiscls}%this is where the cls file goes
\pdfoutput=1
\usepackage{verbatim}
\usepackage[dvipdfmx]{graphicx} \usepackage{bmpsize}
\usepackage{url} % typeset URL's reasonably
\usepackage{listings}
\usepackage{natbib}
\usepackage{multirow}
\usepackage{bigstrut}
\usepackage{subfigure}
\usepackage{booktabs}
\usepackage{float}
\usepackage{epigraph}
\usepackage{rotating}
\usepackage{longtable}
\usepackage{nccmath}
\usepackage{amssymb}
\usepackage{amsmath}
\usepackage{dblfloatfix}
\usepackage{textcomp}
\usepackage{dsfont}
\usepackage[toc,nogroupskip,style=super3col]{glossaries}
\usepackage[htt]{hyphenat}
\usepackage[multiple]{footmisc}
\usepackage[font={small}, labelfont=bf, margin=1cm]{caption}
\usepackage[shortlabels]{enumitem}
\usepackage[dvipsnames,table]{xcolor}
\usepackage{afterpage}

\usepackage{fancyhdr}
\makeglossaries
\loadglsentries{glossary}

\usepackage{lscape}

\begin{document}

% Uncomment the following lines to leave out list of figures, tables
% and copyright until final printing
%\figurespagefalse
%\tablespagefalse
%\copyrightfalse

\title{Correlations between multiple tracers\\of the cosmic web}
\author{Tracey Friday}
%\principaladviser{SUPERVISORS NAME}
%\firstreader{SECOND SUPERVISOR}

\setlength{\footskip}{2cm}

\let\cleardoublepage\clearpage
\beforeabstract

{
\topskip0pt
\vspace*{\fill}

\begin{center}\textit{``I was, am and will be enlightened instantaneously with the Universe.''}\end{center}
\begin{flushright} - Keizan Zenji\end{flushright}

\begin{center}\textit{``People can come up with statistics to prove anything, Kent.\\
40\% of people know that!''}\end{center}
\begin{flushright} - Homer Simpson\end{flushright}
\vspace*{\fill}
}

\prefacesection{Abstract}
\input{abstract}
\afterabstract

\prefacesection{Acknowledgements}

I find myself completing this thesis and writing these acknowledgements during the Covid-19 global pandemic. I am so grateful to the NHS, and all the other people keeping us safe, educated, connected, and fed. Thank you is not enough.

\vspace{5mm}\noindent I thank my supervisors, Roger Clowes, Gerry Williger, and Srini Raghunathan, for the freedom to pursue my hunches, for bringing academic rigour to my work, and for guidance presenting the results of my research.

\vspace{5mm}\noindent I appreciate the support of JHI staff and fellow students, particularly the weekly therapy sessions over tea and doughnuts. I gratefully acknowledge financial support from the Science and Technology Facilities Council.

\vspace{5mm}\noindent I am grateful to Vincent Pelgrims for encouraging me, probably without realising, to investigate an intriguing `cosmic curiosity', and for his assistance with the correlation tests. I also thank Damien Hutsem{\'e}kers for sharing data.

\vspace{5mm}\noindent I finally extend very special thanks to my family and friends for their unwavering encouragement and utter faith in me, and for never questioning me undertaking this endeavour at an age when I really should know better. Thank you.

\printglossary[title={Abbreviations}]

\clearpage\mbox{}\clearpage

\afterpreface

\raggedbottom % needed if using twopage in thesiscls cos otherwise it tries to have same bottoms for facing pages which introduces vertical spacing

% These include the actual text
\include{chapterIntro}
\include{chapterSNeVsCMBT}
\include{chapterLQGSample}
\include{chapterLQGStatsMethods}
\include{chapterLQGResults}
\include{chapterDiscussion}
\include{chapterConclusion}

\pagestyle{fancy}
\lhead{} % left head
%\lhead{\MakeUppercase{\chaptername \ \thechapter}}  % left head
%\chead{Draft: \today} % centre head
\lfoot{}
\cfoot{\thepage}
\rfoot{}
%\bibliography{thesis}             % this causes the references to be
                                  % listed
%\bibliographystyle{alpha}
\bibliographystyle{aa}       % this determines the style in which
                                % the references are printed, other
                                % possible values are plain and abbrv
\bibliography{thesis}             % this causes the references to be
                                  % listed
                                  
% Appendices start here
\appendix
\include{appendixISW}
\include{appendixSZ}
\include{appendixFields1to7}
\include{appendixVariances}
\include{appendixAltPlanckMaps}
\include{appendixComovingDistance}
\include{appendixLQGData}
\include{appendixLQGs3DPerspectives}
\include{appendixVoronoi}
\include{bibliography}
\end{document}

%% file: glossary.tex
\newglossaryentry{HR2}
{name=HR2,
description={Horizon Run 2 (cosmological simulation)\dotfill}}

\newglossaryentry{HOD}
{name=HOD,
description={Halo occupation distribution\dotfill}}

\newglossaryentry{AGN}
{name=AGN,
description={Active galactic nuclei\dotfill}}

\newglossaryentry{SMBH}
{name=SMBH,
description={Supermassive black hole\dotfill}}

\newglossaryentry{KDE}
{name=KDE,
description={Kernel density estimation\dotfill}}

\newglossaryentry{ICM}
{name=ICM,
description={Intra-cluster medium\dotfill}}

\newglossaryentry{A1}
{name=A1,
description={Alignment region 1 (of quasar optical polarization alignment)\dotfill}}

\newglossaryentry{RN1}
{name=RN1,
description={Radio north region 1 (of quasar radio polarization alignment)\dotfill}}

\newglossaryentry{RN2}
{name=RN2,
description={Radio north region 2 (of quasar radio polarization alignment)\dotfill}}

\newglossaryentry{CDM}
{name=CDM,
description={Cold dark matter\dotfill}}

\newglossaryentry{LCDM}
{name=$\Lambda$CDM,
description={$\Lambda$ (dark energy) cold dark matter\dotfill}}

\newglossaryentry{LSS}
{name=LSS,
description={Large-scale structure\dotfill}}

\newglossaryentry{BAO}
{name=BAO,
description={Baryon acoustic oscillations\dotfill}}

\newglossaryentry{BCG}
{name=BCG,
description={Brightest cluster galaxy\dotfill}}

\newglossaryentry{LRG}
{name=LRG,
description={Luminous red galaxy\dotfill}}

\newglossaryentry{SDSS}
{name=SDSS,
description={Sloan Digital Sky Survey\dotfill}}

\newglossaryentry{LQG}
{name=LQG,
description={Large quasar group\dotfill}}

\newglossaryentry{CDF}
{name=CDF,
description={Cumulative distribution function\dotfill}}

\newglossaryentry{EDF}
{name=EDF,
description={Empirical distribution function (aka ECDF)\dotfill}}

\newglossaryentry{ODR}
{name=ODR,
description={Orthogonal distance regression\dotfill}}

\newglossaryentry{FoF}
{name=FoF,
description={Friends-of-friends\dotfill}}

\newglossaryentry{GoF}
{name=GoF,
description={Goodness-of-fit\dotfill}}

\newglossaryentry{MST}
{name=MST,
description={Minimal spanning tree\dotfill}}

\newglossaryentry{CHMS}
{name=CHMS,
description={Convex hull of member spheres\dotfill}}

\newglossaryentry{PA}
{name=PA,
description={Position angle\dotfill}}

\newglossaryentry{HWCI}
{name=HWCI,
description={Half-width confidence interval\dotfill}}

\newglossaryentry{QSO}
{name=QSO,
description={Quasi-steller object (quasar)\dotfill}}

\newglossaryentry{NCP}
{name=NCP,
description={North celestial pole\dotfill}}

\newglossaryentry{SCP}
{name=SCP,
description={South celestial pole\dotfill}}

\newglossaryentry{EQP}
{name=EQP,
description={Equatorial `pole' (for parallel transport toy model)\dotfill}}

\newglossaryentry{NGC}
{name=NGC,
description={North Galactic cap\dotfill}}

\newglossaryentry{SGC}
{name=SGC,
description={South Galactic cap\dotfill}}

\newglossaryentry{HR}
{name=HR,
description={Hermans-Rasson (test of uniformity on the circle)\dotfill}}

\newglossaryentry{KS}
{name=KS,
description={Kolmogorov–Smirnov test\dotfill}}

\newglossaryentry{BC}
{name=BC,
description={Bimodality coefficient ($\mathrm{BC}_{crit} \approx 0.555$)\dotfill}}

\newglossaryentry{HDS}
{name=HDS,
description={Hartigans' dip statistic (test of unimodality)\dotfill}}

\newglossaryentry{SL}
{name=SL,
description={Significance level\dotfill}}

\newglossaryentry{SNe}
{name=SNe,
description={Supernovae\dotfill}}

\newglossaryentry{ISW}
{name=ISW,
description={Integrated Sachs-Wolfe effect\dotfill}}

\newglossaryentry{RS}
{name=RS,
description={Rees-Sciama effect\dotfill}}

\newglossaryentry{CMB}
{name=CMB,
description={Cosmic microwave background\dotfill}}

\newglossaryentry{SZ}
{name=SZ,
description={Sunyaev-Zel\textquotesingle dovich effect\dotfill}}

\newglossaryentry{HEALPix}
{name=HEALPix,
description={Hierarchical Equal Area iso-Latitude Pixelisation scheme\dotfill}}

\newglossaryentry{SAI}
{name=SAI,
description={Sternberg Astronomical Institute\dotfill}}

\newglossaryentry{OSC}
{name=OSC,
description={Open Supernova Catalogue\dotfill}}

\newglossaryentry{FWHM}
{name=FWHM,
description={Full-width at half-maximum\dotfill}}

\newglossaryentry{HRHD}
{name=HRHD,
description={Half-ring half-difference\dotfill}}

\newglossaryentry{HMHD}
{name=HMHD,
description={Half-mission half-difference\dotfill}}

\newglossaryentry{MWU}
{name=MWU,
description={Mann-Whitney U test (aka Mann–Whitney–Wilcoxon test etc.)\dotfill}}

\newglossaryentry{OLS}
{name=OLS,
description={Ordinary least squares linear regression\dotfill}}

\newglossaryentry{WLS}
{name=WLS,
description={Weighted least squares linear regression\dotfill}}

%% file: abstract.tex
% START

The large-scale structure of the Universe is a `cosmic web' of interconnected clusters, filaments, and sheets of matter. This work comprises two complementary projects investigating the cosmic web using correlations between three different tracers: the cosmic microwave background (CMB), supernovae (SNe), and large quasar groups (LQGs). In the first project we re-analyse the apparent correlation between CMB temperature and SNe redshift reported by Yershov, Orlov and Raikov. In the second we investigate for the first time whether LQGs exhibit coherent alignment.

\textbf{Project 1:} The cosmic web leaves a detectable imprint on CMB temperature. Evidence presented by Yershov, Orlov and Raikov showed that the \textit{WMAP}/\textit{Planck} CMB pixel-temperatures at supernovae (SNe) locations tend to increase with increasing redshift. They suggest this could be caused by the Integrated Sachs-Wolfe effect and/or by residual foreground contamination.

We assess the correlation independently using \textit{Planck} 2015 SMICA R2.01 data and, following Yershov et al., a sample of 2,783 SNe from the Sternberg Astronomical Institute. Our analysis supports the prima facie existence of the correlation but attributes it to a composite selection bias caused by the chance alignment of seven deep survey fields with CMB hotspots. These seven fields contain 9.2\% of the SNe sample (256 SNe). Spearman's rank-order correlation coefficient indicates the correlation present in the whole sample (p-value $= 6.7 \times 10^{-9}$) is insignificant for a sub-sample of the seven fields together (p-value $= 0.2$) and entirely absent for the remainder of the SNe (p-value $= 0.6$). We demonstrate the temperature and redshift biases of these deep fields, and estimate the likelihood of their falling on CMB hotspots by chance is $\sim6.8\%-8.9\%$ (approximately 1 in $11-15$). We show that a sample of 7,880 SNe from the Open Supernova Catalogue exhibits the same effect, and conclude that it is an accidental but not unlikely selection bias.

\textbf{Project 2:} The cosmic web influences structure formation and evolution; galaxy and quasar spins correlate with their host large-scale structures. We investigate for the first time whether the LQGs hosting the quasars themselves exhibit coherent alignment. We use the position angle (PA) of 71 LQGs in the redshift range \mbox{$1.0\leq z\leq1.8$} as the tracer here. We find that their PAs do not follow a random distribution (p-values $0.008 \lesssim p \lesssim 0.07$). Their distribution is bimodal with modes at the centre of the A1 region of $\bar{\theta}\sim52\pm2^{\circ}, 137\pm3^{\circ}$. The median location of the modes at all 71 LQG locations is $\bar{\theta}\sim45\pm2^{\circ}, 136\pm2^{\circ}$. We find that the LQG PAs are correlated, specifically aligned and orthogonal, with a maximum significance of $\simeq 0.8\%$ ($2.4\sigma$) at typical angular (comoving) separations of $\sim 30^{\circ}$ ($\sim1.6$ Gpc).

This finding suggests an explanation for the Gpc-scale coherent orientation of quasar polarization, first reported by Hutsem{\'e}kers in 1998, and since widely studied. Our results are remarkably close to the radio polarization angles of $\bar{\theta}\simeq42^{\circ}, 131^{\circ}$ reported by Pelgrims and Hutsem{\'e}kers. The origin of these correlations is still an open question, but is often attributed to quasars' intrinsic alignment. More recently, evidence has emerged that quasar polarization vectors are preferentially parallel and orthogonal to LQG axes. Our results integrate these two findings.

The statistical significance of our results is marginal, and we cannot exclude the LQG correlation being a chance anomaly. However, spatial coincidence between our LQG sample and regions of quasar polarization alignment, and the similarity between LQG PAs and radio polarization angles, suggest an interesting result. The tendency of LQGs to be either aligned or orthogonal warrants further investigation, and we suggest future work to evaluate this intriguing correlation further.

%% file: chapterIntro.tex
% START

\chapter{Introduction} \label{chaptIntro}
\pagenumbering{arabic}

The widely accepted lambda cold dark matter (\gls{LCDM}) cosmological model of a Universe comprising $\sim 69\%$ dark energy (denoted by $\Lambda$), $\sim 26\%$ cold dark matter (\gls{CDM}), and $\sim 5\%$ ordinary baryonic matter is the fundamental foundation of most modern cosmology. Precision measurements of the cosmic microwave background (CMB) temperature, polarization, and lensing anisotropies from the South Pole Telescope, \textit{Planck} satellite, and Atacama Cosmology Telescope \citep[e.g.,][and references therein]{Story2013, Planck2016b, Louis2017} have strongly reinforced the preference for $\Lambda$CDM as the concordance model of cosmology. $\Lambda$CDM describes the evolution of the Universe from the big bang through 13.8 billion years to today. During this time the Universe has expanded from a hot, dense, almost entirely smooth primordial state to the cool, rich, hierarchical large-scale structure (LSS) of galaxy clusters, superclusters, walls, sheets, filaments and voids of the `cosmic web' we see today.

This work investigates correlations between multiple tracers of the cosmic web, up to redshifts $z \sim 1.8$. At high redshifts, direct tracers need to be highly luminous, such as supernovae and quasars as used in this work. In this chapter, we introduce the cosmic web and report on evidence that galaxies and, in the case of active galaxies their quasars, trace and align with the cosmic web. We give an overview of the CMB, and the imprints large-scale structure leaves on the CMB, which can be used as a probe of matter distribution along the line-of-sight.

\section[The large-scale structure of the Universe: the `cosmic web']{The large-scale structure of the Universe:\\the `cosmic web'} \label{secCosmicWeb}

\begin{figure} [t]
    \centering
    \includegraphics[width=0.85\columnwidth]{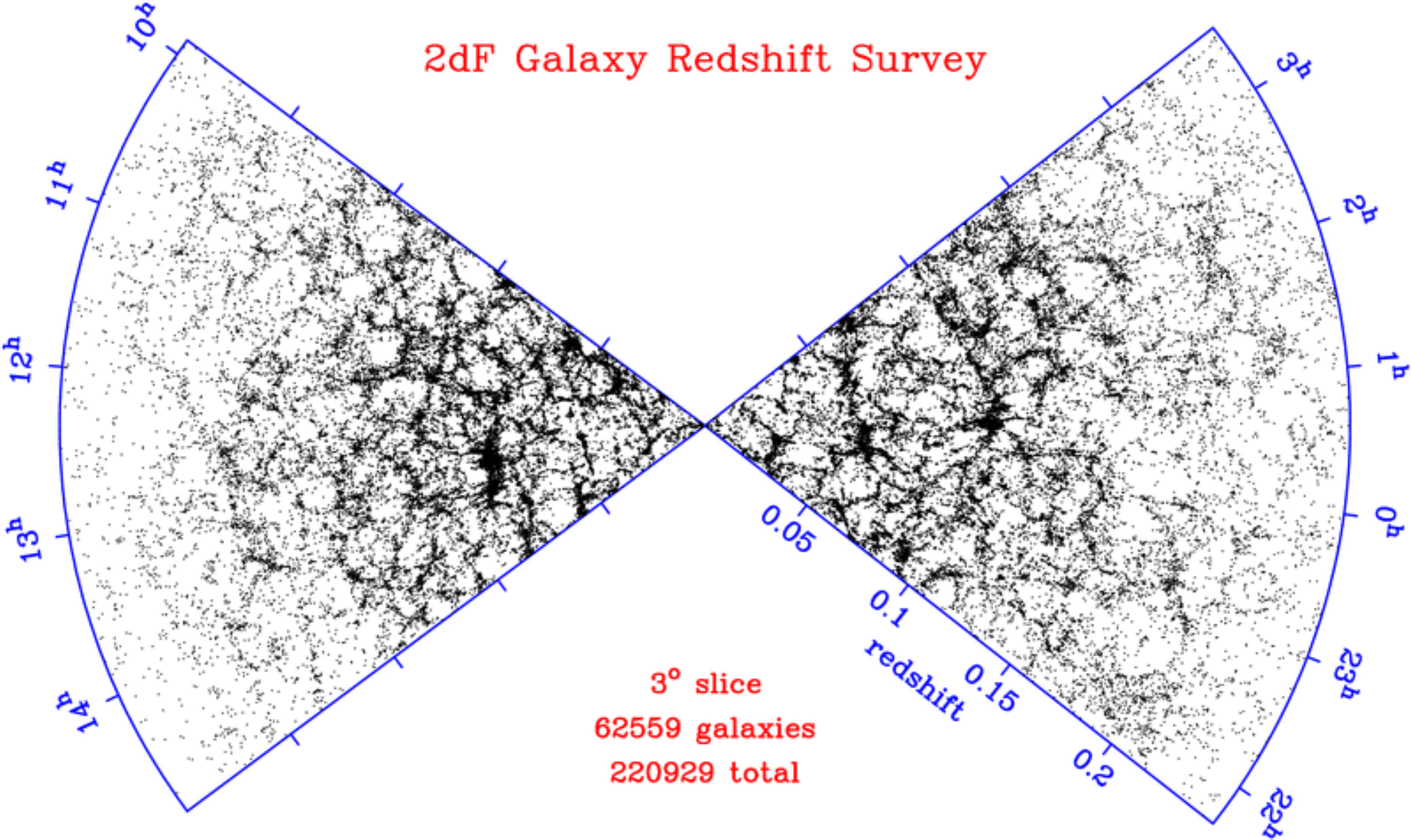}
    \caption[The cosmic web of galaxies from the 2dF Galaxy Redshift Survey]{The spatial distribution of galaxies from the Two-Degree Field Galaxy Redshift Survey \protect\citep{Colless2001}, as a function of redshift and right ascension, for a $3^{\circ}$ slice in declination. Image credit: the 2dFGRS team.}
    \label{fig2dF}
\end{figure}

The large-scale spatial distribution of galaxies in the Universe is far from uniform, as demonstrated by wide-area surveys such as the Sloan Digital Sky Survey \citep[SDSS,][]{York2000}, the Two-Degree Field Galaxy Redshift Survey \citep[2dFGRS, Fig.~\ref{fig2dF},][]{Colless2001}, the Galaxy And Mass Assembly survey \citep[GAMA,][]{Driver2009}, the Dark Energy Survey \citep[DES,][]{DES2016}, and the VIMOS Public Extragalactic Redshift Survey \citep[VIPERS,][]{Scodeggio2018}.

\begin{figure} [h!]
    \centering
    \includegraphics[width=0.65\columnwidth]{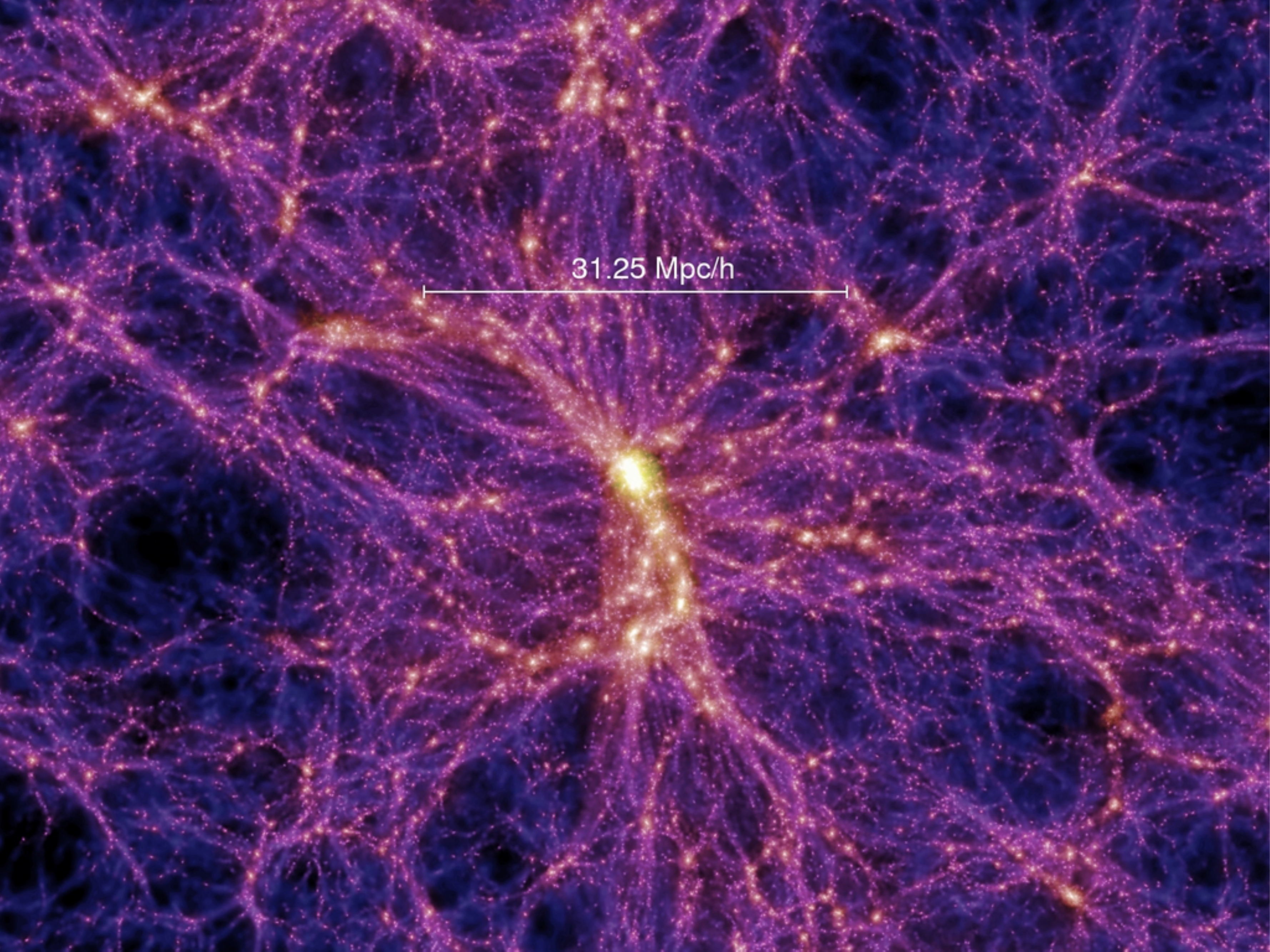}
    \caption[The cosmic web of dark matter from the Millennium Simulation]{The cosmic web of dark matter modelled by the Millennium Simulation \protect\citep{Springel2005}. Visualisation is of a $15\:h^{-1}$ Mpc thick slice through the density field at $z = 1.4$. Image credit: Springel et al. (Virgo Consortium), Max-Planck-Institut f{\"u}r Astrophysik.}
    \label{figMillenium}
\end{figure}

Rather, the large-scale structure (\gls{LSS}) of the Universe forms a `cosmic web' \citep{Bond1996} of filaments, sheets, and voids, which evolved via gravitational instability from small density perturbations in the near-homogeneous early Universe. Using numerical simulations \citet{Bond1996} showed the final web of filaments is present in the initial density fluctuations. More recent cosmological simulations \citep[e.g.\ Fig.~\ref{figMillenium},][]{Springel2005,Dubois2014} now model the Universe with impressive detail and accuracy.

Filaments and walls of galaxies have been observed on $\sim 100\:h^{-1}$ Mpc scales for some time \citep[e.g.][]{deLapparent1986,Geller1989,Gott2005}. More recently, the mass distribution of filaments has been detected by weak lensing of background galaxies \citep{Jauzac2012,Clampitt2016,Epps2017,Yang2020} and of the CMB \citep{He2018}. Filaments have also been detected in ionized gas \citep{deGraaff2019} by the Sunyaev-Zel\textquotesingle dovich (SZ) effect \citep{SunyaevZeldovich1970}.

Analysis of galaxy filaments in Las Campanas Redshift Survey \citep[LCRS,][]{Shectman1996} by \citet{Bharadwaj2004} indicates the length at which filaments are statistically significant is $\lesssim 70-80\:h^{-1}$ Mpc (comoving). On larger scales they find that filaments connect by statistical chance and are not real structures. However, using CMB lensing, \citet{He2018} observe that filament length increases as a function of redshift, up to $\sim 260$ Mpc (comoving, $h = 0.673$) by $z = 0.7$.

\citet{Bond2010} identify filaments in the SDSS galaxy distribution and in cosmological simulations, and find the backbone of the filamentary structure is in place by $z \sim 3$. Indeed, \citet{Moller2001} detect a filament at $z = 3.04$. Further, \citet{Choi2010} show that the distribution of filament lengths has remained roughly constant between redshifts $0.1 \lesssim z \lesssim 0.8$, although filaments become narrower at lower redshift. The statistical properties of filaments (e.g.\ length), and the characteristics of their constituent galaxies (including alignment), strongly depend on which algorithm is used to identify them \citep{Rost2019}.

A key assumption of $\Lambda$CDM is that the Universe is homogeneous and isotropic on scales large enough for the cosmic web to be treated as perturbations, with a statistical distribution independent of position. The transition from non-uniform to homogeneous has been measured by various galaxy and quasar surveys on comoving scales $\sim 70-150\:h^{-1}$ Mpc \citep[e.g.][]{Hogg2005,Scrimgeour2012,Pandey2015}, although other authors claim spatial inhomogeneity of galaxies on comoving scales up to $300\:h^{-1}$ Mpc \citep[e.g.][]{SylosLabini2011,Park2017}. Using a fractal analysis of $\Lambda$CDM $N$-body simulations \citet{Yadav2010} estimate the upper limit to the homgeneity scale is $\sim 260\:h^{-1}$ Mpc (comoving).

\subsection{Galaxies align with their host large-scale structure} \label{secGalaxiesAlign}

It has long been known that galaxy clusters are elongated \citep[e.g.][]{Carter1980}, tend to be orientated towards neighbouring clusters on scales $\sim 30$ Mpc \citep{Binggeli1982}, and that the brightest cluster galaxy (\gls{BCG}) tends to be preferentially aligned with the major axis of its parent cluster \citep[][and references therein]{Sastry1968,Binggeli1982,Plionis2003}.

Most studies of BCG alignment focus on relatively low redshifts, e.g.\ \citet{NiedersteOstholt2010}, who find that clusters at `high' redshift ($z > 0.26$) show less alignment than those at lower redshift. However, \citet{West2017} show that BCG alignments are found in a sample of the most massive galaxy clusters at redshifts $0.19 < z < 1.8$. They suggest the most likely mechanisms for BCG alignments are

\begin{enumerate}[(a)]
\item anisotropic infall of matter along filaments,
\item primordial alignment with the surrounding matter density field,
\item gravitational torques, or
\item some combination of these.
\end{enumerate}

Considering other galaxy types, alignment of luminous red galaxies (\gls{LRG}s) is detected at separations of $\sim 100\:h^{-1}$ Mpc up to redshift $z \lesssim 0.7$ \citep[e.g.][]{Joachimi2011,Singh2015}, and alignment of nearby ($z \leq 0.02$) spirals is reported at separations of $\sim 2\:h^{-1}$ Mpc \citep{Lee2011}. Using the Horizon-AGN \citep{Dubois2014} cosmological simulation, \citet{Bate2020} show that the alignment of elliptical galaxies begins at $z \sim 3$, leading to significant alignment by $z \sim 1$.

On smaller scales (hundreds of kpc), coherent planes of satellite galaxies are observed throughout the local Universe \citep[e.g.][]{Pawlowski2013, Ibata2013, Tully2015}. \citet{Pawlowski2018} suggests these may be caused by accretion along filaments, infall of satellite groups, or the formation of tidal dwarfs. The filamentary accretion hypothesis is supported by \citet{Libeskind2015}, who find that satellite planes are preferentially aligned with the local large-scale structure.

The orientation of galaxies is influenced by the gravitational potential and angular momentum of the surrounding large-scale structure. In the absence of other dynamical factors (e.g.\ mergers) the shape of elliptical galaxies will tend to be perturbed to align with the LSS gravitational potential (tidal distortion), whilst spiral galaxies will tend to align their spin with the LSS angular momentum (tidal torque) \citep{Kiessling2015}.

Cosmological simulations such as Horizon-AGN \citep{Dubois2014} and S\textsc{imba} \citep{Dave2019} indicate the spin of haloes (and galaxies) are preferentially aligned with filaments and sheets at low masses (mainly spirals) and orthogonal at high masses (mainly ellipticals) \citep[e.g.][]{Dubois2014,Codis2018,Kraljic2020}. Using the \textit{Planck} Millennium simulation \citep{Baugh2019}, \citet{GaneshaiahVeena2018} demonstrate this is a result of accretion history, with low-mass haloes tending to accrete mass from orthogonal to their host filament and thus orientating their spins along the filaments. In contrast, they find high-mass haloes tend to accrete along their host filament and have spins orthogonal to them.

These predicted alignments between the orientation of galaxies and their surrounding large-scale structure have been observed. For example, \citet{Zhang2013} find that the major axes of SDSS DR7 galaxies are preferentially aligned with the direction of filaments and within the plane of sheets, and \citet{Tempel2015} find that orientation of SDSS DR10 galaxy pairs is aligned with their host filaments. Recently \citet{Welker2020} detected the mass-dependent transition of galaxy spin alignments with filaments, from parallel at low-mass to orthogonal at high-mass. They found that this shift occurred at $10^{10.4-10.9}$ M\textsubscript{\(\odot\)}, consistent with Horizon-AGN predictions \citep{Dubois2014,Codis2018}.

\subsection{Quasars align with their host large-scale structure} \label{secQuasarsAlign}

Quasars are the most luminous active galactic nuclei (\gls{AGN}), powered by accretion onto supermassive black holes \citep[\gls{SMBH}s, e.g.][]{Salpeter1964}. The fraction of the SMBH lifetime it is actively accreting, or equivalently the fraction of quasars in the population of SMBHs, is known as the quasar `duty cycle'. For a typical duty cycle of $0.1-1\%$ the mean quasar lifetime is $10-100$ Myr \citep{Marinello2016}.

High-redshift quasars are believed to exist within the most massive host galaxies of large dark matter haloes \citep{Turner1991}. These are associated with extreme over-densities in the matter density field, which are predicted to become progressively more rare with increasing redshift \citep{Efstathiou1988}. Quasars are biased tracers of the most over-dense regions in the early Universe, which simulations indicate are the progenitors of galaxy clusters we see in the present epoch \citep{Springel2005,DiMatteo2008,Angulo2012}. Indeed, there are observations of galaxies clustering around some high-redshift ($z \gtrsim 3$) quasars \citep[e.g.][]{Husband2013,McGreer2014,Husband2015}.

\begin{figure}
    \centering
    \includegraphics[width=\columnwidth]{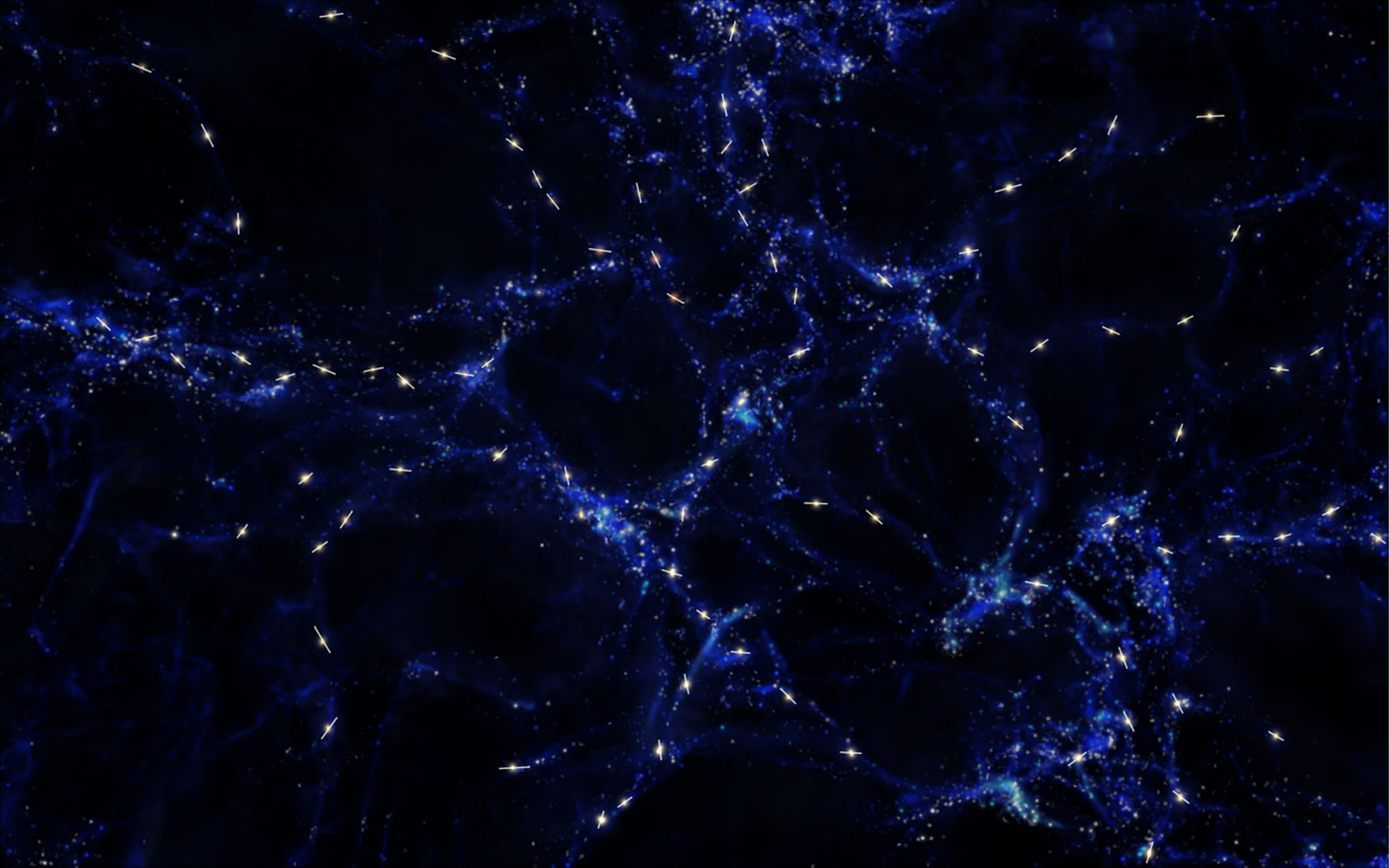}
    \caption[Alignment of quasar polarization with LSS (artist's impression)]{Artist's impression of the alignment of quasar polarization with large-scale structure, reported by \protect\citet{Hutsemekers2014}. The cosmic web is shown in blue, quasars in white, and the rotation axes of their black holes are indicated with a line. Image credit: ESO/M. Kornmesser.}
    \label{figQSOAlignment}
\end{figure}

Large-scale anisotropies, up to Gpc scales, are observed in the spatial distribution of quasars; see discussion of large quasar groups in section~\ref{secLQGs}. There are also anisotropies in quasar polarization, potentially on even larger scales (Fig.~\ref{figQSOAlignment}). This was first reported by \citet{Hutsemekers1998}, who found the polarization of optical light from quasars was coherently orientated on scales $\sim 1,000\:h^{-1}$ Mpc at redshifts of \mbox{$1 \lesssim z \lesssim 2$}. This was then confirmed at higher significance levels by further polarization observations at optical wavelengths \citep{Hutsemekers2001,Cabanac2005,Hutsemekers2005}, the introduction of coordinate-invariant statistics by \citet{Jain2004}, analysis using a new and completely independent statistical method proposed by \citet{Pelgrims2014}, and polarization measurements at radio wavelengths \citep{Tiwari2013,Pelgrims2015}.

\citet{Joshi2007} find no such evidence for large-scale polarization alignment at radio wavelengths. However, they do not use the same source type restriction as \citet{Pelgrims2015}, who only find alignment with radio quasars, but not all radio sources. Nor do they examine different redshift intervals, although from the first detection \citep{Hutsemekers1998} the alignment is reported to be redshift dependent. \citet{Tiwari2019} also report isotropic radio polarizations with no evidence of alignment on scales $\gtrsim 800$ Mpc, although they are prevented from examining smaller scales due to the number density of their radio source sample. The claims of quasar polarization alignment, although widely reported, therefore remain somewhat controversial.

\begin{table} [t]
    \centering
    \small
    \caption[Regions of quasar optical polarization alignment]{Regions of quasar optical polarization alignment \protect\citep[e.g.][]{Hutsemekers1998,Hutsemekers2005}. RA and Dec (B1950) ranges delineate region boundaries. A1\textsubscript{(low)}, A3\textsubscript{(low)}, and A3\textsubscript{(high)} are the low/high-redshift counterparts to \gls{A1} and A3. No departure from random was detected for A3\textsubscript{(low)} and A3\textsubscript{(high)}. A2 was dropped and most objects incorporated in A1\textsubscript{(low)} when the authors increased the sample size between 1998 and 2005. Average polarization directions $\bar{\theta}$ are taken from \citet{Hutsemekers2005}.}
    \label{tabQSORegionsOptical}
    \begin{tabular}{lcccr}
        \hline
        Region & RA ($^{\circ}$) & Dec ($^{\circ}$) & Redshift & $\bar{\theta}$ ($^{\circ}$) \\
        \hline
        A1\textsubscript{(low)} & $168 \leq \alpha \leq 217$ & $\delta \leq 50$ & $0.0 \leq z \leq 1.0$ & 79 \\
        A1 & $168 \leq \alpha \leq 217$ & $\delta \leq 50$ & $1.0 \leq z \leq 2.3$ & 8 \\
        A2 & $150 \leq \alpha \leq 250$ & $\delta \leq 50$ & $0.0 \leq z \leq 0.5$ & - \\
        A3\textsubscript{(low)} & $320 \leq \alpha \leq 360$ & $\delta \leq 50$ & $0.0 \leq z \leq 0.7$ & - \\
        A3 & $320 \leq \alpha \leq 360$ & $\delta \leq 50$ & $0.7 \leq z \leq 1.5$ & 128 \\
        A3\textsubscript{(high)} & $320 \leq \alpha \leq 360$ & $\delta \leq 50$ & $1.5 \leq z \leq 3.0$ & - \\
        \hline
    \end{tabular}
\end{table}

\begin{table} [h!]
    \centering
    \small
    \caption[Regions of quasar radio polarization alignment]{Regions of quasar radio polarization alignment \protect\citep{Pelgrims2015}. RA and Dec (B1950) cooordinates identify the region centre, $\bar{\xi}$ and $\xi_{max}$ are the mean and maximum angular separations of objects from the region centre, and $\bar{\theta}$ is the average polarization direction for the region. No redshift limits, although correlation is more significant for $z > 1$.}
    \label{tabQSORegionsRadio}
    \begin{tabular}{lccccr}
        \hline
        Region & RA ($^{\circ}$) & Dec ($^{\circ}$) & $\bar{\xi}$ ($^{\circ}$) & $\xi_{max}$ ($^{\circ}$) & $\bar{\theta}$ ($^{\circ}$) \\
        \hline
        \gls{RN1} & 163 & 12 & 12 & 21 & 131 \\
        \gls{RN2} & 206 & 38 & 14 & 25 & 42 \\
        RS1 & 340 & 18 & 15 & 25 & 57 \\
        \hline
    \end{tabular}
\end{table}

The alignments in quasar optical and radio polarization are grouped in distinct regions, two in the north Galactic cap and one in the south Galactic cap, each with different preferred directions (Tables~\ref{tabQSORegionsOptical} and \ref{tabQSORegionsRadio}). The regions identified at optical wavelengths differ to those found at radio wavelengths, although there is significant overlap. In the south $\gtrsim 85 \%$ of the objects within the RS1 region are also within A3, and in the north $\gtrsim 70 \%$ of those within RN2 overlap with A1 \citep{Pelgrims2015}. Region RN2 also coincides with the main cluster of aligned polarized radio sources identified independently by \citet{Shurtleff2014} using an alternative statistical method.

A plausible physical mechanism for the alignment of quasar polarization on such large scales is not obvious, unless the polarization is induced along the line-of-sight. From the first detection, interstellar polarization was a concern but deemed unlikely \citep{Hutsemekers1998}. Most recently, \citet{Pelgrims2019} revisits this possibility using \textit{Planck} 353 GHz maps to estimate the contamination of quasar polarization data by Galactic dust \citep{Planck2015a,Planck2015b}. He finds the alignments are robust against Galactic dust contamination.

Another potential line-of-sight mechanism widely discussed is exotic particles, such as axion-photon mixing in external magnetic fields \citep[e.g.][]{Cabanac2005,Hutsemekers2005,Payez2008,Hutsemekers2011}, although this is now disfavoured using constraints from circular polarization measurements \citep{Hutsemekers2010,Payez2011}.

If polarization is not induced along the line-of-sight, we must consider instrinsic alignment of the quasar spin axes \citep[e.g.][]{Hutsemekers1998,Cabanac2005,Pelgrims2016thesis}. \citet{Hutsemekers2014} report that optical quasar polarization is preferentially either aligned or orthogonal to the host large-scale structure. They propose that this bimodality is due to the orientation of the accretion disk with respect to the line-of-sight, and conclude that quasar spin axes are likely parallel to their host large-scale structures. \citet{Pelgrims2016} report a similar result using radio wavelengths and large quasar groups (LQGs). They also conclude that the quasar spin axes are preferentially parallel to the LQG major axis for LQGs with at least 20 members, although they suggest this becomes orthogonal with fewer members ($10 < m < 20$).

Several studies report radio jets are aligned over large scales \citep{Taylor2016,Contigiani2017}, supporting the intrinsic alignment explanation independently of polarization measurements. However, using a different jet scale, \citet{Blinov2020} find no such alignment, although they do report a weak correlation between jet direction and polarization orientation. The intrinsic alignment interpretation of coherent quasar polarization orientations remains open.

\subsection{Large quasar groups (LQGs)} \label{secLQGs}

\begin{figure} [h!]
    \centering
    \includegraphics[width=0.7\columnwidth]{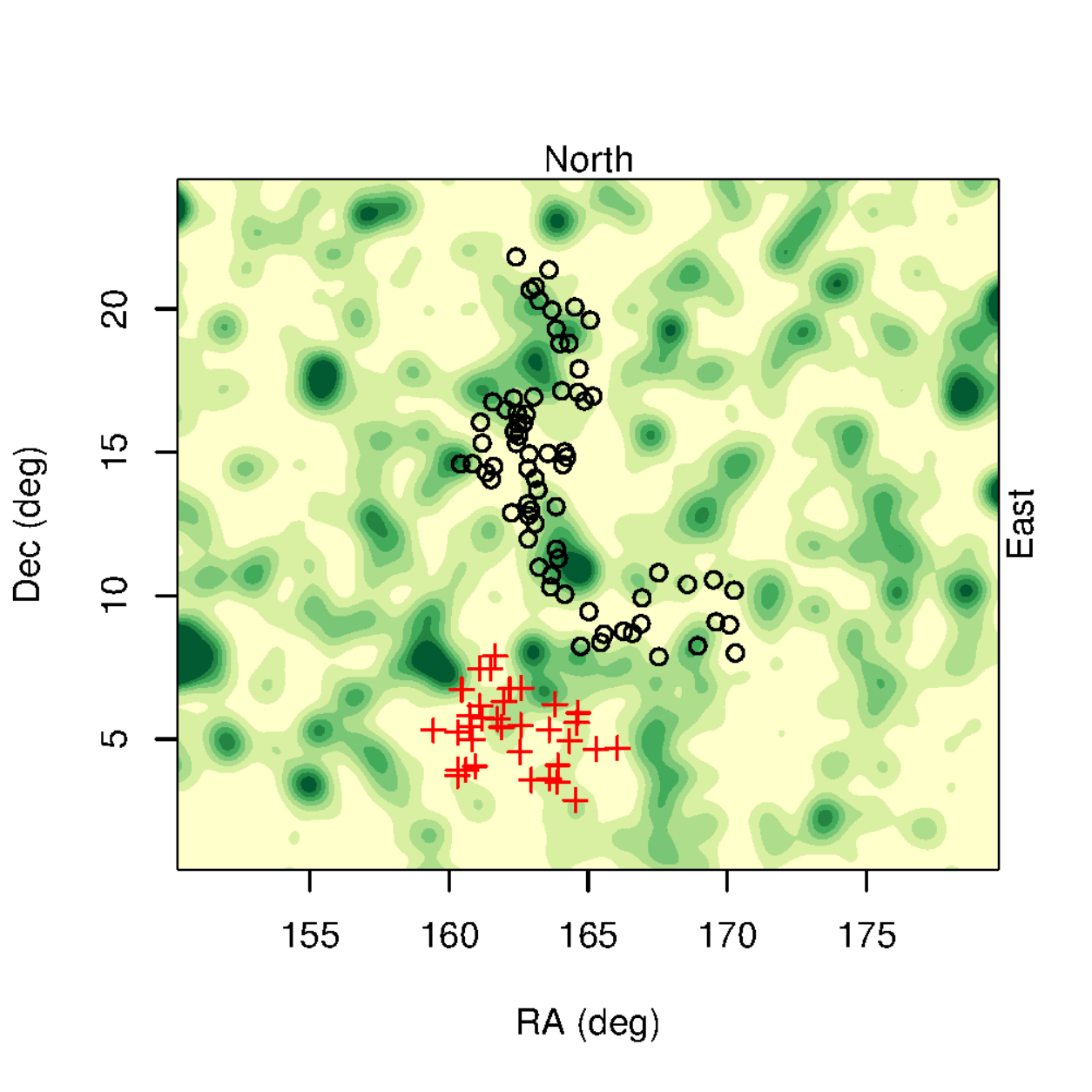}
    \caption[The Huge-LQG and Clowes-Campusano LQG]{Quasars comprising the Huge-LQG (black circles) and the Clowes-Campusano LQG (red crosses). Green contours represent a kernel-smoothed intensity map of MgII absorbers in the joint redshift range of the LQGs ($1.17 \leq z \leq 1.42$). Image credit: \protect\citet{Clowes2013}.}
    \label{figHugeLQG}
\end{figure}

Large quasar groups (LQGs) are collections of quasars that represent peaks in the quasar density field. The first \gls{LQG} observed \citep{Webster1982} contained 4 quasars at $z \simeq 0.37$ and had a size $\sim 100 \ \mathrm{Mpc}$. The largest LQG so far identified is known as the Huge-LQG \citep[][Fig.~\ref{figHugeLQG}]{Clowes2013}, contains 73 quasars at $\bar{z} = 1.27$, and has a longest dimension $\sim 1,240 \ \mathrm{Mpc}$.

\clearpage

We are not aware of any work investigating the duty cycle of quasars in LQGs. If it is similar to other quasars ($0.1-1\%$), LQGs will also contain many quiescent SMBHs, with the potential for their activation to illuminate the LQG with different quasars at different epochs. However, if LQGs are triggered by some environmental factor, causing a greater fraction of quasars to activate at a particular epoch, then LQGs could be phenomena as transient as their constituent quasars ($10-100$ Myr).

Categorisation of LQGs as huge `structures' that could potentially challenge the cosmological principle is controversial \citep{Nadathur2013,Pilipenko2013,Park2015}. However, the presence of the Huge-LQG and Clowes-Campusano LQG \citep{Clowes1991}, in particular, are supported by their association with MgII absorbers \citep{Williger2002,Clowes2013}, and correlation with quasar polarization \citep{Hutsemekers2014,Pelgrims2016}. The Huge-LQG may also be coincident with a CMB temperature anomaly \citep{Romano2015}.

It has been suggested LQGs are the high-redshift precursors to the galaxy superclusters we see today \citep{Komberg1996,Pilipenko2007}. \citet{Pilipenko2007} found that quasars tend to lie two-dimensionally, in sheets. At low redshifts ($z \sim 0.3$) \citet{Sochting2002,Sochting2004} show that quasars follow the large-scale structure traced by galaxy clusters, being preferentially located on their peripheries.

It is likely that LQGs trace the regions of enhanced matter density of the cosmic web, although their origin and significance are still somewhat disputed.

\section{The cosmic microwave background} \label{secCMB}

The cosmic microwave background (\gls{CMB}) was first detected by \citet{Penzias1965} who, while investigating the source of radio `interference', detected an isotropic signal from the sky. In a companion paper \citet{Dicke1965} attributed this emission to the remnant radiation from the hot big bang, as earlier postulated by \citet{Alpher1948}.

In the $\Lambda$CDM cosmological model the early Universe was a hot dense plasma (mostly electrons, protons, neutrons and photons). This tightly coupled photon-baryon\footnote{Here `baryon' includes electrons, which are carried along with the more massive baryons} fluid \citep{Peebles1970}, opaque to radiation, subsequently cooled as it expanded. After $\sim$ 380,000 years, at a redshift of $z \sim 1,100$ it had cooled sufficiently ($\sim$ 3,000K) for free electrons to combine with protons and form neutral hydrogen \citep[`recombination', ][]{Peebles1968}. The Universe became transparent to photons, which were then able to free-stream across the Universe to be observed by us arriving isotropically from all directions on the sky.

Over the intervening 13.8 billion years the Universe continued expanding and cooling, and the CMB we see today has an effective temperature of $T = 2.726 \pm 0.001$K \citep{Fixsen2009}. The CMB is almost perfect black-body radiation and its temperature is highly uniform, with only tiny fluctuations (anisotropies) of the order of one part in $10^5$ first identified by COBE \citep{Smoot1992}.

\begin{figure}[t] % CMB_SMICA_UT78
	\centering
	\includegraphics[width=0.7\linewidth]{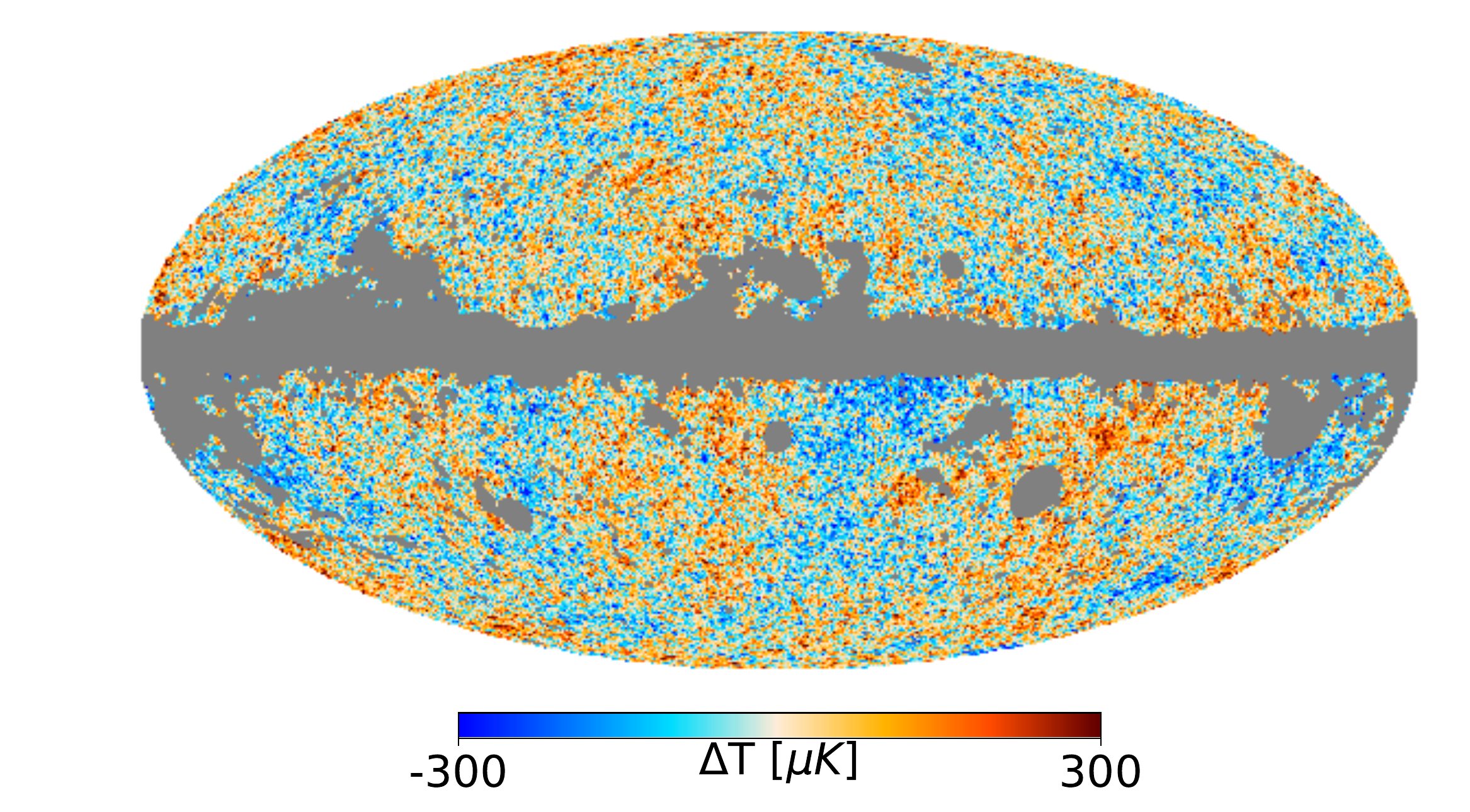}
	\caption[\textit{Planck} 2015 SMICA component separated CMB temperature map]{\textit{Planck} 2015 SMICA component separated CMB temperature map at full resolution of FWHM $5'$ with $1.7'$ pixels. Grey areas are the \textit{Planck} UT78 confidence mask \protect\citep{Planck2016c}. Mollweide projection in Galactic coordinates centred on the Galactic centre.}
	\label{figPlanckSMICAmap}
\end{figure}

The CMB has since been surveyed by a multitude of different telescopes, including the Wilkinson Microwave Anisotropy Probe \citep[WMAP, ][]{Hinshaw2013}, the \textit{Planck} satellite \citep{Planck2016a}, the Atacama Cosmology Telescope \citep[ACT, e.g.][]{Das2014}, and the South Pole Telescope \citep[SPT, ][]{Schaffer2011}. See Fig.~\ref{figPlanckSMICAmap} for a map of the CMB temperature anisotropies.

It is estimated that $\sim$ 90\% of the CMB photons we observe have travelled unimpeded across the Universe to us. This represents the oldest primordial electromagnetic signal we can observe, and it carries with it information about the properties and physics of the Universe at the time of recombination. Information about the density, velocity and temperature of the primordial plasma are all encoded in the primary anisotropies.

\begin{figure}[h]
	\centering
	\includegraphics[width=0.8\columnwidth]{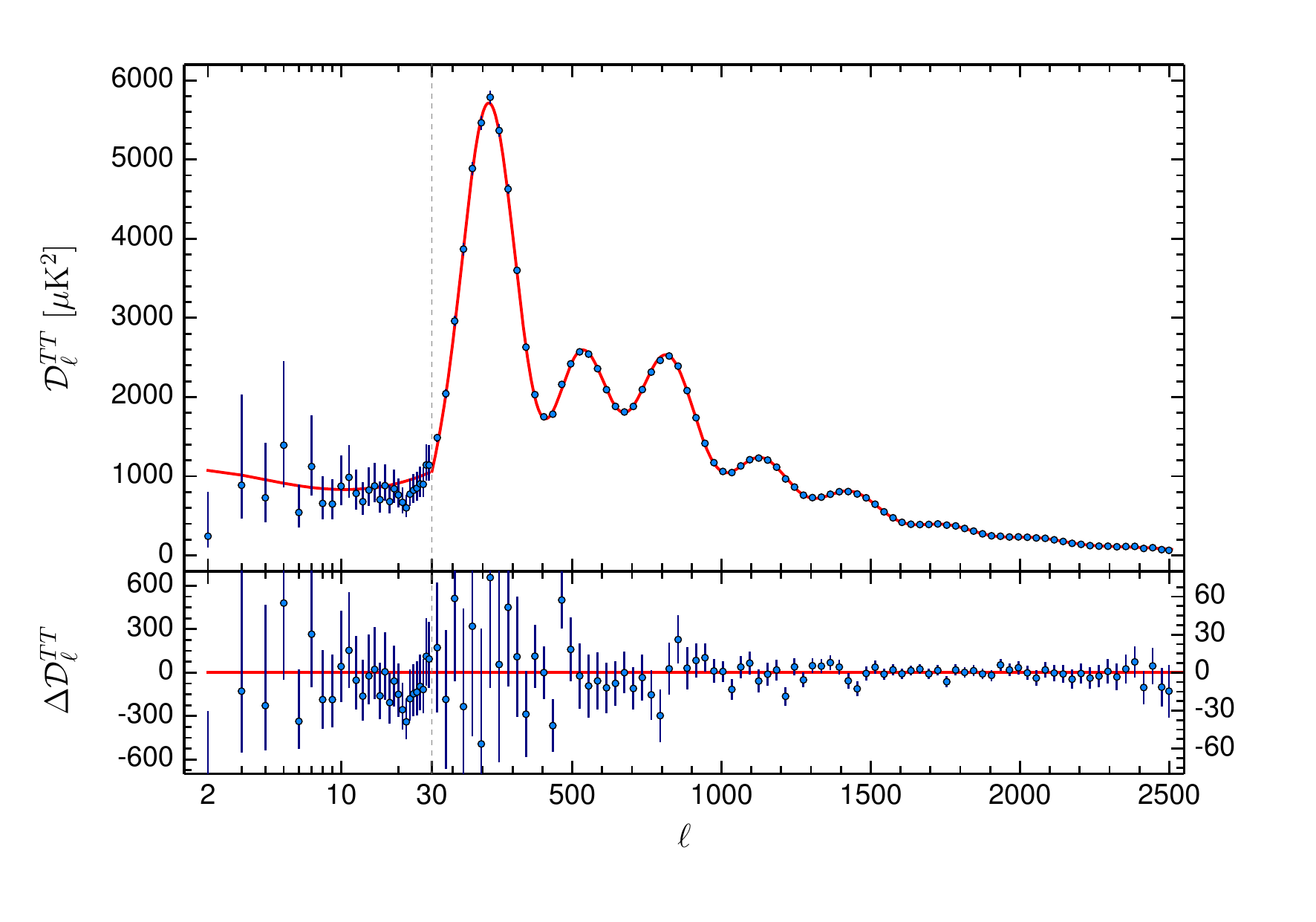}
	\caption[\textit{Planck} 2015 CMB power spectrum]{\textit{Planck} 2015 CMB power spectrum with $\Lambda$CDM fit (red line). The upper panel shows the spectrum and the lower panel shows the residuals. Vertical axis is power shown as $D_\ell$ (Eq.~\ref{eqCMBDell}) and horizontal axis is multipole moment ($\ell$). Image credit: \protect\citet{Planck2016d}.}
	\label{figPlanckPower}
\end{figure}

\clearpage

The other $\sim$ 10\% of the CMB photons bear the imprints of the processes and environments that they have travelled through, e.g.\ reionization, hot intra-cluster gas, structure formation, and gravitational lensing. These can all induce secondary anisotropies in the CMB, which tell us about the Universe during the epochs through which the photons have travelled.

The CMB angular power spectrum (Fig.~\ref{figPlanckPower}) refers to the angular coherence of fluctuations in the CMB temperature, and it is this that is generally used for statistical analysis of the CMB. Observation of the CMB power spectrum and CMB anisotropies has revolutionised our understanding of cosmology, and the excellent fit between theory and observations has cemented $\Lambda$CDM as our fiducial cosmological model \citep{Hu2002}.

The CMB temperature signal is a function over a sphere, so analogous to a Fourier expansion, the contributions from different angular scales can be separated by expanding it as a spherical harmonic series

\begin{equation}
T(\theta,\phi) = \sum_{\ell = 0}^{\infty} \sum_{m = -\ell}^{\ell} a_{lm} Y_{lm}(\theta,\phi) \ ,
\end{equation}

\noindent where indices $\ell = 0,...,\infty$ and $-\ell \leq m \leq \ell$ are the degree and order of Legendre polynomials, with $\ell$ (the multipole moment) related to angular size and $m$ related to orientation/shape, $a_{lm}$ are the spherical harmonic coefficients (i.e.\ the amplitude of the contribution from each spherical harmonic) and $Y_{lm}(\theta,\phi)$ are the spherical harmonics themselves. The power spectrum ($C_\ell$) is defined by the average variance of the spherical harmonic coefficients as

\begin{equation}
C_\ell \equiv \langle \mid a_{lm} \mid^2 \rangle = \frac{1}{2\ell + 1} \sum_{m = -\ell}^{\ell} \mid a_{lm} \mid^2 \ .
\end{equation}

The observed power spectrum is intrinsically limited by the number of $m$-modes we can measure, since there are only ($2 \ell + 1)$ of these for each multipole ($\ell$). This leads to an unavoidable error in the estimation of any $C_\ell$ of

\begin{equation}
\Delta C_\ell = \sqrt{2/(2\ell + 1)} \ ,
\end{equation}

\noindent which is usually termed cosmic variance and gives rise to the larger uncertainties at the low multipole end of the power spectrum. Note that it has become customary to plot the observed power spectrum as

\begin{equation} \label{eqCMBDell}
D_\ell = \ell(\ell+1) \times C_\ell/(2\pi) \ ,
\end{equation}

\noindent because this changes more slowly as a function of $\ell$, gives a flat plateau at large angular scales, and emphasises the structure at smaller scales.

The distinctive peaks in the CMB power spectrum (Fig.~\ref{figPlanckPower}) occur as a result of cosmological perturbations creating sound (pressure) waves in the tightly coupled photon-baryon fluid \citep{Peebles1970}. Matter falling into gravitational potential wells was compressed and heated, increasing radiation pressure and causing the plasma to rebound and rarefy before gravity once again dominated and the oscillation repeated. The propagation of these waves effectively ended at the epoch of recombination and the pattern of oscillations was `frozen in' to the CMB.

The acoustic peaks reveal a harmonic series of oscillations with various modes corresponding to the number of oscillations completed before recombination. The longest wavelength mode, corresponding to collapse but ending before rebound, is the fundamental mode and was measured first \citep{Miller1999}. Since then, the acoustic peaks have emerged as a powerful cosmological probe \citep[e.g.][]{deBernardis2000, Benoit2003} measuring the contents and curvature of the Universe \citep[e.g.][]{Efstathiou2002, Spergel2003, Tegmark2004}.

We observe the imprint of these acoustic signatures in the large-scale clustering of matter, known as baryon acoustic oscillations (\gls{BAO}), which give more opportunities to test $\Lambda$CDM \citep[e.g.][]{Eisenstein2005, Percival2010, Beutler2011}. BAO analysis has been extended to higher redshifts using quasars \citep[e.g.][]{Delubac2015, Ansarinejad2018}. The fit with $\Lambda$CDM so far is exquisite.

%\subsection{LSS imprints on CMB temperature and lensing} \label{secCMBImprints}
\subsection{LSS imprints on CMB temperature} \label{secCMBImprints}

The $\Lambda$CDM cosmological model predicts that large-scale structure will leave an imprint on the cosmic microwave background (CMB). The CMB temperature is almost uniform, with fluctuations of the order of one part in $10^5$ originating from the epoch the CMB was emitted (primary anisotropies), and even smaller fluctuations from the physical processes that have affected it since then (secondary anisotropies). Measurements of these anisotropies by the \textit{Planck} satellite are in excellent agreement with the $\Lambda$CDM cosmological model \citep{Planck2016a}. Cross-correlation of CMB anisotropies with the cosmic web can constrain cosmological parameters \citep[e.g.][]{Giannantonio2016, DES2017}.

Secondary anisotropies in the CMB can generally be grouped into those caused by gravitational effects and those that result from scattering of CMB photons by free electrons. In a $\Lambda$CDM Universe gravitational effects such as the integrated Sachs-Wolfe \citep[ISW,][]{SachsWolfe1967} and Rees-Sciama \citep[RS,][]{Rees1968} effects are expected to dominate at large angular scales. Scattering effects such as the thermal and kinetic Sunyaev-Zel\textquotesingle dovich effects \citep[SZ,][]{SunyaevZeldovich1970, SunyaevZeldovich1972} are expected to dominate at small angular scales. See Appendices~\ref{appISW} and \ref{appSZ} for summaries of the ISW (and RS) and SZ effects respectively. The amplitude of the anisotropies caused by both groups of effects are predicted to be tiny, at least an order of magnitude smaller than the primary anisotropies (one part in $10^5$) of the CMB.

There is evidence that LSS does indeed imprint on the CMB as hot and cold spots via the ISW and SZ effects \citep[e.g.][]{Granett2008, Cai2014, Kovacs2017}. These are generally statistical studies of many superclusters and voids across large areas of the sky. \citet{Granett2008} report a $\sim 4 \sigma$ detection of the ISW imprint using stacked supervoids and superclusters at $\bar{z} \sim 0.5$ identified from SDSS LRGs. They find mean ISW signals of $\Delta T_v \sim -11 \pm 3 \mu K$ and $\Delta T_c \sim 8 \pm 3 \mu K$ respectively, $\gtrsim 3 \sigma$ above $\Lambda$CDM. Using a similar stacking method, \citet{Kovacs2017} detect the ISW imprint of Dark Energy Survey (DES) super-structures on the CMB using 102 superclusters at redshifts 0.2 \textless \ $z$ \textless \ 0.65. They reported a cumulative hot imprint of superclusters of $\Delta T \sim 5.1 \pm 3.2 \mu K$, which they find to be $\sim 1.2 \sigma$ higher than expected in $\Lambda$CDM. The SZ effect has also been used to probe nearby galaxy clusters \citep[e.g.][]{Bourdin2015}.

However, not all apparent imprints on CMB temperature have proved to be real. \citet{Yershov2012, Yershov2014} reported that the \textit{WMAP}/\textit{Planck} CMB pixel-temperatures at SNe locations tend to increase with increasing redshift. The signal they find does not exhibit the frequency dependence of the SZ effect. They suggest it could be caused by the ISW effect instead. \citet[][chapter~\ref{chaptSNeCMBT}]{Friday2018} independently assess this apparent correlation and offer a different interpretation. Our analysis supports the prima facie existence of the correlation but attributes it instead to a composite selection bias due to the chance alignment of seven deep survey fields with CMB hotspots. These fields contain just 9.2\% of the SNe sample (256 out of 2783 SNe); the `signal' is entirely absent without them. We estimate the likelihood of this alignment occurring by chance is $6.8\%-8.9\%$.

One individual anisotropy that has been a particular source of tension with $\Lambda$CDM is the CMB Cold Spot \citep{Vielva2004}, a $\sim 5^{\circ}$ radius feature with a temperature decrement $\Delta T \sim -150 \mu K$. Various mechanisms to explain the Cold Spot have been suggested, including rotation of the Universe \citep{Jaffe2005} and the imprint of a supervoid via the ISW effect \citep{Inoue2006}. However, insufficient voids have been found in the direction of the Cold Spot to explain the comparatively large drop in temperature \citep[e.g.][]{Bremer2010}. \citet{Mackenzie2017} reach similar conclusions and suggest a primordial origin.

% Srini suggests explaining this more thoroughly with references (Planck, ACT, SPT) or removing
%In addition to imprinting on CMB temperature, LSS also deflects CMB photons by gravitational lensing. This leaves statistical imprints in the temperature anisotropies, for example over-densities magnify the angular size of hot and cold spots. This can be used to reconstruct the lensing potential (predominantly due to dark matter) along the line-of-sight. Again, the fit with $\Lambda$CDM is so far excellent.

%The cross-correlation of LSS and CMB lensing has been detected, first using multiple tracers of LSS including luminous red galaxies (LRGs), quasars, and radio sources \citep[e.g.][]{Ho2008,Hirata2008} and more recently with single tracers \citep[e.g., using quasars,][]{Sherwin2012,Han2019}. \citet{He2018} report the first detection of the imprint of filaments, specifically, on CMB lensing.

\section{Summary and thesis structure}

The large-scale structure (LSS) of the cosmic web leaves a detectable imprint on the cosmic microwave background (CMB). This induces anisotropies in the temperature (hot and cold spots) via gravitational and scattering effects such as the Integrated Sachs-Wolfe (ISW, Appendix~\ref{appISW}) and Sunyaev-Zel\textquotesingle dovich (SZ, Appendix~\ref{appSZ}) effects. The gravitational lensing of these anisotropies can be used to reconstruct the lensing potential (predominantly due to dark matter) along the line-of-sight. Both CMB temperature and lensing therefore correlate with LSS.

In chapter~\ref{chaptSNeCMBT} we investigate the cross-correlation of supernovae (\gls{SNe}) redshift and CMB temperature, reported by \citet{Yershov2012,Yershov2014} and attributed by them to the ISW effect and/or to some unrelated foreground emission. The SNe we analyse are predominately at redshifts $z \lesssim 1.5$, so it is plausible large-scale structure hosting their SNe sample could leave an imprint on the CMB via the ISW effect. However, the signal reported by \citet{Yershov2012,Yershov2014} increases with redshift in a manner unexpected for the ISW effect and unexplained by them. We therefore examine their result further in order to explain the correlation and assess the feasibility of using SNe as tracers of the cosmic web at high redshifts. However, we conclude that the apparent correlation is a composite selection bias (high CMB T $\times$ high SNe $z$) caused by the chance alignment of seven deep survey fields with CMB hotspots. This work was originally presented in \citet{Friday2018}.

We include supplementary information to chapter~\ref{chaptSNeCMBT} in Appendices~\ref{appFields1to7}, \ref{appVariances}, and \ref{appAltPlanckMaps}. In Appendix~\ref{appFields1to7} we illustrate the \textit{Planck} CMB temperature in the vicinity of the seven SNe deep survey fields. Appendices \ref{appVariances} and \ref{appAltPlanckMaps} show the apparent correlation of CMB temperature as a function of SNe redshift produced using different estimates of \textit{Planck} pixel variance (Appendix~\ref{appVariances}) and different \textit{Planck} maps (Appendix~\ref{appAltPlanckMaps}); the results are consistent.

The filaments and walls of the cosmic web are both predicted by cosmological simulations and are observed on scales of several $100$ Mpc up to redshift $z \sim 3$. Galaxies are expected to align with these structures, and this is observed on comparable scales. The spin axes of quasars exhibit similar physical alignment, and there is also evidence quasars' optical and radio polarization is coherently orientated on Gpc scales, and that they are aligned with their host LSS.

In this work we investigate for the first time whether LQGs exhibit coherent orientation, and whether this can explain the reported alignments of quasar polarization from \citet{Hutsemekers1998} to \cite{Pelgrims2019}. This examines scales larger than those so far analysed, and potentially offers corroborating evidence for, and enhancement of, the intrinsic alignment interpretation of the results from many quasar polarization studies \citep[e.g.][]{Hutsemekers2001,Jain2004,Cabanac2005,Tiwari2013,Pelgrims2014,Pelgrims2015}. If true, it would represent LSS alignments over $\gtrsim$ Gpc scales, larger than those predicted by cosmological simulations and larger than any so far observed. It could potentially challenge the cosmological assumptions of isotropy and homogeneity, although the existence of an individual fluctuation (whatever it might be) cannot necessarily be used to make inferences about the large-scale statistical isotropy and homogeneity of the Universe.

In chapter~\ref{chaptLQGSample} we present the large quasar group (LQG) sample we use to trace the cosmic web at redshifts $1.0 \leq z \leq 1.8$. We describe the methods for determining LQG morphology and orientation, discuss the methods for analysis of axial data, and explain the need for coordinate-invariance and how parallel transport provides this. For three-dimensional analyses we calculate radial comoving distance as detailed in Appendix~\ref{appComovingDistance}. We later show that the distribution of LQG position angles (PAs) is bimodal. This bimodality, together with the axial nature of PAs, and their wide distribution on the celestial sphere, makes statistical analysis somewhat challenging. We devote chapter~\ref{chaptLQGStatsMethods} to describing and evaluating several statistical methods to analyse the uniformity, bimodality, and correlation of LQG PAs.

We present the results of our analysis of LQG PAs in chapter~\ref{chaptLQGOrientation}, with all LQG locations, memberships, and position angles listed in Appendix~\ref{appLQGData}, and their positions in three-dimensional comoving space illustrated in Appendix~\ref{appLQGs3DPerspectives}. We find marginal evidence for non-uniformity, although we do find the PA distribution is bimodal with 97\% confidence. We identify modes (peaks) at $\sim52\pm2^{\circ}$ and $\sim137\pm3^{\circ}$, remarkably close to the mean angles of radio quasar polarization, $\bar{\theta}\simeq42^{\circ}$ and $\bar{\theta}\simeq131^{\circ}$, reported by \cite{Pelgrims2015}\footnote{Mean angles derived by the \citet{Hawley1975} method; errors were not reported}. We examine the correlation in LQG PAs and find they are aligned and orthogonal across Gpc scales, with the typical likelihood of this occurring by chance of $\sim 1.5 - 3.3 \%$. The correlation is most significant (significance level $\sim 0.8 \%, 2.4\sigma$) for groups of $\sim45$ LQGs, corresponding to typical angular (comoving) separations of $\sim 30^{\circ}$ (1.6 Gpc).

In chapter~\ref{chaptDiscussion} we discuss our work and speculate about the physical origin of the LQG orientation correlation (see also Appendix~\ref{appVoronoi} regarding Voronoi tessellation), and in chapter~\ref{chaptConclusion} we offer our conclusions and suggest future work.

\afterpage{\null\newpage} % because this ended on an odd page so the next chapter started on an even page

% END

%% file: chapterSNeVsCMBT.tex
% START

\chapter{Accidental deep field bias in CMB $T$ and SNe $z$ correlation} \label{chaptSNeCMBT}

This chapter presents our first project and was published in \citet{Friday2018}.

\section{Abstract} \label{secAbstract}

Evidence presented by Yershov, Orlov and Raikov apparently showed that the \textit{WMAP}/\textit{Planck} cosmic microwave background (CMB) pixel-temperatures ($T$) at supernovae (SNe) locations tend to increase with increasing redshift ($z$). They suggest this correlation could be caused by the Integrated Sachs-Wolfe effect and/or by some unrelated foreground emission. Here, we assess this correlation independently using \textit{Planck} 2015 SMICA R2.01 data and, following Yershov et al., a sample of 2,783 SNe from the Sternberg Astronomical Institute. Our analysis supports the prima facie existence of the correlation but attributes it to a composite selection bias (high CMB $T$ $\times$ high SNe $z$) caused by the accidental alignment of seven deep survey fields with CMB hotspots. These seven fields contain 9.2\% of the SNe sample (256 SNe). Spearman's rank-order correlation coefficient indicates the correlation present in the whole sample ($\rho_s = 0.5$, p-value $= 6.7 \times 10^{-9}$) is insignificant for a sub-sample of the seven fields together ($\rho_s = 0.2$, p-value $= 0.2$) and entirely absent for the remainder of the SNe ($\rho_s = 0.1$, p-value $= 0.6$). We demonstrate the temperature and redshift biases of these seven deep fields, and estimate the likelihood of their falling on CMB hotspots by chance is at least $\sim$ 6.8\% (approximately 1 in 15). We show that a sample of 7,880 SNe from the Open Supernova Catalogue exhibits the same effect and we conclude that the correlation is an accidental but not unlikely selection bias.

\section{Introduction} \label{secIntro}

Observations of the cosmic microwave background (CMB) and Type Ia supernovae (SNIa) are exceptional probes of cosmological parameters. The measurements of the CMB by the \textit{WMAP} satellite \citep{Bennett2013} and SNIa by the high-$z$ supernova search team \citep{Riess1998} and the supernova cosmology project \citep{Perlmutter1999} have established the six parameter $\Lambda$CDM cosmological model. Precision measurements of the CMB temperature, polarization, and lensing anisotropies from the South Pole Telescope, \textit{Planck} satellite, and Atacama Cosmology Telescope \citep[e.g.,][and references therein]{Story2013, Planck2016b, Louis2017} have strongly reinforced the preference for $\Lambda$CDM as the concordance model of cosmology.

Cross-correlation of CMB observations with the large-scale structure (LSS) of the Universe, revealed by surveys such as the \textit{Sloan Digital Sky Survey} \citep[SDSS,][]{York2000} and the \textit{Dark Energy Survey} \citep[DES,][]{DES2016}, provide powerful tests of $\Lambda$CDM \citep[e.g.,][]{Giannantonio2016, DES2017}. Increased efforts are currently being made for accurate and precise calibration of the distance-redshift relation using SNIa up to high redshifts \citep[e.g.,][]{LSST2009, Kessler2015} in order to understand the expansion history and late-time ($z \le 1$) accelerated expansion of the Universe attributed to dark energy.

\begin{figure} % A1_SNe_all_fields_remainder
	\centering
	\includegraphics[width=0.5\columnwidth]{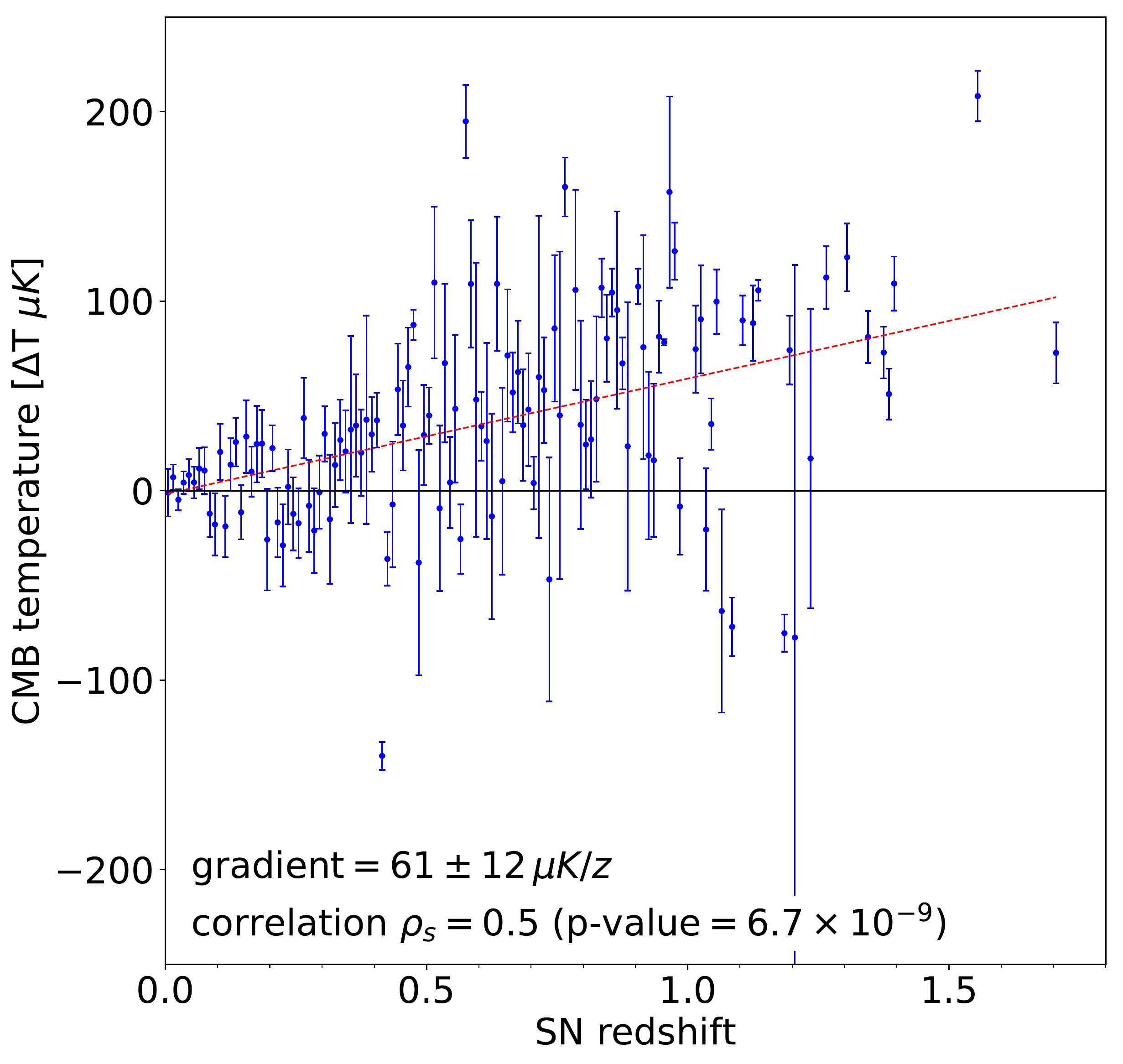}
	\caption[CMB $T$ at SNe locations as a function of SNe redshift]{Plot of CMB temperature at SNe locations versus SNe redshift binned with bin sizes $\Delta z=0.01$. Data are restricted by Galactic latitude ($|b|>40^\circ$), redshift ($z>0.005$), and \textit{Planck} UT78 confidence mask. Error bars are the standard error on the bin mean. The dashed line indicates ordinary least squares linear regression, representing the correlation reported by \citet{Yershov2012, Yershov2014}.}
	\label{figBinnedScatter}
\end{figure}

\citet{Yershov2012, Yershov2014} combined CMB and supernovae (SNe) data and detected a correlation between the CMB temperature anisotropies ($T$) and the redshift ($z$) of the SNe (CMB $T$ $\times$ SNe $z$). High-$z$ SNe appear to be preferentially associated with hotter CMB temperatures (Fig.~\ref{figBinnedScatter}). This effect was particularly strong for the SNIa sample. They concluded that the correlation is not caused by the Sunyaev-Zel\textquotesingle dovich (SZ) effect \citep{SunyaevZeldovich1970} and suggested it may instead be caused by the Integrated Sachs-Wolfe (ISW) effect \citep{SachsWolfe1967} or some remnant contamination in the CMB data, possibly from low redshift foreground \citep{Yershov2014}.

In this chapter we re-analyse the SNe samples of \citet{Yershov2012, Yershov2014} and offer an alternative explanation for the correlation, namely that it is a composite selection bias caused by the chance alignment of certain deep survey fields with CMB hotspots. This bias (high CMB $T$ $\times$ high SNe $z$) is the combined result of a selection bias (high-$z$ SNe in deep fields) and the chance alignment of those deep fields with CMB hotspots.

The remainder of this chapter describes our analyses of the reported correlation. In section~\ref{secDataMethod} we describe the data and summarise the variety of methods used to demonstrate the prima facie existence of the correlation in these data. We present the results from re-analysing the SNe sample of \citet{Yershov2012, Yershov2014} in section~\ref{secBias}. Specifically, we identify SNe fields with a high surface density of SNe (\ref{subsecFieldID}) and show that these cause the apparent correlation (\ref{subsecContribution}) due to their bias to hotter CMB temperature and higher redshift than the remainder of the SNe sample (\ref{subsecTempRedshift}). We quantify the likelihood of this bias occurring by chance: at least 6.8\%, or approximately 1 in 15 (\ref{subsecLikelihood}). We present corroborating results from analysing alternative data in section~\ref{secAlternativeData}. In section~\ref{secConclusions} we conclude that the correlation reported by \citet{Yershov2012, Yershov2014} is actually an accidental but not exceptionally unlikely composite selection bias and we briefly speculate on further potential implications for cosmology.

\section{Data and methods} \label{secDataMethod}

\subsection{SNe and CMB data} \label{subsecData}

We used CMB data from the \textit{Planck} 2015 \citep{Planck2016a} maps\footnote{\url{http://irsa.ipac.caltech.edu/data/Planck/release_2/all-sky-maps/}}, specifically SMICA R2.01 with $N_{side}=2048$. These maps are provided and analysed using the Hierarchical Equal Area iso-Latitude Pixelisation scheme \citep[\gls{HEALPix}\footnote{\url{http://healpix.jpl.nasa.gov/}}, ][]{Gorski2005}. Our temperature distribution was consistent with that previously determined by \citet{Yershov2014} using the \textit{Planck} 2013 \citep{Planck2014a} SMICA R1.20 map. 

The \textit{Planck} component-separated CMB maps were produced using four techniques: Commander \citep{Eriksen2008}, NILC \citep{Delabrouille2009}, SEVEM \citep{FernandezCobos2012} and SMICA \citep{Cardoso2008}. \citet{Planck2016c} provide a critical analysis of the applicability of the resultant four 2015 maps, and confirm that, as in 2013 \citep{Planck2014b}, SMICA is preferred for high-resolution temperature analysis.

As recommended by \citet{Planck2016c}, for analysing component-separated CMB temperature maps, we used the \textit{Planck} UT78 common mask. This is the union of the Commander, SEVEM, and SMICA confidence masks. UT78 excludes point sources, some of the Galactic plane (note also our subsequent Galactic latitude restriction), and some other bright regions. It has a fraction of unmasked pixels of $f_{sky} = 77.6\%$. Note that our results were consistent using an alternative mask (UTA76), and without masking.

Our initial analysis used SNe data provided by \citet{Yershov2014} as supplementary data\footnote{\url{https://academic.oup.com/mnras/article-lookup/doi/10.1093/mnras/stu1932}}, derived from the Sternberg Astronomical Institute (\gls{SAI}) Supernova Catalogue\footnote{\url{http://www.sai.msu.su/sn/sncat/}} \citep{Bartunov2007} as of October 2013. This provides a sample of 6359 SNe of all types. To avoid contamination from the Galactic plane we restricted this sample to high Galactic latitude $|b|>40^\circ$, the same conservative restriction used by \citet{Yershov2014}. We could not reproduce the identical sample for the redshift restriction apparently used by \citet{Yershov2014}, so we adopted a restriction of $z>0.005$, which yielded a similar sample size to theirs. We excluded SNe on masked (\textit{Planck} UT78) HEALPix pixels. The resultant SAI sample contained 2783 SNe.

Above redshift $z\sim1.2$, SNe in the SAI sample become rather sparse and are predominantly associated with hotter than average CMB temperature. To analyse this high redshift region further, we obtained $z>0.005$ data from the Open Supernova Catalogue\footnote{\url{https://sne.space/}} \citep[\gls{OSC},][]{Guillochon2017} as of June 2017. We removed SNe without co-ordinate information, those not yet confirmed as SNe ($Type=Candidate$) and gamma ray bursts ($Type=LGRB$). These selections provide a sample of 12879 SNe of all types. We also restricted this sample to high Galactic latitude $|b|>40^\circ$ and excluded SNe on masked (\textit{Planck} UT78) HEALPix pixels. The resultant OSC sample contained 7880 SNe.

Unless specified otherwise, all analysis in this chapter was performed on the SAI sample after the restrictions on Galactic latitude ($|b|>40^\circ$), redshift ($z>0.005$), and \textit{Planck} UT78 confidence mask. We repeated our analyses using the OSC sample (section~\ref{subsecOSC}) with the same restrictions to verify that our results are not specific to the SAI sample.

\subsection{Methods} \label{subsecMethod}

We constructed Fig.~\ref{figBinnedScatter} broadly following \citet{Yershov2014}. We determined the temperature\footnote{We follow the practice of referring to temperature anisotropies ($\Delta T$) as temperature ($T$)} of the CMB map pixel at each SN location. The data were grouped into redshift bins of width $\Delta z = 0.01$ and the weighted mean CMB temperature of each bin ($\overline{T}$) was calculated as

\begin{ceqn}
\begin{equation}
\overline{T} = \sum_{i=1}^{n} w_iT_i / \sum_{i=1}^{n} w_i\ ,
\end{equation}
\end{ceqn}

\noindent where $T_i$ is the individual pixel temperature, $w_i$ is the weight of each pixel, and $n$ is the number of SNe per bin. Error bars are the standard error on the weighted bin mean ($\sigma_{\overline{T}}$), calculated from the weighted variances as

\begin{ceqn}
\begin{equation}
\sigma_{\overline{T}} = \sqrt{\sum_{i=1}^{n} w_i^2\sigma_i^2 / \left(\sum_{i=1}^{n} w_i\right)^2}\ ,
\end{equation}
\end{ceqn}

\noindent where $\sigma_i^2$ is our variance estimate for each pixel. Note that for bins with only one SN, error bars are $\sigma_{\overline{T}} = \sigma_i$.

Individual pixel variances ($\sigma_i^2$) were estimated\footnote{Following private communications with \textit{Planck} Legacy Archive and NASA/IPAC Archive} by producing a squared, smoothed ($0.5^\circ$ \gls{FWHM}\footnote{Full-width at half-maximum}) half-mission half-difference (\gls{HMHD}\footnote{Find the difference between the half-mission maps; halve, square, and smooth the result}) map from the two \textit{Planck} 2015 SMICA R2.01 half-mission maps. Weights ($w_i$) are thus

\begin{ceqn}
\begin{equation}
w_i = 1 / \sigma_i^2\ .
\end{equation}
\end{ceqn}

The choice of smoothing scale and the method of estimating pixel variance affects the resultant weights. We tested a number of these and our results are consistent. See Appendix~\ref{appVariances} for versions of Fig.~\ref{figBinnedScatter} produced using different smoothing scales (none, $5'$, 0.5$^\circ$, and 5$^\circ$ FWHM) and different estimates of individual pixel variance (\textit{Planck} 2013 R1.20 SMICA map noise, \textit{Planck} 2015 R2.01 SMICA HMHD and \gls{HRHD} maps, and \textit{Planck} 2015 R2.02 143GHz and 217GHz frequency maps).

We fitted an ordinary least squares (\gls{OLS}) linear regression to the binned data (dashed line) using Python's\footnote{Statistical models library \citep{Seabold2010}} \texttt{statsmodels.api.OLS}, and calculated its slope plus the standard error on the gradient. We followed the same method to calculate the OLS linear regression gradient throughout this chapter. Note that results using weighted least squares (\gls{WLS}) linear regression (weighted by $\sigma_{\overline{T}}$ or $n$), and results using OLS and WLS linear regression of unbinned data, were consistent.

Several of our analyses compared various SNe sub-samples using the parametric independent 2-sample Welch's t-test \citep[or unequal variance t-test,][]{Welch1938} and the non-parametric 1-sided Mann-Whitney U (\gls{MWU}) test. We used the unequal variance t-test to account for the different angular extent of the deep survey fields, and hence difference in variance of their CMB temperature. It also accommodates the wide variation in the number of SNe per redshift bin, and the resultant differing variance of both their CMB temperature and their redshift. The MWU test does not assume that the population follows any specific parameterised distribution, unlike the t-test which assumes a normal distribution, and it is less sensitive to outliers than the t-test.

The t-test gives the probability (p-value) of obtaining SNe sub-samples with differences in mean CMB temperature (or SNe redshift) at least as extreme as those observed, assuming the null hypothesis is true

\begin{ceqn}
\begin{equation}
H_{0}: \mu_{T1} = \mu_{T2} \ (H_{0}: \mu_{z1} = \mu_{z2})\ .
\end{equation}
\end{ceqn}

In other words, it tests whether the means of their populations differ. The 2-sample Welch's t-test was implemented using Python's\footnote{SciPy (scientific computing in Python) community project \citep{Virtanen2019}} \texttt{scipy.stats.ttest\_ind} with \texttt{equal\_var $=$ False}.

MWU combines the sub-samples of CMB temperature (or SNe redshift), ranks the combined sample, and determines the mean of the ranks ($\overline{R}$) for each sub-sample. It gives the probability (p-value) of obtaining SNe samples with differences in mean ranks at least as extreme as those observed, assuming the null hypothesis is true

\begin{ceqn}
\begin{equation}
H_{0}: \overline{R}_{T1} = \overline{R}_{T2} \ (H_{0}: \overline{R}_{z1} = \overline{R}_{z2})\ .
\end{equation}
\end{ceqn}

In practice, this is generally interpreted as whether the distributions of the sub-samples differ, since the ranks of the sub-samples will differ if so. We used the 1-sided MWU to test the alternative hypothesis that the CMB temperature (or SNe redshift) distribution of one sub-sample was greater than that of the other

\begin{ceqn}
\begin{equation}
H_{1}: \overline{R}_{T1} \ge \overline{R}_{T2} \ (H_{1}: \overline{R}_{z1} \ge \overline{R}_{z2})\ .
\end{equation}
\end{ceqn}

The 1-sided MWU test was implemented using Python's \texttt{scipy.stats.mann-whitneyu} with \texttt{alternative $=$ `greater'}.

To analyse which modes dominate the apparent correlation (section~\ref{subsubAngularScales}) we filtered the map to remove large angular scales. We used Python's\footnote{Python wrapper \citep{Zonca2019} for HEALPix \citep{Gorski2005}} \texttt{healpy.spht-func.almxfl} to apply a high pass filter to the spherical harmonic coefficients ($a_{\ell m}$) of the \textit{Planck} 2015 SMICA map. We then computed the filtered map from these filtered $a_{\ell m}$ values. We repeated the OLS linear regression and Spearman's rank-order correlation coefficient analyses for each filtered map.

In our analysis of likelihood (section~\ref{subsecLikelihood}) we performed randomisations of SNe locations within fields and of SNe field centres on the sky. We used Python's \texttt{random.uniform} to select random HEALPix pixels, subject to the same Galactic latitude restriction ($|b|>40^\circ$) as the original sample. After each randomisation the masking (\textit{Planck} UT78) was re-applied, the CMB temperature and variance were re-sampled, and the weights were re-calculated before the OLS linear regression was re-fitted.

We also assessed the likelihood by creating simulations of the CMB. We used Python's \texttt{healpy.sphtfunc.anafast} to compute the power spectrum ($C_\ell$) of the original (unmasked) \textit{Planck} 2015 SMICA map. Note that this extracts $C_\ell$ values from the given map and does not assume a particular power spectrum or underlying cosmology. We then used \texttt{healpy.sphtfunc.synfast} to generate new synthetic maps from these $C_\ell$ values, at full resolution $5'$ FWHM, $N_{side} = 2048$ \citep{Planck2016c}, to match our fiducial \textit{Planck} 2015 SMICA map. After each simulation the CMB temperature was re-sampled. The SNe mask (\textit{Planck} UT78), variances, and weights were left unchanged (as the SNe had not moved) and the OLS linear regression was re-fitted.

We apply these methods to the data and various randomisations in the following sections.

\section{Results: deep field bias} \label{secBias}

\setlength{\tabcolsep}{3pt} % default is 6pt
\begin{table}
    \small
	\centering
	\caption[SNe fields identified in the Sternberg Astronomical Institute sample]{SNe fields identified in the SAI sample. `No. SNe' is the number of SNe in each field after restrictions by Galactic latitude ($|b|>40^\circ$), redshift ($z>0.005$), and \textit{Planck} UT78 confidence mask. RA ($\alpha$, J2000) and Dec ($\delta$, J2000) are of the approximate field centres. Angular size is of a square (rectangle) in RA-Dec orientation encompassing each field 1-7 (Stripe 82). The equivalent surface density of SNe per deg$^2$ has been calculated. `No. pixels' is the number of CMB map pixels ($N_{side} = 2048$) whose centres are within the field and which are not masked by \textit{Planck} UT78. `Deep survey field(s)' lists examples coincident with the SNe fields. Values for the `remainder' sample (after removing fields 1-7 and Stripe 82), where applicable, have been shown for comparison.}
	\label{tabSNFields}
	\begin{tabular}{lrrrcrrl}
		\hline
		\multirow{2}{*}{Field} & No. & $\alpha$ (J2000) & $\delta$ (J2000) & Angular & SNe & No. & Deep survey \\
		& SNe & h m s & $^\circ\ '\ ''$ & size & deg$^{-2}$ & pixels & field(s) \\
		\hline
		Field 1 & 50 & 14:19:28 & 52:40:28 & $1.1^\circ \times 1.1^\circ$ & 41.3 & 1599 & SNLS D3, EGS \\
		Field 2 & 29 & 12:36:55 & 62:16:40 & $0.4^\circ \times 0.4^\circ$ & 181.3 & 221 & HDF-N, \\
		&&&&&&& GOODS-N \\
		Field 3 & 54 & 02:31:41 & -08:24:43 & $2.0^\circ \times 2.0^\circ$ & 13.5 & 4944 & ESSENCE wdd \\
		Field 4 & 22 & 02:25:55 & -04:30:58 & $1.1^\circ \times 1.1^\circ$ & 18.2 & 1596 & SNLS D1 \\
		Field 5 & 64 & 02:07:53 & -04:19:04 & $2.0^\circ \times 2.0^\circ$ & 16.0 & 5000 & ESSENCE wcc, \\
		&&&&&&& NDWFS \\
		Field 6 & 21 & 22:15:36 & -17:42:17 & $1.1^\circ \times 1.1^\circ$ & 17.4 & 1593 & SNLS D4 \\
		Field 7 & 16 & 03:32:26 & -27:38:25 & $0.4^\circ \times 0.4^\circ$ & 100.0 & 219 & CDF-S, \\
		&&&&&&& GOODS-S \\
		Stripe 82 & 665 & 00:55:00 & 00:00:00 & $90^\circ \times 2.8^\circ$ & 2.8 & 352871 & SDSS Stripe 82 \\
		&&&&&&&\\
		Remainder & 1862 & n/a & n/a & 14474 deg$^2$ & 0.2 & 16971444 & n/a \\
		\hline
	\end{tabular}
\end{table}
\setlength{\tabcolsep}{6pt}

\citet{Yershov2012, Yershov2014} detected a correlation between SNe redshifts and CMB temperature using OLS linear regression and Pearson's correlation coefficient. We verified this correlation using the independent 2-sample Welch's t-test, the 1-sided MWU test, and Spearman's rank-order correlation coefficient, so we do not dispute that a correlation is found.

However, we do not believe that there is any astrophysical origin of the correlation and conclude that it is a composite selection bias caused by the chance alignment of certain deep survey fields with CMB hotspots. In this section we describe our identification of the deep survey fields in question and determine the significance of their contribution to the correlation. We show how their temperature and redshift biases cause this composite selection bias and we quantify the likelihood of it occurring by chance.

\subsection{Identification of fields} \label{subsecFieldID}

SNe are transient objects, historically detected both by chance and by repeated observations of specific fields, galaxy clusters etc. Reliably detecting and following the lightcurves of SNe at high redshift requires particularly targeted approaches \citep[e.g.,][]{Filippenko1998, Dawson2009}. As a result SNe are not evenly detected across the sky and most SNe datasets are not spatially uniform. This situation is changing with wide and time-domain surveys such as DES \citep{DES2016}.

We analysed the SAI sample to identify regions with a high surface density of SNe. Visual inspection of a \textit{Topcat} Sky Plot suggested that defining these as regions containing $\geqslant 20$ SNe with an average surface density of $\geqslant 20$ SNe per square degree would be appropriate and productive. We placed no constraint on the overall angular size of the region. Our algorithm identified 7 SNe fields meeting these criteria, plus 2 additional fields within SDSS Stripe 82.

Table~\ref{tabSNFields} lists the 7 SNe fields plus Stripe 82. These fields contain a total of 921 SNe (33.1\%), with Stripe 82 containing 665 SNe (23.9\%) and fields 1-7 containing 256 SNe (9.2\%) of the sample. For each field 1-7 we identified corresponding deep survey fields coincident with the SNe field. There were 3 fields from the Supernova Legacy Survey \citep[SNLS,][]{Astier2006}, 2 from the ESSENCE supernova survey \citep{Miknaitis2007}, the Hubble Deep Field North \citep[HDF-N,][]{Williams1996}, and the Chandra Deep Field South \citep[CDF-S,][]{Giacconi2001}.

\begin{figure} % SNe_plot_field_v2
    \centering
	\includegraphics[width=0.5\columnwidth]{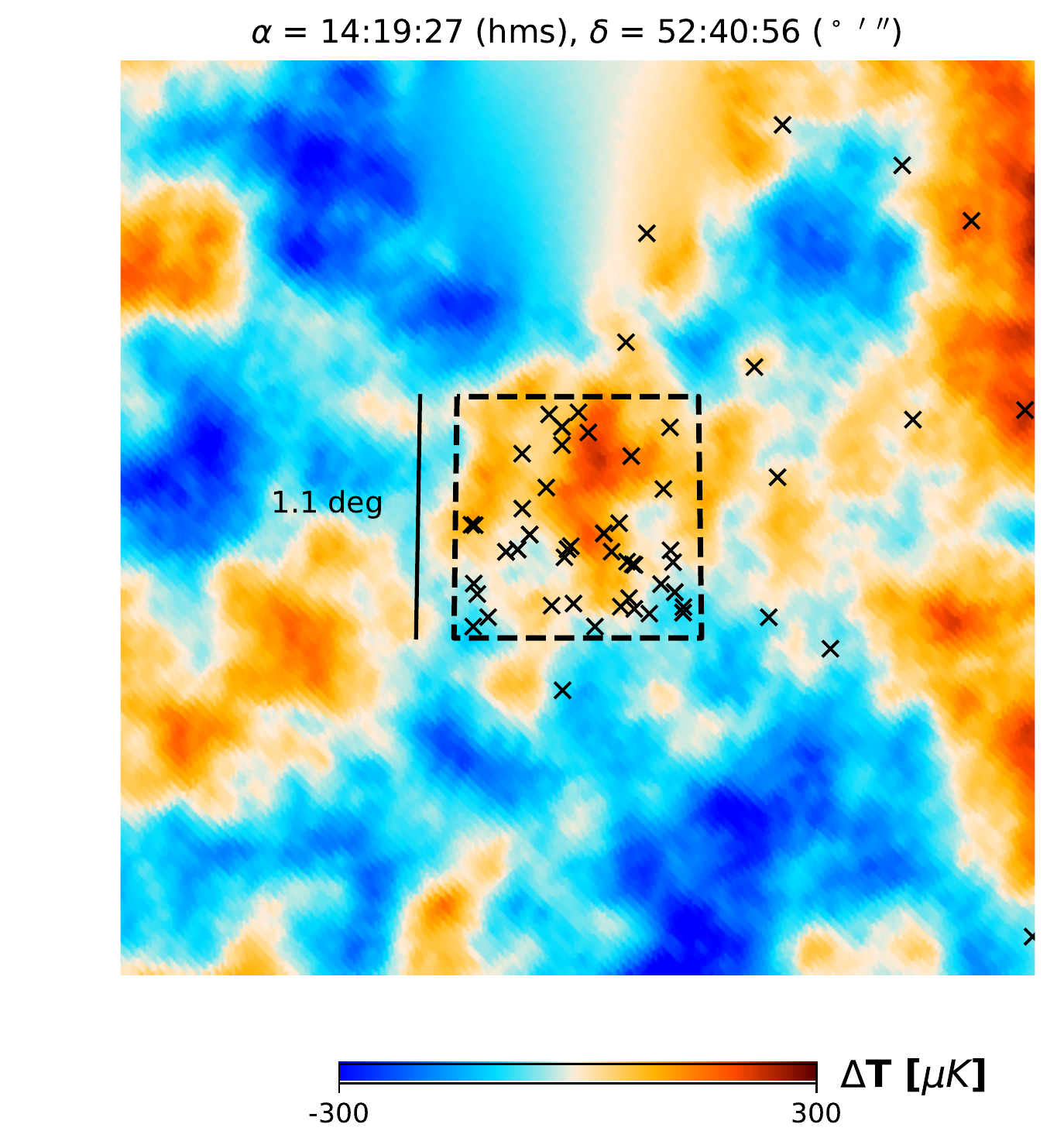}
	\caption[CMB $T$ in the vicinity of SNe field 1]{Gnomic projection of the \textit{Planck} 2015 SMICA CMB map in the vicinity of field 1. The location of SAI sample SNe are plotted with crosses ($\times$). The dashed square is the boundary of field 1, centred on the SNLS D3 deep survey field at the specified coordinates ($\alpha$ and $\delta$, J2000). See Appendix~\ref{appFields1to7} for all fields 1-7.}
	\label{figField1}
\end{figure}

For each field 1-7 we defined a square in RA and Dec orientation consistent with the deep survey field footprint and encompassing the bulk of the SNe identified by our algorithm. For the majority of the fields the best fit was to increase the deep survey field edge lengths by 10\%. For field 2 and field 7, coincident with HDF-N and CDF-S respectively, the best fit was to rotate (to RA and Dec orientation) a square enclosing the deep survey field and increase the deep survey field edge lengths by 20\%. Note that localising our SNe fields to the corresponding deep survey fields in this way generally reduced their angular size and the number of SNe they contained, which in some cases reduced the number of SNe and/or their surface density below the initial detection thresholds used.

Fig.~\ref{figField1} shows the resultant boundary of field 1 (dashed box), with SNe positions plotted with crosses ($\times$) and CMB temperature shown by the colour-bar (red indicating hotter than average, blue indicating colder than average). For all fields 1-7 see Appendix~\ref{appFields1to7}.

\subsection{Contribution to correlation} \label{subsecContribution}

To test whether these fields contribute to the correlation we compared the OLS linear regression both with and without them in the sample. We created `remainder' samples containing SNe from the SAI sample minus those in fields 1-7 combined, minus those in Stripe 82, and minus those in fields 1-7 and Stripe 82 together. We calculated the OLS linear regression gradient and Spearman's rank-order correlation coefficient of these remainders and compared them with those of the whole sample. Note that results using WLS linear regression, and values of Pearson's correlation coefficient, were consistent.

\setlength{\tabcolsep}{1pt} % default is 6pt
\begin{table} % SNe_CORE_all_versus_fields_v4, uncomment stripe 82 section for those data, SNe_correlation_coefficients
	\centering
	\small
	\caption[Gradient and Spearman's correlation coefficient for SAI sub-samples]{OLS gradient (with uncertainty of standard error on the gradient) and Spearman's rank-order correlation coefficient for the SAI sample after removing subsets of SNe fields. `No. SNe' is the number of SNe in each `remainder' sample.}
	\label{tabGradients}
	\begin{tabular}{llcrrrclcrll}
		\hline
		\multirow{2}{*}{Fields removed} && No. && \multicolumn{4}{c}{Gradient} && \multicolumn{3}{c}{Corr. Coeff.}\\
		&& SNe && \multicolumn{4}{c}{($\mu K / z$)} && $\rho_s$ && p-value\\
		\hline
		None &~~~& 2783 &~~~&~~~& 61 & $\pm$ & 12 &~~~& 0.5 &~~~& $6.7 \times 10^{-9}$\\
		Fields 1-7 && 2527 &&& -2 & $\pm$ & 22 && 0.1 && 0.6\\
		Stripe 82 && 2118 &&& 54 & $\pm$ & 14 && 0.4 && $4.9 \times 10^{-7}$\\
		Fields 1-7 \& Stripe 82 && 1862 &&& -2 & $\pm$ & 25 && 0.0 && 0.7\\
		\hline
	\end{tabular}
\end{table}
\setlength{\tabcolsep}{6pt}

Table~\ref{tabGradients} shows the gradient of the OLS linear regression slope for each remainder sample in units of $\mu K$ per unit redshift, plus the standard error on the gradient. In these units the gradient of the whole sample is $61 \pm 12 \,\mu K / z$, significantly above zero. Spearman's rank-order correlation coefficient for the whole sample shows a moderate correlation ($\rho_s = 0.5$) which is statistically significant (p-value $= 6.7 \times 10^{-9}$).

SDSS Stripe 82 is the largest field we identified, both in terms of the number of SNe (665) and angular size. Therefore the SNe in Stripe 82 cover a wider variety of CMB pixels and any statistical contribution from them should be much less prone to selection bias. Indeed, removing Stripe 82 from the sample does not significantly affect the OLS linear regression slope ($54 \pm 14 \,\mu K / z$) or Spearman's rank-order correlation coefficient ($\rho_s = 0.4$, p-value $= 4.9 \times 10^{-7}$). However, removing fields 1-7 (256 SNe) reduces the gradient dramatically to $-2 \pm 22 \,\mu K / z$, consistent with zero, and there is no correlation evident ($\rho_s = 0.1$, p-value $= 0.6$) in the remainder. Removing both fields 1-7 and Stripe 82 together has a similar effect.

\begin{figure} % A1_SNe_all_fields_remainder
	\centering
	\subfigure[Whole sample (2783 SNe)]{\includegraphics[width=0.45\columnwidth]{binned_scatter_fields_1to7_all_square.pdf} \label{subBinnedScatterFields1to7All}}
	\subfigure[Fields 1-7 sample (256 SNe)]{\includegraphics[width=0.45\columnwidth]{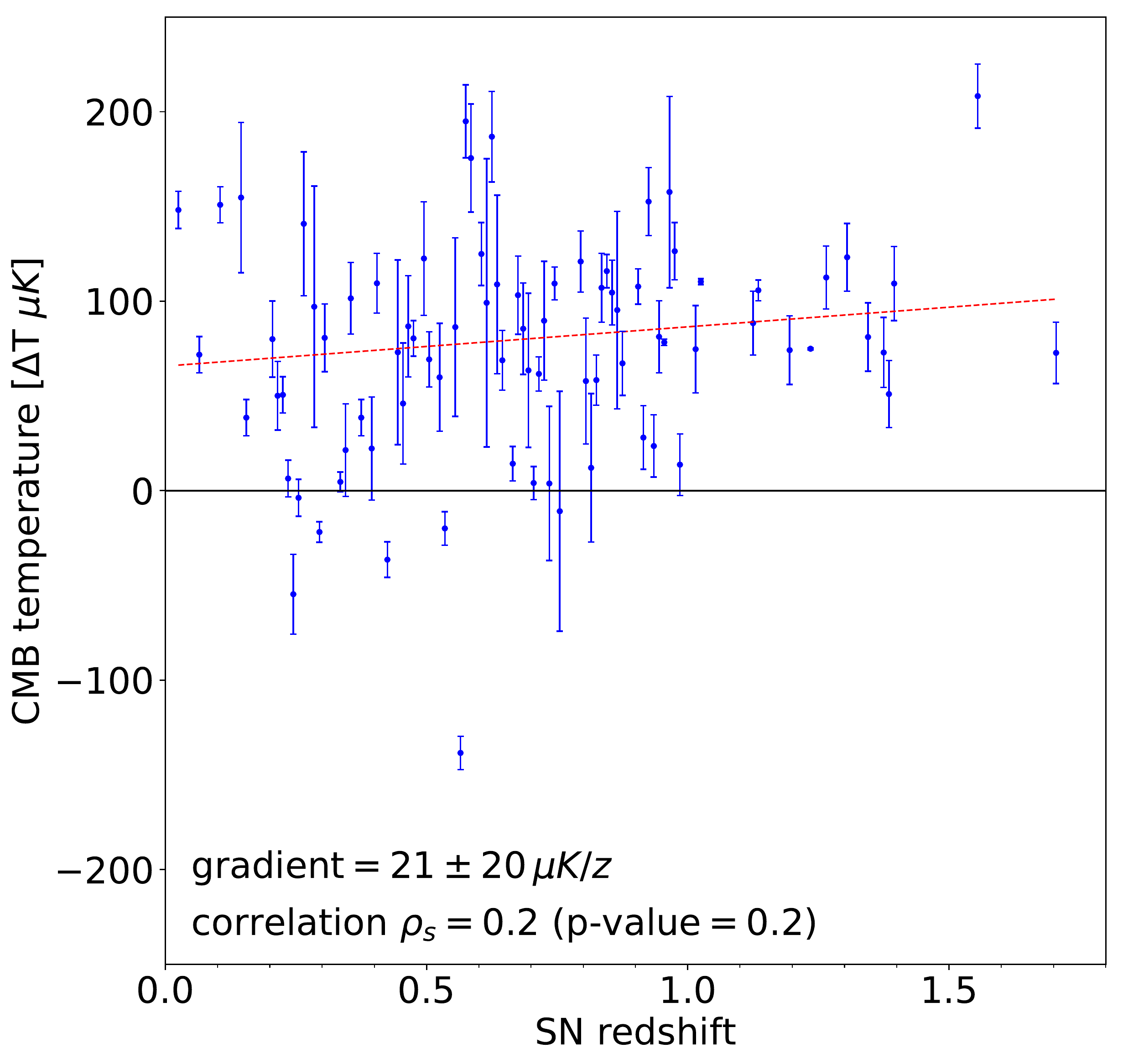} \label{subBinnedScatterFields1to7Fields}}
	\\
	\subfigure[Remainder sample (2527 SNe)]{\includegraphics[width=0.45\columnwidth]{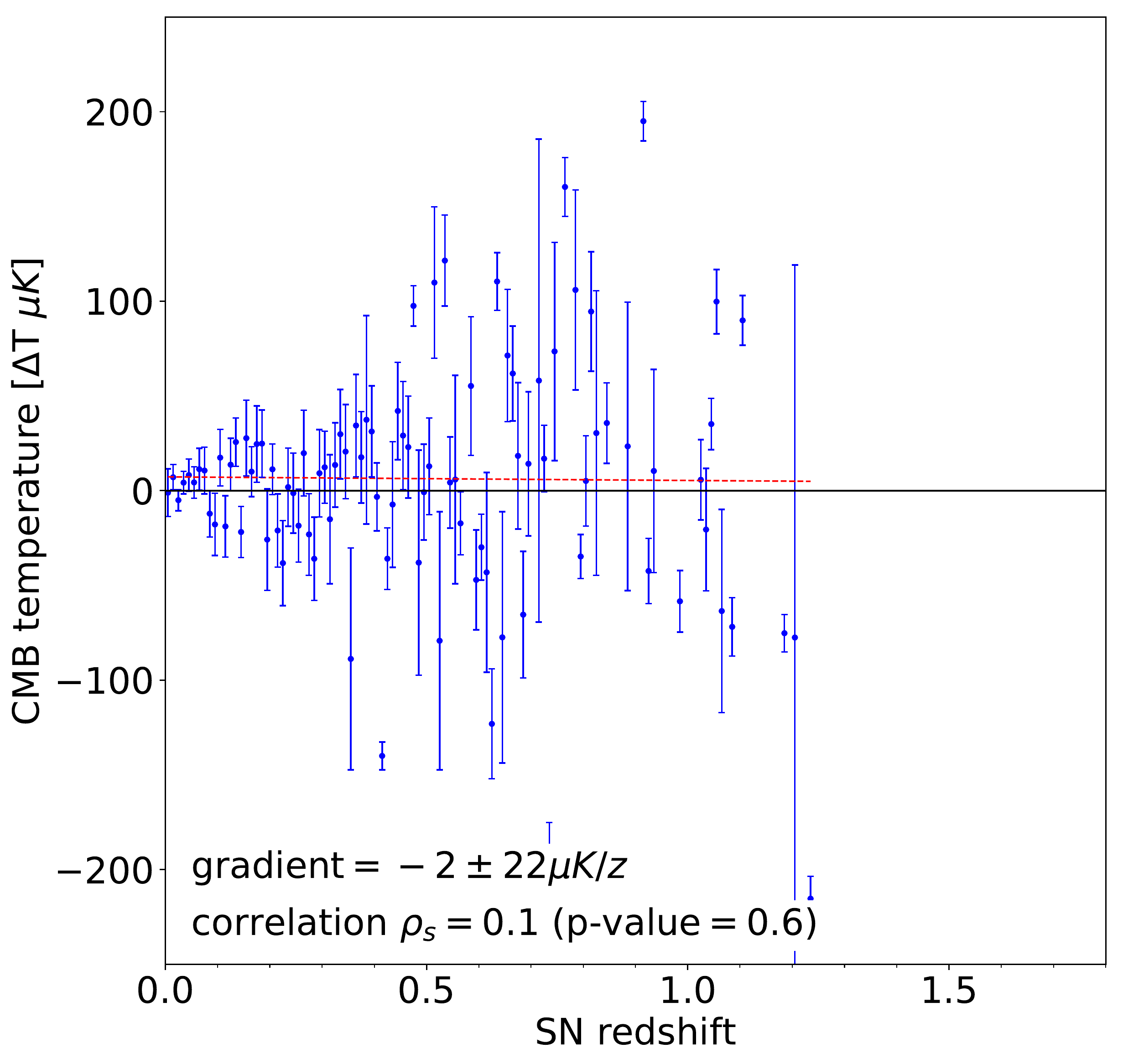} \label{subBinnedScatterFields1to7Remainder}}
	\caption[CMB $T$ at SNe locations as a function of SNe redshift (sub-samples)]{Plot of CMB temperature at SAI sample SNe locations versus SNe redshift binned with bin sizes $\Delta z=0.01$. Data are restricted by Galactic latitude ($|b|>40^\circ$), redshift ($z>0.005$), and \textit{Planck} UT78 confidence mask. Sub-figures are \protect\subref{subBinnedScatterFields1to7All} the whole sample, \protect\subref{subBinnedScatterFields1to7Fields} fields 1-7 sample, and \protect\subref{subBinnedScatterFields1to7Remainder} remainder (fields 1-7 removed) sample. Error bars are the standard error on the bin mean. The dashed line indicates ordinary least squares linear regression.}
	\label{figBinnedScatterFields1to7}
\end{figure}

The result of removing fields 1-7 from the SAI sample is illustrated in Fig.~\ref{figBinnedScatterFields1to7}, which plots the weighted mean CMB temperature at SNe locations in redshift bins of $\Delta z = 0.01$. This plot is repeated for the whole sample (\ref{subBinnedScatterFields1to7All}), fields 1-7 only (\ref{subBinnedScatterFields1to7Fields}) and the remainder of the sample after fields 1-7 are removed (\ref{subBinnedScatterFields1to7Remainder}). Note that OLS linear regression gradients of unbinned data were consistent.

The OLS gradient and correlation present in the whole sample (gradient $= 61 \pm 12 \,\mu K / z$, $\rho_s = 0.5$, p-value $= 6.7 \times 10^{-9}$) are entirely absent in the remainder (gradient $= -2 \pm 22 \,\mu K / z$, $\rho_s = 0.1$, p-value $= 0.6$). The OLS gradient of the fields 1-7 sample is slightly positive ($21 \pm 20 \,\mu K / z$) but Spearman's rank-order correlation coefficient indicates that there is no significant correlation ($\rho_s = 0.2$, p-value $ = 0.2$).

The data clearly indicate that the correlation is caused by fields 1-7 and that SDSS Stripe 82 does not contribute significantly.

\subsubsection{Angular scales} \label{subsubAngularScales}

We checked whether large-scale hot/cold spots, or anisotropies on scales of $\sim 1^\circ$ ($\ell \sim 100$), dominate the apparent correlation. We filtered the \textit{Planck} 2015 SMICA map to remove large angular scales ($\ell < 10$, $\ell < 50$, and $\ell < 100$) and repeated the OLS linear regression and Spearman's rank-order correlation coefficient analyses. In all cases there was no significant correlation evident (e.g., $\ell < 50$, $\rho_s = 0.1$, p-value $= 0.1$). The large angular scales are dominating, as expected, indicating that the CMB map pixels at SNe locations contribute no more than any other pixels within these scales.

This supports our likelihood results (section~\ref{subsecLikelihood}, Fig.~\ref{subShuffleInFields}), which indicate that the CMB map pixels at SNe locations are no more relevant than any other pixels within fields 1-7 (angular size from $0.4^\circ \times 0.4^\circ$ to $2.0^\circ \times 2.0^\circ$).

\subsection{Temperature and redshift} \label{subsecTempRedshift}

We investigated whether the CMB temperature at SNe locations and/or the redshift of SNe within fields 1-7 and Stripe 82 differ from those in the rest of the sample. We calculated the mean CMB temperature ($\overline{T} \pm \sigma_{\overline{T}}$) and mean SNe redshift ($\overline{z} \pm \sigma_{\overline{z}}$) for each sample compared with the remainder. We also analysed the CMB temperature and SNe redshift distributions using the independent 2-sample Welch's t-test and 1-sided MWU test. Note that results using the median CMB temperature at SNe locations, and median SNe redshift, were consistent.

\subsubsection{Mean CMB $T$ and mean SNe $z$} \label{subsubMeanTempRedshift}

\begin{figure} % A6_SNe_field_1_to_7_z_and_T
    \centering
	\includegraphics[width=0.6\textwidth]{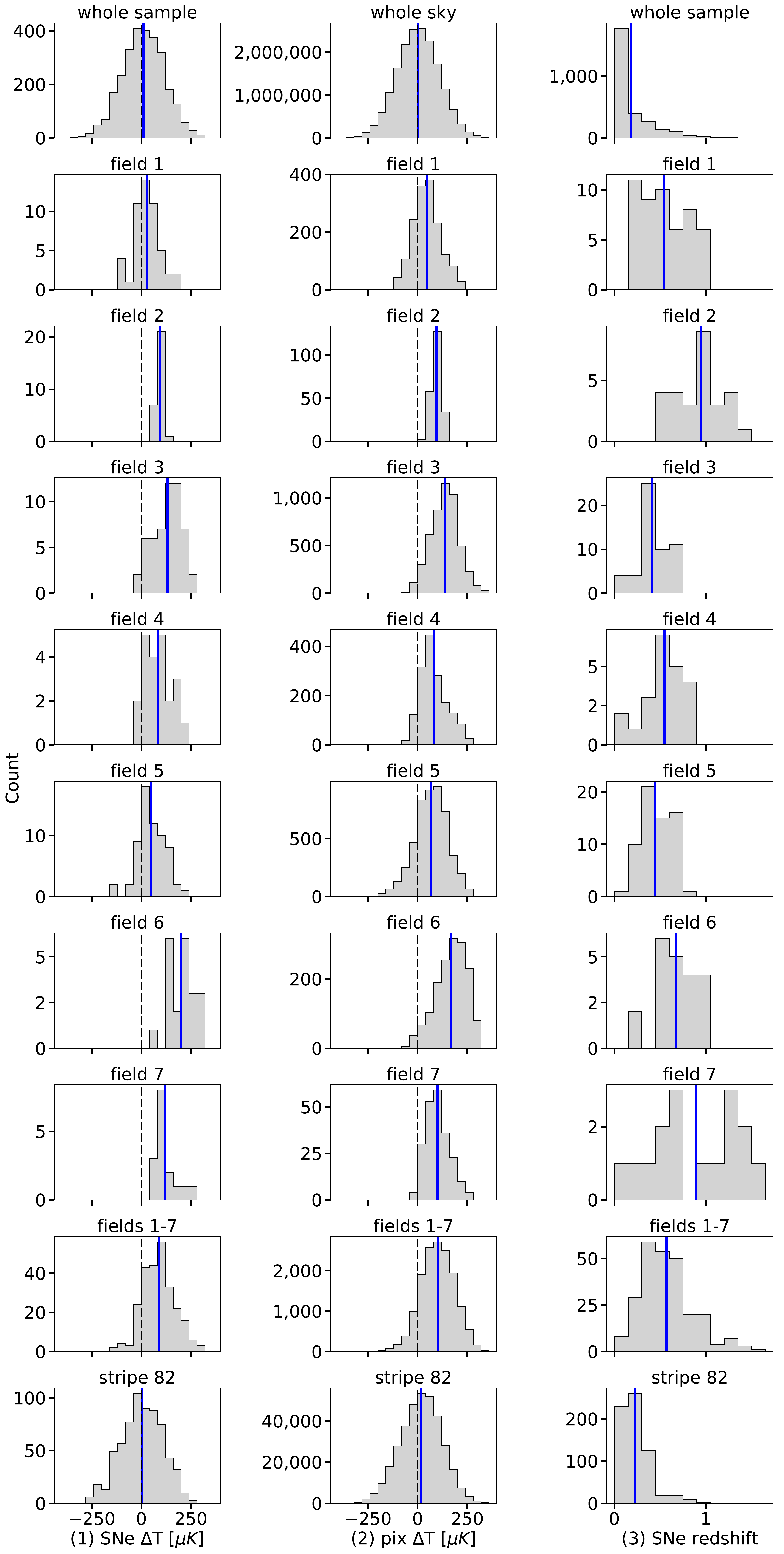}
	\caption[CMB $T$ and SNe $z$ distributions for SNe/pixel sub-samples]{SAI sample CMB $T$ and SNe redshift distributions for fields 1-7, Stripe 82, and the whole sample or sky. Column (1) shows CMB $T$ at SNe locations, $\Delta T = 40 \,\mu K$ bins. Column (2) shows CMB $T$ of all HEALPix pixels, $\Delta T = 40 \,\mu K$ bins. Column (3) shows redshift of SNe, $\Delta z = 0.15$ bins. SNe and pixels are restricted as described in the text. Solid vertical lines are arithmetic (unweighted) mean values for each distribution. Dashed vertical lines in columns (1) and (2) are at $\Delta T = 0$.}
	\label{figHistograms}
\end{figure}

The mean CMB temperature ($\overline{T} \pm \sigma_{\overline{T}}$) at SNe locations, and of all CMB map HEALPix pixels within each sample, and mean SNe redshift ($\overline{z} \pm \sigma_{\overline{z}}$) are specified in Table~\ref{tabHistograms}. The samples are fields 1-7 individually, fields 1-7 combined, Stripe 82, and the whole sample. Fig.~\ref{figHistograms} illustrates these distributions, namely CMB temperature at SNe locations (1), CMB temperature of all CMB map HEALPix pixels within each sample (2), and SNe redshift (3) in these samples. In both Table~\ref{tabHistograms} and Fig.~\ref{figHistograms} SNe are restricted by Galactic latitude ($|b|>40^\circ$), redshift ($z>0.005$), and \textit{Planck} UT78 confidence mask and CMB map HEALPix pixels are restricted by Galactic latitude ($|b|>40^\circ$) and \textit{Planck} UT78 confidence mask.

SNe in all fields 1-7 are biased to CMB temperatures hotter than the mean of the whole sample ($10.3 \pm 2.0 \,\mu K$). Fields 3, 6 and 7 are particularly extreme, with mean CMB temperatures at SNe locations of $130.8 \pm 9.2 \,\mu K$, $199.3 \pm 13.6 \,\mu K$, and $120.5 \pm 12.0 \,\mu K$ respectively. SNe in SDSS Stripe 82 are not biased to CMB temperatures hotter than the mean of the whole sample. For all the fields, fields 1-7 and Stripe 82, the CMB temperature  distribution (and mean) at SNe locations is generally representative of the CMB map HEALPix pixel temperature distribution (and mean) to within $\pm \sim 30 \,\mu K$.

SNe in all fields 1-7 are also biased to higher redshift than the mean of the whole sample ($z = 0.18 \pm 0.00$). Fields 2, 6 and 7 are particularly extreme, with mean SNe redshifts of $z = 0.94 \pm 0.06$, $z = 0.67 \pm 0.04$ and $z = 0.89 \pm 0.11$ respectively. SNe in SDSS Stripe 82 are biased to slightly higher redshift of $z = 0.23 \pm 0.01$ than the whole sample. However, although Stripe 82 is deeper it is not hotter, which explains why it does not significantly contribute to the correlation.

\setlength{\tabcolsep}{1pt} % default is 6pt
\begin{table} % A4_SNe_find_field_z_and_T & A5_SNe_find_field_pixels
	\centering
	\small
	\caption[CMB $\overline{T}$ and SNe $\overline{z}$ distributions for SNe sub-samples]{Arithmetic (unweighted) mean CMB temperature, at SNe locations within each field sample and of all HEALPix pixels within each field sample area, and mean SNe redshift for each field sample and the whole sample or sky. SNe are restricted by Galactic latitude ($|b|>40^\circ$), redshift ($z>0.005$), and \textit{Planck} UT78 confidence mask. Pixels are restricted by Galactic latitude ($|b|>40^\circ$) and \textit{Planck} UT78 confidence mask. The uncertainty is the standard error of the mean.}
	\label{tabHistograms}
	\begin{tabular}{lrclcrclcrcl}
		\hline
		\multirow{2}{*}{Field} & \multicolumn{7}{c}{CMB temperature ($\mu K$)} && \multicolumn{3}{c}{Redshift} \\
		& \multicolumn{3}{c}{SNe} && \multicolumn{3}{c}{pixels} && \multicolumn{3}{c}{SNe} \\
		\hline
		Field 1 & 29.1 & $\pm$ & 9.1 &~~~& 47.5 & $\pm$ & 1.7 &~~~& 0.55 & $\pm$ & 0.03  \\
		Field 2 & 92.9 & $\pm$ & 3.3 && 94.3 & $\pm$ & 1.7 && 0.94 & $\pm$ & 0.06 \\
		Field 3 & 130.8 & $\pm$ & 9.2 && 137.5 & $\pm$ & 1.0 && 0.41 & $\pm$ & 0.02 \\
		Field 4 & 85.3 & $\pm$ & 14.3 && 81.7 & $\pm$ & 1.7 && 0.55 & $\pm$ & 0.05 \\
		Field 5 & 49.7 & $\pm$ & 8.5 && 68.2 & $\pm$ & 1.2 && 0.45 & $\pm$ & 0.02 \\
		Field 6 & 199.3 & $\pm$ &13.6 && 168.9 & $\pm$ & 1.9 && 0.67 & $\pm$ & 0.04 \\
		Field 7 & 120.5 & $\pm$ & 12.0 && 100.8 & $\pm$ & 4.0 && 0.89 & $\pm$ & 0.11 \\
		Fields 1-7 & 87.4 & $\pm$ & 5.0 && 101.5 & $\pm$ & 0.7 && 0.57 & $\pm$ & 0.02 \\
		Stripe 82 & 3.8 & $\pm$ & 4.0 && 17.2 & $\pm$ & 0.2 && 0.23 & $\pm$ & 0.01 \\
		Whole sample & 10.3 & $\pm$ & 2.0 && 3.1 & $\pm$ & 0.02 && 0.18 & $\pm$ & 0.00 \\
		\hline
	\end{tabular}
\end{table}
\setlength{\tabcolsep}{6pt}

\subsubsection{MWU and t-test}

We analysed whether the CMB temperature distribution at SNe locations and/or SNe redshift distribution within fields 1-7 and Stripe 82 differ from the rest of the sample using the independent 2-sample Welch's t-test and 1-sided MWU test. To recap from section~\ref{subsecMethod}, Welch's t-test (or unequal variance t-test) accommodates both the different angular extent of the deep survey fields, and hence difference in variance of the CMB temperature, and the variation in the number of SNe per redshift bin. The MWU test makes fewer assumptions (in particular, the t-test assumes a normal distribution) and is less sensitive to outliers than the t-test. Clearly not all the samples we tested are normally distributed (see Fig.~\ref{figHistograms}) but we have included all the results for completeness.

We performed all the analysis using a constant `remainder' sample created by removing fields 1-7 and Stripe 82 from the sample. This was tested against samples containing SNe from fields 1-7 individually, fields 1-7 combined, and Stripe 82. See Table~\ref{tabFieldMWU} for the results (p-values). Note that comparison between the very small p-values is unlikely to be meaningful.

For CMB temperature both p-values for SDSS Stripe 82 and the MWU p-value for field~1 are above the $\alpha = 1$\% significance level. Therefore we cannot reject the null hypotheses that Stripe 82 has the same mean temperature and same temperature distribution as the remainder of the sample, nor the null hypothesis that field 1 has the same temperature distribution as the remainder of the sample.

However, for all other tests the p-values indicate that the individual field samples do not have the same mean temperature as the remainder, and that the temperature distribution of the fields is significantly hotter than that of the remainder.

For SNe redshift all the p-values of all the samples indicate that the individual fields do not have the same mean redshift as the remainder, and that the redshift distribution of the fields is significantly higher than that of the remainder.

\begin{table} % A3_SNe_MWU_ttest_temp_z
	\centering
	\small
	\caption[Results of Welch's t-test and MWU test of CMB $T$ and SNe $z$]{Results (p-values) from independent 2-sample Welch's t-tests and 1-sided (greater) MWU tests of CMB temperature and SNe redshift for each field sample. All tests are against a constant remainder sample after removing fields 1-7 and Stripe 82.}
	\label{tabFieldMWU}
	\begin{tabular}{lllll}
		\hline
		\multirow{3}{*}{Field} & \multicolumn{4}{c}{p-values} \\
		& \multicolumn{2}{c}{CMB temperature} & \multicolumn{2}{c}{SN redshift} \\
		& t-test & MWU & t-test & MWU \\
		\hline
		Field 1 & $5.8 \times 10^{-3}$ & $1.8 \times 10^{-2}$ & $5.8 \times 10^{-17}$ & $3.0 \times 10^{-26}$ \\
		Field 2 & $2.9 \times 10^{-31}$ & $9.9 \times 10^{-9}$ & $9.0 \times 10^{-15}$ & $1.7 \times 10^{-19}$ \\
		Field 3 & $9.6 \times 10^{-20}$ & $1.8 \times 10^{-18}$ & $2.3 \times 10^{-20}$ & $3.9 \times 10^{-25}$ \\
		Field 4 & $1.2 \times 10^{-5}$ & $3.5 \times 10^{-5}$ & $9.6 \times 10^{-9}$ & $1.4 \times 10^{-12}$ \\
		Field 5 & $1.0 \times 10^{-6}$ & $1.7 \times 10^{-5}$ & $1.5 \times 10^{-23}$ & $1.9 \times 10^{-29}$ \\
		Field 6 & $3.8 \times 10^{-12}$ & $1.0 \times 10^{-12}$ & $2.9 \times 10^{-11}$ & $1.2 \times 10^{-13}$ \\
		Field 7 & $6.1 \times 10^{-8}$ & $7.6 \times 10^{-7}$ & $6.0 \times 10^{-6}$ & $8.2 \times 10^{-11}$ \\
		Fields 1-7 & $3.8 \times 10^{-42}$ & $1.6 \times 10^{-36}$ & $3.7 \times 10^{-71}$ & $1.2 \times 10^{-112}$ \\
		Stripe 82 & 0.7 & 0.2 & $8.3 \times 10^{-49}$ & $1.8 \times 10^{-118}$ \\
		\hline
	\end{tabular}
\end{table}

We have demonstrated that fields 1-7 are biased to hotter CMB temperatures, specifically at SNe locations but also at all CMB map HEALPix pixels within the fields. We believe this is the result of the chance alignment of those fields with CMB hotspots. This would not on its own be sufficient to lead to the correlation reported by \citet{Yershov2012, Yershov2014}. However, fields 1-7 are also biased to higher redshifts because they are the result of deep survey fields. The remainder of the SNe are generally lower redshift and are spread more uniformly across the sky, so they have a mean CMB temperature closer to the mean of the whole CMB map.

The composite effect is to introduce enough high-redshift SNe at locations of sufficiently high CMB temperature to skew all the analyses we have performed to demonstrate the presence of the correlation, namely OLS linear regression, Welch's t-test, MWU test, and Spearman's rank-order correlation coefficient. This effect was caused by 256 SNe, comprising 9.2\% of the restricted SAI sample of 2783 SNe.

\subsection{Likelihood} \label{subsecLikelihood}

We quantified the likelihood of this selection bias happening by chance by analysing the effect on the OLS gradient of moving SNe to random positions within fields 1-7, and by moving fields 1-7 to random positions on the sky. In both analyses the SNe were not moved between fields. We also analysed the effect on the OLS gradient of simulating the CMB sky, without moving the SNe at all.

Within each field 1-7 we moved SNe to 1000 random positions within the field boundaries defined in section~\ref{subsecFieldID}. We also moved each field 1-7 to 10000 random positions on the sky, compliant with the Galactic latitude restriction ($|b|>40^\circ$), whilst keeping the field size and shape constant and the SNe in approximately the same position within each field (within small angle approximation). In both cases all other SNe outside fields 1-7 were left in their original positions. After each move (within each field or of each field on the sky) the masking (\textit{Planck} UT78) was re-applied, the CMB temperature and variance were re-sampled, and the weights were re-calculated.

We created 10000 simulations of the CMB sky from the power spectrum of our fiducial \textit{Planck} 2015 SMICA map, as described in section~\ref{subsecMethod}. All SNe were left in their original positions. After each simulation the CMB temperature was re-sampled. The masking (\textit{Planck} UT78) and variances were left unchanged as the SNe remained on their original CMB map HEALPix pixel.

Following each move or simulation and subsequent derivations/calculations we binned the data, re-calculated the weighted mean CMB temperature of each bin, fitted an OLS linear regression, and determined the gradient of the slope as previously described (section~\ref{subsecMethod}).

\begin{figure} % shuffle_SNe_within_fields.py, shuffle_SNe_fields_on_sky.py, sim_CMB_behind_SNe_fields.py
    \centering
	\subfigure[Move SNe within fields 1-7]{\includegraphics[width=0.65\columnwidth]{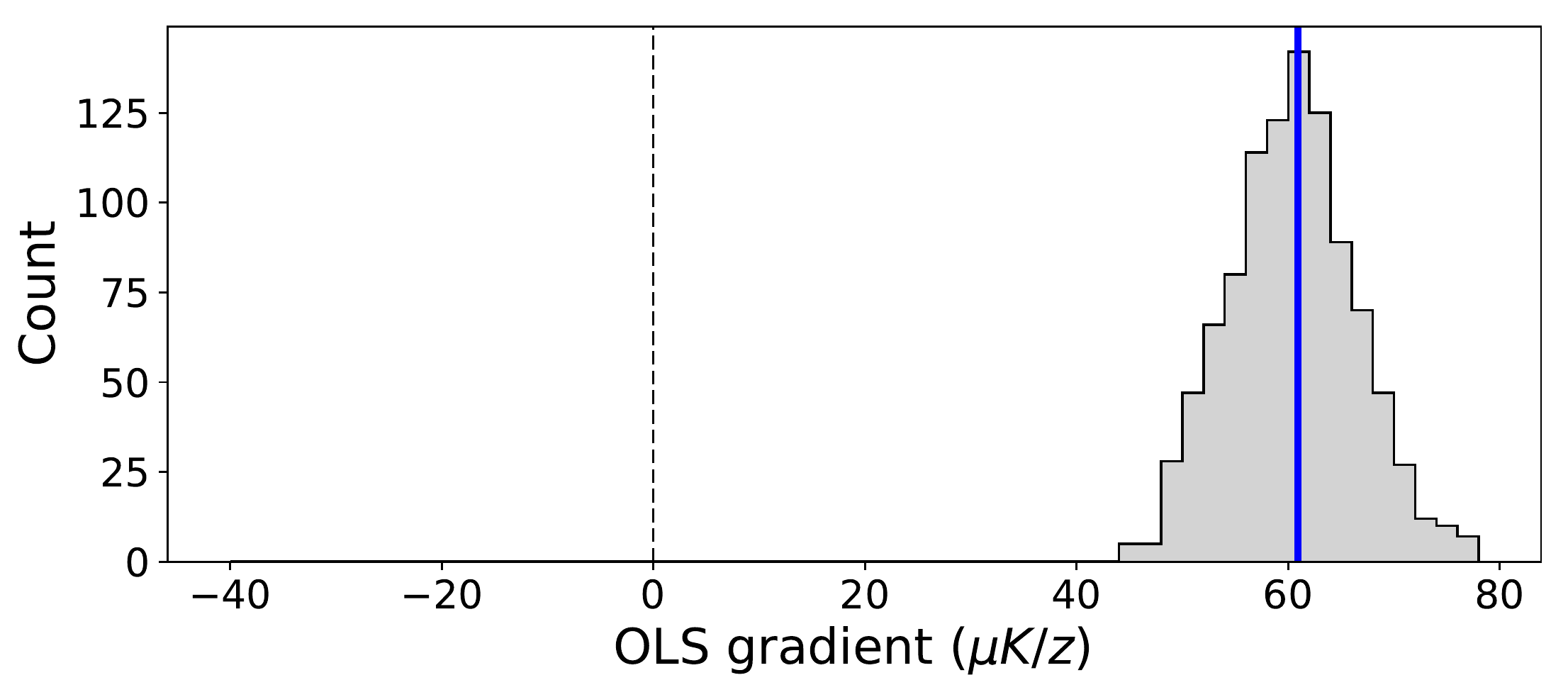} \label{subShuffleInFields}}
	\\
	\subfigure[Move fields 1-7 on sky]{\includegraphics[width=0.65\columnwidth]{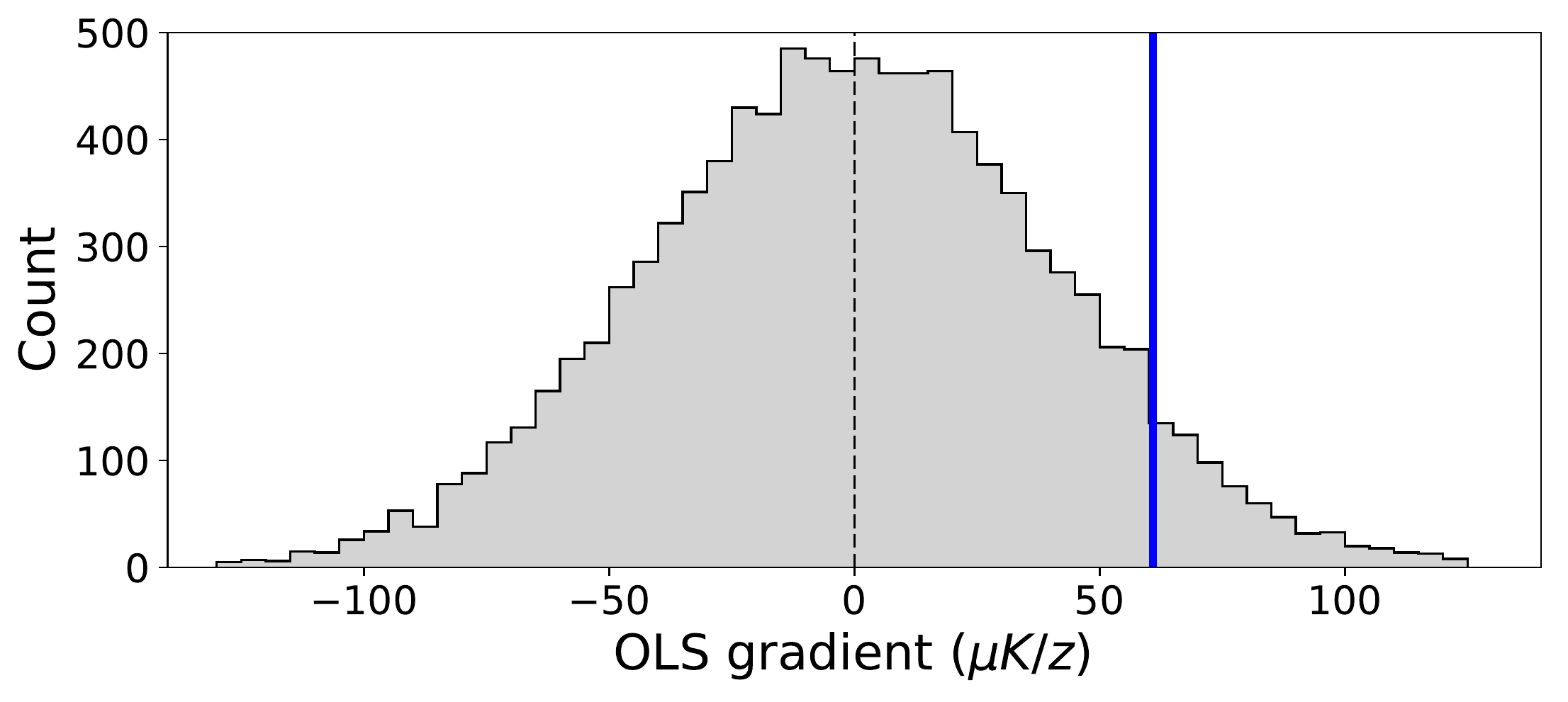} \label{subShuffleOnSky}}
	\\
	\subfigure[Simulate CMB]{\includegraphics[width=0.65\columnwidth]{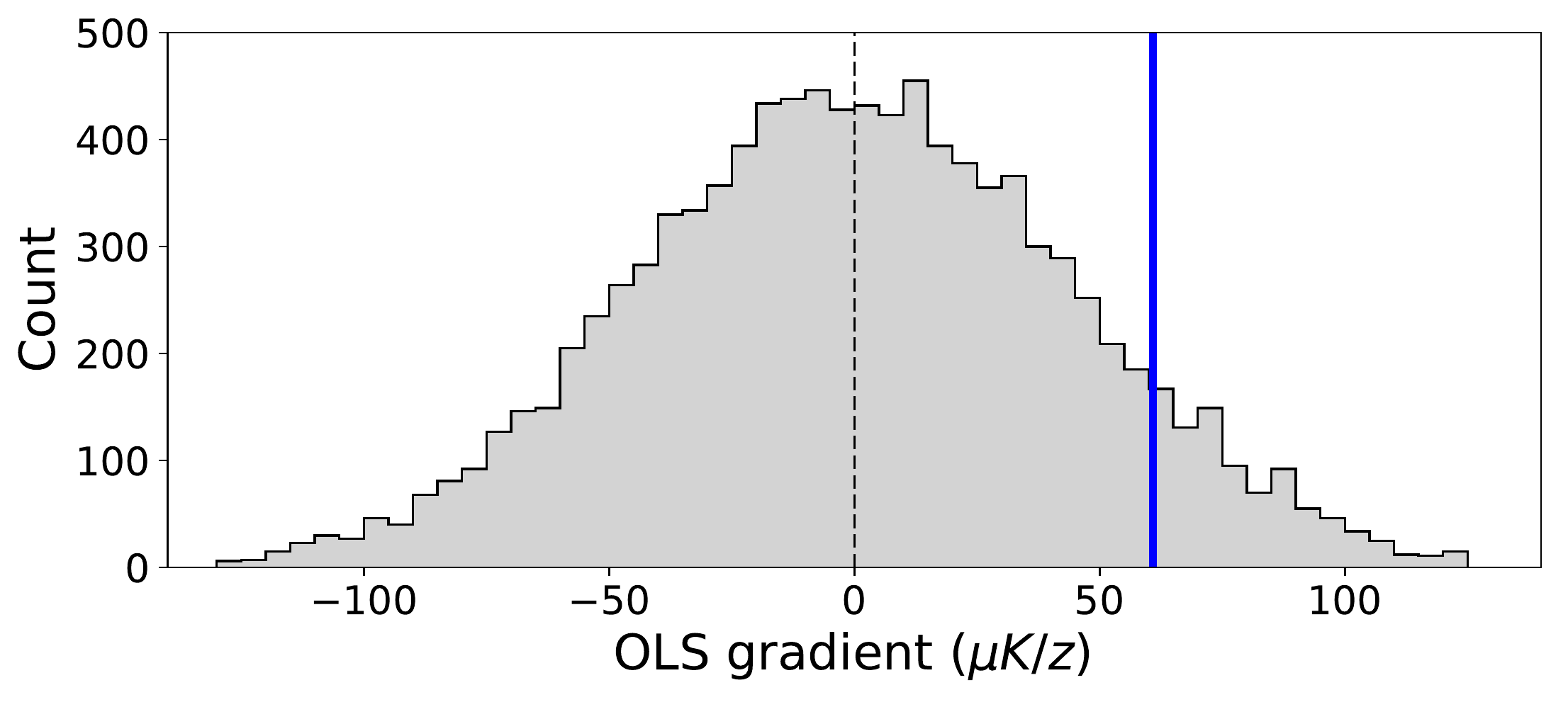} \label{subSimCMB}}
	\caption[Distribution of gradient after randomisations (shuffling \& simulation)]{Histogram of OLS gradient for the SAI sample after \protect\subref{subShuffleInFields} moving SNe to 1000 random positions within each field 1-7, \protect\subref{subShuffleOnSky} moving fields 1-7 to 10000 random positions on the sky ($|b|>40^\circ$), and \protect\subref{subSimCMB} 10000 simulations of the CMB sky. SNe positions are restricted by Galactic latitude ($|b|>40^\circ$) and \textit{Planck} UT78 confidence mask. Dashed vertical lines are at zero gradient. Solid vertical lines are original value of the gradient.}
	\label{figShuffle}
\end{figure}

Fig.~\ref{figShuffle} shows the distribution of OLS gradients after these random moves and simulations. For comparison, the original gradient ($61 \pm 12 \,\mu K / z$) is shown as a solid vertical line. Note that the uncertainty in the original gradient is the standard error on the gradient as calculated by the OLS linear regression, whereas the uncertainties in the means of the distributions, described below, are the standard errors of the means.

After moving SNe within each field 1-7 to 1000 random positions within the fields, the mean of the OLS gradient distribution (\ref{subShuffleInFields}) is $60 \pm 0 \,\mu K / z$. The distribution is narrow and consistent with the original gradient, indicating that the position of SNe within fields 1-7 does not significantly affect the correlation.

After moving each field 1-7 to 10000 random positions on the sky, the mean of the OLS gradient distribution (\ref{subShuffleOnSky}) is $-1 \pm 0 \,\mu K / z$. The distribution is wide, centred near zero, and inconsistent with the original gradient, unsurprisingly indicating that the position of fields 1-7 on the sky is responsible for the correlation.

After 10000 simulations of the CMB sky, the mean of the OLS gradient distribution (\ref{subSimCMB}) is $-0 \pm 1 \,\mu K / z$. The distribution is consistent with the results from moving each field 1-7 to 10000 random positions on the sky.

Assuming a standard normal distribution we calculated the z-score (standard score) of the original OLS gradient (X) as

\begin{ceqn}
\begin{equation}
z = (X - \mu) / \sigma \ ,
\end{equation}
\end{ceqn}

\noindent where $\mu$ is the mean of the gradient distribution and $\sigma$ is its standard deviation. We then used the standard normal distribution table to provide the probability of observing a gradient at least as extreme as X within our gradient distributions. For moving fields 1-7 on the sky (\ref{subShuffleOnSky}) the probability is 6.8\% ($\sim$ 1 in 15) and for simulating the CMB (\ref{subSimCMB}) it is 8.9\% ($\sim$ 1 in 11). Therefore the chance alignment of fields 1-7 with CMB hotspots is not an exceptionally unlikely event.

\section{Results: alternative data} \label{secAlternativeData}

\subsection{SNe types} \label{subsecSNTypes}

\citet{Yershov2014} demonstrated that the correlation between SNe redshifts and CMB temperature was particularly strong for the SNIa sub-sample, whereas for the rest of the SNe it vanished. Is this consistent with our assertion that the correlation is the result of a composite selection bias caused by the chance alignment of certain deep survey fields (fields 1-7) with CMB hotspots?

\begin{table}
	\centering
	\small
	\caption[Number and proportion of SNIa within sub-samples]{Number and proportion (per cent) of SNIa within the whole SAI sample, fields 1-7 sample, and remainder (fields 1-7 removed) sample.}
	\label{tabSNIa}
	\begin{tabular}{lrrr}
		\hline
		\multirow{2}{*}{Sub-sample} & No. & No. & \% \\
		& SNe & SNIa & SNIa \\
		\hline
		Whole sample & 2783 & 1,749 & 62.8\% \\
		Fields 1-7 & 256 & 235 & 91.8\% \\
		Remainder & 2527 & 1,514 & 59.9\% \\
		\hline
	\end{tabular}
\end{table}

Table~\ref{tabSNIa} shows the number and proportion of Type Ia SNe in our SNe samples. Supernova surveys such as SNLS and ESSENCE primarily targeted SNIa \citep{Pritchet2005, Miknaitis2007}, so it is unsurprising that fields 1-7 contain predominantly SNIa (91.8\%). As expected in the whole sample, a little over half the SNe are SNIa. Removing fields 1-7 from the sample does not significantly decrease the proportion of SNIa, which drops from 62.8\% in the whole sample to 59.9\% in the remainder. However, we have shown that the correlation present in the whole sample (Fig.~\ref{subBinnedScatterFields1to7All}) is entirely absent in this remainder (Fig.~\ref{subBinnedScatterFields1to7Remainder})

Fields 1-7 together comprise 9.2\% of the whole sample, but when the sample is restricted to SNIa only this increases to 13.4\%. Thus, restricting the sample to SNIa increases the influence of fields 1-7. We suggest that the correlation is not caused by SNIa themselves, but that it is inadvertently enhanced by restricting the sample to SNIa due to the dominance of SNIa in fields 1-7. And further, removing SNIa from the sample effectively removes fields 1-7, causing the correlation to vanish.

\subsection{SNe catalogues} \label{subsecOSC}

We have demonstrated that the correlation reported by \citet{Yershov2012, Yershov2014} is a composite selection bias caused by the chance alignment of certain deep survey fields with CMB hotspots. \citet{Yershov2012, Yershov2014} analysed the Sternberg Astronomical Institute \citep[SAI,][]{Bartunov2007} SNe catalogue, but it seems reasonable that other SNe catalogues could show a similar effect.

We repeated our analyses from section~\ref{secBias} using the Open Supernova Catalogue \citep[OSC,][]{Guillochon2017}. Data were obtained, restricted and weighted as described in section~\ref{secDataMethod}, yielding a sample of 7880 SNe. We found that the OSC sample does indeed exhibit a similar apparent correlation (with a gradient of $42 \pm 7 \,\mu K / z$) to the SAI sample.

We applied the same SNe field detection algorithm with the same detection thresholds described in section~\ref{subsecFieldID} to the OSC sample.

\begin{table}
	\centering
	\small
	\caption[Number of SNe within fields identified in SAI and OSC sub-samples]{SNe fields identified in the SAI and OSC samples and the number of SNe within each. See Table~\ref{tabSNFields} for field positions, sizes and further information.}
	\label{tabSNFieldsOSC}
	\begin{tabular}{lrr}
		\hline
		\multirow{2}{*}{Field} & \multicolumn{2}{c}{No. SNe} \\
		& SAI & OSC \\
		\hline
		Field 1 & 50 & 152 \\
		Field 2 & 29 & 91 \\
		Field 3 & 54 & 79 \\
		Field 4 & 22 & 116 \\
		Field 5 & 64 & 91 \\
		Field 6 & 21 & 92 \\
		Field 7 & 16 & 55 \\
		Stripe 82 & 665 & 2445 \\
		\hline
	\end{tabular}
\end{table}

Our algorithm identified the same 7 SNe fields with the same boundaries, but with somewhat increased SNe membership, plus 13 fields within Stripe 82. These fields (Table~\ref{tabSNFieldsOSC}) contain a total of 3,121 SNe (39.6\%), with Stripe 82 containing 2,445 SNe (31.0\%) and fields 1-7 containing 676 SNe (8.6\%) of the OSC sample. We compared the OLS linear regression both with and without these fields in the sample and calculated Spearman's rank-order correlation coefficient of these samples, as described in section~\ref{subsecContribution}.

Table~\ref{tabGradientsOSC} shows the gradient of the OLS linear regression slope for each OSC remainder sample in units of $\mu K$ per unit redshift, plus the standard error on the gradient. In these units the gradient of the whole sample is $42 \pm 7 \,\mu K / z$, which as for the SAI sample is significantly above zero. Spearman's rank-order correlation coefficient for the whole sample shows a moderate correlation ($\rho_s = 0.6$) which is statistically significant (p-value $= 2.0 \times 10^{-15}$). These results are consistent with those for the whole SAI sample (gradient $= 61 \pm 12 \,\mu K / z$, $\rho_s = 0.5$ and p-value $= 6.7 \times 10^{-9}$).

SDSS Stripe 82 is again the largest field we identified, both in terms of the number of SNe (2,445) and angular size. Removing Stripe 82 from the OSC sample does not significantly affect the OLS linear regression slope ($38 \pm 7 \,\mu K / z$) or Spearman's rank-order correlation coefficient ($\rho_s = 0.5$, p-value $= 1.1 \times 10^{-11}$). However, removing fields 1-7 (676 SNe) reduces the gradient dramatically to $10 \pm 11 \,\mu K / z$, and there is no correlation evident in the remainder ($\rho_s = 0.1$, p-value $= 0.5$). Removing both fields 1-7 and Stripe 82 together has a similar effect.

\setlength{\tabcolsep}{1pt} % default is 6pt
\begin{table} % SNe_CORE_all_versus_fields_v4, uncomment stripe 82 section for those data, SNe_correlation_coefficients
	\centering
	\small
	\caption[Gradient and Spearman's correlation coefficient for OSC sub-samples]{OLS gradient (with uncertainty of standard error on the gradient) and Spearman's rank-order correlation coefficient for the OSC sample after removing subsets of SNe fields. `No. SNe' is the number of SNe in each `remainder' sample.}
	\label{tabGradientsOSC}
	\begin{tabular}{llcrrrclcrll}
		\hline
		\multirow{2}{*}{Fields removed} && No. && \multicolumn{4}{c}{Gradient} && \multicolumn{3}{c}{Corr. Coeff.}\\
		&& SNe && \multicolumn{4}{c}{($\mu K / z$)} && $\rho_s$ && p-value\\
		\hline
		None &~~~& 7880 &~~~&~~~& 42 & $\pm$ & 7 &~~~& 0.6 &~~~& $2.0 \times 10^{-15}$\\
		Fields 1-7 && 7204 &&& 10 & $\pm$ & 11 && 0.1 && 0.5\\
		Stripe 82 && 5435 &&& 38 & $\pm$ & 7 && 0.5 && $1.1 \times 10^{-11}$\\
		Fields 1-7 \& Stripe 82 && 4759 &&& 11 & $\pm$ & 13 && -0.0 && 0.1\\
		\hline
	\end{tabular}
\end{table}
\setlength{\tabcolsep}{6pt}

\begin{figure} % A1_SNe_all_fields_remainder
	\centering
	\subfigure[Whole sample (7880 SNe)]{\includegraphics[width=0.45\columnwidth]{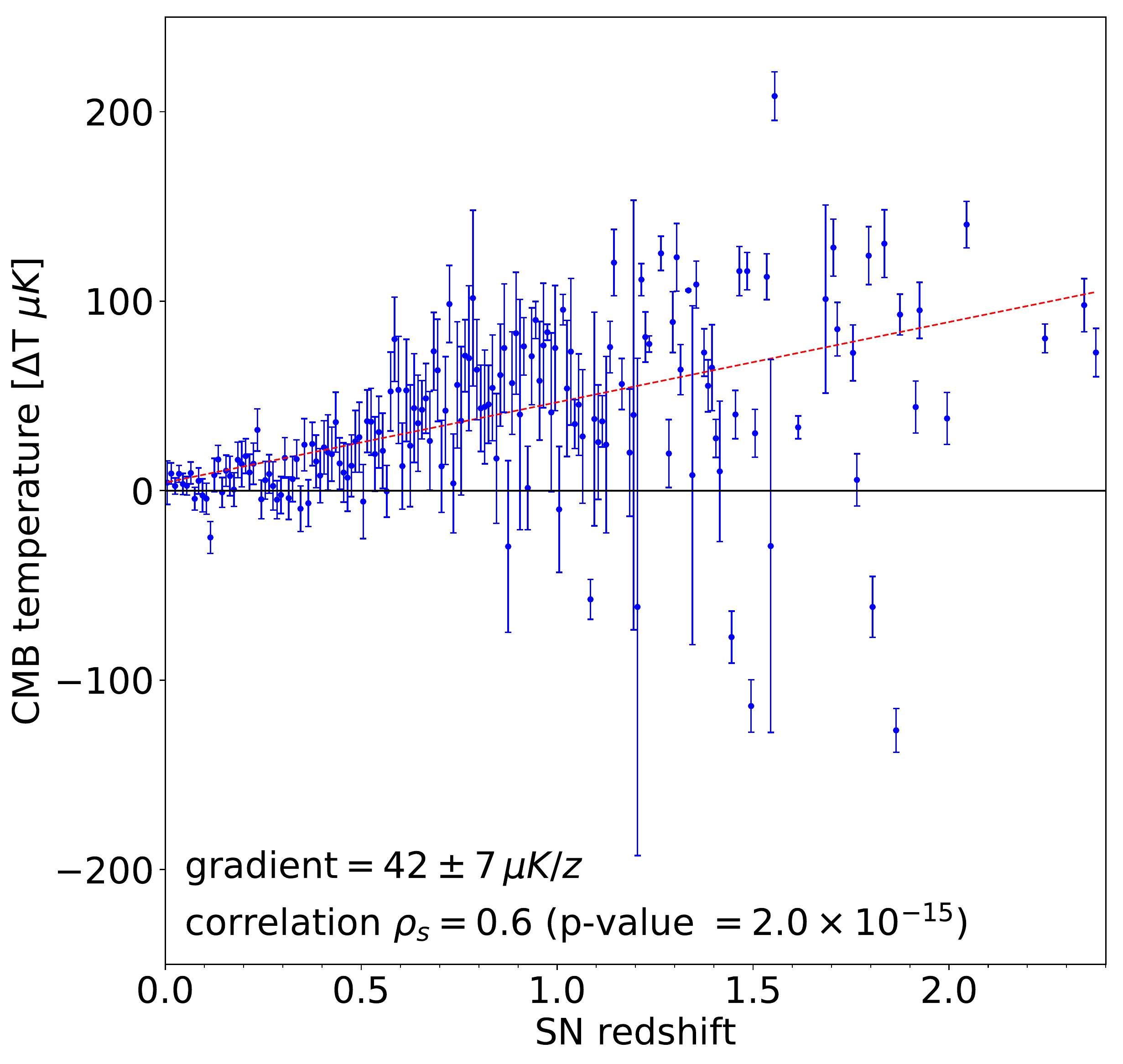} \label{subBinnedScatterFields1to7AllOSC}}
	%\\
	\subfigure[Fields 1-7 sample (676 SNe)]{\includegraphics[width=0.45\columnwidth]{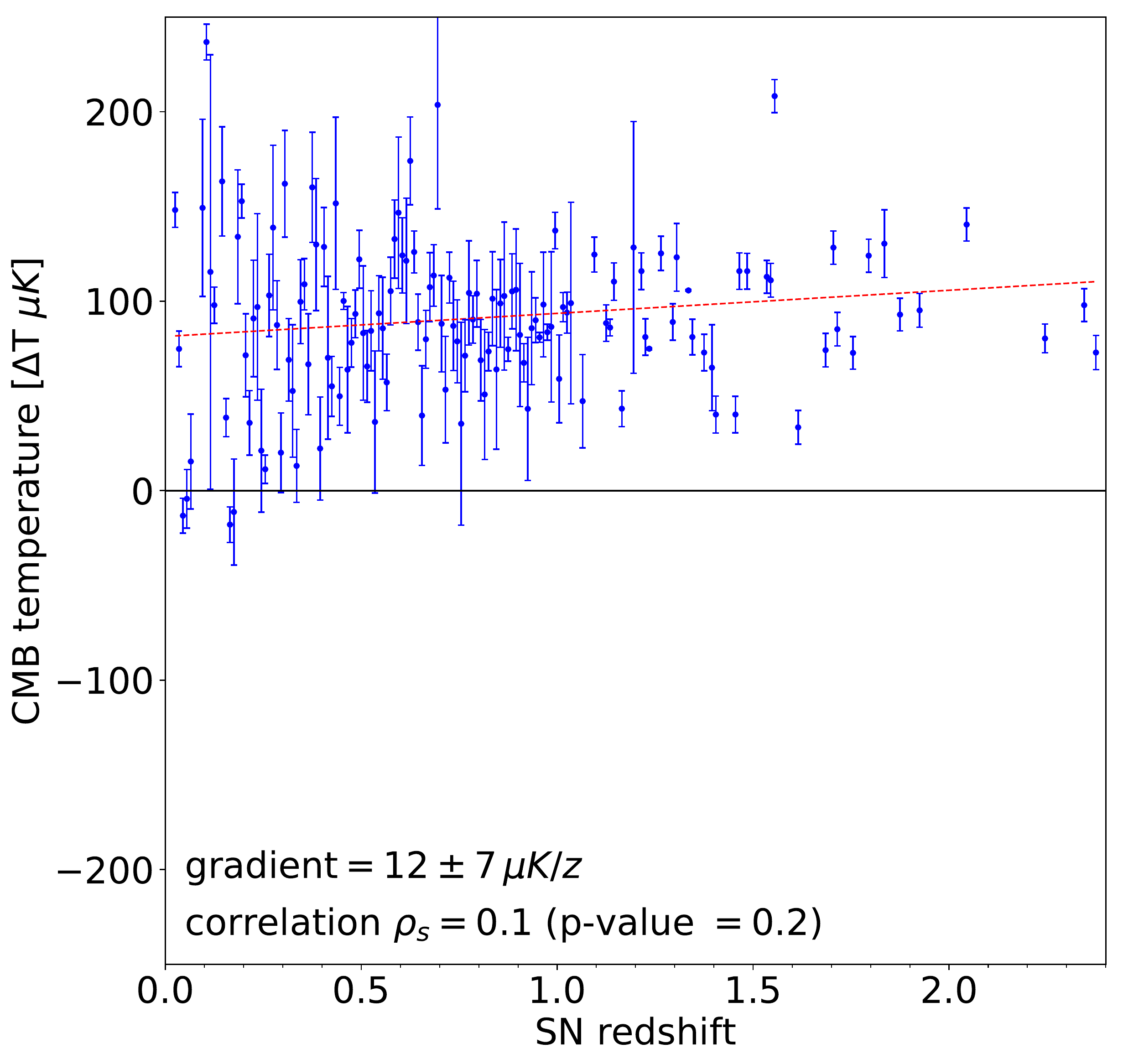} \label{subBinnedScatterFields1to7FieldsOSC}}
	\\
	\subfigure[Remainder sample (7204 SNe)]{\includegraphics[width=0.45\columnwidth]{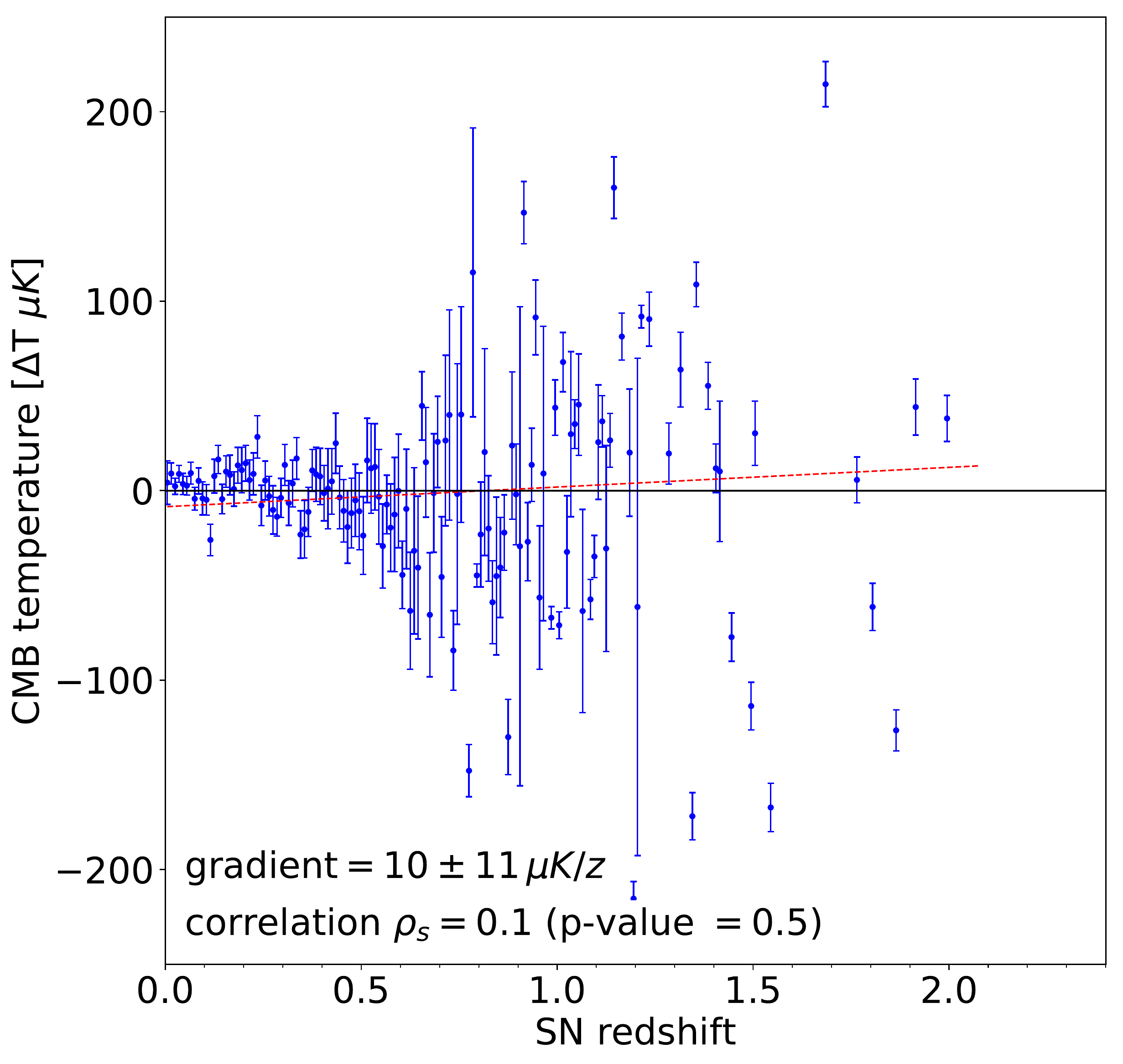} \label{subBinnedScatterFields1to7RemainderOSC}}
	\caption[As Fig.~\ref{figBinnedScatterFields1to7} but for OSC sub-samples]{Same as Fig.~\ref{figBinnedScatterFields1to7} but for the OSC sample.}
	\label{figBinnedScatterFields1to7OSC}
\end{figure}

The result of removing fields 1-7 from the OSC sample is illustrated in Fig.~\ref{figBinnedScatterFields1to7OSC}, which plots the weighted mean CMB temperature at SNe locations in redshift bins of $\Delta z = 0.01$. This plot is repeated for the whole sample (\ref{subBinnedScatterFields1to7AllOSC}), fields 1-7 only (\ref{subBinnedScatterFields1to7FieldsOSC}) and the remainder of the sample after fields 1-7 are removed (\ref{subBinnedScatterFields1to7RemainderOSC}).

\clearpage

The results for the OSC sample indicate that the correlation is caused by fields 1-7 and that SDSS Stripe 82 does not contribute significantly, which is consistent with those for the SAI sample.

The results of our final two OSC sample analyses, namely determining the temperature and redshift biases of fields 1-7 and quantifying the likelihood of the selection bias happening by chance, are entirely consistent with those for the SAI sample (sections~\ref{subsecTempRedshift} and~\ref{subsecLikelihood} respectively).

\subsection{\textit{Planck} CMB maps} \label{subsecMaps}

Both our analysis and that of \citet{Yershov2014} used maps produced by the \textit{Planck} SMICA component separation pipeline (\textit{Planck} 2015 SMICA R2.01 and \textit{Planck} 2013 SMICA R1.20 respectively). To check our results are consistent across all four of the \textit{Planck} component separation pipelines we repeated selected analyses from section~\ref{secBias} using the \textit{Planck} 2015 Commander, NILC, and SEVEM CMB maps. In all cases the pixel variance estimates were calculated from the corresponding HMHD maps as described in section~\ref{subsecMethod}.

We repeated the OLS linear regression gradient and Spearman's rank-order correlation coefficient analyses from section~\ref{subsecContribution}. For all four maps (Commander, NILC, SEVEM, and SMICA) the OLS gradient and correlation present in the whole sample are entirely absent in the remainder once fields 1-7 are removed. For SMICA the contribution of fields 1-7 to the correlation was illustrated in Fig.~\ref{figBinnedScatterFields1to7}. For Commander, NILC, and SEVEM see Appendix~\ref{appAltPlanckMaps} Figs.~\ref{figBinnedScatterFields1to7Commander}, \ref{figBinnedScatterFields1to7nilc}, and \ref{figBinnedScatterFields1to7sevem} respectively.

We repeated the mean CMB temperature analysis from section~\ref{subsubMeanTempRedshift}. For all four maps SNe in fields 1-7, and all HEALPix pixels within each sample, are biased to CMB temperatures hotter than the mean of the whole sample. In all cases fields 3, 6, and 7 are particularly extreme (Appendix~\ref{appAltPlanckMaps} Table~\ref{tabHistogramsAltPlanckMaps}).

Our results are entirely consistent across all four \textit{Planck} maps.

\section{Discussion and conclusions} \label{secConclusions}

We have shown that the apparent correlation of CMB temperature and SNe redshift reported by \citet{Yershov2012, Yershov2014} using OLS linear regression, Pearson's correlation coefficient, and an SAI SNe sample, is also evident using Spearman's rank-order correlation coefficient, Welch's t-test, and MWU test, and it is discernible in at least one other SNe sample (OSC).

Whilst our analysis supports the prima facie existence of the apparent correlation, the data indicate that it is actually a composite selection bias (high CMB $T$ $\times$ high SNe $z$) caused by the accidental alignment of seven deep survey fields (fields 1-7) with CMB hotspots. These fields include 3 from the Supernova Legacy Survey, 2 from the ESSENCE supernova survey, HDF-N and CDF-S. These comprise 9.2\% of the SAI sample and 8.6\% of the OSC sample. These deep fields by their very nature contain SNe at higher redshift than the remainder of the samples. We have shown that the SNe within fields 1-7 are also biased to hotter CMB temperature than the remainder of the samples. Our results are consistent across all four of the \textit{Planck} maps.

We have quantified the likelihood of fields 1-7 falling on CMB hotspots by chance and have found this to be at least 6.8\%, or approximately 1 in 15. We conclude that the correlation reported by \citet{Yershov2012, Yershov2014} is a composite selection bias caused by the chance alignment of certain deep survey fields with CMB hotspots. This bias (high CMB $T$ $\times$ high SNe $z$) is the combined result of both a selection bias (high-$z$ SNe in deep fields) and the chance alignment of those deep fields with CMB hotspots.

This selection bias results in heteroscedastic data, where the variance of CMB temperature at SNe locations is unequal across the range of redshifts. We have shown that high-redshift SNe tend to be in deep survey fields which, given the chance alignments, generally give hot \textit{Planck} pixel temperatures. Low-redshift SNe are more uniformly scattered across the sky and thus have much wider variance of hot and cold \textit{Planck} pixel temperatures. This heteroscedasticity was hidden by binning the data.

This chapter shows that deep survey fields have biased SNe cross-correlation with CMB temperature, but the implications could extend further. Deep fields could potentially bias any cross-correlation between astronomical objects (e.g., SNe, galaxies, GRBs, quasars) and the CMB. It is conceivable that deep fields could, by chance, also be aligned with distant large-scale structures, voids, cosmic bulk flows, or even regions of anisotropic cosmic expansion (should they exist).

% END

%% file: chapterLQGSample.tex
% START

\chapter{Data and methods to detect LQGs and analyse their morphology} \label{chaptLQGSample}

Chapters~\ref{chaptLQGSample}-\ref{chaptLQGOrientation} present our second project, investigating for the first time whether large quasar groups (LQGs) exhibit coherent alignment. In this chapter we introduce our sample of 71 LQGs, summarise the method used to detect them, explain calculation of their orientation, including uncertainties and weighting, and describe the parallel transport method used for coordinate invariance.

\section{Cosmography and cosmological parameters}

Several of the methods in this and subsequent chapters require relative distances or three-dimensional positions, for which we use comoving coordinates which remain the same with epoch. In this section we summarise our calculation of comoving positions, and we specify the cosmological parameters used throughout this work.

It is conventional to express the energy density of the Universe at time $t$ as dimensionless parameters $\Omega_{i}(t)$ ($i = r, m, k, \Lambda$ for radiation, matter, curvature, and dark energy respectively). For the concordance cosmological model, at the present epoch ($t=t_0$), $\Omega_{k,0} = 0$, $\Omega_{r,0} \simeq 0$, and $\Omega_{m,0} + \Omega_{\Lambda,0} = 1$. Further, we use $H_0 = 67.8 \mathrm{\ km\ s^{-1}Mpc^{-1}}$ and $\Omega_{m,0} = 0.308$ \citep{Planck2016b}. These parameters ($H_0$, $\Omega_{m,0}$, $\Omega_{k,0}$, $\Omega_{r,0}$) are used as the working cosmological model throughout this work. When reporting other work, we sometimes use the dimensionless equivalent to $H_0$, `little $h$' \citep{Croton2013}, where $h = H_0 / (100\mathrm{\ km\ s}^{-1} \mathrm{Mpc}^{-1})$.

\subsection{Calculation of comoving positions} \label{subComovingPositions}

The three-dimensional comoving position ($x$, $y$, $z$) of an object is calculated as

\begin{equation} \label{eqCartCoords}
\begin{aligned}
    x &= r\cos(\delta)\cos(\alpha)\ , \\
    y &= r\cos(\delta)\sin(\alpha)\ , \\
    z &= r\sin(\delta)\ ,
\end{aligned}
\end{equation}

\noindent where $\alpha$ and $\delta$ are right ascension and declination and $r$ is comoving radial distance. Here $z$ denotes the third Cartesian coordinate, not redshift.

To calculate comoving radial distance $r(z)$ of an object at redshift $z$ we follow \citet{Hogg1999}, who in turn follows \citet{Peebles1993}, and use

\begin{equation} \label{eqRadialDist}
    r(z) = \frac{c}{H_0}\int_{0}^{z}\frac{dz}{E(z)}\ ,
\end{equation}

\noindent where $H_0$ is the Hubble parameter at the present epoch ($t = t_0$) and $c$ is the speed of light. The function $E(z)$ is derived from the Friedmann expansion equation (Appendix~\ref{appComovingDistance}), and for the concordance cosmological model is

\begin{equation} \label{eqDimensionlessFriedmann}
    \frac{H}{H_0} = E(z) = \sqrt{1 + \Omega_{m,0}((1 + z)^3 -1)}\ ,
\end{equation}

\noindent where $\Omega_{m,0}$ is the matter density parameter of the Universe at $t_0$. $\Omega_{i}(t)$ is the energy density $\rho_i(t)$ as a fraction of the critical density $\rho_c(t) = 3 H^2(t) / 8 \pi G$, so $\Omega_m(t) = \rho_m(t) / \rho_c(t)$. See Appendix~\ref{appComovingDistance} for more details.

The Cartesian coordinates (Eq.~\ref{eqCartCoords}) are used to compute relative positions and distances, both to determine LQG orientation and to define groups of nearest neighbours. In both cases, the Hubble parameter is a scaling factor of the radial distance (Eq.~\ref{eqRadialDist}), and LQG orientation and membership of nearest neighbour groups do not depend on the chosen value of $H_0$. The same is not true of $\Omega_m$ (Eq.~\ref{eqDimensionlessFriedmann}), and results may depend on the value chosen. We calculate $r(z)$ using Python's\footnote{Astropy community project \citep{Astropy2013,Astropy2018}} \texttt{astropy.cosmology}, which follows \citet{Hogg1999}.

\section{Detecting large quasar groups} \label{secLQGMethods}

Our LQG sample is from the work of \citet{Clowes2012,Clowes2013}. The LQGs were detected using quasars from the Sloan Digital Sky Survey \citep[\gls{SDSS},][]{York2000}, specifically Quasar Redshift Survey Data Release 7 \citep[DR7QSO,][]{Schneider2010}. The DR7\gls{QSO} catalogue of 105,783 quasars covers a region of $\sim 9,380 \ \mathrm{deg}^2$, with its main contiguous area of $\sim 7,600 \ \mathrm{deg}^2$ in the north Galactic cap (\gls{NGC}).

\citet{Clowes2012,Clowes2013} restrict their quasar sample to low-redshift ($z \leq 2$) quasars with apparent magnitude $i \leq 19.1$ in order to achieve an approximately spatially uniform sample \citep{VandenBerk2005,Richards2006}. They further restrict their sample to a redshift range of $1.0 \leq z \leq 1.8$, within which the comoving number density of quasars as a function of redshift is sufficiently flat for clustering analysis.

\subsection{Large quasar group finder} \label{subsecIDLQGs}

\citet{Clowes2012,Clowes2013} detect LQG candidates using a three-dimensional single-linkage hierarchical clustering algorithm, also known as friends-of-friends (\gls{FoF}). This is equivalent to a three-dimensional minimal spanning tree (\gls{MST}). These type of methods are widely used to detect galaxy clusters, superclusters, voids, and filaments \citep[e.g.][]{Press1982,Einasto1997,Park2012,Pereyra2019}, as well as LQGs \citep[e.g.][]{Clowes2012,Clowes2013,Nadathur2013,Einasto2014,Park2015}. They make no assumptions about cluster morphology.

Using the terminology of \citet{Barrow1985}, MST treats the dataset as a graph made up of vertices (nodes, in this case quasars) which are connected by edges (straight lines). This is a `tree' when it has no closed paths (sequence of edges) and a `spanning' tree when it contains all the vertices. For any graph, there are multiple possible spanning trees; the `minimal' spanning tree is that of minimal length (sum of edge lengths). To identify clusters within the MST it may be separated, where edges exceeding a certain length are removed, leaving groups of objects with mutual separations less than this `linkage length'.

The choice of linkage length is crucial - too long and clusters merge to fill the entire volume, too short they break up into pairs and triplets \citep{Graham1995}. \citet{Pilipenko2007} categorises the criteria for making this choice as physical or formal. With a priori knowledge of physical parameters (e.g.\ size, membership, density) of the clusters, it is possible to choose the scale that maximises the fraction of clusters with those parameters. The criterion used by \citet{Graham1995} is an example of a physical approach; they choose the scale that maximises the number of clusters of a minimum membership. An example formal approach would be to choose a scale based on the mean nearest-neighbour separation.

\citet{Clowes2012,Clowes2013} use this latter approach. Their quasar sample has a mean nearest-neighbour separation of $\sim 74 \ \mathrm{Mpc}$. They also account for uncertainties in their edge lengths (i.e.\ comoving distances) due to redshift errors and peculiar velocities \citep[for estimates, see][]{Clowes2012} and choose a linkage length of $100 \ \mathrm{Mpc}$.

\citet{Clowes2012,Clowes2013} estimate the overdensity and statistical significance of their LQGs using a convex hull of member spheres (\gls{CHMS}) method. This is described in detail in \citet{Clowes2012}, but briefly, for each LQG they calculate the volume of a convex hull of spheres of radius half the mean edge length at each vertex (member quasar location). The CHMS volume of an LQG of $m$ members is then compared with the distribution of CHMS volumes of clusters of $m$ points in Monte Carlo simulations of the same size and density as their control area, in order to estimate overdensity and statistical significance. We note this method of estimating statistical significance is not universally accepted \citep{Nadathur2013,Pilipenko2013,Park2015}.

The Huge-LQG and Clowes-Campusano LQG, amongst others, have been independently detected using different FoF algorithms \citep{Nadathur2013,Einasto2014,Park2015}. Indeed, our LQG sample has many objects in common with the publicly available catalogue\footnote{\url{https://vizier.u-strasbg.fr/viz-bin/VizieR?-source=J/A+A/568/A46}} of \citet{Einasto2014}, also detected from DR7QSO. Note that their catalogue does not include estimates of overdensity and statistical significance, which makes quality-assured sample selection non-trivial.

\subsection{Large quasar group sample} \label{subsecLQGSample}

Our LQG sample is taken\footnote{Clowes (2016), private communication} from the work of \citet{Clowes2012,Clowes2013}. In order to confidently determine the geometric properties of the LQGs (e.g.\ orientation and morphology) we restrict their original sample of 398 LQGs to those with membership $m \geq 20$, giving a sample of 89 LQGs.

We select the most convincing of these using the significance estimates of \citet{Clowes2012,Clowes2013}. Whilst the absolute values of these may be contentious \citep{Nadathur2013,Pilipenko2013}, they provide a legitimate relative order for ranking based on confidence. As a compromise between sample size and confidence, we restrict our sample to LQGs with `significance'\footnote{We attribute no significance to the value of 2.8; it is used as a relative threshold only} $\geq 2.8 \sigma$, yielding 72 LQGs.

We finally exclude one LQG in the south Galactic cap, giving our final sample of 71 LQGs, of varied and generally irregular morphologies, shown in Fig.~\ref{figAllLQGsPt1}.

\clearpage

\begin{figure} % plot_all_LQGs
    \centering
    \includegraphics[width=0.8\columnwidth]{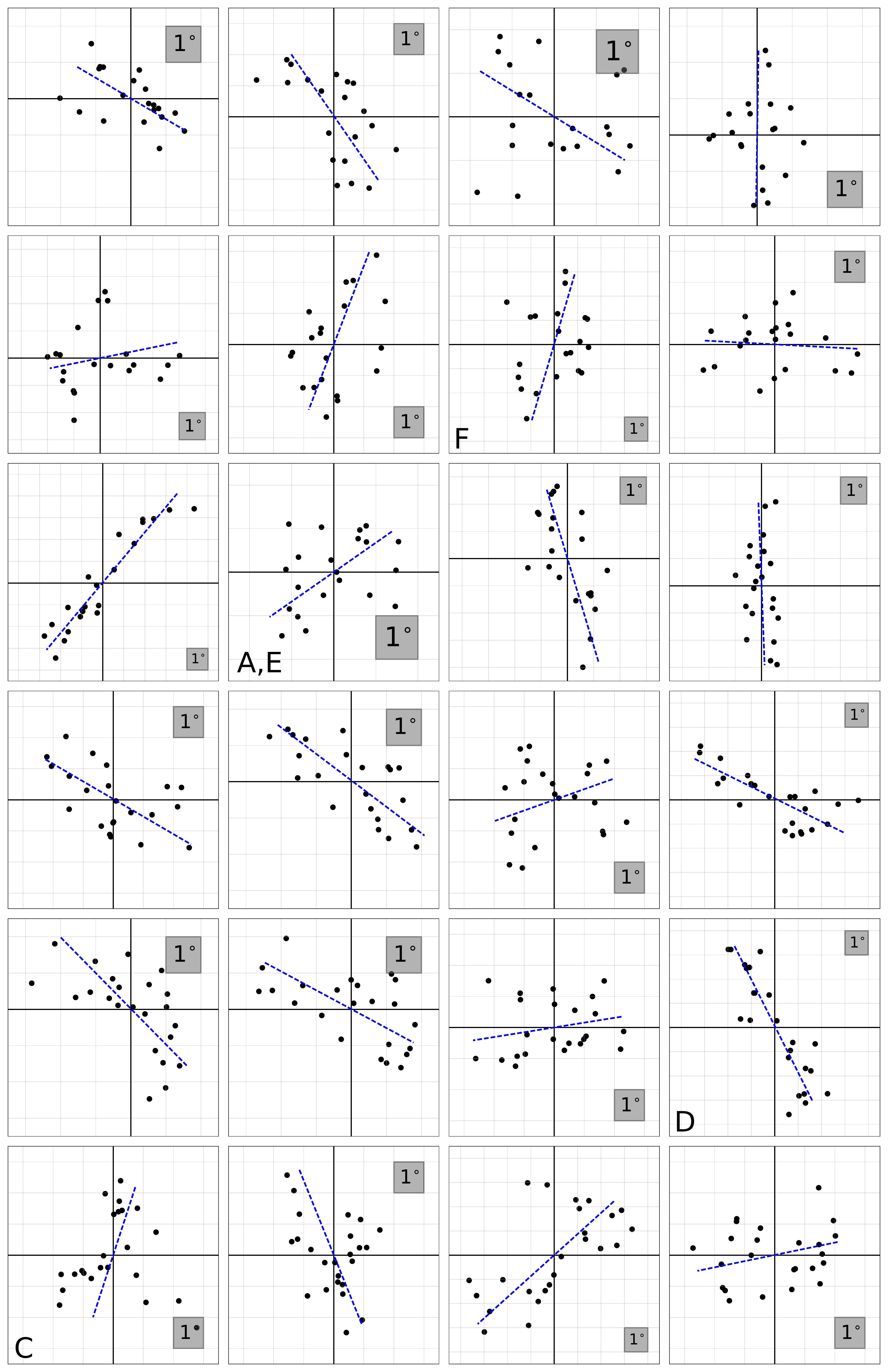}
    \caption[Tangent plane projection of the 71 LQGs in our sample]{(a) Tangent plane projection of 1-24 (across then down) of the 71 LQGs in our sample, in Cartesian coordinates. Member quasars shown as black dots, orthogonal distance regression fit shown as dashed blue line. Solid black lines indicate $x=0$ and $y=0$, and grey square illustrates scale ($1^{\circ}$). LQGs labelled A, C (D, E, F) are discussed later in Fig.~\ref{figPA2D3D} (Fig.~\ref{figBootstraps}).}
    \label{figAllLQGsPt1}
\end{figure}

\begin{figure}
    \ContinuedFloat
    \centering
    \includegraphics[width=0.8\columnwidth]{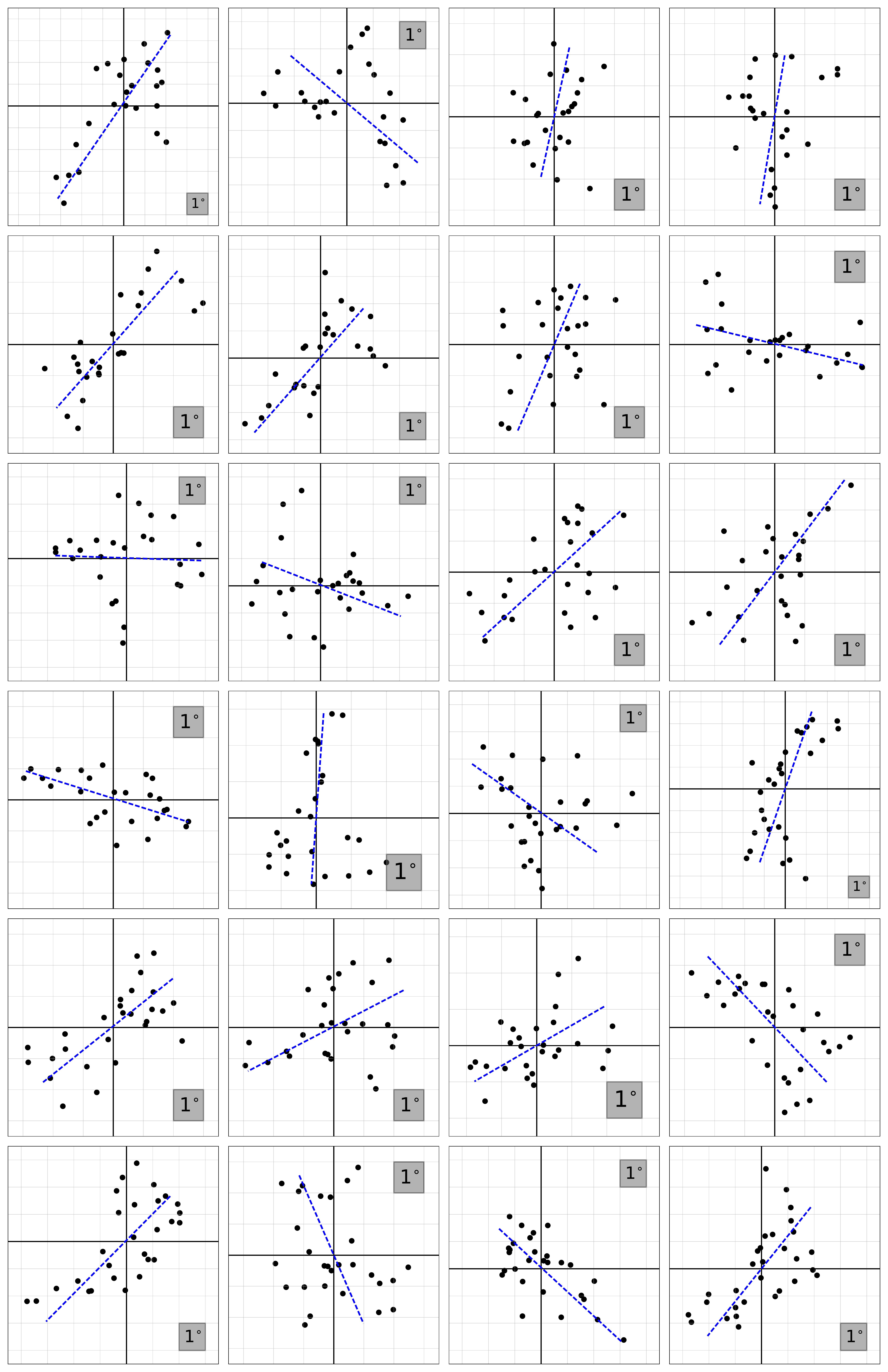}
    \caption[]{(b) Tangent plane projection of 25-48 (across then down) of the 71 LQGs in our sample, in Cartesian coordinates. Member quasars shown as black dots, orthogonal distance regression fit shown as dashed blue line. Solid black lines indicate $x=0$ and $y=0$, and grey square illustrates scale ($1^{\circ}$).}
    \label{figAllLQGsPt2}
\end{figure}

\begin{figure}
    \ContinuedFloat
    \centering
    \includegraphics[width=0.8\columnwidth]{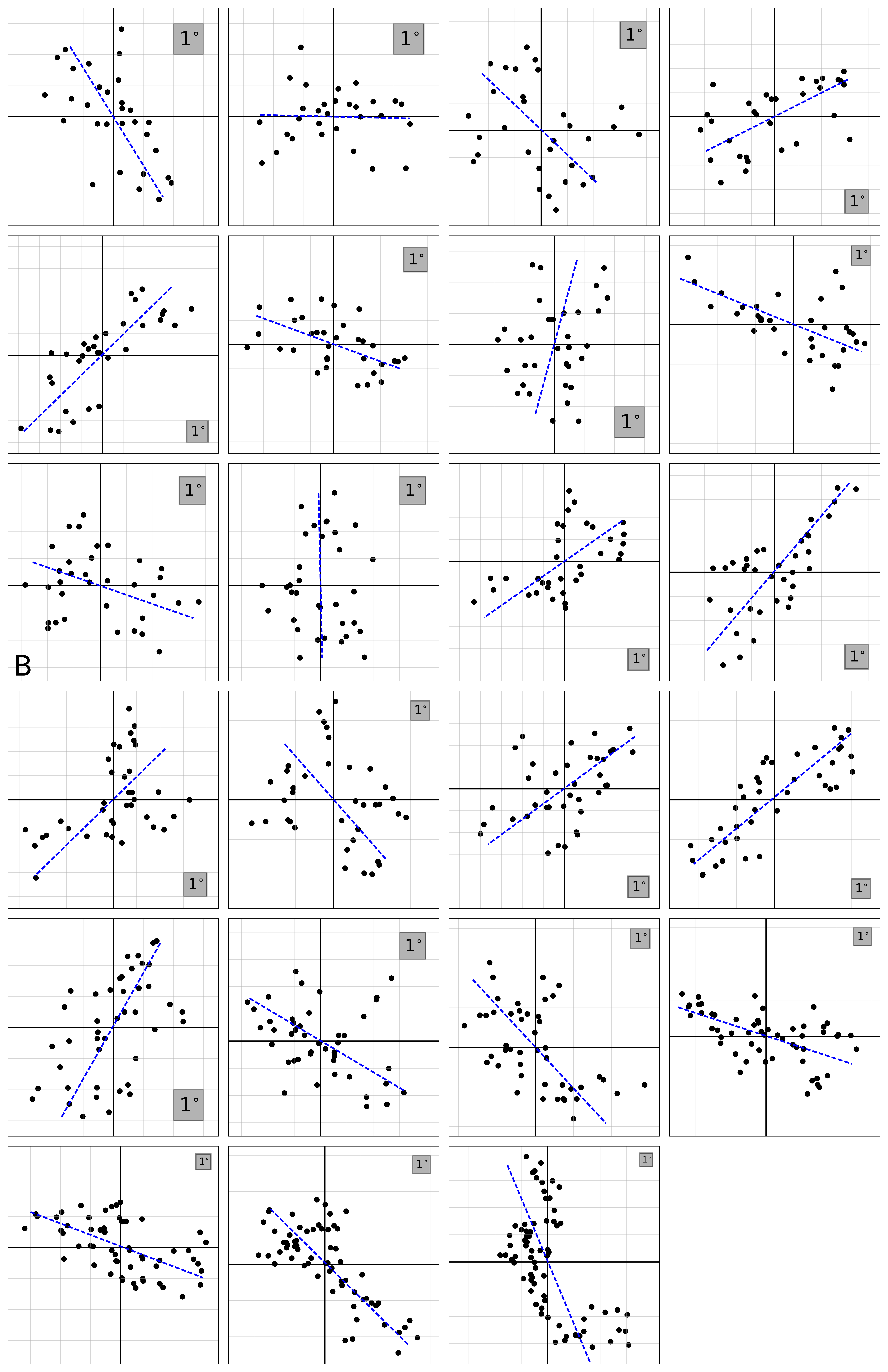}
    \caption[]{(c) Tangent plane projection of 49-71 (across then down) of the 71 LQGs in our sample, in Cartesian coordinates. Member quasars shown as black dots, orthogonal distance regression fit shown as dashed blue line. Solid black lines indicate $x=0$ and $y=0$, and grey square illustrates scale ($1^{\circ}$). LQG labelled B is discussed later (Fig.~\ref{figPA2D3D}).}
    \label{figAllLQGsPt3}
\end{figure}

\clearpage

For data about each LQG and their individual morphologies see Tables~\ref{tabLQGDataPart} (example LQGs) and \ref{tabLQGDataFull} (full sample). Methods used to determine position angles and their uncertainties, to calculate two-dimensional orthogonal distance regression goodness-of-fit weights, and to fit three-dimensional ellipsoids are described later in this chapter. The three-dimensional morphology of LQGs (e.g.\ ellipsoid axis ratios) is discussed in chapter~\ref{chaptLQGOrientation}.

\section{Determining large quasar group orientation} \label{secLQGOrientations}

The position angle (\gls{PA}) of a large quasar group can be calculated in either two or three dimensions. In two dimensions we treat the quasars as points on the celestial sphere, whereas in three dimensions we take into account their comoving radial distances. For either approach we measure PAs with respect to celestial north\footnote{$0^{\circ}$, $90^{\circ}$, and $180^{\circ}$ correspond axes pointing north, east, and south respectively}. This angle is clearly not coordinate invariant; it depends on the location of the celestial north pole. See section \ref{secParallelTransport} for details of how we ensure our analysis methods are co-ordinate invariant using parallel transport.

\begin{figure} % LQG_PAs_2D_3D_v2 for LQG:23:209.53:34.29:1.65
    \centering
    \includegraphics[width=0.4\columnwidth]{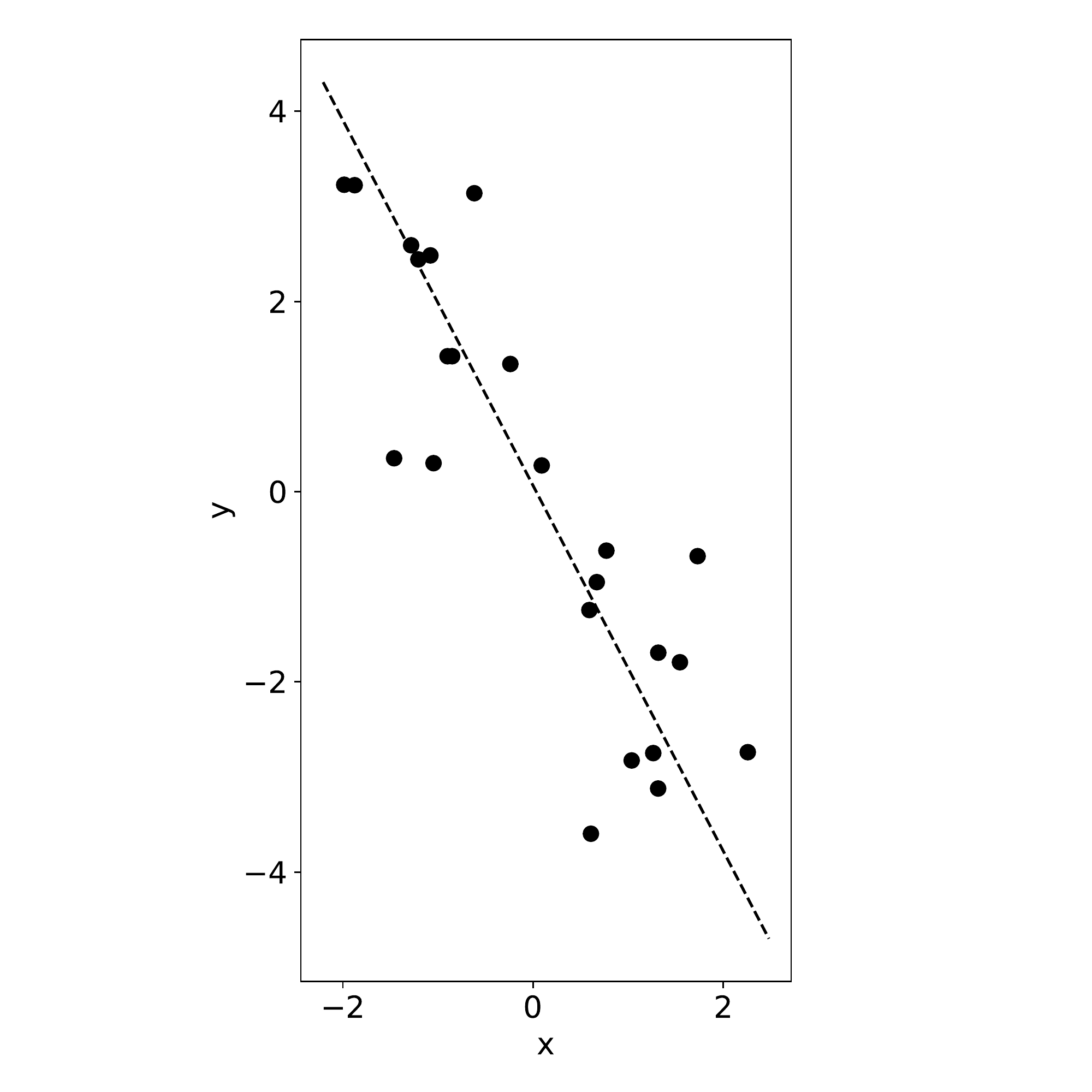}
    \caption[Example position angle calculated by 2D approach]{Tangent plane projection of an LQG, with coordinates ($\alpha$, $\delta$) projected to ($x$, $y$). Member quasars are shown as black dots and orthogonal distance regression fit shown as dashed line. LQG is also shown in Figs.~\ref{figAllLQGsPt1} and \ref{figBootstraps}, labelled D. Using the 2D PA approach, PA = $152.5^{\circ}$.}
    \label{fig2DPA}
\end{figure}

For the 2D approach quasar positions (in right ascension and declination) are projected onto the tangent plane as Cartesian ($x$, $y$) points. This plane meets the celestial sphere at the centre of gravity of the LQG, calculated assuming quasars are point-like unit masses. To determine LQG orientation we use linear regression of these projected points. Ordinary least-squares regression is inappropriate; we cannot categorise either $x$ or $y$ as `independent' or `dependent' variables, and must treat them symmetrically. We compute the orthogonal distance regression (\gls{ODR}) of the points (see Fig.~\ref{fig2DPA} for an example). This minimises the sum of the squares of the orthogonal residuals between the points and the line \citep[for a discussion of OLS, ODR, and other regression methods, see][]{Isobe1990}.

\clearpage

\begin{figure}
    \centering
    \includegraphics[width=0.6\columnwidth]{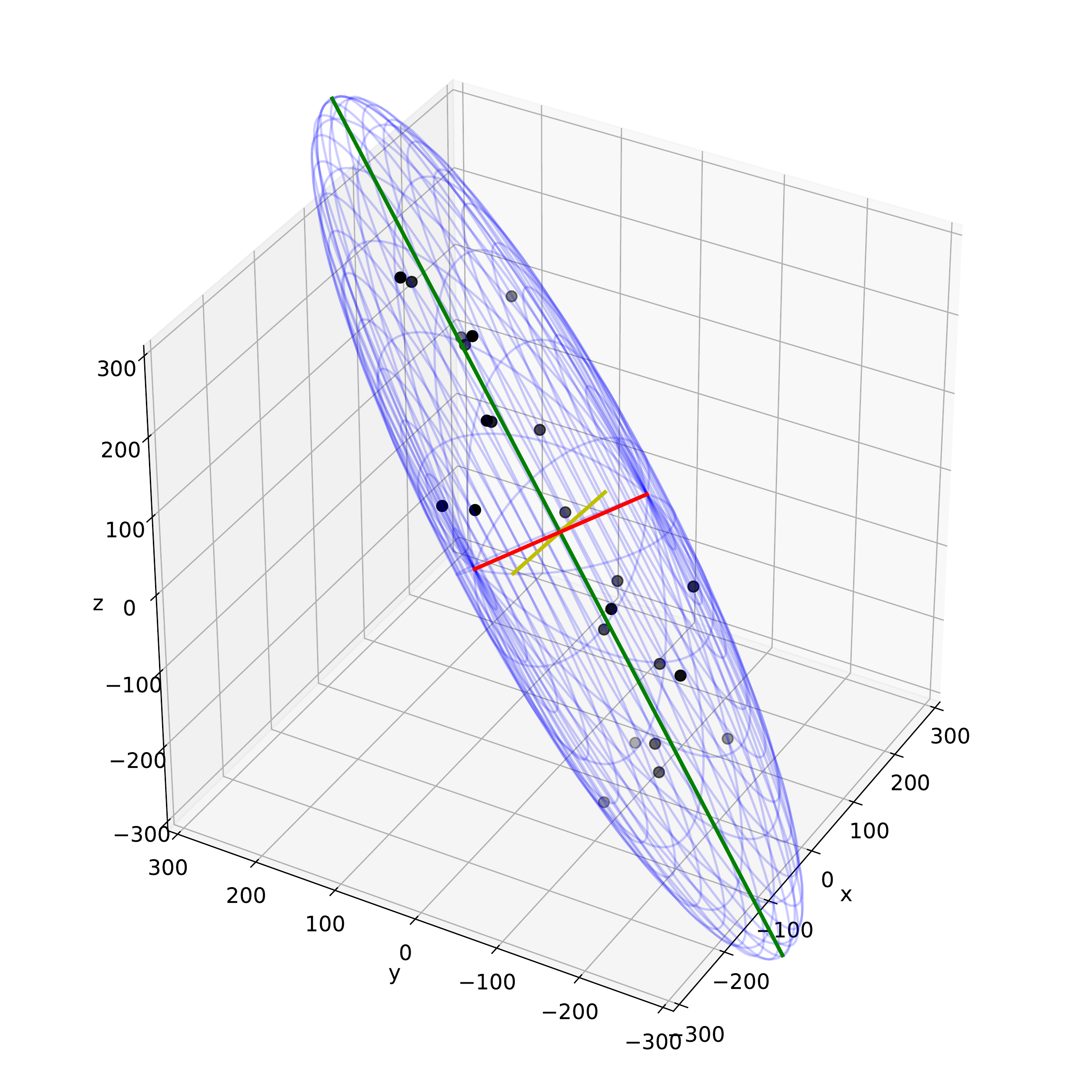}
    \caption[Example position angle calculated by 3D approach]{Same LQG as Fig.~\ref{fig2DPA}, but shown in 3D comoving coordinates orientated with line-of-sight orthogonal to the page. Axes of enclosing ellipsoid (green = \textit{a}-axis, yellow = \textit{b}-axis, red = \textit{c}-axis) constructed from eigenvectors and eigenvalues. Using the 3D PA approach, 2D projected PA = $152.3^{\circ}$.}
    \label{fig3DPA}
\end{figure}

\clearpage

For the 3D approach the covariance matrix of the quasar comoving positions is decomposed into its eigenvectors and eigenvalues. Again, quasars are assumed to be point-like unit masses. The axes of a confidence ellipsoid (e.g.\ Fig.~\ref{fig3DPA}) are constructed from the eigenvectors and eigenvalues; each axis is in the direction of its eigenvector and its length $\ell$ is a function of its eigenvalue $\lambda$, specifically $\ell \propto \sqrt{\lambda}$. The first principal component (ellipsoid major axis) is given by the eigenvector with the largest eigenvalue. Finally, this is projected onto the plane orthogonal to the line-of-sight to define the (2D projected) PA.

\subsection{Axial position angle data} \label{subsecAxial}

We measure large quasar group orientation as the position angle from celestial north. It is important to recognise that the PA data are axial [$0^{\circ}$, $180^{\circ}$), more specifically 2-axial; $0^{\circ}$ and $180^{\circ}$ are equivalent. This can be visualised by considering PAs as axes on the celestial sphere. When analysing alignment, the orientation of the axis is important, but its direction (i.e.\ which end is the `head' and which is the `tail') is arbitrary and has no physical meaning. Axial data are synonymous with undirected data, unlike vectors which are directed.

Statistical analysis of directed data typically uses vector algebra (known as Fisher statistics). However, non-directed data cannot be treated as vectors. \citet{Fisher1993} recommends a statistically valid solution for axial (2-axial) or, generally, $p$-axial data. First transform the angles to vector (circular) data as

\begin{equation} \label{eqpAxialCirc}
    \Theta\ [0^{\circ},\ 360^{\circ}) =
    \begin{cases}
        2 \times \theta \qquad \mathrm{for\ axial\ data} & [0^{\circ},\ 180^{\circ})\ ,\\
        p \times \theta \qquad \mathrm{for\ }p\mathrm{\text{-}axial\ data} & [0^{\circ},\ 360^{\circ}/p)\ ,
    \end{cases}
\end{equation}

\noindent then analyse the data as required and back-transform the results. The final step, back-transformation, is generally only required to find direction \citep{Fisher1993}. In the case of axial data, back-transformation is simply halving any resultant angles, e.g.\ to determine the direction of a mean resultant vector.

\citet{Hutsemekers2014} and \citet{Pelgrims2016thesis} test for alignment and, simultaneously, for orthogonality (which they describe as `anti-alignment'), by converting from 2-axial $\theta$ [$0^{\circ}$, $180^{\circ}$) to 4-axial $\theta_{4ax}$ [$0^{\circ}$, $90^{\circ}$) data using

\begin{equation} \label{eq4Axial}
   \theta_{4ax}\ [0^{\circ},\ 90^{\circ}) = \mathrm{mod}(\theta, 90^{\circ})\ .
\end{equation}

\noindent For vector algebra, this is transformed to circular data  $\Theta$ using Eq.~\ref{eqpAxialCirc}, specifically

\begin{equation} \label{eq4AxialCirc}
   \Theta_{4ax}\ [0^{\circ},\ 360^{\circ}) = 4 \times \theta_{4ax}\ ,
\end{equation}

\noindent where back-transformation, if required, would be to quarter any resultant angles. Conversion to 4-axial data is a statistically legitimate method, and has been used in the literature \citep[e.g.][]{Hutsemekers2014}. However, we urge caution interpreting any results manifest only when analysing 4-axial data, unless this conversion is physically well motivated (e.g.\ orthogonality is expected).

Throughout this work we specify which construct of PA data we use, i.e.\ raw 2-axial ($\theta$), circular 2-axial ($\Theta_{2ax} = 2\theta$, Eq.~\ref{eqpAxialCirc}), 4-axial ($\theta_{4ax}$, Eq.~\ref{eq4Axial}), or circular 4-axial ($\Theta_{4ax} = 4\theta_{4ax}$, Eq.~\ref{eq4AxialCirc}).

\subsection{Position angle uncertainties} \label{subsecPAUncertainties}

To estimate the measurement uncertainties in the LQG position angles we use bootstrap re-sampling with replacement \citep{Efron1979}. For an LQG with $m$ members, the dataset consists of $m$ observed quasar positions (right ascension, declination, and redshift). We construct $n$ bootstrap LQGs, each with the same number of $m$ members which are drawn at random from the original dataset. Each draw is made from the entire dataset, and each member is replaced in the dataset before the next draw. Thus, each bootstrap LQG is likely to miss some members and have duplicates (or triplicates or more) of others.

These boostraps are used to estimate the uncertainty on parameters (e.g.\ PAs) derived from the dataset, without any assumption about the underlying population \citep{Feigelson2012}. The position angle of each bootstrap LQG is calculated through the same 2D and 3D approaches used for the observed LQGs.

For a sample of $n$ bootstrap LQGs we can determine the mean PA and its associated uncertainty. Using the linear mean $\bar{\theta} = (\sum_{i = 1}^{n}\theta_i) / n$ is inappropriate for axial data [$0^{\circ}$, $180^{\circ}$), where $0^{\circ}$ and $180^{\circ}$ are equivalent. For example, a sample of PAs centred on $0^{\circ}$, with around half in the range $0^{\circ} \lesssim \theta \lesssim 10^{\circ}$ and half in the range $170^{\circ} \lesssim \theta \lesssim 180^{\circ}$, has a linear mean $\bar{\theta} \sim 90^{\circ}$ rather than the correct answer $\bar{\theta} \sim 0^{\circ}$ (or, equivalently, $\bar{\theta} \sim 180^{\circ}$).

Therefore, instead of linear mean, we calculate the `circular' mean \citep{Fisher1993} of axial PAs. First, following \citet{Mardia2000}, let

\begin{equation} \label{eqCosSin}
    \bar{C} = \frac{1}{n} \sum_{i = 1}^{n} \cos 2\theta_i\ , \ \ \bar{S} = \frac{1}{n} \sum_{i = 1}^{n} \sin 2\theta_i\ ,
\end{equation}

\noindent where $\theta_i$ is the PA of the $i^{\mathrm{th}}$ bootstrap LQG and $n$ is the total number of boostraps for this LQG. The factor of two accounts for the axial (rather than circular) nature of the data. The mean direction $\bar{\theta}$ is given by

\begin{equation} \label{eqCircMean}
    \bar{\theta} =
    \begin{cases}
        \dfrac{1}{2} \arctan \left( \dfrac{\bar{S}}{\bar{C}} \right) & \mathrm{if}\ \bar{C} \geq 0\ ,\\
        \dfrac{1}{2} \arctan \left( \dfrac{\bar{S}}{\bar{C}} \right) + \pi & \mathrm{if}\ \bar{C} < 0\ ,
    \end{cases}
\end{equation}

\noindent where the factor $\frac{1}{2}$ converts from vector algebra back to axial data (i.e.\ `back-transformation').

\begin{figure} % LQG_PA_bootstrap
    \centering
    \subfigure[Example 1 - narrow LQG]{\includegraphics[width=0.45\columnwidth]{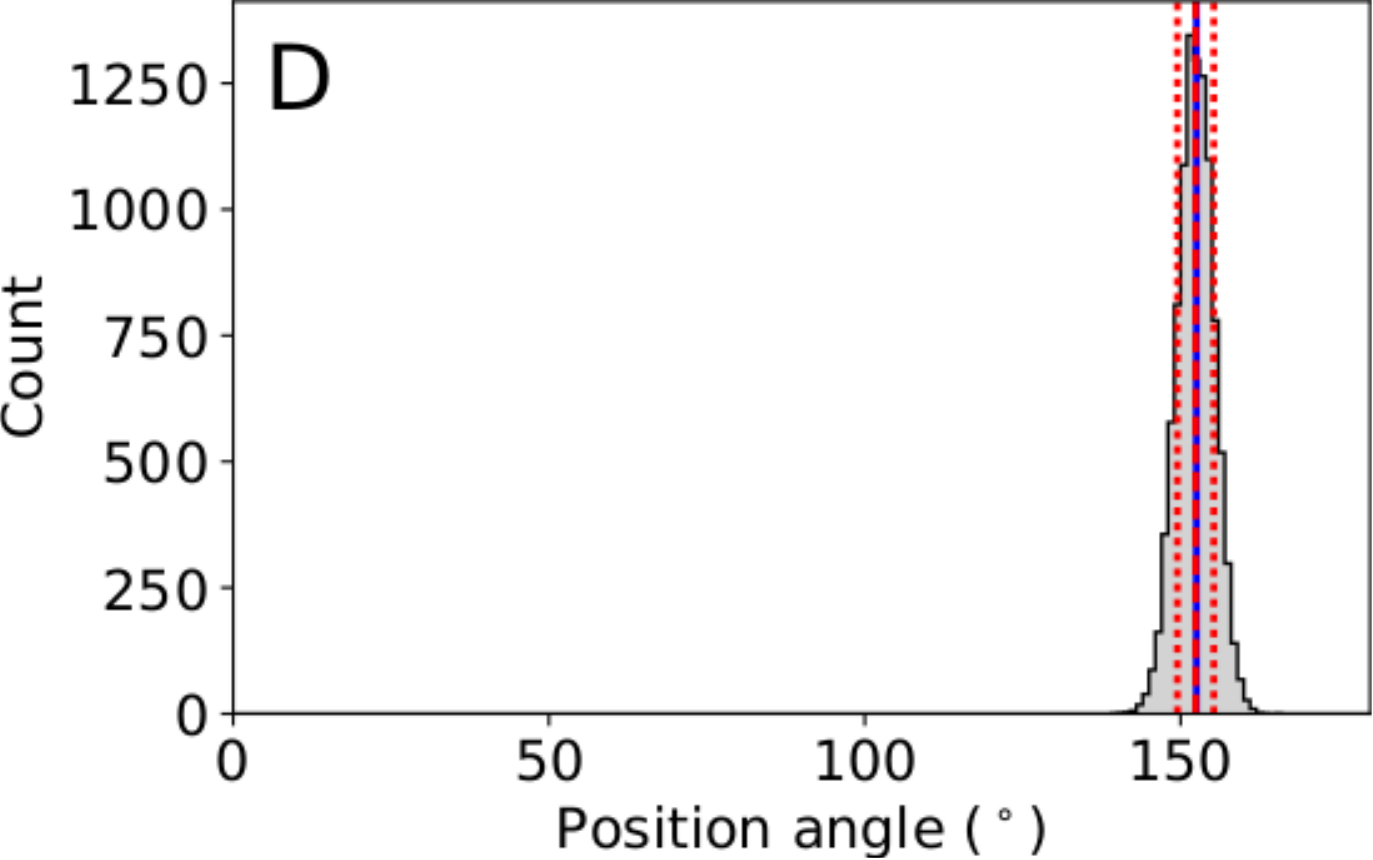} \label{subBSNarrow}}\\ % LQG:23:209.53:34.29:1.65, step = 1 deg
    \subfigure[Example 2 - broad LQG]{\includegraphics[width=0.45\columnwidth]{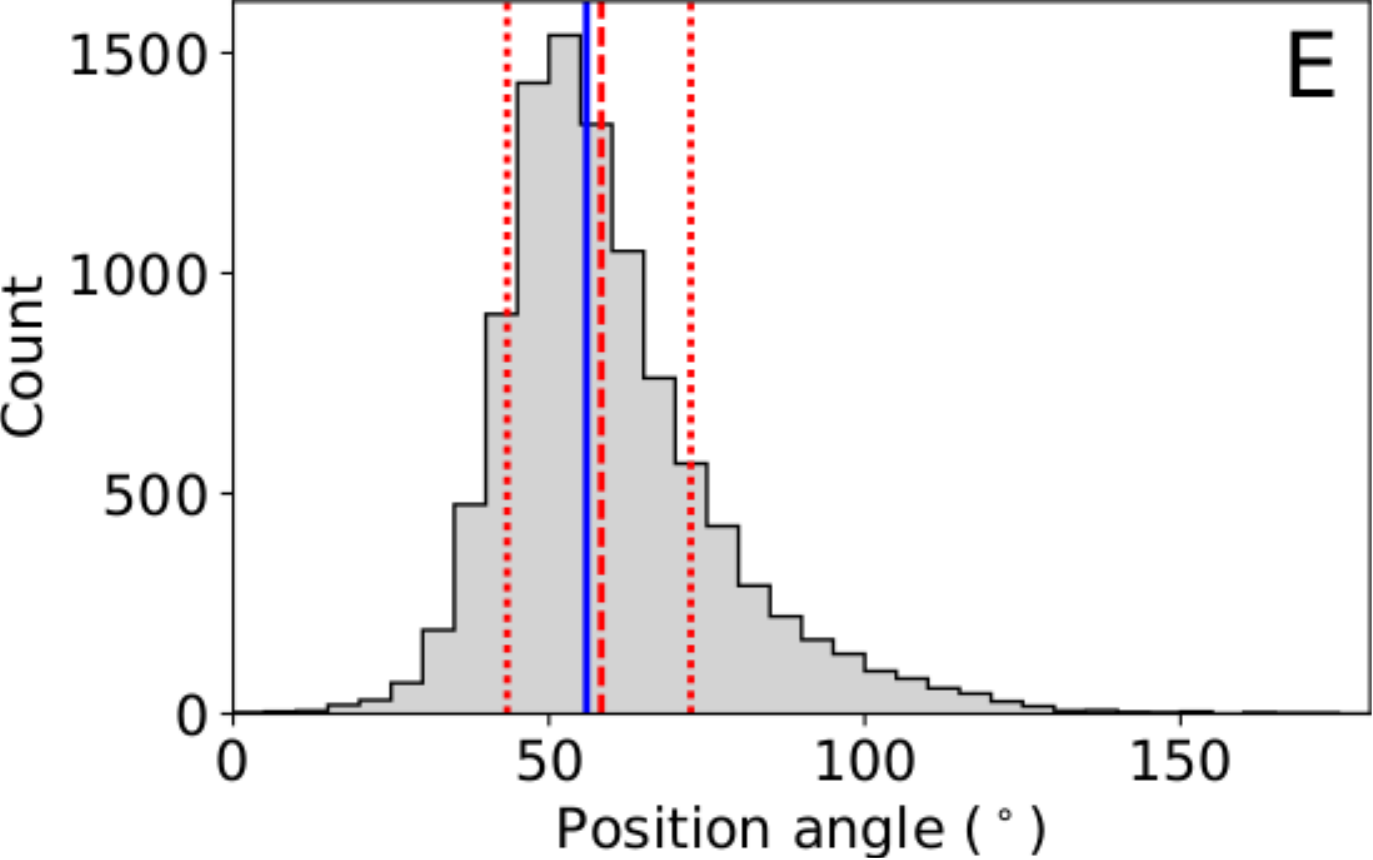} \label{subBSWide}}\\ % LQG:21:209.13:3.21:1.56, step = 5 deg
    \subfigure[Example 3 - intermediate LQG]{\includegraphics[width=0.45\columnwidth]{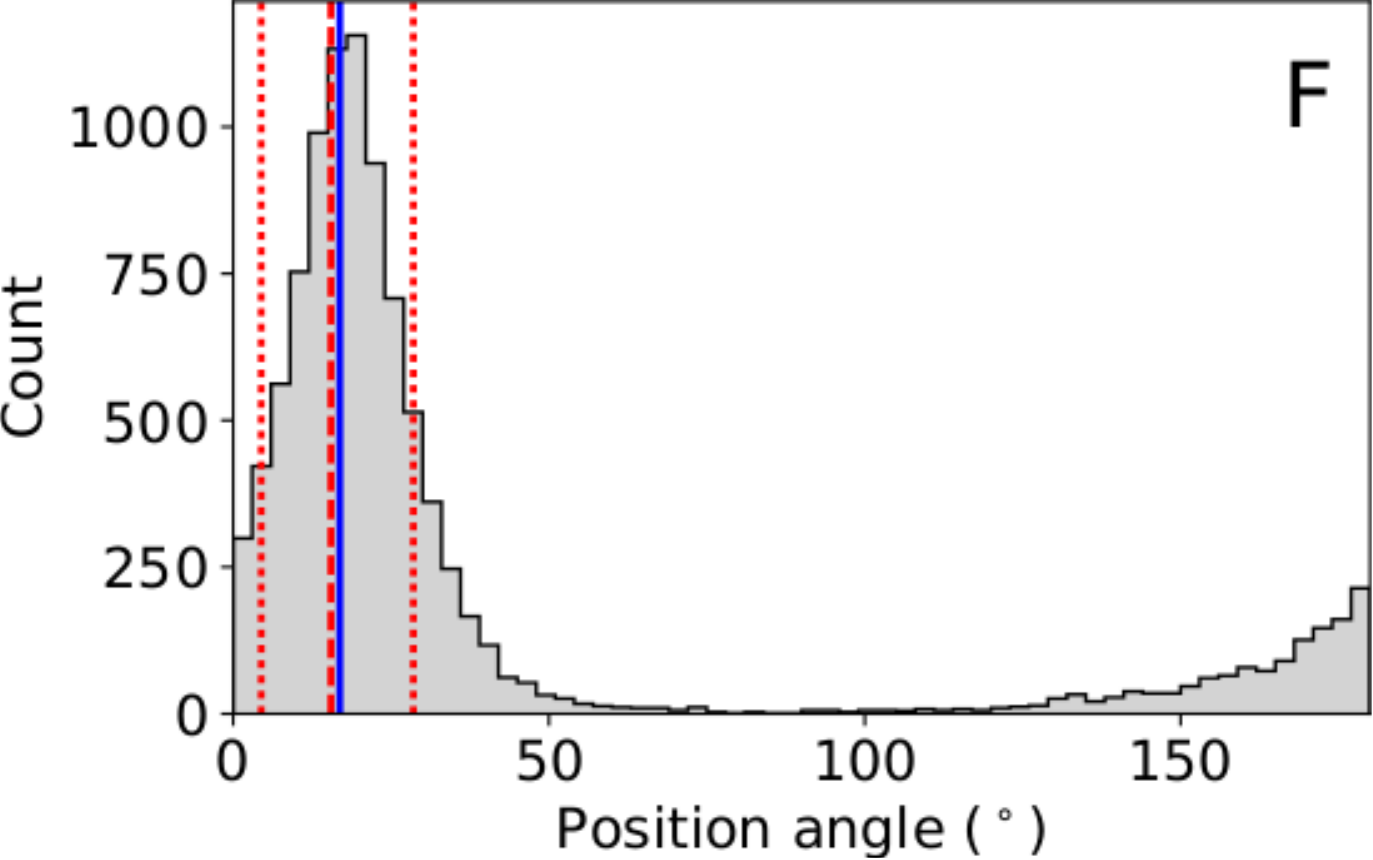} \label{subBSWrap}} % LQG:21:133.48:41.17:1.4, step = 3 deg
    \caption[Position angle distributions from bootstraps of three example LQGs]{LQG position angles from 10,000 bootstraps for three example LQGs, labelled D, E, F in Fig.~\ref{figAllLQGsPt1}, and visually classified as \protect\subref{subBSNarrow} narrow, \protect\subref{subBSWide} broad, and \protect\subref{subBSWrap} intermediate morphology. All PA calculations use the 2D approach with no parallel transport. Dashed orange line shows the circular mean of the bootstraps $\bar{\theta}$, dotted orange lines show the 68\% confidence interval, solid blue line shows the observed PA.}
    \label{figBootstraps}
\end{figure}

For each LQG, to estimate the uncertainty in the mean $\bar{\theta}$, we calculate its confidence interval following \citet{Pelgrims2016thesis}, who in turn follows \cite{Fisher1993}, and for each individual bootstrap $i$ out of a total of $n$ we define the residual as

\begin{equation} \label{eqCircGammas}
    \gamma_i = \frac{1}{2} \arctan \left( \frac{\sin(2(\theta_i - \bar{\theta}))}{\cos(2(\theta_i - \bar{\theta}))}\right)\ ,
\end{equation}

\noindent where $\theta_i$ is the PA of the $i^{\mathrm{th}}$ bootstrap LQG. For $i = 1,...,n$ we sort the $\gamma_i$ in ascending order to give an ordered list $\gamma_{(1)} \leq ... \leq \gamma_{(n)}$. To determine the confidence interval at the $100(1 - \alpha)\%$ level\footnote{Confidence level $= 1 - \alpha$, where $\alpha$ is the significance level} we find the $\gamma_i$ list elements at lower index $l$ which is the integer part of $(n\alpha + 1) / 2$ and upper index $u = n - l$. The confidence interval for $\bar{\theta}$ is then $[\bar{\theta} + \gamma_{(l + 1)}, \bar{\theta} + \gamma_{(u)}]$. We calculate the confidence interval at the 68\% level ($\alpha = 0.32$) and define the half-width of the confidence interval (HWCI) as $\gamma_h = (\gamma_{(u)} - \gamma_{(l + 1)}) / 2$.

\begin{figure} % LQG_PA_bootstrap
    \centering
    \includegraphics[width=0.5\columnwidth]{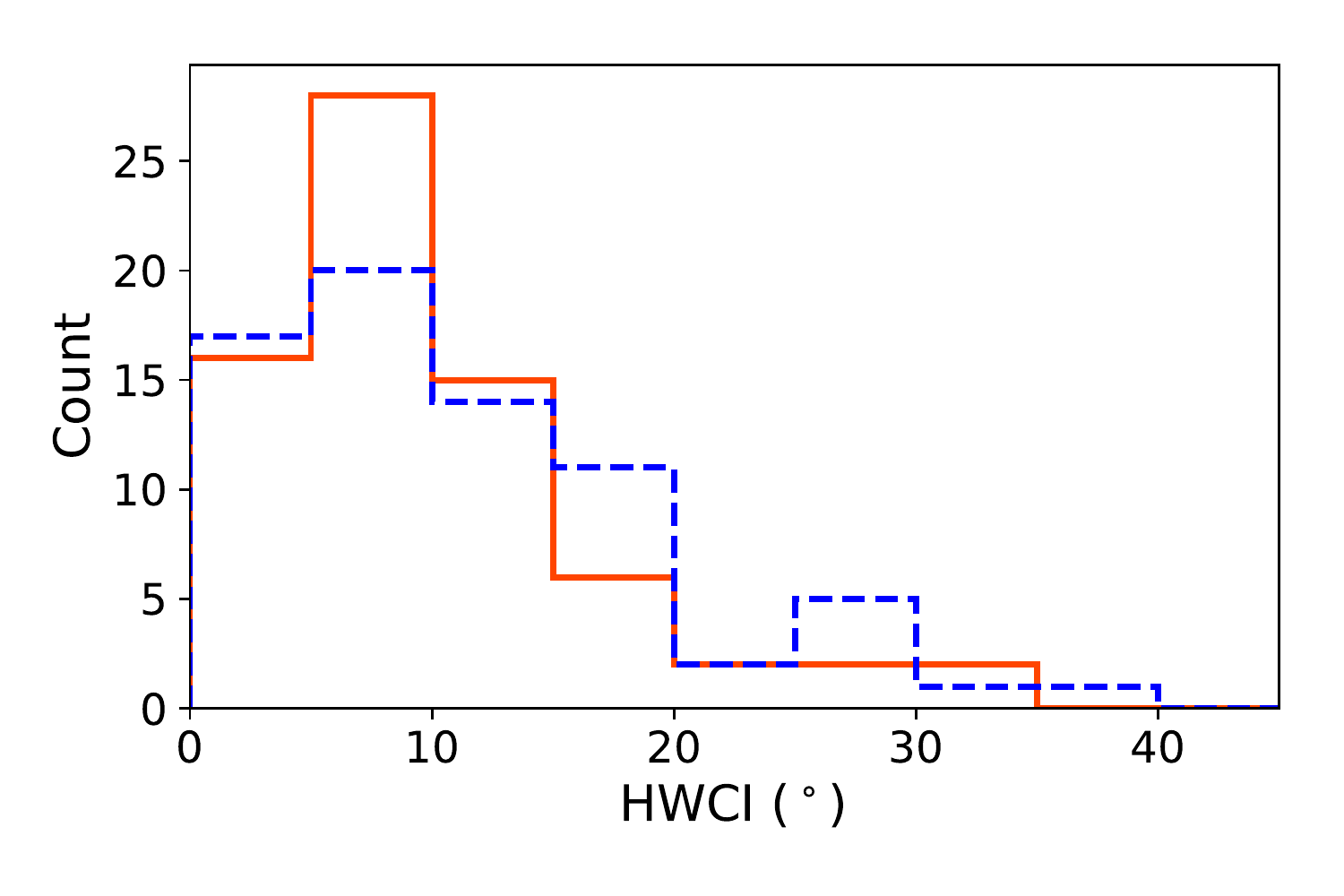}
    \caption[Uncertainties in position angles calculated by 2D and 3D approaches]{Half-width confidence interval (HWCI) calculated for the 71 large quasar group position angles (PAs) using bootstrap re-sampling. PAs calculated by the 2D approach (solid orange) have smaller uncertainties than those calculated by the 3D approach (dashed blue).}
    \label{figHWCI}
\end{figure}

For each of our sample of 71 LQGs, we create $n = 10,000$ bootstraps, calculate their PAs $\theta_i$ using the 2D or 3D approach, and then determine the mean $\bar{\theta}$ and confidence interval. Fig.~\ref{figBootstraps} shows the distribution of bootstrap PAs $\theta_i$ calculated using the 2D approach for three example LQGs. The circular mean of the bootstraps $\bar{\theta}$ (dashed orange line) generally agrees well with the observed PA (solid blue line).

As expected, the 68\% confidence interval (Fig.~\ref{figBootstraps}, dotted orange lines) is smaller for (a) a `narrow' LQG than (b) a `broad' sheet-like LQG. Example (c) illustrates why the linear mean is inappropriate, since the distribution of the bootstrap PAs may `wrap' from $180^{\circ}$ back to $0^{\circ}$.

%As expected, the 68\% confidence interval (dotted orange line) is smaller for `narrow' LQGs (Fig.~\ref{subBSNarrow}; the same LQG as shown in Figs.~\ref{fig2DPA} and \ref{fig3DPA}) than `broad' sheet-like LQGs (Fig.~\ref{subBSWide}). LQGs where the $\theta_i$ distribution `wraps' from $180^{\circ}$ back to $0^{\circ}$ (Fig.~\ref{subBSWrap}) illustrate why the linear mean is inappropriate.

The distribution of the half-width confidence interval (\gls{HWCI}) for 10,000 bootstraps of our sample of 71 LQGs, with PAs $\theta_i$ calculated using both the 2D and 3D approaches, is shown in Fig.~\ref{figHWCI}. In general, the confidence intervals are slightly lower for the 2D approach (solid orange) than the 3D approach (dashed blue). The mean (median) HWCI for the 2D approach is $\sim10^{\circ}$ ($\sim8^{\circ}$), and for the 3D approach it is $\sim11^{\circ}$ ($\sim9^{\circ}$). The highest HCWIs ($\gtrsim 20^{\circ}$) are due to LQG morphology and/or orientation and also have low goodness-of-fit weighting (section~\ref{subsecGoF}). The HWCIs for individual LQG PAs, calculated using both the 2D and 3D approaches, are shown as error bars in Fig.~\ref{figPA2D3D}.

\subsection{Evaluation of 2D and 3D position angle approaches} \label{subsec2D3DEval}

PAs determined by the two approaches may differ, for example the 2D approach may be susceptible to projection effects and the 3D approach may be susceptible to redshift-space distortions. The orientation of the LQG with respect to the line-of-sight and its morphology may also induce differences. We expect PAs determined by the two approaches to agree well when the LQG is linear and orthogonal to the line-of-sight, but they may differ significantly when the LQG is broad, crooked, curved or aligned along the line-of-sight.

\begin{figure} % LQG_PA_bootstrap
    \centering
	\includegraphics[width=0.5\columnwidth]{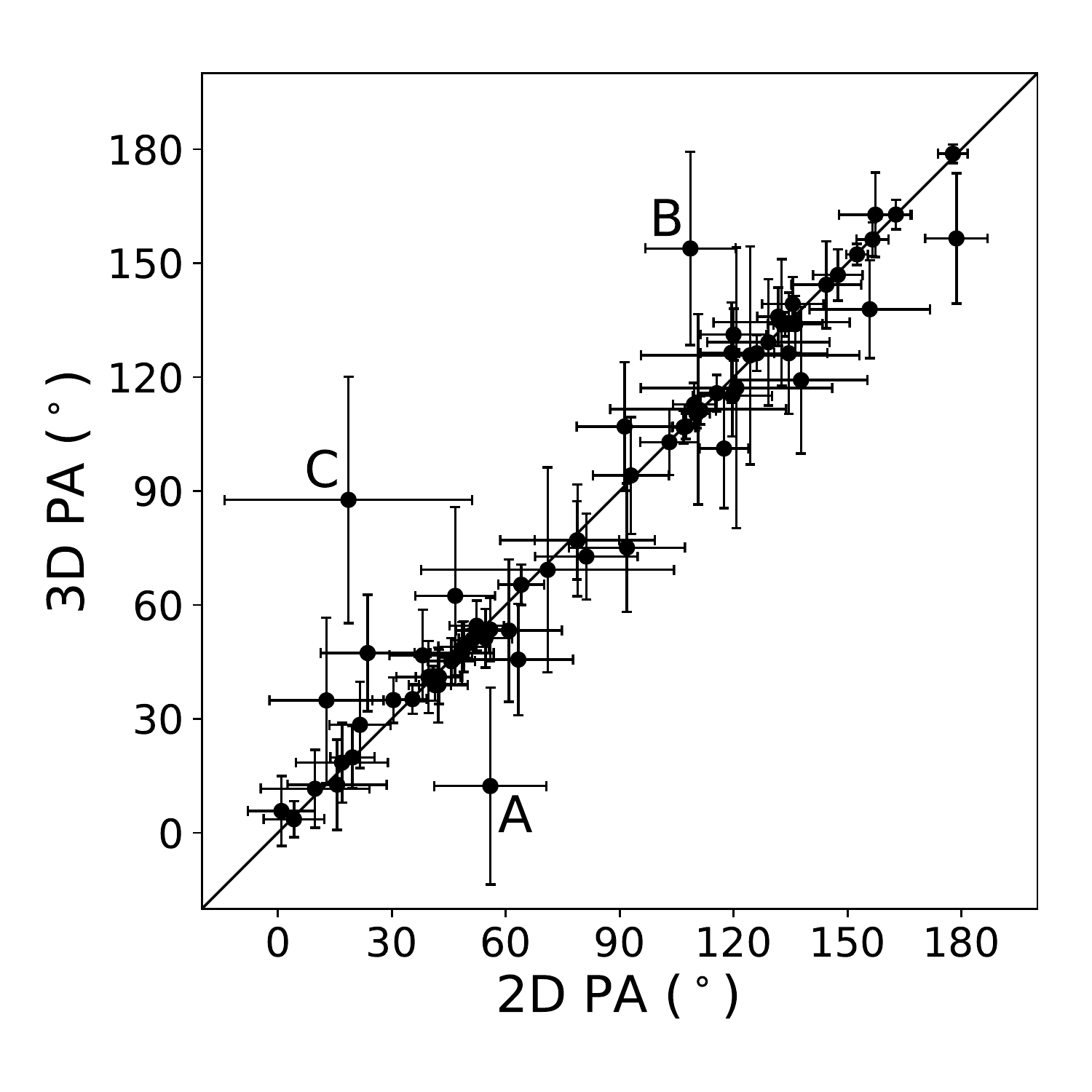}
	\caption[Position angles calculated by 2D and 3D approaches (comparison)]{Position angles of 71 LQGs calculated by 2D and 3D approaches. The two generally agree well, with the three widest outliers (A, B, C) due to the geometry of those particular LQGs (see text and Fig.~\ref{figAllLQGsPt1}). Error bars show half-width confidence intervals estimated using bootstrap re-sampling.}
	\label{figPA2D3D}
\end{figure}

For our sample of 71 LQGs we find that the two approaches are generally consistent. Fig.~\ref{figPA2D3D} shows the PAs calculated using both the 2D and 3D approaches. Note that the PAs have not been parallel transported, but these angles serve as a useful comparison between the two approaches. The error bars are the half-width confidence intervals estimated using bootstrap re-sampling. Note that, usually, the measurement uncertainties are slightly larger for the 3D approach.

The three widest outliers from the 1:1 diagonal line in Fig.~\ref{figPA2D3D} are due to the geometry of these particular LQGs (A, B, C; also labelled in Fig.~\ref{figAllLQGsPt1}). Two of these LQGs (A and B) have their major axes orientated towards the line-of-sight, and not significantly longer than their first minor axes. Indeed, for both of these, the 2D approach fits a regression comparable to the first minor axes rather than the major axes. One LQG (C) is very irregular so linear fits are poor, and corresponding PAs are uncertain, in both the 2D and 3D approaches. The PAs of all three of these LQGs are given little weight by goodness-of-fit weighting (section~\ref{subsecGoF}).

\begin{figure}[t!] % LQG_PAs_2D_3D
    \centering
	\subfigure[2D approach]{\includegraphics[width=0.48\columnwidth]{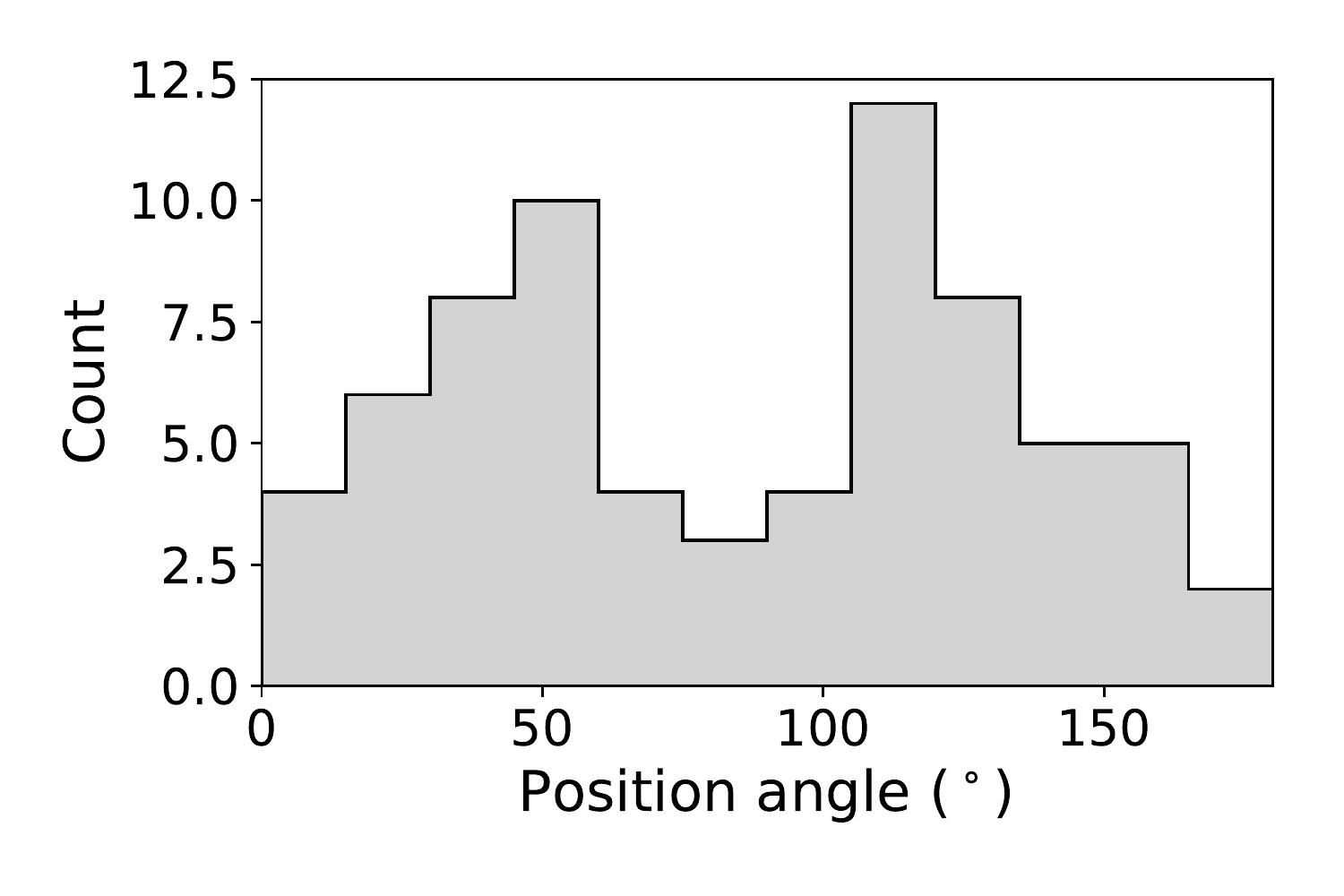} \label{subHist2D}}
	\subfigure[3D approach]{\includegraphics[width=0.48\columnwidth]{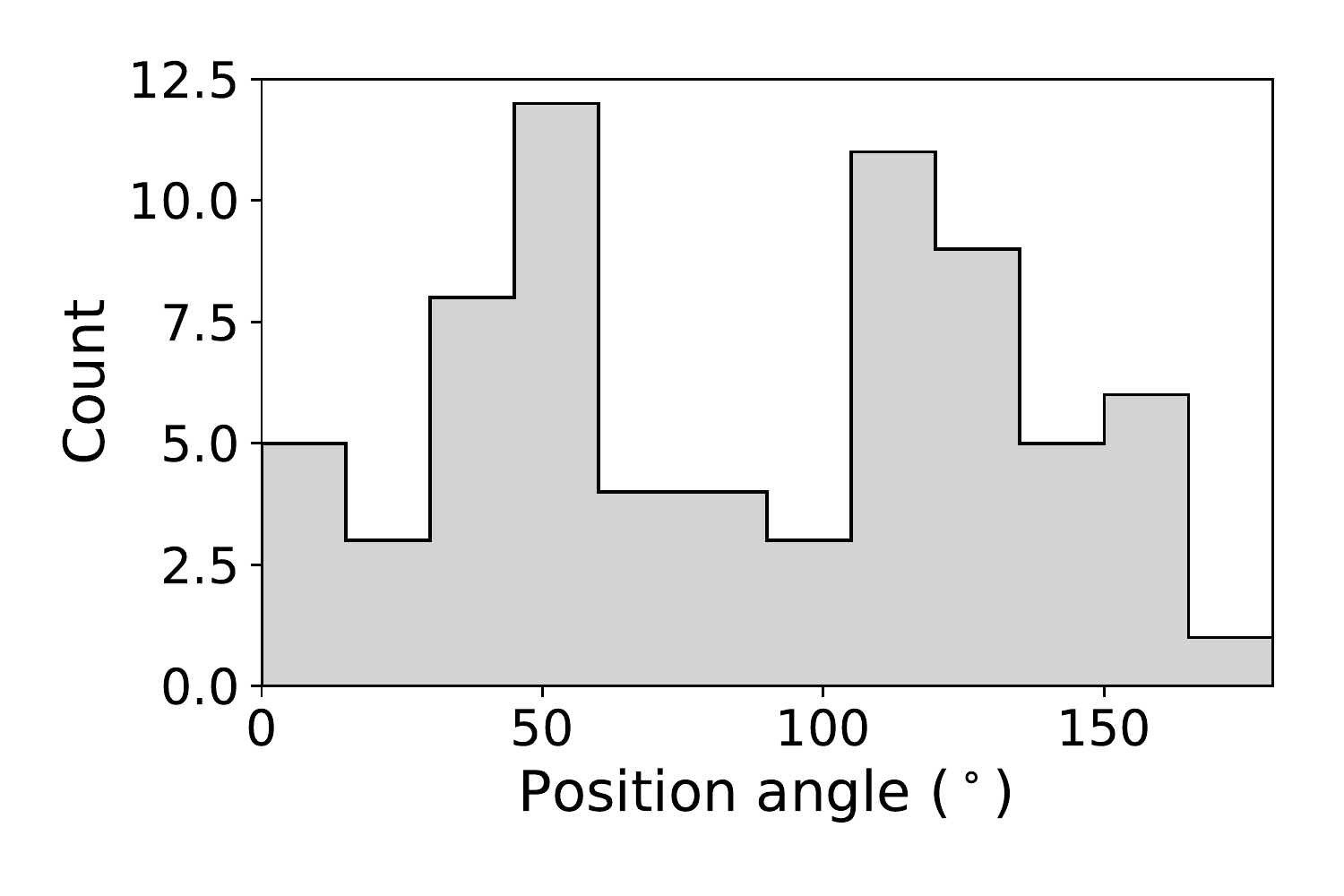} \label{subHist3D}}
	\caption[Position angles calculated by 2D and 3D approaches (distribution)]{LQG position angles, with no parallel transport, using the \protect\subref{subHist2D}~2D and \protect\subref{subHist3D} 3D approaches, both with $15^{\circ}$ bins. The bimodal distribution is robust to which approach is taken.}
	\label{figPAHist2D3D}
\end{figure}

The PA distributions determined using the two and three-dimensional approaches are shown in Fig.~\ref{figPAHist2D3D}, and the apparent bimodality is robust to which approach is chosen. Note the PAs have not been parallel transported, and neither histogram is goodness-of-fit weighted. Results incorporating these are reported in chapter~\ref{chaptLQGOrientation}.

Both the 2D and 3D approaches have been used to determine the PAs of LQGs. \citet{Hutsemekers2014} use the 2D approach to demonstrate alignment of quasars' optical linear polarization with LQG axes, whilst \citet{Pelgrims2016} use the 3D approach to evidence alignment of quasars' radio polarization with more LQG axes. The latter derived eigenvectors and eigenvalues from the inertia tensor rather than covariance matrix; results are equivalent. \citet{Pelgrims2016} report that for the 2D approach PAs calculated using ODR are consistent (within $1^{\circ}$) with those determined using the inertia tensor.

\citet{Pelgrims2016} show that both approaches usually agree well, and argue that the 3D approach is more physically motivated. We agree, but note that any 3D analysis is susceptible to redshift errors. The quasars in our sample are from SDSS DR7QSO, which typically has quoted redshift errors of $\Delta z \sim 0.004$ \citep{Schneider2010}. There is also evidence for systematic errors of $\Delta z \sim 0.003$ \citep{Hewett2010}. Using Monte Carlo simulations we find that these redshift errors introduce uncertainty in the PA, generally of a few degrees, but up to $\sim 30^{\circ}$ for LQGs particularly orientated along the line-of-sight (e.g.\ LQGs A and B).

Furthermore, in the 3D approach, there is also the potential for errors due to redshift-space distortions from the quasars' peculiar velocities, causing their real-space distribution to be either elongated \citep{Jackson1972} or squashed \citep{Kaiser1987} along the line-of-sight. We find that the measurement uncertainties (section~\ref{subsecPAUncertainties}) are slightly larger for the 3D approach. We therefore base our analysis on the two-dimensional approach; tangent plane projection of the LQG and orthogonal distance regression of the projected quasars.

\subsection{Position angle confidence weighting} \label{subsecGoF}

Due to the filamentary nature of LQGs, the orthogonal distance regression (ODR) of the 2D approach gives a better linear fit for some LQGs than others. Therefore, we have higher confidence in some PAs than others. We can weight the PA of each LQG according to its ODR goodness-of-fit (\gls{GoF}) by

\begin{equation} \label{eqWeight}
    w = \ell / \sigma^2\ ,
\end{equation}

\noindent where $\ell$ is the length of the ODR line fitted to the LQG, and $\sigma^2$ is the residual variance of the $m$ quasars in the LQG, calculated as

\begin{equation}
    \sigma^2 = \frac{1}{m-1} \sum_{q=1}^{m} e_q^2\ ,
\end{equation}

\noindent where $e_q$ is the orthogonal residual of the $q^{\mathrm{th}}$ quasar from the ODR line. Note that this definition of weight (Eq.~\ref{eqWeight}) is only dimensionless after normalization.

We found weighting by both length $\ell$ and inverse residual variance $1/\sigma^2$ combined to be optimal; it gives long, linear LQGs the highest weighted PAs. Inverse variance weighting is standard. But without also considering length, a `short' LQG with low residual variance would have the same high weight as an LQG twice as long with the same variance and membership, despite us having higher confidence in the PA of the latter. It also makes physical sense; we are weighting by inverse residual variance per unit length.

In chapter~\ref{chaptLQGStatsMethods} we describe the statistical methods we use to analyse the uniformity, bimodality and correlation of the PA distribution. Where possible these analyses are performed using both unweighted and weighted PAs. In addition to unweighted analyses, we apply the weighting described here (Eq.~\ref{eqWeight}) to the $\chi^2$ test, Kuiper's test of uniformity, and the bimodality coefficient. It is not used for the Fourier-type Hermans-Rasson test for uniformity on the circle, Hartigans' dip statistic of unimodality, or for the S test for PA correlation.

\section{Coordinate invariance: parallel transport} \label{secParallelTransport}

The position angles calculated by either the two-dimensional or three-dimensional method, are dependent on the coordinate system in which they are measured, and in particular the position of the pole used to define $\theta = 0^{\circ}$.

\begin{figure} % LQG_plot_ortho
	\centering
    \includegraphics[width=0.75\columnwidth]{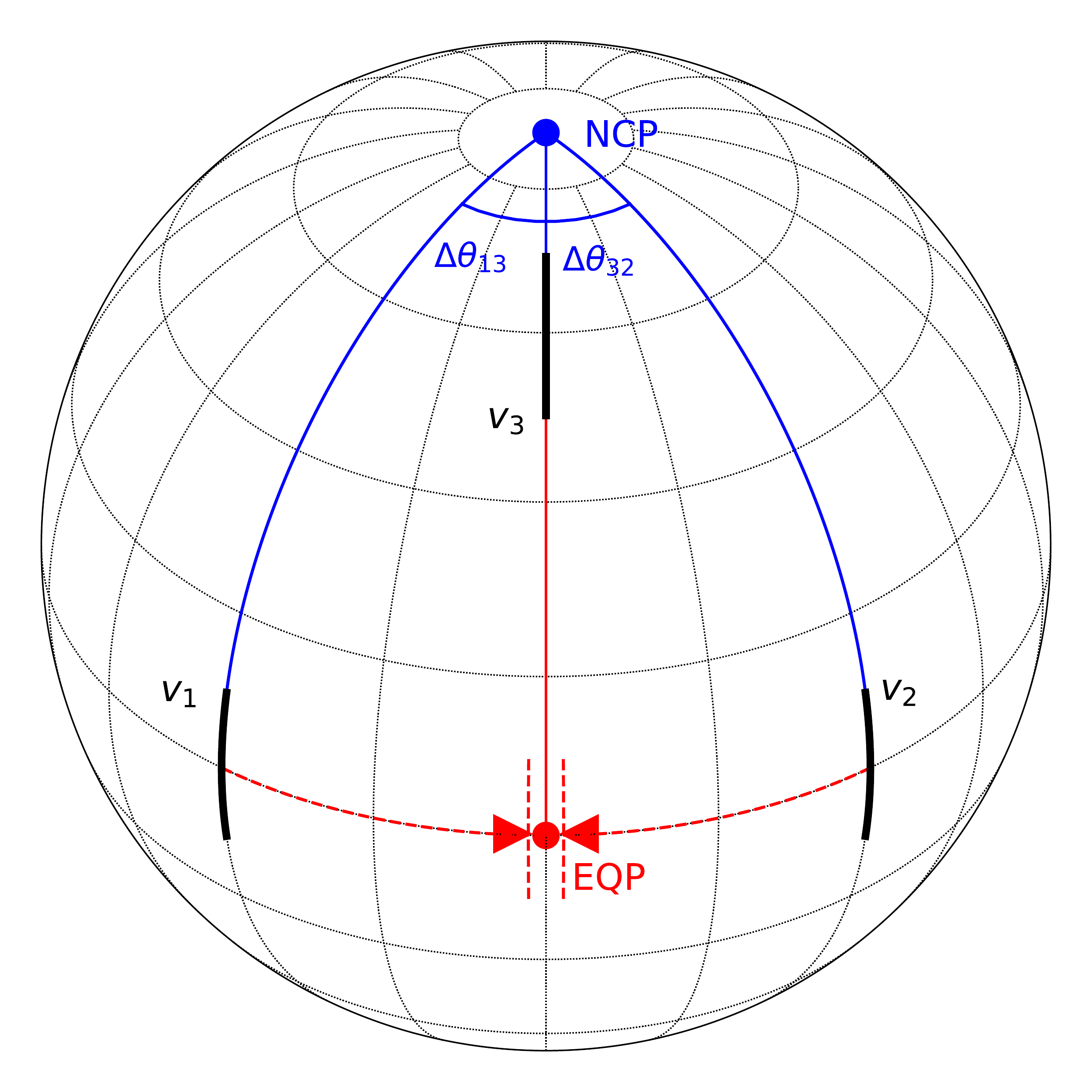}
    \caption[Illustration of PA coordinate system dependence (toy model)]{Illustration of position angle coordinate system dependence. Vectors $v_1$, $v_2$, and $v_3$ (solid black lines) appear aligned when measured in the NCP (blue dot) coordinate system, but $v_3$ is orthogonal when measured in the EQP (red dot) coordinate system. Conversely, transporting vectors to NCP (blue lines) indicates they are not aligned, whilst transporting them to EQP (red lines) indicates they are. See text for explanation.}
    \label{figPTModel}
\end{figure}

To illustrate this, using Fig.~\ref{figPTModel} consider the two vectors $v_1$ and $v_2$ (solid black lines), located at two different positions on the celestial equator, and both aligned with their own meridians. Their PAs measured with respect to the north celestial pole (\gls{NCP}, blue dot) are both $\theta_{ncp} = 0^{\circ}$. However, their PAs measured with respect to an alternative (bogus) equatorial `pole' (\gls{EQP}, red dot) are both $\theta_{eqp} = 90^{\circ}$. In both cases $v_1$ and $v_2$ have the same PA, so we might consider them to be aligned.

Consider now a third vector $v_3$, at a different location, also aligned with its own meridian, which this time passes through EQP. Again, its PA measured with respect to NCP is $\theta_{ncp} = 0^{\circ}$, but its PA measured with respect to EQP is now $\theta_{eqp} = 0^{\circ}$. Assuming that two vectors are considered `aligned' if they have the same PA $\theta$, is $v_3$ aligned with $v_1$ and $v_2$? In the NCP coordinate system all three vectors are aligned ($\theta_{ncp} = 0^{\circ}$), but in the EQP coordinate system $v_1$ and $v_2$ are aligned ($\theta_{eqp} = 90^{\circ}$) whilst $v_3$ is orthogonal ($\theta_{eqp} = 0^{\circ}$).

Furthermore, suppose the three vectors $v_1$, $v_2$, and $v_3$ are extended along their meridians to the NCP (solid blue lines). They are not parallel when they intersect at the NCP ($\Delta\theta_{13} \neq 0$ and $\Delta\theta_{32} \neq 0$). Similarly consider the vectors at EQP; $v_1$ and $v_2$ are moved along the equator to EQP (whilst maintaining $90^{\circ}$ between vector and equator), and $v_3$ is extended along its meridian to EQP (dashed and solid red lines respectively). The three vectors are parallel when all considered at EQP.

The results from this toy model, both comparing PAs in situ with respect to a common origin (NCP or EQP), and after being moved to a common point (again, NCP or EQP), are clearly not invariant under coordinate transformation. This coordinate system dependence is clearly undesirable.

\subsection{Parallel transport method} \label{secParallelTransportMethod}

\citet{Jain2004} developed a method to overcome this coordinate dependence, specifically in relation to particular statistical tests, the S and Z tests (section~\ref{subsecCorrelationMethods}), but transferable to other analyses. Rather than analysing the PAs as measured with respect to their own meridians, which are coordinate system dependent, they introduce corrections to the PAs based on their relative locations on the celestial sphere. These corrections are calculated using parallel transport.

Parallel transport is the concept of moving a vector along a path whilst preserving its orientation with respect to that path  (e.g.\ transporting $v_1$ and $v_2$ along the equator in our example above). It had already been used for studies of galaxy spin alignment \citep[e.g.][]{Pen2000} and CMB polarization \citep[e.g.][]{Challinor2002}.

For two objects at locations $P_1$ and $P_2$ on the celestial sphere, \citet{Jain2004} proposed parallel transporting the vector at the location of one object to the location of the other before comparing them. The path they use is the geodesic (great circle) between the objects. Parallel transport preserves the angle between the PA vector and the vector tangent to this geodesic. The correction to apply between $P_1$ and $P_2$ is the difference between the angles the geodesic makes with one of the basis vectors at each location. That is, if the tangent plane to the sphere has local basis vectors ($\hat{\theta}_1$, $\hat{\phi}_1$) at location $P_1$, and the tangent unit vector to the geodesic at this point is given by $\hat{t}_1$, then the angle $\xi_1$ between $\hat{t}_1$ and $\hat{\phi}_1$ is given by \citep{Pelgrims2016thesis}

\begin{equation}
    \xi_1 = \tan^{-1}(-\hat{t}_1 \cdot \hat{\theta}_1,\ \hat{t}_1 \cdot \hat{\phi}_1) \ ,
\end{equation}

\noindent with angle $\xi_2$ at location $P_2$ being similarly obtained.

The parallel transport correction between locations $P_1$ and $P_2$, i.e.\ the angle by which a vector rotates during parallel transport from $P_1$ to $P_2$, is given by the difference between angles $\xi_1$ and $\xi_2$ \citep{Jain2004}. So, to parallel transport the position angle $\theta_k$ of object $k$ to the location of object $i$ we compute

\begin{equation} \label{eqPTCorrections}
\begin{aligned}
    \theta_k^{(i)} &= \theta_k + \Delta_{k \rightarrow i} \ , \\
    &= \theta_k + \xi_k - \xi_i \ ,
\end{aligned}
\end{equation}

\noindent where $\theta$ refers to position angle, not spherical coordinates. Applying these corrections results in coordinate-invariant statistics \citep{Jain2004, Hutsemekers2005}. The result of parallel transporting a vector from $P_1$ to $P_2$ depends on the path taken between them. If a different path was chosen the parallel transport correction ($\Delta_{k \rightarrow i}$) would differ.

\subsection{Parallel transport variants} \label{secParallelTransportVariants}

In its simplest form, we can parallel transport all PAs to a single point on the celestial sphere to compare them. We choose the point indicated by a black cross in Fig.~\ref{figLQGsSphere} ($\alpha = 193.6^{\circ}$, $\delta = 24.7^{\circ}$, J2000). This is the centre of the A1 region \citep{Hutsemekers1998} of large-scale alignment of the optical polarization of quasars, so if LQGs are also aligned it is plausible any correlation may be centred here. It is also near the centroid of our LQG sample, so it minimises the distance over which we parallel transport, and the amount by which PAs are rotated.

This is the method of parallel transport we use for all subsequent analysis of uniformity and bimodality, which require either categorical data ($\chi^2$ test and bimodality coefficient) or the empirical cumulative distribution function (Kuiper's test and Hartigans' dip statistic). It is also used for the Fourier-type Hermans-Rasson test for uniformity on the circle.

To analyse alignment correlation we use the S test, specifically the coordinate invariant version of \citet{Jain2004}. This incorporates parallel transport corrections into the measure of dispersion of groups of nearest neighbours as detailed in section~\ref{subSTest}. With this form of parallel transport we minimise the distance over which we transport, and the size of the correction required.

\section{Summary and conclusions} \label{secLQGSampleConclusions}

In this chapter we present our sample of 71 LQGs, and summarise the minimal spanning tree method \citet{Clowes2012,Clowes2013} used to detect them. We explain the two-dimensional and three-dimensional methods of determining their orientation, and demonstrate they are generally consistent. Due to potential systematics and larger measurement errors of the 3D approach, we choose the 2D approach for this work, namely tangent plane projection of the LQG and orthogonal distance regression (ODR) of the projected quasars.

Position angle data are 2-axial [$0^{\circ}$, $180^{\circ}$); we explain the implications of this for statistical analysis and present statistically valid solutions, including conversion to 4-axial [$0^{\circ}$, $90^{\circ}$) data. The mean measurement uncertainty of 2-axial PAs is $\sim10^{\circ}$, although this varies considerably between different LQGs. We account for this by goodness-of-fit weighting.

Finally, we demonstrate the need for coordinate-invariant statistics and explain how parallel transport provides this. For many tests we transport all PAs to a single point on the celestial sphere, namely the centre of the A1 region of large-scale quasar optical polarization alignment \citep{Hutsemekers1998}. Parallel transport corrections depend on the path taken, and therefore on the destination. We later show our results are robust to choice of destination (section~\ref{secHDSResults}), but note the potential ambiguity which choosing a destination could introduce. We use the coordinate-invariant version of the S test \citep{Jain2004} to analyse alignment correlation; this incorporates parallel transport corrections by design and has no `arbitrary' parallel transport destination.

\afterpage{\null\newpage} % because this ended on an odd page so the next chapter started on an even page

% END

%% file: chapterLQGStatsMethods.tex
% START
\addtocontents{toc}{\protect\pagebreak} % forces toc entry onto next toc page

\chapter{Methods for the statistical analysis of LQG position angles} \label{chaptLQGStatsMethods}

Statistical analysis of large quasar group (LQG) position angle (PA) data requires appropriate methods. LQGs are widely and non-uniformly distributed, both on the celestial sphere (separation $\lesssim 120^{\circ}$) and in redshift ($1 \le z \le 1.8$), and their PAs are axial data. Furthermore, the PA distribution may be bimodal, with PAs both aligned and orthogonal \citep{Hutsemekers2014,Pelgrims2016thesis}. Many statistical methods lack discriminatory power in multimodal cases.

In this chapter we evaluate several statistical methods which may be appropriate to analyse the LQG PA data. We draw on literature from astrophysics, applied statistics, psychology, and biology and apply some of these methods to astrophysical data for the first time (e.g.\ the Hermans-Rasson test). Furthermore, we introduce weighting and/or 4-axial data to other methods (e.g.\ Kuiper's test, S test).

%We use methods for statistical analysis of the uniformity, bimodality and correlation of LQG PAs, specifically to determine:
We analyse the uniformity, bimodality and correlation of LQG PAs to determine:

\begin{itemize}
    \item Are they likely to be drawn from a uniform distribution? (section~\ref{subsecUniformityMethods})
    \item Is their distribution bimodal, and where are the peaks? (section~\ref{subsecBimodalityMethods})
    \item Are they more aligned/orthogonal than random simulations? (section~\ref{subsecCorrelationMethods})
\end{itemize}

\section{Uniformity tests} \label{subsecUniformityMethods}

For coordinate invariance, we perform uniformity tests on PAs parallel transported to the centre ($\alpha = 193.6^{\circ}$, $\delta = 24.7^{\circ}$, J2000) of the A1 region \citep{Hutsemekers1998} of large-scale alignment of the polarization of quasars.

Many uniformity tests suffer reduced power with multimodal distributions. In some cases, where we expect perfect \emph{f}-fold symmetry, it is legitimate to convert multimodal \emph{f}-fold symmetric raw data into unimodal data \citep{Fisher1993}, e.g.\ for circular [$0^{\circ}$, $360^{\circ}$) data

\begin{equation}
    \Theta = \mathrm{mod}(f \times \theta, 360^{\circ})\ .
\end{equation}

However, situations where this is valid are limited; the alternative hypothesis must predict that the multiple peaks are the same size, symmetrical, and evenly distributed around the circle. This is close to what we observe (Fig.~\ref{figPAHist}), but we cannot \emph{a priori} make this prediction for the PA data, except in the special case where we are explicitly testing for both alignments and orthogonality, i.e.\ where $\Delta\theta\sim90^{\circ}$. However, we have independent empirical motivation for suspecting this type of distribution from \cite{Pelgrims2015}, who identify large-scale alignment and orthogonality of the polarization vectors of radio quasars.

We test the PA distribution for departure from uniformity using Kuiper's test, the Hermans-Rasson (HR) test, and the $\chi^2$ test. Kuiper's test and the HR test are both nonparametric tests that are applied to continuous data. In contrast, the $\chi^2$ test is applied to categorical data. We introduce and evaluate each of these tests in this section.

\subsection{Kuiper's test} \label{secKuipers}

Circular [$0^{\circ}$, $360^{\circ}$) and axial [$0^{\circ}$, $180^{\circ}$) data are relatively common in the fields of geology and biology, where the most common test of departure from uniformity is the Rayleigh test \citep{Rayleigh1880}. However, \citet{Upton1989} note that this is only appropriate for unimodal distributions.

A popular alternative to the Rayleigh test is Kuiper's test \citep{Kuiper1960}, which is a rotationally invariant version of the better-known Kolmogorov–Smirnov (\gls{KS}) test. To compare an empirical cumulative distribution function (\gls{EDF}) to a theoretical cumulative distribution function (\gls{CDF}) Kuiper's test quantifies the maximum positive \emph{and} negative differences between the EDF and CDF as

\begin{equation} \label{eqCDFDiffs}
\begin{split}
    D_+ &= \max_{0^{\circ}\le x \le 180^{\circ}} |F_{EDF}(x) - F_{CDF}(x)|\\
    D_- &= \max_{0^{\circ}\le x \le 180^{\circ}} |F_{CDF}(x) - F_{EDF}(x)|\ ,
\end{split}
\end{equation}

\noindent where $F_{CDF}$ is the theoretical cumulative distribution function being tested against (e.g.\ a uniform distribution) and $F_{EDF}$ is the empirical cumulative distribution function of the observations (in this case LQG PAs). Kuiper's test statistic is

\begin{equation} \label{eqKuiperV}
    V = D_+ + D_-\ .
\end{equation}

We apply Kuiper's test to both unweighted and weighted EDFs. To incorporate weighting we compute a weighted EDF, where for any measurement $x$, $F^w_{EDF}(x)$ is equal to the sum of the normalized weights of all measurements less than or equal to $x$. Following \citet{Monahan2011}, that is

\begin{equation} \label{eqKuiperWeighted}
    F^w_{EDF}(x) = \sum_{i=1}^{n_x} w_i \bigg/ \sum_{i=1}^{n} w_i\ ,
\end{equation}

\noindent where $n$ is the total sample size, $n_x$ is the number of measurements up to and including $x$, and $w_i$ are their goodness-of-fit weights ($w = \ell/\sigma^2$). Kuiper's test is then computed normally (Eqs.~\ref{eqCDFDiffs} and \ref{eqKuiperV}) using $F^w_{EDF}(x)$ in place of $F_{EDF}(x)$.

Kuiper's test is implemented using Python's \texttt{astropy.stats.kuiper}, which for unweighted data evaluates the test statistic $V$ plus an approximate p-value. To incorporate weighting we also coded a manual version (Eqs.~\ref{eqCDFDiffs}, \ref{eqKuiperV}, \ref{eqKuiperWeighted}) and checked the unweighted results are consistent between the two methods. The p-values are evaluated by simulation. We generate 10,000 samples of $n$ random PAs, drawn from a uniform distribution, apply the same weighting, and calculate the fraction of samples with Kuiper's test statistic $V$ at least as extreme as the observations. Note that using alternative weights (e.g.\ $w_i^2$, $w = 1/\gamma$, $w = 1/\gamma^2$ where $\gamma$ is the half-width confidence interval) does not significantly affect the p-value.

The rotational invariance of Kuiper's test makes it independent of the `origin' PAs are measured against (in this case celestial north). If we choose a different origin, $D_+$ and $D_-$ would change individually, but their sum remains constant. This symmetry makes Kuiper's test appropriate for circular and axial data that `wrap' between one end of the distribution and the other, and also gives it equal sensitivity at all values of $x$.

Kuiper's test is evaluated for both 2-axial and 4-axial PAs. For 2-axial we analyse raw PAs ($\theta$) and PAs of the form $\Theta_{2ax} = 2\theta$; results are consistent, as expected for a method using the EDF. For 4-axial we evaluate Kuiper's test for PAs of the form $\theta_{4ax}$ and $\Theta_{4ax} = 4\theta_{4ax}$; again results are consistent.

\subsection{Hermans-Rasson test} \label{secHR}

\citet{Landler2018} test the performance of the Rayleigh and Kuiper’s tests (amongst others) with a variety of multimodal distributions. They show that these tests lack statistical power in most multimodal cases, and find the Hermans-Rasson (\gls{HR}) test for uniformity on the circle \citep{Hermans1985} significantly out-competes the alternatives. The HR method is a family of tests, based on decomposing a circular distribution using Fourier series \citep{Landler2019}. Variants of the HR test are controlled by the parameter $\beta$, with $\beta = 2.895$ being recommended by both \citet{Hermans1985} and \citet{Landler2018}\footnote{Recommendation in electronic supplementary material 2} as offering power in both unimodal and multimodal cases. In this case, the HR statistic $T$ of $n$ measurements $\theta_1,...,\theta_n$ is defined \citep{Landler2018} as

\begin{equation} \label{eqHR}
    T = \frac{1}{n} \sum_{i=1}^{n} \sum_{j=1}^{n} \pi - \lvert\pi - \lvert\theta_i - \theta_j\rvert\rvert + 2.895\lvert\sin(\theta_i - \theta_j)\rvert\ ,
\end{equation}

\noindent where $\theta_i$ is the PA of the $i^{\mathrm{th}}$ LQG and $\theta_j$ is the PA of the $j^{\mathrm{th}}$ LQG, from a sample of $n$ LQGs. The smaller the $T$ statistic the greater the departure from uniformity (opposite to KS and Kuiper's tests).

To the best of our knowledge, the HR test is not yet provided in a Python library, so we coded our own version (Eq.~\ref{eqHR}). Our implementation of the HR test does not currently incorporate weighting. The p-values are evaluated by simulation. We generate 10,000 samples of $n$ random PAs, drawn from a uniform distribution, and calculate the fraction of samples with the HR statistic $T$ at least as extreme as the observations. Note smaller $T$ statistics are more significant.

The HR test requires circular data so we apply it to 2-axial and 4-axial PAs of the form $\Theta_{2ax} = 2\theta$ and $\Theta_{4ax} = 4\theta_{4ax}$ respectively.

\subsection{$\chi^2$ test} \label{secChi2}

The final test we review is the very well-known $\chi^2$ test which, unlike Kuiper's test and the HR test, is applied to categorical data. It tests whether the observed frequency distribution is consistent with a theoretical distribution (in this case uniform). It is equally sensitive across the whole distribution, but binning reduces its discriminatory power compared with tests applied to continuous data, like Kuiper's test and the HR test. Unlike those tests it does not account for the `wrap-around' nature of circular/axial data.

For a histogram comprising $m$ bins, the $\chi^2$ statistic is

\begin{equation} \label{eqChi2}
    \chi^2 = \sum_{i=1}^{m} \frac{(O_i - E_i)^2}{E_i}\ ,
\end{equation}

\noindent where $O_i$ is the observed frequency and $E_i$ is the expected frequency per bin $i$. For a total of $n$ measurements in $m$ bins $E_i = n/m$.

To incorporate weighting into the $\chi^2$ test we compute the frequencies of a weighted histogram. Either $O_i$ and $E_i$ must both be normalized, or, more simply and equivalently, the weighted frequencies $O_{i,w}$ must be scaled such that

\begin{equation} \label{eqChi2Weights}
    \sum_{i=1}^{m} O_{i,w} = n\ ,
\end{equation}

\noindent where $n$ is the total number of measurements in all $m$ bins.

The $\chi^2$ test is implemented using Python's\footnote{SciPy (scientific computing in Python) community project \citep{Virtanen2019}} \texttt{scipy.stats.chisquare}, which for unweighted data evaluates the $\chi^2$ statistic plus a p-value. It can also be applied to weighted histograms after appropriate scaling of the observed frequency $O_{i,w}$ (Eq.~\ref{eqChi2Weights}). The expected frequency $E_i$ is uniform and unweighted.

The $\chi^2$ test is evaluated for both 2-axial and 4-axial PAs. For 2-axial we analyse raw PAs ($\theta$), and for 4-axial we evaluate the $\chi^2$ test for PAs of the form $\theta_{4ax}$. In all cases the bin width is chosen to ensure $E_i > 5$.

\section{Bimodality tests} \label{subsecBimodalityMethods}

\citet{Freeman2013} compare three different measures of bimodality: the bimodality coefficient \citep[BC,][]{Yamada2009}, Hartigans' dip statistic \citep[HDS,][]{Hartigan1985}, and Akaike's information criterion difference \citep[AIC$_{diff}$,][]{Akaike1974}. They report HDS has the highest sensitivity, followed by BC, and that both methods are generally convergent. They found that AIC$_{diff}$ behaves quite differently to the other two methods, and erroneously identifies bimodality in their simulations and experimental data. We thus focus on BC and HDS only.

As for uniformity tests, we perform bimodality tests on PAs after parallel transport to the centre of the A1 region \citep{Hutsemekers1998}.

\subsection{Bimodality coefficient} \label{secBC}

The bimodality coefficient \citep[\gls{BC},][]{Yamada2009} is calculated from categorical data as

\begin{equation} \label{eqBC}
    \mathrm{BC} = \frac{m_3^2+1}{m_4 + 3\left(\frac{(n-1)^2}{(n-2)(n-3)}\right)}\ ,
\end{equation}

\noindent where $n$ is the sample size (number of LQGs), $m_3$ is the skewness of the distribution, and $m_4$ is its excess kurtosis. Values for $m_3$ and $m_4$ are calculated using Python's \texttt{scipy.stats.skew} and \texttt{scipy.stats.kurtosis} respectively.

A uniform distribution has $\mathrm{BC}_{crit} = 5/9 \approx 0.555$. The BC of an empirical distribution is compared to this critical value; $\mathrm{BC} > \mathrm{BC}_{crit}$ indicates bimodality, $\mathrm{BC} < \mathrm{BC}_{crit}$ indicates unimodality.

The BC is evaluated for 2-axial PAs, using both raw PAs ($\theta$) and PAs of the form $\Theta_{2ax} = 2\theta$; results are not consistent, probably due to sensitivity to bin size. The BC is not calculated for 4-axial PAs, since this conversion yields unimodal data, for which BC is not meaningful.

\begin{figure} [h] % illustrative_figs_BC
    \centering
    \includegraphics[width=0.65\columnwidth]{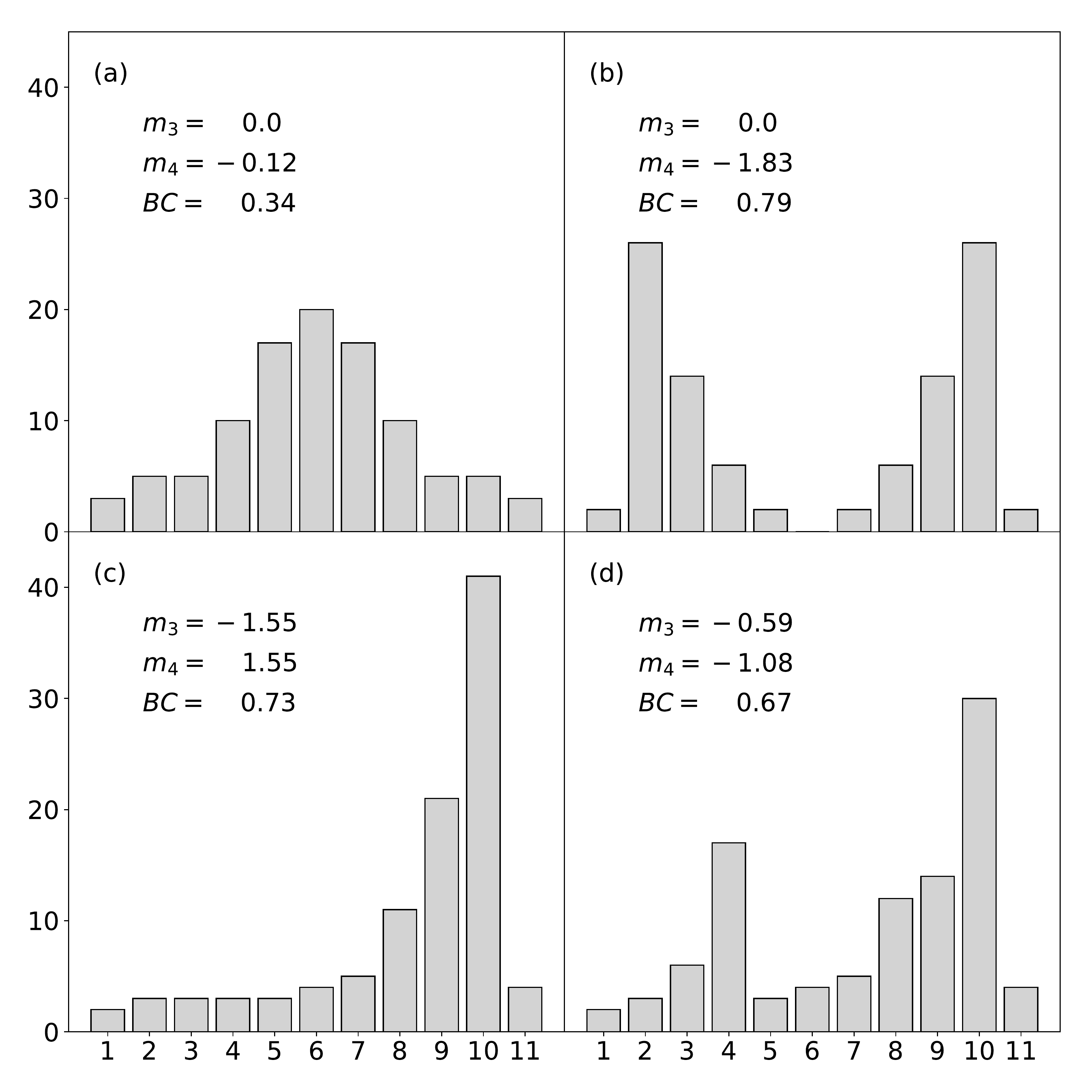}
    \caption[Bimodality coefficient - example distributions]{Histograms for four example distributions, showing skewness ($m_3$), kurtosis ($m_4$), and bimodality coefficient (BC), adapted from \citet{Pfister2013}. $\mathrm{BC} > 0.555$ indicates bimodality; (a), (b) and (d) are correctly classified by the BC, (c) is erroneously classified as bimodal due to its skewness.}
    \label{figBCExamples}
\end{figure}

The BC is dependent on both skewness and kurtosis. High BCs result from high skewness (whether left or right; negative or positive) and/or low or negative kurtosis (light-tailed distributions). \citet{Pfister2013} show that this, particularly the influence of skewness, can result in undesired behaviour of the BC, including erroneous identification of bimodality (Fig.~\ref{figBCExamples}). Furthermore, they explain that the inability to derive a BC probability density function \citep{Knapp2007} precludes using BC for hypothesis significance testing. 

We find that the BC is extremely sensitive to the choice of bin size, which is also undesirable. Therefore we look beyond BC to detect potential multimodality in the PA data.

\subsection{Hartigans' dip statistic} \label{secHDS} % illustrative_figs_HDS

Hartigans' dip statistic \citep[\gls{HDS},][]{Hartigan1985} is a non-parametric test of the unimodality of continuous data. A distribution is categorised as unimodal if its cumulative distribution function (CDF) is convex up to its maximum gradient (which corresponds to the peak in the distribution) and concave afterwards, i.e.\ with a single inflection point. HDS analyses the empirical cumulative distribution function (EDF) and measures the minimum achievable vertical distance $2d$ between the two vertically offset copies of the EDF $F$ and a set of piecewise-linear unimodal (but otherwise arbitrary) cumulative distribution functions (CDFs) $\mathds{G}$. The `dip' statistic $d$ is half the difference between the possible copies of $F$, which permit a unimodal $G$, and where the choice of $G$ minimises that difference.

\begin{figure}
    \centering
    \includegraphics[width=0.6\columnwidth]{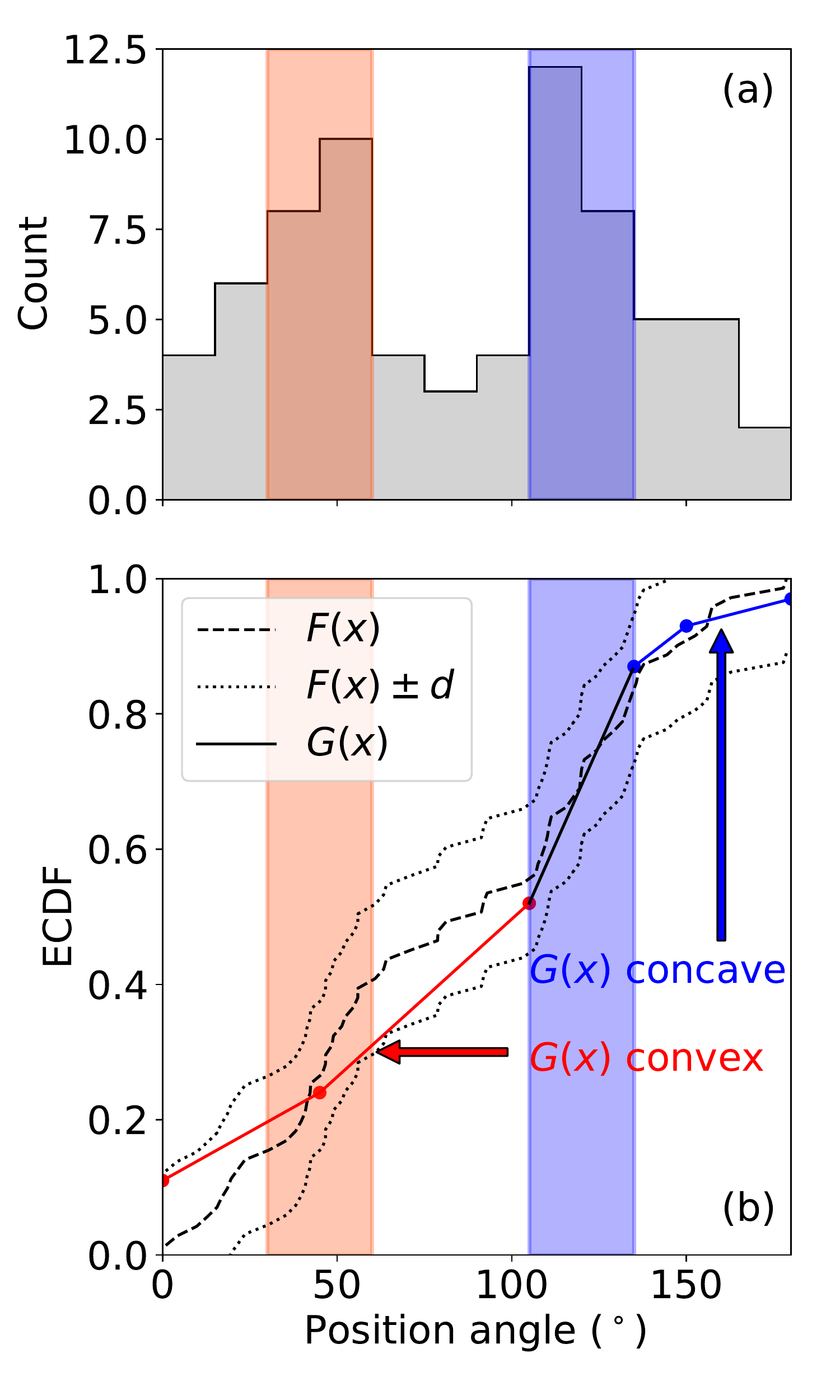}
    \caption[Illustration of Hartigans' Dip Statistic method]{Illustration of Hartigans' Dip Statistic method. Shaded regions indicate peaks in the data. (a) Distribution of position angle data, with no parallel transport, and $15^{\circ}$ bins. (b) Empirical cumulative distribution function (EDF, $F(x)$, black dashed line), two vertically offset copies of EDF ($F(x) \pm d$, grey dotted lines), and unimodal distribution ($G(x)$, solid line). The EDF is based on real data, but choice of offset $d$ and unimodal $G(x)$ are arbitrary and for illustration only. Note convex (red), concave (blue), and inflection (black) regions of $G(x)$; see text for explanation.}
    \label{figHDSexample}
\end{figure}

To illustrate HDS, see Fig.~\ref{figHDSexample}, loosely following \citet{Maurus2016}.\linebreak
Fig.~\ref{figHDSexample}(a) shows a histogram of LQG position angles for illustration only; HDS uses the EDF, not binned data. Fig.~\ref{figHDSexample}(b) shows the EDF for the PA data ($F(x)$, black dashed line), two vertically offset copies of the EDF ($F(x) \pm d$, grey dotted lines), with a piecewise-linear unimodal CDF between them ($G(x)$, solid line). Note that the EDF is based on real data, but the choice of offset $d$ and unimodal fit $G(x)$ are arbitrary and chosen solely for demonstration purposes.

The $F(x)$ gradient is larger where there are peaks in the data (shaded regions). The unimodal $G(x)$ is convex (red), reaches maximum gradient (black), and is then concave (blue). The further $F(x)$ is from unimodality, the larger $d$ will need to be in order to accommodate a unimodal $G(x)$. HDS compares $d$ to a reference unimodal distribution in order to calculate the p-value, i.e.\ the probability of a unimodal CDF having a `dip' greater than the multimodal EDF $F(x)$.

HDS thus gives a measure of how far $F(x)$ departs from unimodality. It also indicates the location of this departure (i.e.\ the peak); this is the location of the maximum gradient of $G(x)$ (i.e.\ the inflection point between convex and concave). Identification of one peak can then be exploited to recursively find the remainder of the peaks in a multimodal distribution \citep{Maurus2016}.

We evaluate HDS using Benjamin Doran's Python port of \texttt{unidip.UniDip}\footnote{\url{https://github.com/BenjaminDoran/unidip}}, which follows \citet{Maurus2016}. The sensitivity of this test is controlled by the parameter $\alpha$; we use $\alpha = 0.03$ to isolate peaks with at least $97\%$ signal-to-noise confidence. This implementation does not currently accommodate weighted data; enhancing it, or coding a weighted version, is beyond the scope of this work. The approach to weighting used for Kuiper's test, following \citet{Monahan2011}, should be considered in the future to extend the usefulness of this test.

\section{Correlation tests} \label{subsecCorrelationMethods}

Uniformity and bimodality tests do not quantify the degree of alignment in the data; for this we need specific statistical methods appropriate to axial data on the celestial sphere. We briefly review two tests, the S test and the Z test, that have been widely used to analyse polarization alignments (e.g.\ \citet{Hutsemekers1998}, \citet{Jain2004}, \citet{Pelgrims2015}), but are appropriate to analyse the alignment of any vectors on the celestial sphere (e.g.\ \cite{Contigiani2017}).

\subsection{S test} \label{subSTest}

The S test was developed by \citet{Hutsemekers1998} and analyses the dispersion of vectors with respect to their nearest $n_v$ neighbours. For each vector~$i$ a measure of the dispersion $d_i$ is calculated as

\begin{equation} \label{eqDispOrig}
    d_i(\theta) = 90 - \frac{1}{n_v} \sum_{k=1}^{n_v}|90 - |\theta_k - \theta||\ ,
\end{equation}

\noindent where $\theta_k$ are the PAs [$0^{\circ}$, $180^{\circ}$) of the neighbouring $n_v$ vectors, including central vector $i$. The value of $\theta$ that minimises the function $d_i(\theta)$ is a measure of the average PA at the location of $i$. Use of absolute values accounts for the axial nature of the data \citep{Fisher1993}.

For vector $i$ the mean dispersion $D_i$ of its $n_v$ nearest neighbours is calculated to be the minimum value of $d_i(\theta)$, which will be small for coherently aligned vectors. The measure of alignment within the whole sample of $n$ vectors is given by the S test statistic

\begin{equation} \label{eqSD}
    S_D = \frac{1}{n} \sum_{i=1}^{n}D_i\ ,
\end{equation}

\noindent with one free parameter $n_v$. If the vectors are aligned, the value of $S_D$ will be smaller than if they are uniformly distributed. So, the significance level for this version of the S test test is evaluated as the probability that a random numerical simulation has a lower $S_D$ than that observed \citep{Cabanac2005}. We will discuss the $S_D$ distribution in detail in section~\ref{subSTestSDDist}.

\citet{Jain2004} introduce a coordinate invariant version of the S test, similar to the original except that, instead of the dispersion measure in Eq.~\ref{eqDispOrig}, they use

\begin{equation} \label{eqDispJain}
    d_i(\theta) = \frac{1}{n_v} \sum_{k=1}^{n_v}\cos[2\theta - 2(\theta_k + \Delta_{k \rightarrow i})]\ ,
\end{equation}

\noindent where $\Delta_{k \rightarrow i}$ is the angle by which the PA $\theta_k$ changes during parallel transport from position $k$ to position $i$. Here, the factor two accounts for the axial nature of the data. The measure of dispersion is given by the \emph{maximum} value of Eq.~\ref{eqDispJain} (as opposed to the minimum value of Eq.~\ref{eqDispOrig}). The S statistic is calculated as previously (Eq.~\ref{eqSD}). \citet{Pelgrims2016thesis} notes that the same value of $\theta$ that maximises Eq.~\ref{eqDispJain} at the same time minimises Eq.~\ref{eqDispOrig}, so the two versions are fully equivalent.

\citet{Jain2004} show the maximisation of $d_i(\theta)$ is calculated analytically as

\begin{equation} \label{eqJainDiMax}
    d_i\Bigr\rvert_{max} = \frac{1}{n_v}\left[\left(\sum_{k=1}^{n_v}\cos\theta_k^{'}\right)^2 + \left(\sum_{k=1}^{n_v}\sin\theta_k^{'}\right)^2\right]^{1/2}\ ,
\end{equation}

\noindent where $\theta_k^{'} = 2(\theta_k + \Delta_{k \rightarrow i})$ is the circular version of $\theta_k$ after parallel transport to position $i$. We can similarly apply this to the 4-axial version of $\theta_k$ by using a factor of 4. This calculation is straightforward to code and avoids the time-consuming trials of the original version \citep{Hutsemekers1998}. A large value of $d_i\Bigr\rvert_{max}$ indicates small dispersion, and hence a large value of $S_D$ indicates strong alignment.

The significance level for the \citet{Jain2004} version of S test is conventionally evaluated as the probability that a random numerical simulation has a higher $S_D$ than that observed (i.e.\ opposite to the \citet{Hutsemekers1998} version).

The S test is evaluated for both 2-axial and 4-axial PAs. For 2-axial we analyse PAs of the form $\Theta_{2ax} = 2\theta$, and for 4-axial we analyse PAs of the form $\Theta_{4ax} = 4\theta_{4ax}$. In both cases we apply parallel transport corrections before transforming the angles.

\subsection{Z test} \label{subZtest}

The Z test was originally developed by Andrews and Wasserman \citep{Bietenholz1986} and subsequently modified by \citet{Hutsemekers1998}. It compares the PA of vectors to the mean resultant vector of their nearest $n_v$ neighbours. For each vector $j$ the mean resultant vector $\boldsymbol{Y}_j$ is calculated as

\begin{equation}
    \boldsymbol{Y}_j = \frac{1}{n_v}\left(\sum_{k=1}^{n_v}\cos 2\theta_k, \sum_{k=1}^{n_v}\sin 2\theta_k\right)\ ,
\end{equation}

\noindent where $\theta_k$ are the PAs [$0^{\circ}$, $180^{\circ}$) of the neighbouring $n_v$ vectors, this time \emph{excluding} central vector $j$, and the factor of two takes into account the axial nature of the PAs \citep{Fisher1993}. The mean direction $\bar{\theta}_j$ is given by the normalized mean resultant vector $\boldsymbol{\bar{Y}}_j$ through

\begin{equation}
    \boldsymbol{\bar{Y}}_j = (\cos 2\bar{\theta}_j, \sin 2\bar{\theta}_j)\ .
\end{equation}

A measure of the closeness of vector $i$ ($\theta_i$) to the mean resultant vector of the $n_v$ nearest neighbours of vector $j$ ($\bar{\theta}_j$) is given by the inner product $D_{ij} = \boldsymbol{y}_i \boldsymbol{\cdot} \boldsymbol{\bar{Y}}_j$, where $\boldsymbol{y}_i = (\cos 2\theta_i, \sin 2\theta_i)$. If the PAs are correlated \emph{with their positions}, then on average $D_{ij}$ will be larger for $i = j$ than for $i \neq j$.

\citet{Hutsemekers1998} introduced a modification to the Z test, computing instead $D_{ij} = \boldsymbol{y}_i \boldsymbol{\cdot} \boldsymbol{Y}_j$. This retains the length of the mean resultant vector $\boldsymbol{Y}_j$, which will be larger if the dispersion of the angles is smaller \citep{Fisher1993}. This modification is therefore expected to give more weight to sets of nearest neighbours with coherent alignments.

To evaluate the Z statistic, for each vector $i$ the inner products ($D_{ij}$'s) are sorted in ascending order and the rank $r_i$ of $D_{i,j=i}$ is determined. The statistic for the whole sample of $n$ vectors is then given by

\begin{equation} \label{eqZc}
    Z_c = \frac{1}{n} \sum_{i=1}^{n}\frac{r_i - (n+1)/2}{\sqrt{n/12}}\ ,
\end{equation}

\noindent where again $n_v$ is a free parameter. $Z_c$ will be larger when PAs are non-uniformly distributed and position-dependent \citep{Hutsemekers1998}.

This statistic is not coordinate invariant, because it compares vectors at different locations without performing parallel transport. \citet{Pelgrims2016thesis} shows that there are at least two different mechanisms to introduce the parallel transport corrections of \citet{Jain2004} into the Z test:

\begin{enumerate}
    \item Calculate $\boldsymbol{Y}_j$ using $\theta_k$'s that have been parallel transported to position $j$, then parallel transport $\boldsymbol{Y}_j$ to position $i$ to evaluate the coordinate-invariant $D_{ij}$ as\\
    \begin{equation} \label{eqDijNotUsed}
        D_{ij} = \frac{1}{n_v}\sum_{k=1}^{n_v}\cos[2(\theta_i - (\theta_k + \Delta_{k \rightarrow j} + \Delta_{j \rightarrow i}))]\ .
    \end{equation}
    \item Calculate $\boldsymbol{Y}_j$ using $\theta_k$'s that have been parallel transported direct to position $i$, then evaluate the coordinate-invariant $D_{ij}$ as\\
    \begin{equation} \label{eqDijCoordInv}
        D_{ij} = \frac{1}{n_v}\sum_{k=1}^{n_v}\cos[2(\theta_i - (\theta_k + \Delta_{k \rightarrow i}))]\ .
    \end{equation}
\end{enumerate}

\citet{Pelgrims2016thesis} notes that despite both these calculations of $D_{ij}$ being coordinate invariant, they are generally different. This is because the result of parallel transport depends on the path taken, and $\Delta_{k \rightarrow j} + \Delta_{j \rightarrow i}$ is (usually) different to $\Delta_{k \rightarrow i}$. In this work Eq.~\ref{eqDijCoordInv} is adopted, following the literature \citep{Jain2004,Hutsemekers2005,Pelgrims2015}. Further, \citet{Pelgrims2016thesis} shows that the significance levels derived using Eqs.~\ref{eqDijNotUsed} and \ref{eqDijCoordInv} agree with each other within less than a factor of two, which is close enough for hypothesis testing.

The significance level for the Z test is the probability that a random numerical simulation has a higher $Z_c$ than that observed \citep{Cabanac2005}.

\section{Applying the S test to LQG orientations} \label{secSTestApplying}

%This section might also be applicable to the Z test if I can ever get it working...

We find that LQG PAs are not correlated with position, therefore the S test is more appropriate than the Z test to quantify their alignment (section~\ref{secStatsAppropriateness}). In this section we elaborate on the use and interpretation of the S test, particularly when applying it to 2-axial or 4-axial data, both with and without parallel transport corrections. We discuss evaluation of the S~test significance level when data have both aligned and orthogonal modes, since this differs from the conventional interpretation discussed in section~\ref{subSTest}.

\subsection{Nearest neighbours free parameter} \label{subSTestNeighbours}

The S test quantifies the coherence of PA alignment by measuring the dispersion of groups of $n_v$ nearest neighbours, where $n_v$ is a free parameter. We explore a range of $n_v$ values; we do not choose a specific value. For each LQG, its nearest neighbours can be determined either in two dimensions (angular separation) or three dimensions (comoving separation).

In 2D, nearest neighbours are identified by calculating the angular separation $\theta$ on the celestial sphere between LQG 1 and LQG 2 as

\begin{equation}
    \theta = \cos^{-1}[\sin \delta_1 \sin \delta_2 + \cos \delta_1 \cos \delta_2 \cos(\alpha_1 - \alpha_2)] \ ,
\end{equation}

\noindent where $\alpha_1$ and $\delta_1$ ($\alpha_2$ and $\delta_2$) are the right ascension and declination of the centroids of LQG 1 (2) respectively. For each LQG, its nearest neighbours are those $n_v$ LQGs separated from it by the smallest angular distances.

In 3D, nearest neighbours are identified by calculating the three-dimensional comoving positions ($x$, $y$, $z$) of each LQG centroid. For each LQG, its nearest neighbours are those $n_v$ LQGs separated from it by the smallest comoving distances.

The 2D and 3D approaches of identifying nearest neighbours will return different groups of LQGs. When these groups are used to compute the S statistic $S_D$ we find that the values calculated using the two approaches are remarkably consistent. This is probably due to the geometry of the three-dimensional survey volume; our redshift restriction of $1.0 \le z \le 1.8$ yields a `shell' of LQGs of finite thickness. At low $n_v$ the 3D approach may find neighbours in the radial direction, but at high $n_v$ it can only find them tangentially (on the sky), like the 2D approach. Since the two approaches give very similar results, and the 3D approach is more physically motivated, we use the 3D approach of identifying nearest neighbours in this work.

The number of nearest neighbours $n_v$ is a free parameter which can be adjusted to quantify departure from uniformity over a range of scales. The LQGs are not uniformly distributed, either on the celestial sphere (Fig.~\ref{figLQGsSphere}), or in three dimensions. Therefore, we cannot directly interpret $n_v$ as corresponding to a particular comoving (or angular) scale. Rather, we expect the S test to indicate a significance level (SL) which may be smaller for some values of $n_v$ than others. The lowest value of SL does not indicate the overall significance of any departure from uniformity, but indicates the value of $n_v$ at which the correlation is most significant \citep{Pelgrims2016thesis}.

\subsection{S statistic distribution} \label{subSTestSDDist}

\citet{Jain2004} show that for large sample size $n$ the S test statistic $S_D$ follows a normal distribution while $n_v \ll n$. They find that for small $n$, and when $n_v$ approaches $n$, the $S_D$ peak shifts towards smaller values and deviates from normality. They perform numerical simulations of $n=213$ random PAs drawn from a uniform distribution. Their results are well fitted by $S_D \approx 0.89 / \sqrt{n_v}$ with a standard deviation of $\sigma \approx 0.33 / \sqrt{n}$ while $n_v \ll n$.

\begin{figure} % S_test_SD_dist_v2, figsize = (10., 6.), plt.hist(S_D_pt_2axial1... & plt.hist(S_D_pt_2axial2
    \centering
    \includegraphics[width=0.6\columnwidth]{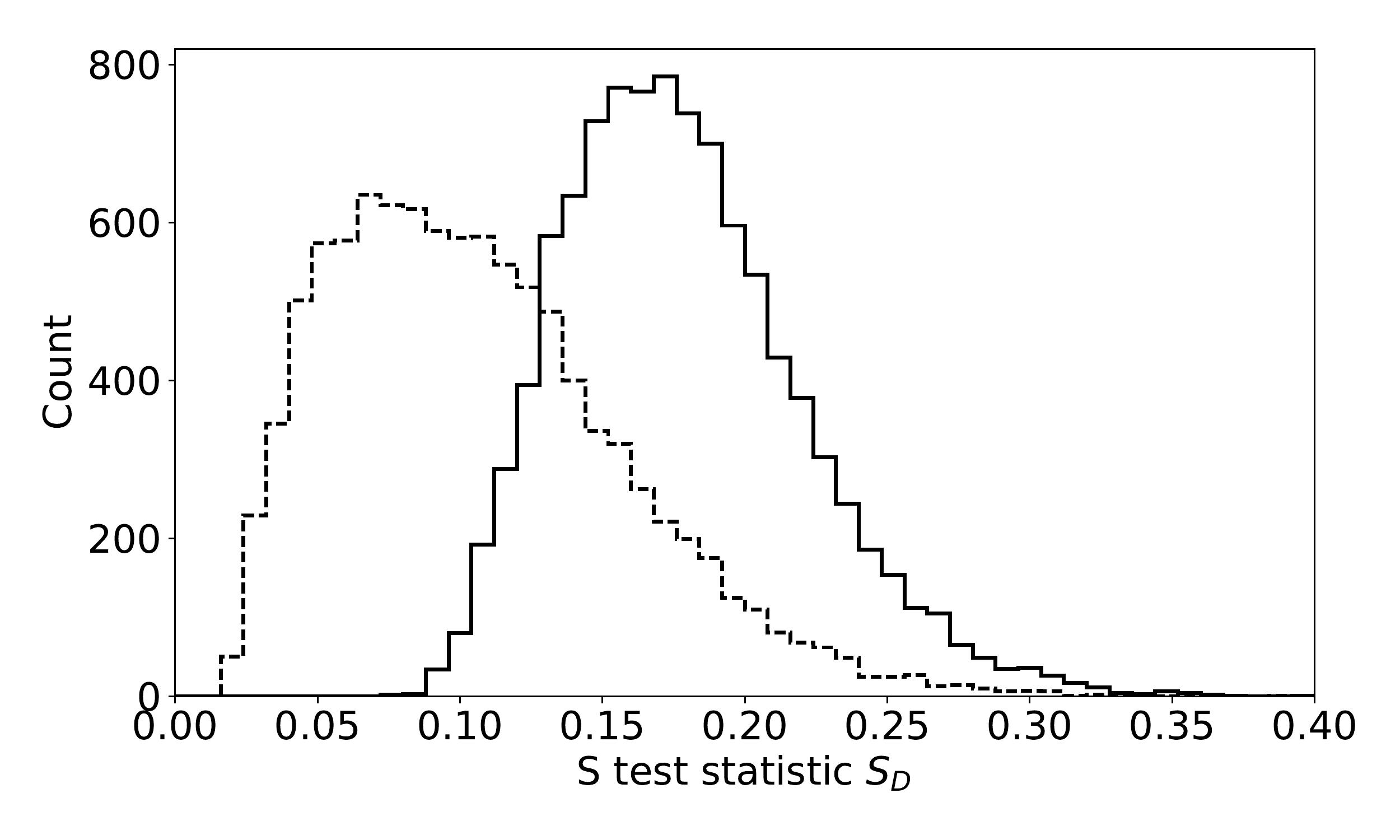}
    \caption[S test statistic distribution for different values of $n_v$]{The S test statistic distribution for 10,000 numerical simulations of $n = 71$ PAs drawn from a uniform distribution. $S_D$ is calculated using 2-axial PAs with parallel transport corrections, and nearest neighbours $n_v$ are determined in 3D. The $S_D$ distribution for $n_v = 25$ ($n_v = 70$) is shown by the solid (dashed) histogram. As $n_v$ increases, the distribution shifts to smaller values of $S_D$ and deviates further from normality.}
    \label{figSD_nv_25_70}
\end{figure}

Following \citet{Jain2004} we perform 10,000 numerical simulations of $n = 71$ random PAs, drawn from a uniform distribution, and confined to LQG locations. Due to our small sample size, and our exploration of a range of nearest neighbours up to $n \sim n_v$ ($10 \le n_v \le 70$), we find that the $S_D$ distribution is skewed from normality, leaving a longer tail of higher $S_D$ values. This is more extreme for higher $n_v$, when the peak of the distribution also shifts to smaller values of $S_D$ (Fig.~\ref{figSD_nv_25_70}). We therefore cannot estimate the mean and standard deviation using the empirical approximations of \citet{Jain2004} and need numerical simulations instead.

\begin{figure}% S_test_SD_dist_v2, figsize = (10., 6.), plt.hist(S_D_inc_pt_2axial..., plt.hist(S_D_inc_pt_4axial..., plt.axvline(x = obs_2axial_pt..., plt.axvline(x = obs_4axial_pt
    \centering
    \includegraphics[width=0.6\columnwidth]{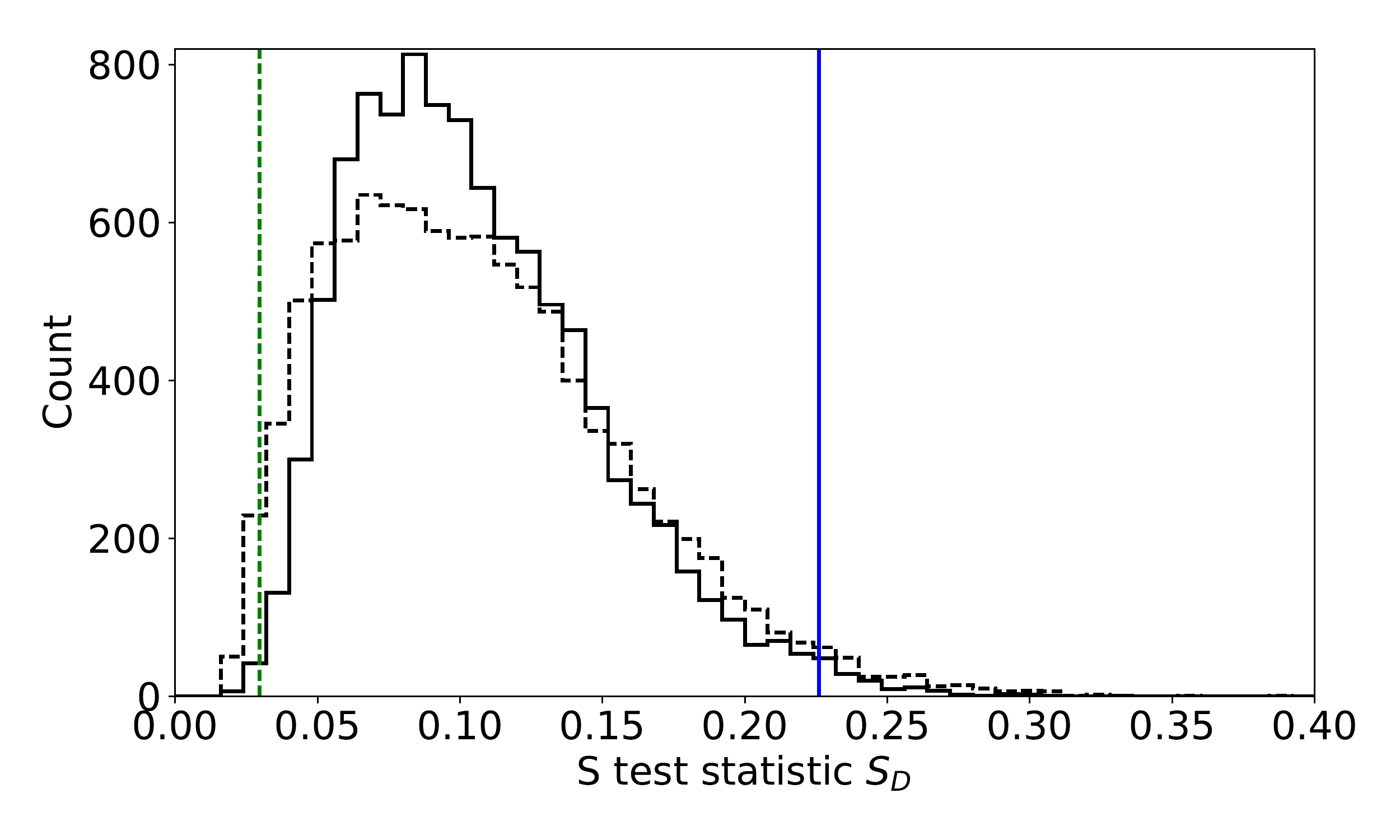}
    \caption[S test statistic distribution for 2-axial and 4-axial position angles]{As Fig.~\ref{figSD_nv_25_70} but for $n_v = 70$, with $S_D$ calculated using 2-axial (dashed) and 4-axial (solid) PAs, both with parallel transport corrections. The $S_D$ distributions of 2-axial and 4-axial numerical simulations for this $n_v$ differ. The vertical lines indicate the values of $S_D$ for our observed sample of 2-axial (dashed green; $\mathrm{SL} \sim 1.9\%$) and 4-axial (solid blue; $\mathrm{SL} \sim 1.2\%$) PAs; these are in the left and right tails of the distributions respectively. See text for interpretation.}
    \label{figSD_nv_70_obs}
\end{figure}

We apply the S test to both 2-axial and 4-axial LQG PAs. The numerical simulations for each of these constructs yields a different $S_D$ distribution (Fig.~\ref{figSD_nv_70_obs}), with the differences being greater for larger values of $n_v$. To calculate the significance level of the S test the values of $S_D$ for our observed sample of 2-axial (dashed green) and 4-axial (solid blue) LQG PAs is compared to the $S_D$ distributions of 2-axial (dashed) and 4-axial (solid) numerical simulations respectively, for each $n_v$.

Conventionally, a large value of the S test statistic $S_D$ indicates alignment. When we apply the S test to 4-axial LQG PAs we find that $S_D$ is in the right-hand tail of the simulated $S_D$ distribution (solid blue line in Fig.~\ref{figSD_nv_70_obs}). This is interpreted as 4-axial PAs being more aligned than random orientations. However, when we apply it to 2-axial LQG PAs we find the opposite, that $S_D$ is in the left-hand tail (dashed green line in Fig.~\ref{figSD_nv_70_obs}). Interpretation of this is less straightforward; how can we argue that 2-axial PAs are less aligned than random orientations?

A position angle distribution with two preferred orientations orthogonal to one another \citep[e.g.][]{Hutsemekers2014,Pelgrims2016thesis} will lead to a small value of $d_i\Bigr\rvert_{max}$ indicating large dispersion, and hence a small value of $S_D$. Therefore, we may interpret values of $S_D$ in the left-hand tail of the distribution as 2-axial PAs being more orthogonal than random orientations.

\subsection{Effect of parallel transport} \label{subSTestPT}

\begin{figure}
    \centering
	\subfigure[$n_v = 25$, 2-axial PAs, \newline SL = 14.6\% (29.3\%) with (out) PT \newline]{\includegraphics[width=0.4\columnwidth]{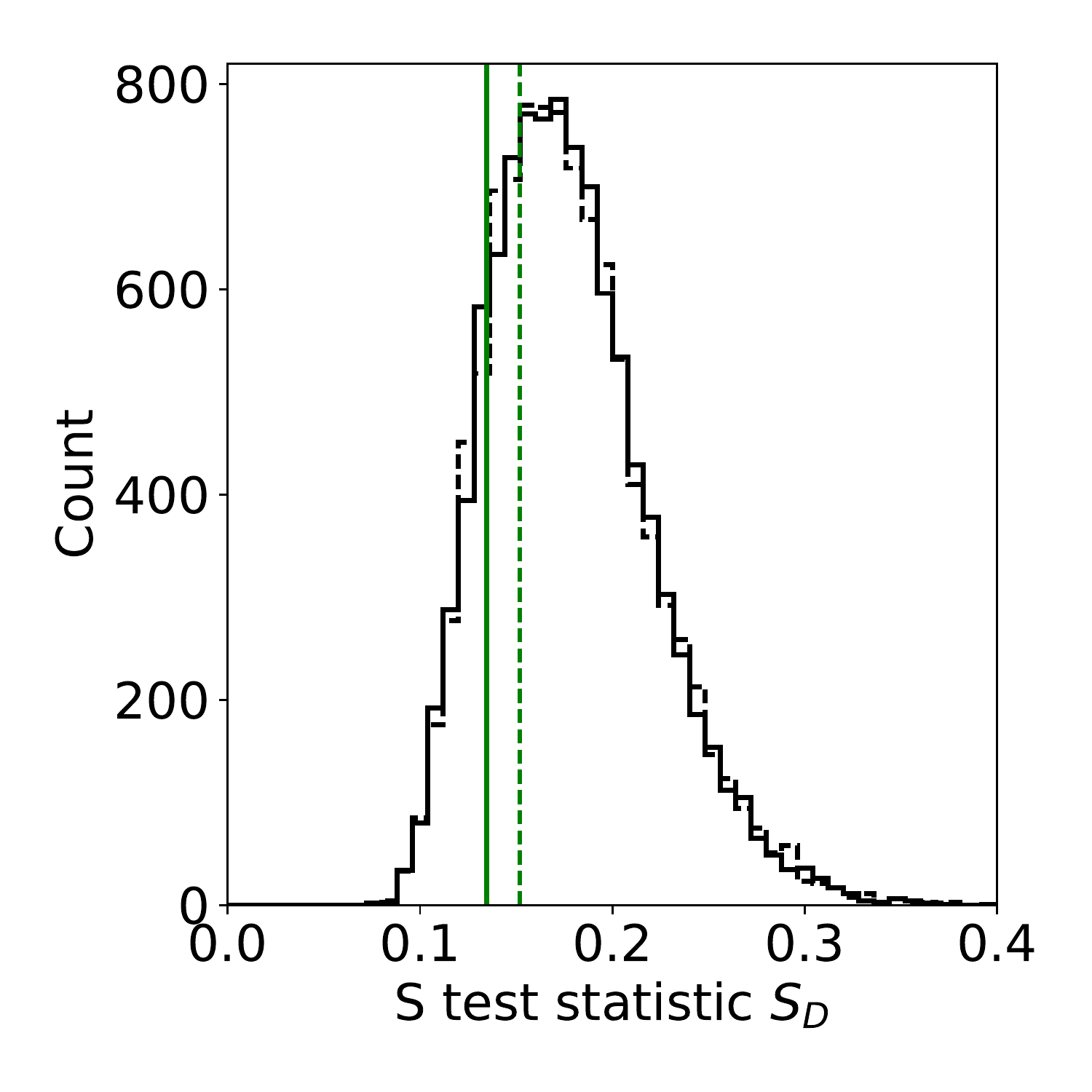} \label{sub2Ax25}}
	\subfigure[$n_v = 25$, 4-axial PAs, \newline SL = 2.6\% (1.0\%) with (out) PT \newline]{\includegraphics[width=0.4\columnwidth]{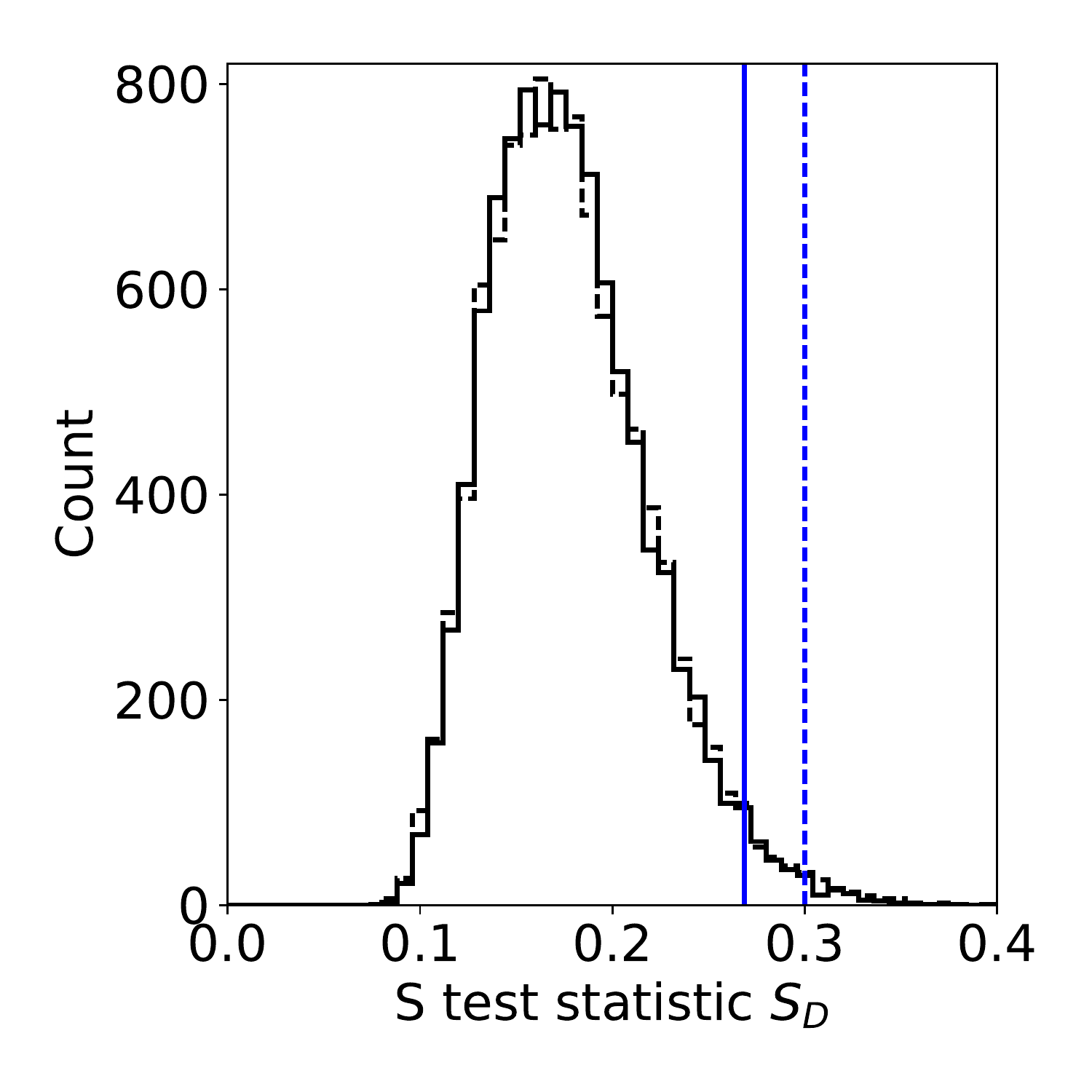} \label{sub4Ax25}} \\
	\subfigure[$n_v = 70$, 2-axial PAs, \newline SL = 1.9\% (24.0\%) with (out) PT]{\includegraphics[width=0.4\columnwidth]{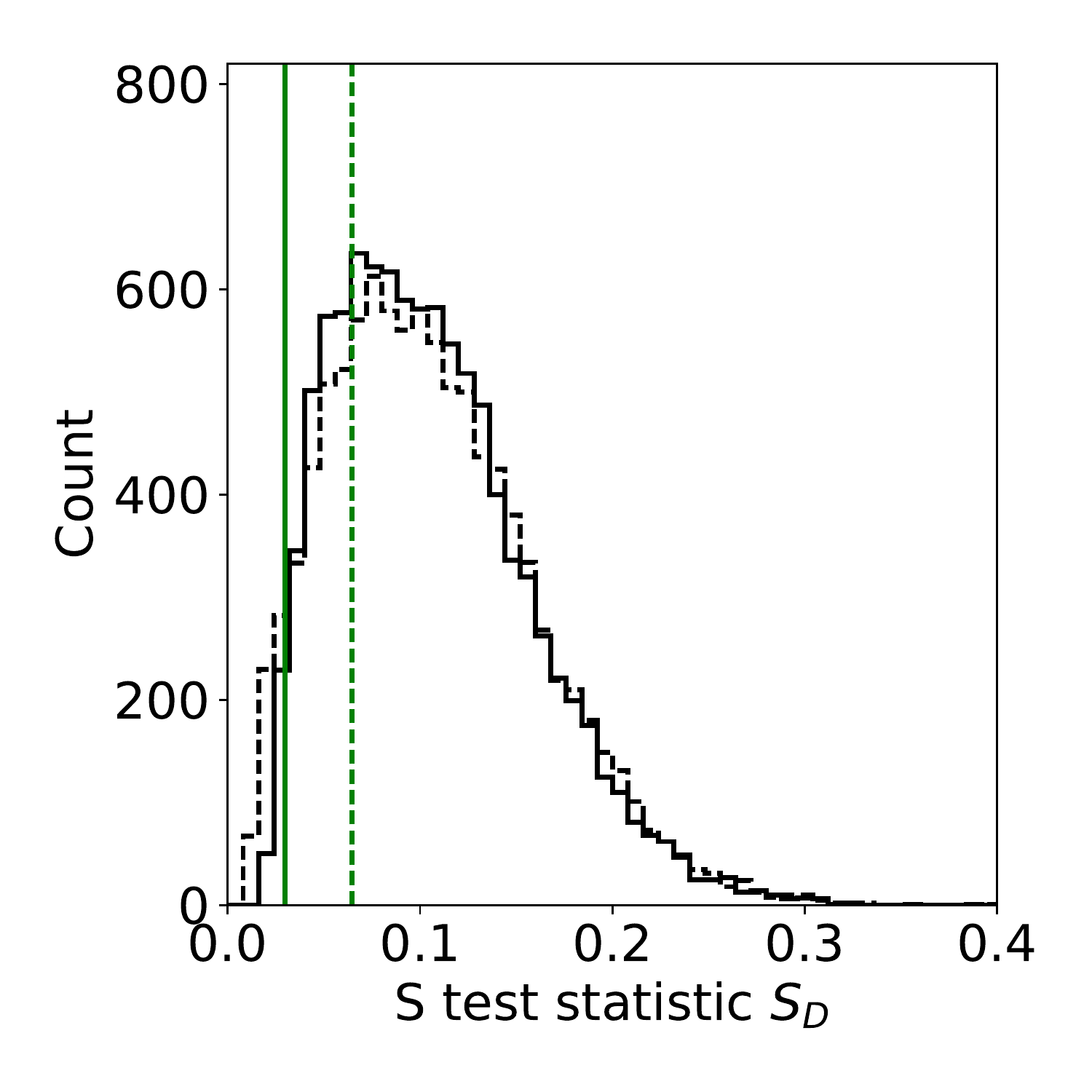} \label{sub2Ax70}}
	\subfigure[$n_v = 70$, 4-axial PAs, \newline SL = 1.2\% (0.4\%) with (out) PT]{\includegraphics[width=0.4\columnwidth]{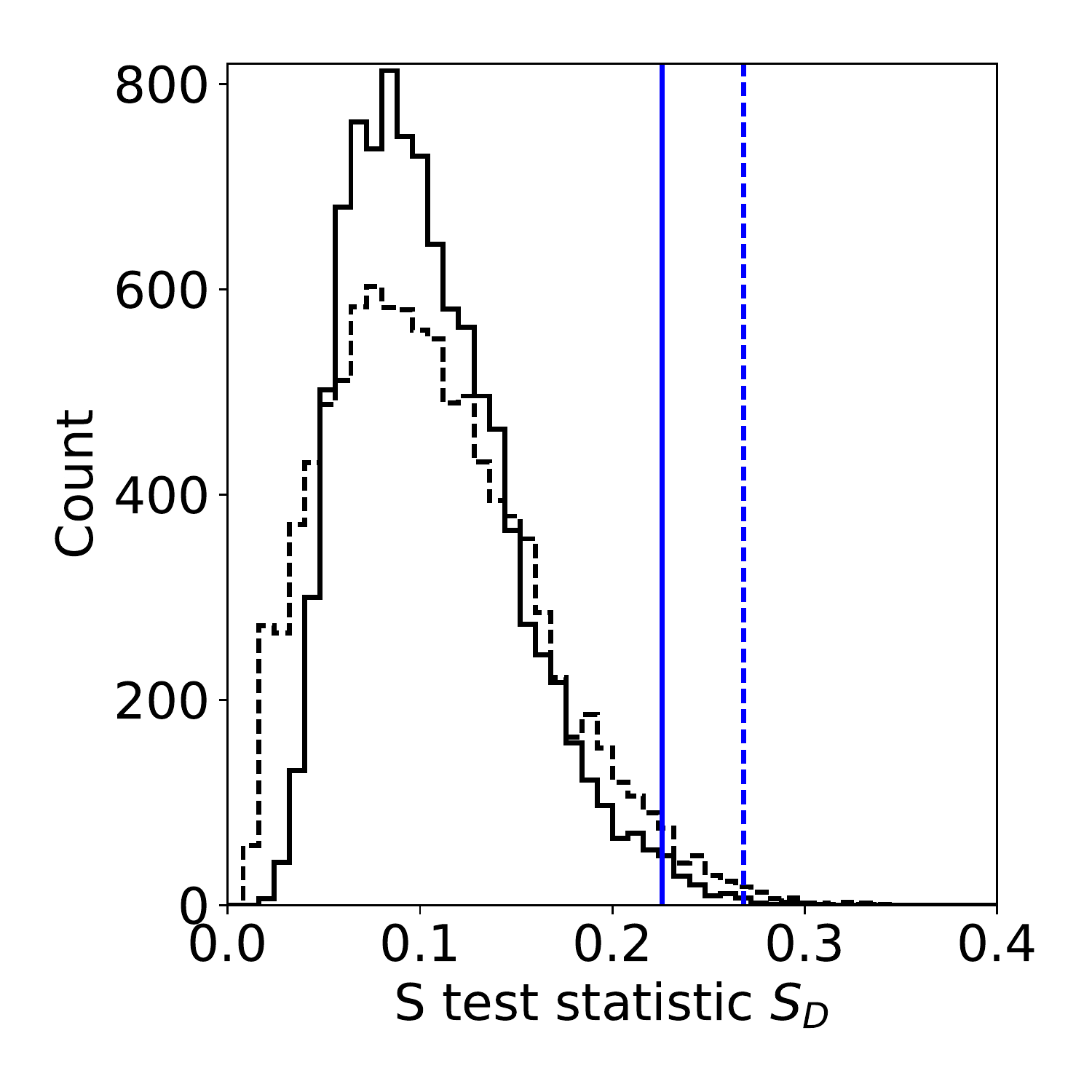} \label{sub4Ax70}}
	\caption[S test statistic distribution with and without parallel transport]{As Fig.~\ref{figSD_nv_25_70} but for  $n_v = 25$ and 70 (row 1 and 2 respectively) and 2-axial and 4-axial PAs (column 1 and 2 respectively). The histograms show the $S_D$ distributions of numerical simulations with (solid) and without (dashed) parallel transport corrections. For $n_v = 70$, particularly for 4-axial PAs, these distributions differ. The vertical lines indicate the values of $S_D$ for our observed sample of 2-axial (green) and 4-axial (blue) LQG PAs, with (solid) and without (dashed) parallel transport (PT) corrections; significance level (SL) of each is given in the caption. The extent to which observed $S_D$ values are in the tails of the distributions is affected by parallel transport.}
	\label{figSDTransport}
\end{figure}

The S test is a coordinate-invariant method to test for alignment of LQG PAs, when it includes parallel transport corrections. In this section we explain the effect of parallel transport on the $S_D$ distribution. Using $n_v = 25$ and 70, Fig.~\ref{figSDTransport} shows the $S_D$ distributions for 10,000 numerical simulations of $n = 71$ PAs randomly drawn from a uniform distribution, calculated using both 2-axial and 4-axial PAs. The solid and dashed histograms show the $S_D$ distributions when the S test includes and excludes parallel transport corrections, respectively. The vertical lines indicate the values of $S_D$ calculated for our observed sample of 2-axial (green) and 4-axial (blue) LQG PAs, both with (solid) and without (dashed) parallel transport corrections.

Considering first $n_v = 25$, the $S_D$ distribution for numerical simulations is similar, whether or not parallel transport corrections are included. For 2-axial PAs (Fig.~\ref{sub2Ax25}) parallel transport reduces the standard deviation of the distribution by $\sim 1\%$, although the mean is consistent (within $\sim 0.1\%$). For 4-axial PAs (Fig.~\ref{sub4Ax25}) the effect is similar ($\sim 4\%$ and $\sim 0.2\%$ respectively). The effect of applying parallel transport corrections to our observed sample of LQG PAs is indicated by the difference between the dashed and solid lines. For 2-axial PAs this increases the significance level of orthogonality from 0.29 to 0.15, and for 4-axial PAs it reduces the significance level of alignment from 0.01 to 0.03.

For $n_v = 70$, the effect of parallel transport is larger. For 2-axial PAs (Fig.~\ref{sub2Ax70}) it reduces the standard deviation of the distribution by $\sim 5\%$, although the mean is consistent (within $\sim 0.3\%$). For 4-axial PAs (Fig.~\ref{sub4Ax70}) it reduces the standard deviation by $\sim 20\%$, although the mean is similarly consistent (within $\sim 0.3\%$). The effect of applying parallel transport corrections to our observed sample of LQG PAs for $n_v = 70$ follows a similar pattern to $n_v = 25$. For 2-axial PAs this increases the significance level of orthogonality from 0.24 to 0.02, and for 4-axial PAs it reduces the significance level of alignment from 0.004 to 0.01.

\citet{Jain2004} report that, in most cases, their results are unchanged by including parallel transport corrections. An exception is if $n_v$ is comparable to $n$, when the $S_D$ distribution of the simulations becomes more sensitive to the precise location of the objects involved. Our results are more sensitive to parallel transport corrections, due to our smaller sample size and our exploration of nearest neighbours up to $n \sim n_v$. Note that as $n_v$ increases, the physical size of the groups of nearest neighbours will also generally increase so parallel transport corrections will be larger. This can be seen in the vertical lines of Fig.~\ref{figSDTransport}; the difference between the values of $S_D$ calculated with (solid) and without (dashed) parallel transport corrections is larger for $n_v = 70$ than $n_v = 25$.

\subsection{Numerical simulations} \label{subSTestRandoms}

The S test yields an $S_D$ statistic for each number of nearest neighbours $n_v$ assessed. The significance level of these values compared to randomness cannot be evaluated analytically, due to: overlaps between groups of nearest neighbours, deviation of the $S_D$ distribution from normality (particularly for small $n$ and when $n_v \sim n$), and the dependence of parallel transport corrections on the precise location of the objects involved. Therefore, numerical simulations are required.

These simulations preserve the three-dimensional locations of the observed LQGs (right ascension, declination, and redshift) and replace observed PAs with `random' PAs at these positions.

\citet{Hutsemekers1998} and \citet{Hutsemekers2001} generate random samples by shuffling the observed PAs randomly between objects, whilst keeping their positions fixed. This has the effect of erasing any correlation between PAs and positions. \citet{Jain2004} enhance this method to make it coordinate-invariant, by including parallel transport corrections during each shuffle. However, there is some ambiguity about which path(s) the parallel transport takes, each of which can lead to different $S_D$ distributions \citep{Pelgrims2016thesis}. Also, as \citet{Jain2004} note, the shuffling method is unable to test for global alignment.

We have no reason to expect LQG PAs to be correlated with LQG positions; alignment could be global. Therefore, we generate random samples from a uniform distribution, again keeping LQG positions fixed. We generate $n = 71$ PAs, randomly drawn from a uniform distribution. These are generated in the ranges [$0^{\circ}$, $180^{\circ}$) for 2-axial PAs and [$0^{\circ}$, $90^{\circ}$) for 4-axial PAs. Note that for 4-axial PAs, initially generating PAs in the range [$0^{\circ}$, $180^{\circ}$), then converting to [$0^{\circ}$, $90^{\circ}$) after parallel transport and before the S test calculation is equivalent.

The significance level (\gls{SL}) of the S test is defined as the percentage of simulations that have an $S_D$ statistic at least as extreme as the one from our observations \citep{Pelgrims2016thesis}. It is computed by comparing the statistic of our observations ($S_{D, obs}$) with the statistic from a large number of numerical simulations ($S_{D, sim}$). 

When we convert 2-axial PAs to 4-axial, we simultaneously test for alignment and orthogonality by combining the modes, so an `alignment plus orthogonality' signal will manifest as alignment only. The SL is therefore the proportion of numerical simulations with statistic $S_D$ higher than that observed. Indeed, for 4-axial PAs we find that $S_{D, obs}$ is in the high $S_{D, sim}$ (right-hand) tail of the distribution (Fig.~\ref{figSD_nv_70_obs}), with the SL being the probability that $S_{D, sim} > S_{D, obs}$. Note that an `alignment only' signal would also be in the right-hand tail, but would be differentiated by its 2-axial result.

For 2-axial PAs we find that $S_{D, obs}$ is in the low $S_{D, sim}$ (left-hand) tail of the distribution. To interpret this, consider the potential bimodality of the LQG PA distribution, with the peaks separated by $\sim 90^{\circ}$. This orthogonality leads to a small value of $d_i\Bigr\rvert_{max}$, indicating large dispersion, and yielding a small value of $S_D$. Therefore, it is legitimate to interpret a result in the left-hand tail as an orthogonal signal, with the SL being the probability that $S_{D, sim} < S_{D, obs}$. Note that the two modes of a combined `alignment plus orthogonality' signal would tend to erase any signal, reducing the power of this test. However, a residual signal in the left (right) tail indicates the orthogonality (alignment) mode dominates.

In terms of interpretation, the SL is the likelihood that the statistic from our observations is a statistical fluke drawn at random from a uniform distribution. For 4-axial PAs this is the probability of a random sample being more aligned than that observed, and for 2-axial PAs it is the probability of a random sample being more orthogonal than that observed.

The $S_D$ distributions of 2-axial and 4-axial numerical simulations differ, particularly as $n_v$ approaches $n$. Therefore, we generate separate numerical simulations for each, and calculate the significance levels as

\begin{equation} \label{eqSL2ax}
    \mathrm{SL_{2ax}} = P(S_{D, sim(2ax)} < S_{D, obs(2ax)})\ ,
\end{equation}

\begin{equation} \label{eqSL4ax}
    \mathrm{SL_{4ax}} = P(S_{D, sim(4ax)} > S_{D, obs(4ax)})\ ,
\end{equation}

\noindent where $P$ indicates probability, $2ax$ and $4ax$ indicate 2-axial and 4-axial PAs, and $sim$ and $obs$ indicate simulations and observations respectively.

\section{Mock LQG catalogues} \label{secMockLQGs}

The methods detailed so far give a thorough statistical evaluation of the LQG PA distribution, but cannot inform about its compatibility with the cosmic web expected in the concordance $\Lambda$CDM cosmological model. We assess this using mock LQG catalogues constructed by \citet{Marinello2016} from the Horizon Run 2 (\gls{HR2}) dark matter only $N$-body cosmological simulation \citep{Kim2011}. HR2 was the largest ($7200h^{-1}\;\mathrm{Mpc}$ box) simulation available with a mass resolution sufficient to reproduce the spatial statistics of the observed quasar distribution.

\citet{Marinello2016} take a snapshot of HR2 at redshift $z = 1.4$, divide the volume into 11 sub-volumes, then create quasar samples by applying a semi-empirical halo occupation distribution (\gls{HOD}) model to each. The HOD model is probabilistic, and can produce separate quasar realizations from the same haloes. They apply the model 10 times to each of the 11 sub-volumes, producing a total of 110 mock quasar catalogues. Then they use the LQG finder \citep[section~\ref{subsecIDLQGs};][]{Clowes2012,Clowes2013} to construct 110 mock LQG catalogues.

We restrict each of these mock catalogues to LQGs with membership $m \geq 20$ and significance $\geq 2.8 \sigma$, to match our observed LQG sample. The mean number of LQGs in our mocks is $\bar{n} = 30\pm 0.4$; numbers in individual mocks vary $20\leq n \leq 42$. The total number of LQGs in all 110 mocks is 3,296.

The sample size of each individual mock catalogue is small, so we stack them to increase the statistical power. The 10 quasar mock catalogues created from each sub-volume are not truly independent; each HOD model realization samples the same set of dark matter haloes. Indeed, \citet{Marinello2015thesis} find that $\sim 99\%$ of LQG regions are detected in at least two realizations, albeit often with different quasars, although $\sim 18\%$ of those are detected with the same quasars. Therefore, for each realization we stack the 11 sub-volumes, which are independent. The mean number of LQGs in each stack is $\bar{n} = 330\pm 3$.

Stacking could average out any LQG correlation, unless it is present throughout the entire HR2 volume. This would make it difficult to differentiate between the null hypothesis of uniformity and the existence of a correlation on scales smaller than HR2. Since we have no \emph{a priori} reason to expect `global' LQG orientations, this is a potential drawback.

Another potential deficiency is use of the quasar duty cycle in the HOD model, since it is unknown whether LQG quasars exhibit a similar duty cycle to other quasars ($0.1-1\%$). \citet{Marinello2016} choose a duty cycle which reproduces the spatial distribution of quasars in SDSS DR7QSO \citep{Marinello2015thesis}, such that their quasar mocks are statistically compatible with observations. Further, \citet{Marinello2016} find their mock LQGs are compatible with their observed LQG sample. For our purposes, we conclude that their choice of duty cycle is reasonable.

We calculate position angles for mock LQGs as for our observed sample, including applying parallel transport corrections, and analyse the distributions of both 2-axial and 4-axial PAs.

\section{Discussion and conclusions} \label{secLQGStatsMethodsConclusions}

Our LQG position angle sample is a challenging dataset: the data are axial, bimodal, and initially coordinate-dependent. The latter factor is particularly significant when the objects under consideration are widely distributed on the celestial sphere, such as our LQGs.

We use parallel transport to ensure our methods are coordinate invariant. For all analysis of uniformity and bimodality we parallel transport all PAs to a single point on the celestial sphere before analysis. For analysis of alignment correlation we directly incorporate parallel transport corrections into the method.

Parallel transport could introduce ambiguity into the tests, since the corrections depend on the direction and distance of the path taken. \citet{Pelgrims2016thesis} recognised this is an issue for the Z test, where there are two possible paths, although they show the difference is small. Another potential parallel transport issue is when generating random samples. If we had chosen to generate these by shuffling the observed PAs randomly between objects \citep{Hutsemekers1998,Hutsemekers2001} then again there are a choice of paths and potential ambiguity. Instead, we generate numerical simulations from a uniform distribution, keeping LQG positions fixed, which requires no additional parallel transport corrections.

When we parallel transport all PAs to a single point on the celestial sphere, however well motivated, the choice of destination obviously determines the path taken. Using different destinations, we show that our results are robust to this choice (section~\ref{secHDSResults}).

\subsection{Test availability and appropriateness} \label{secStatsAppropriateness}

In this chapter we described and evaluated various statistical tests of uniformity, bimodality, and correlation. In this section we assess their availability and appropriateness, specifically for axial and bimodal position angle data.

\subsubsection{Uniformity tests}

To analyse uniformity we use Kuiper's test, the Hermans-Rasson (HR) test, and the $\chi^2$ test. Kuiper's and $\chi^2$ tests are widely used in astrophysics, and are provided in Python libraries. These libraries do not support weighting so we create bespoke versions of these tests to also analyse goodness-of-fit weighted data. We do not find the HR test in a Python library, and to the best of our knowledge it has not previously been used in astrophysics, so in this work we introduce it for the first time.

Kuiper's test is appropriate for axial PA data, but will have its discriminatory power reduced if the distribution is multimodal, as it may be for our data. The HR test is suitable for PA data converted to circular data, and should have higher power than Kuiper's test in case of a potentially multimodal distribution. The $\chi^2$ test will have lower discriminatory power compared with tests applied to continuous data (e.g.\ Kuiper's and HR tests) and is less appropriate for axial or circular data, which may further reduce its power. The $\chi^2$ test is included predominantly due to its familiarity and its ease of computation.

We recommend the HR test as the most appropriate for our LQG PA data.

\subsubsection{Bimodality tests}

To test for bimodality we use the bimodality coefficient (BC) and Hartigans' dip statistic (HDS), both of which are used in astrophysics. The BC is calculated from the skewness and kurtosis of histograms of categorical data, both of which are available in Python libraries. HDS is also available in Python.

BC gives a simple measure of whether or not the PA data are drawn from a bimodal distribution. However, it does not provide any estimate of significance. We note further limitations of this test, particularly its undesirable sensitivity to skewness and bin size. HDS is more robust and can be used for hypothesis testing. It also quantifies the location of any peaks.

We recommend HDS as the most robust and informative for our LQG PA data.

\subsubsection{Correlation tests}

To evaluate the correlation of LQG PA alignments we use the S and Z tests. Both tests have been used to analyse alignments on the celestial sphere, but as far as we know are not publicly available in any programming language. We therefore create Python implementations of these.

The S and Z tests are both, by design, appropriate for axial data. However, their interpretation for bimodal data, particularly when this comprises alignment and orthogonality, is not straightforward. For 2-axial data this would tend to erase any signal, reducing the power of the test, whilst for 4-axial data it could combine the signals (if the PA distribution is 2-fold symmetric), increasing its power. The conventional interpretation is that an observed statistic in the right-hand (high $S_D$) tail of the distribution indicates alignment. We further argue that an observed statistic in the left-hand (low $S_D$) tail indicates orthogonality. This is a new interpretation introduced in this work.

\citet{Hutsemekers2014} report the Z test is better suited to small samples than the S test, because the latter uses a measure of angle dispersion which suffers reduced power with small samples. However, if PA alignment is `global' (i.e.\ correlations are present throughout the survey area), then PAs will not be correlated to positions and the power of the Z test will reduce dramatically (Pelgrims 2019, private communication).

We try both the modified Z test \citep{Hutsemekers1998}, together with the parallel transport corrections, and the \citet{Jain2004} version of the S test. We do not find a signal with the former, probably because the alignments are indeed global rather than local, but we find the S test is appropriate for our data, indicating both alignment and orthogonality.

In section~\ref{secSTestApplying} we discuss some subtleties in the application and interpretation of the S test. We choose to identify the $n_v$ nearest neighbours in three-dimensional comoving space rather than two-dimensional angular separation. We show how the value of $n_v$, and the use of 2-axial or 4-axial data, affects the distribution of the S test statistic $S_D$, and similarly the effects of parallel transport. These factors must all be accounted for when constructing numerical simulations to estimate the significance level of the S test.

% END

%% file: chapterLQGResults.tex
% START

\chapter{Evidence for large-scale correlation of LQG orientations} \label{chaptLQGOrientation}

In this chapter we determine the orientation of large quasar groups, as position angles. We present a sample of 71 LQGs whose PA distribution is apparently bimodal, with modes at $\bar{\theta}\sim52\pm2^{\circ}, 137\pm3^{\circ}$ (with goodness-of-fit weighting) at the centre of the A1 region \citep{Hutsemekers1998}. The median location of the peaks at all 71 LQG locations is $\bar{\theta}\sim45\pm2^{\circ}, 136\pm2^{\circ}$. The peaks are separated by $\Delta\theta\sim90^{\circ}$, indicating that some LQGs have PAs that are preferentially aligned with each other, whilst others are preferentially orthogonal. To the best of our knowledge nothing in the LQG finder could introduce such an aligned plus orthogonal effect. Indeed, it makes no assumptions about LQG morphology or orientation. We apply the statistical methods detailed in chapter~\ref{chaptLQGStatsMethods} to analyse the uniformity, bimodality, and correlation of these PAs.

\textbf{Uniformity}. We find very marginal evidence for non-uniformity of 2-axial PAs, but with 4-axial PAs the evidence is more intriguing ($0.8\% - 7\%$ significance level, $1.5\sigma - 2.4\sigma$). It is not legitimate to convert PA data to 4-axial based on its \textit{a posteriori} near 2-fold symmetry. However, we have \textit{a priori} motivation for suspecting this type of distribution from \cite{Pelgrims2015}, who identify large-scale alignment of the polarization vectors of radio quasars. They report mean angles\footnote{Errors on these means were not reported} of $\bar{\theta}\simeq42^{\circ}, 131^{\circ}$, remarkably close to our PA peaks.

\textbf{Bimodality}. Our analyses indicate that the PA data are bimodal. We find the bimodality coefficient is rather unstable, but for data weighted by goodness-of-fit it generally indicates bimodality. Hartigans' dip statistic is more robust and indicates bimodality with 97\% confidence. Furthermore, it identifies modes (peaks) at $\bar{\theta}\sim52\pm2^{\circ}, 137\pm3^{\circ}$ in PAs parallel transported to the centre of the A1 region.

\textbf{Correlation}. We assess PA correlation using the S test, and compare our observations with numerical simulations to estimate the significance level (SL) of any alignment/orthogonality. On most scales we find 4-axial PAs are more aligned than randoms ($\overline{\mathrm{SL}} = 1.5\%, 2.2\sigma$), and on the largest scales 2-axial PAs are more orthogonal than randoms ($\overline{\mathrm{SL}} = 3.3\%, 1.8\sigma$). We note that the S test, being based on dispersion, is likely to have limited power with small samples.

\section{Results: LQG position angles} \label{LQGPAResults}

\begin{figure} [t] % LQG_plot_ortho
	\centering
    \includegraphics[width=\columnwidth]{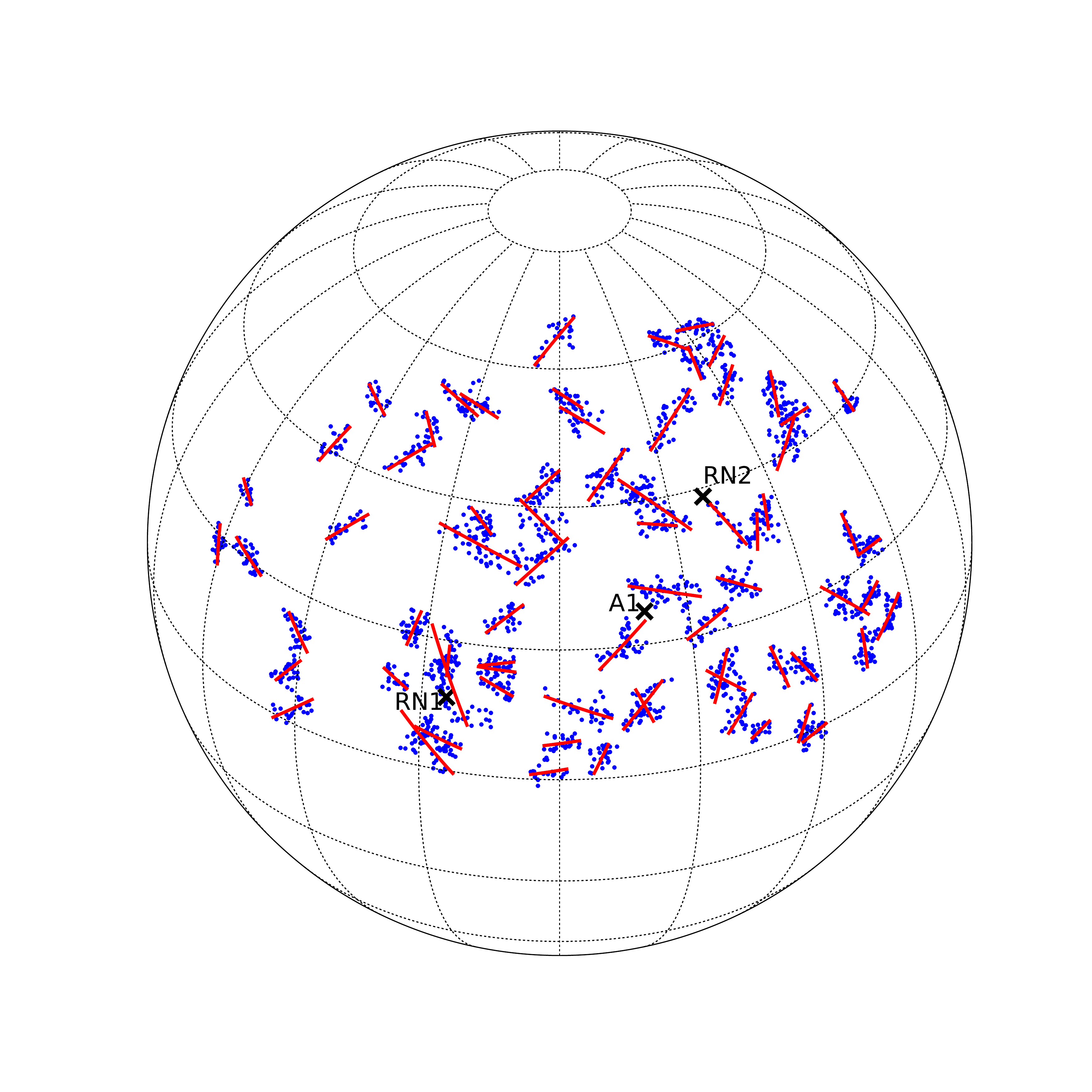}
    \caption[LQGs and their position angles on the celestial sphere]{LQG quasars (blue dots) and ODR axes (red lines) shown on the celestial sphere (east to the right). Also shown, the centres (black crosses) of A1, RN1, and RN2 regions \citep{Hutsemekers1998,Pelgrims2015}. A1 is the parallel transport destination for uniformity and bimodality tests. Projection is centred on $\alpha = 180^{\circ}$, $\delta = 35^{\circ}$ (J2000), parallels and meridians are separated by $20^{\circ}$. Some alignment is apparent by eye; we analyse this statistically in subsequent sections.}
    \label{figLQGsSphere}
\end{figure}

\begin{figure} [h] % LQG_plot_3D
    \centering
    \includegraphics[width=0.9\columnwidth]{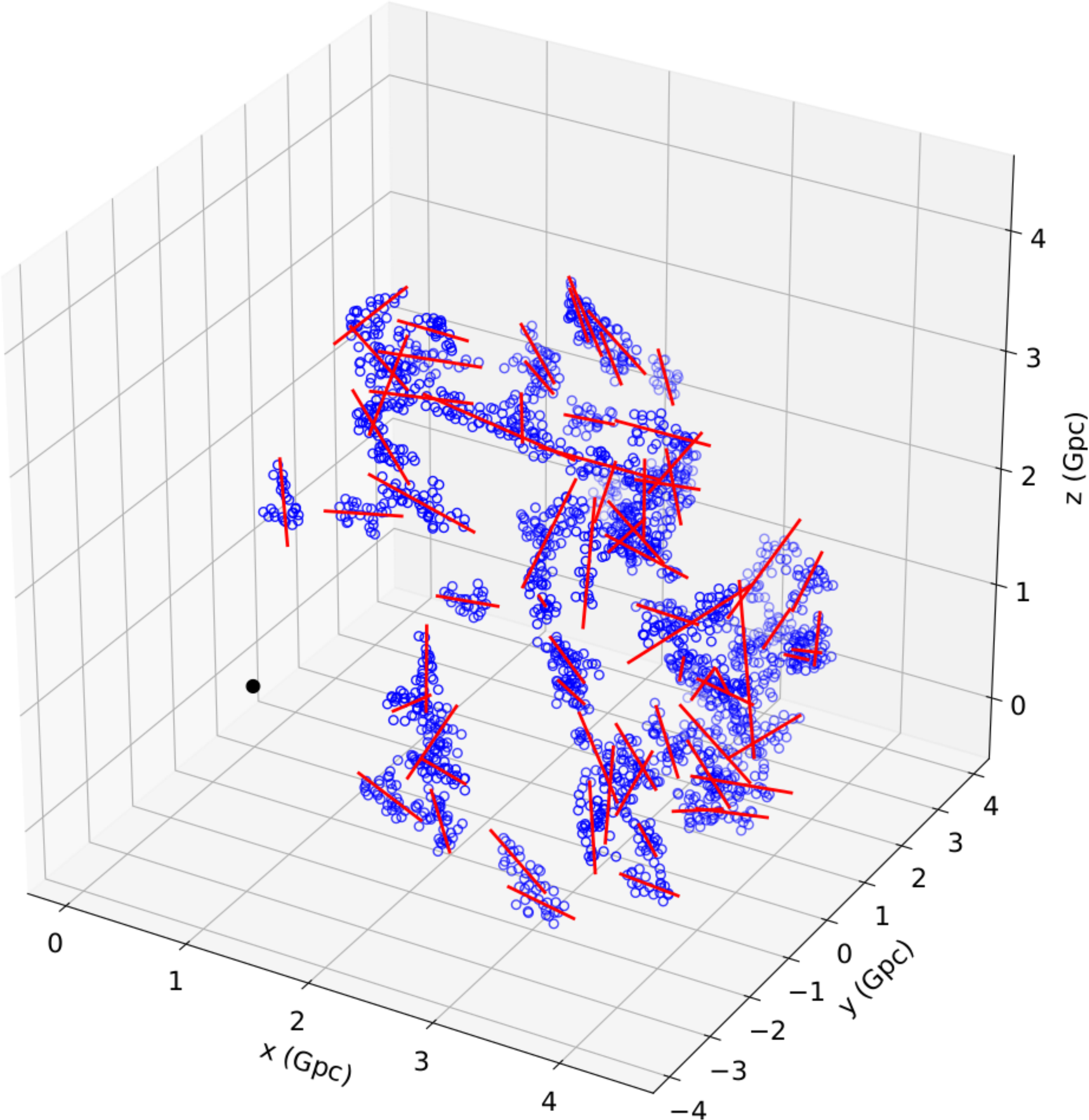}
    \caption[LQGs and their major axes in three-dimensional comoving space]{LQGs in three-dimensional comoving space, showing LQG quasars (blue circles) and LQG major axes (red lines). Quasar markers are shaded to give the appearance of depth, with lighter shades representing more distant quasars. Our location at ($x, y, z$) coordinates (0, 0, 0) is indicated by a black dot. See Appendix~\ref{appLQGs3DPerspectives} for three alternative, orthogonal perspectives.}
    \label{figLQGs3D}
\end{figure}

We identify 71 LQGs of $\ge 20$ quasars and detection significance $\ge 2.8\sigma$. The LQG positions on the celestial sphere, and their orientation as determined by the two-dimensional method, are illustrated in Fig.~\ref{figLQGsSphere}. By eye it appears that the orientations may be somewhat preferentially aligned, but we caution that the orthographic projection may be deceiving. We analyse this potential alignment in subsequent sections. The LQG positions in three-dimensional comoving space, and their orientation as determined by the three-dimensional method, are illustrated in Fig.~\ref{figLQGs3D} with additional perspectives shown in Fig.~\ref{figLQGs3DAspectsPt1} (Appendix~\ref{appLQGs3DPerspectives}).

\clearpage

The orientation of each LQG is quantified as a position angle. We consider two potential methods of determining PAs. To recap, the 2D approach involves tangent plane projection of the LQG quasars, followed by orthogonal distance regression (ODR) of the projected points. These ODR fits are indicated by the solid red lines in Fig.~\ref{figLQGsSphere}. The 3D approach requires determining the comoving coordinates of the LQG quasars, performing principal component analysis on the covariance matrix of these, then tangent plane projection of the resultant major axis. These major axes are indicated by the solid red lines in Figs.~\ref{figLQGs3D} and \ref{figLQGs3DAspectsPt1}. For both approaches, bootstrap re-sampling with replacement is used to estimate the uncertainty in the form of the half-width confidence interval (HWCI) of 10,000 bootstraps.

We analyse PAs determined by the 2D approach in this work, predominantly due to the potential influence of redshift errors, and the increased measurement uncertainties of the 3D approach. However, data from the 3D approach are used for some preliminary analysis of the morphology of LQGs. The results of both approaches are presented in Tables~\ref{tabLQGDataPart} (example LQGs) and \ref{tabLQGDataFull} (full LQG sample, in Appendix~\ref{appLQGData}) and generally agree well. The PAs listed in these tables are measured in situ at the location of each LQG and will have parallel transport corrections applied before statistical analysis.

For methods where we parallel transport all PAs to a single point on the celestial sphere to compare them, we choose the point indicated by a black cross in Fig.~\ref{figLQGsSphere} ($\alpha = 193.1^{\circ}$, $\delta = 24.7^{\circ}$, J2000). This is the centre of the A1 region \citep{Hutsemekers1998} of large-scale alignment of the optical polarization of quasars, which is roughly at the centroid of the LQG distribution.

\begin{figure}[t] % plot_PAs
    \centering
	\subfigure[Unweighted]{\includegraphics[width=0.48\columnwidth]{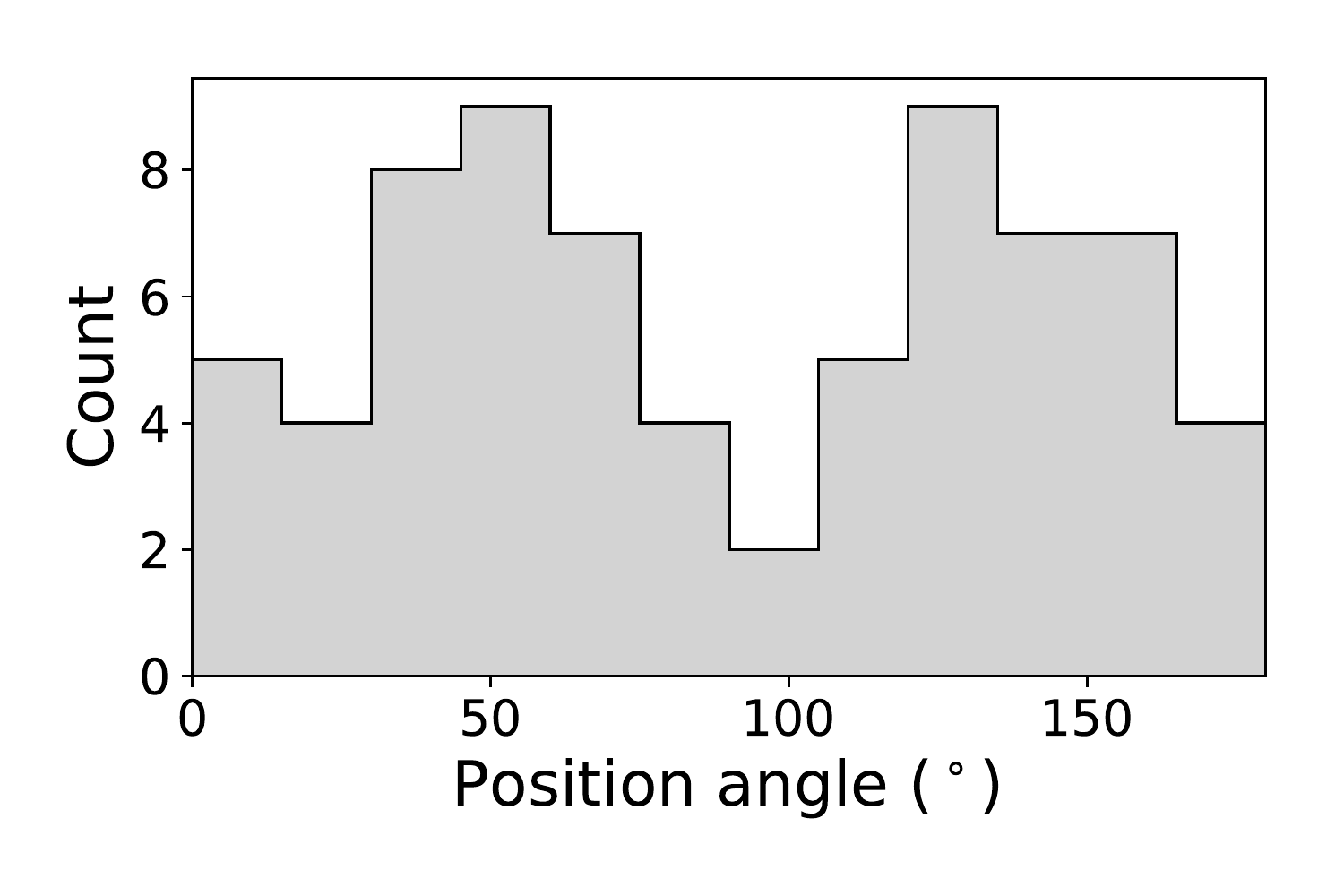} \label{subHistUnweighted}}
	\subfigure[Weighted]{\includegraphics[width=0.48\columnwidth]{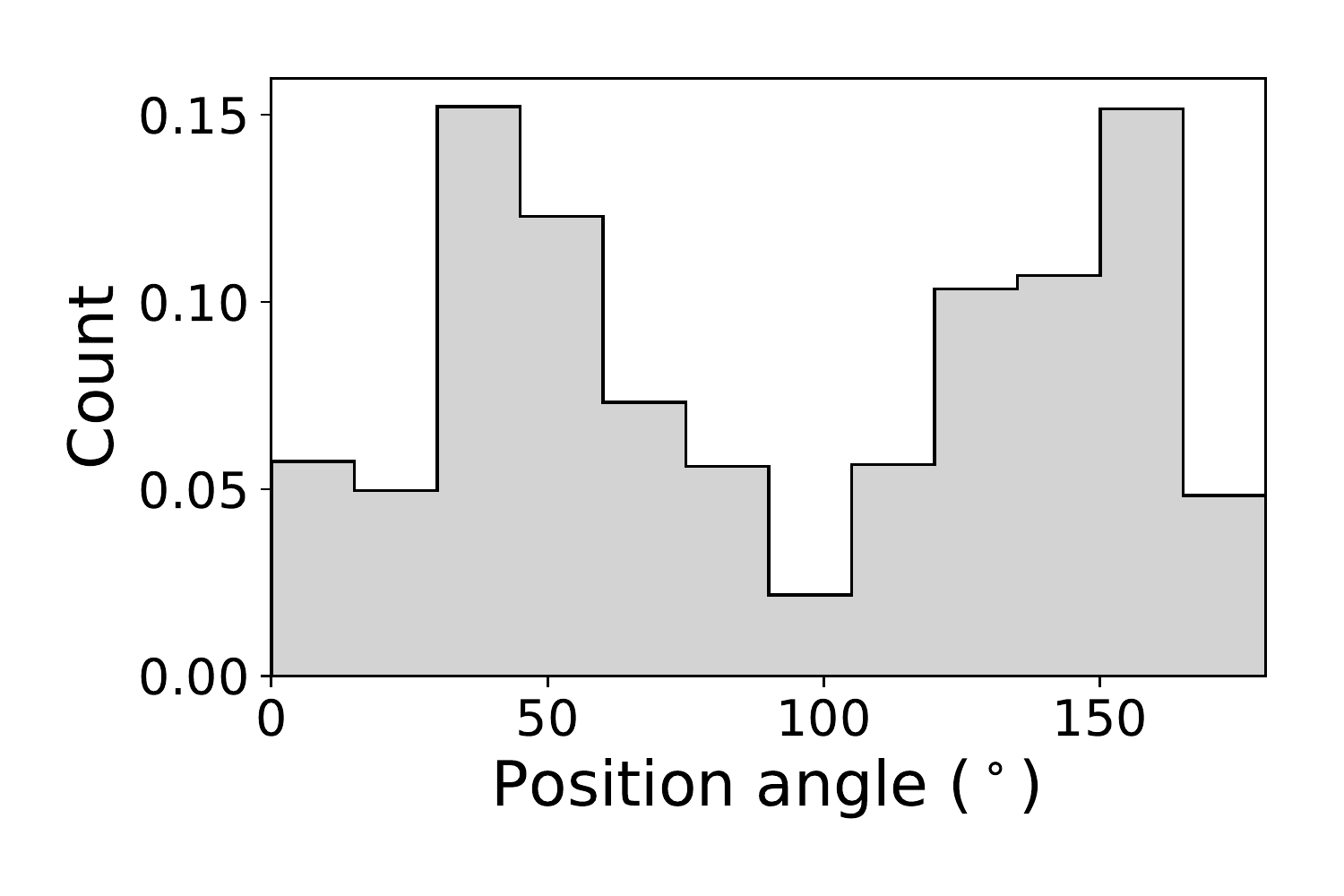} \label{subHistWeighted}}
	\caption[Position angles with and without weighting (histogram)]{LQG position angles, all parallel transported to and measured at the centre of the A1 region \citep{Hutsemekers1998}. \protect\subref{subHistUnweighted} is unweighted and \protect\subref{subHistWeighted} is ODR goodness-of-fit weighted, both with $15^{\circ}$ bins. The bimodal distribution is robust to whether or not ODR goodness-of-fit weighting is used.}
	\label{figPAHist}
\end{figure}

\begin{figure}[h] % plot_PAs
    \centering
	\subfigure[Unweighted]{\includegraphics[width=0.48\columnwidth]{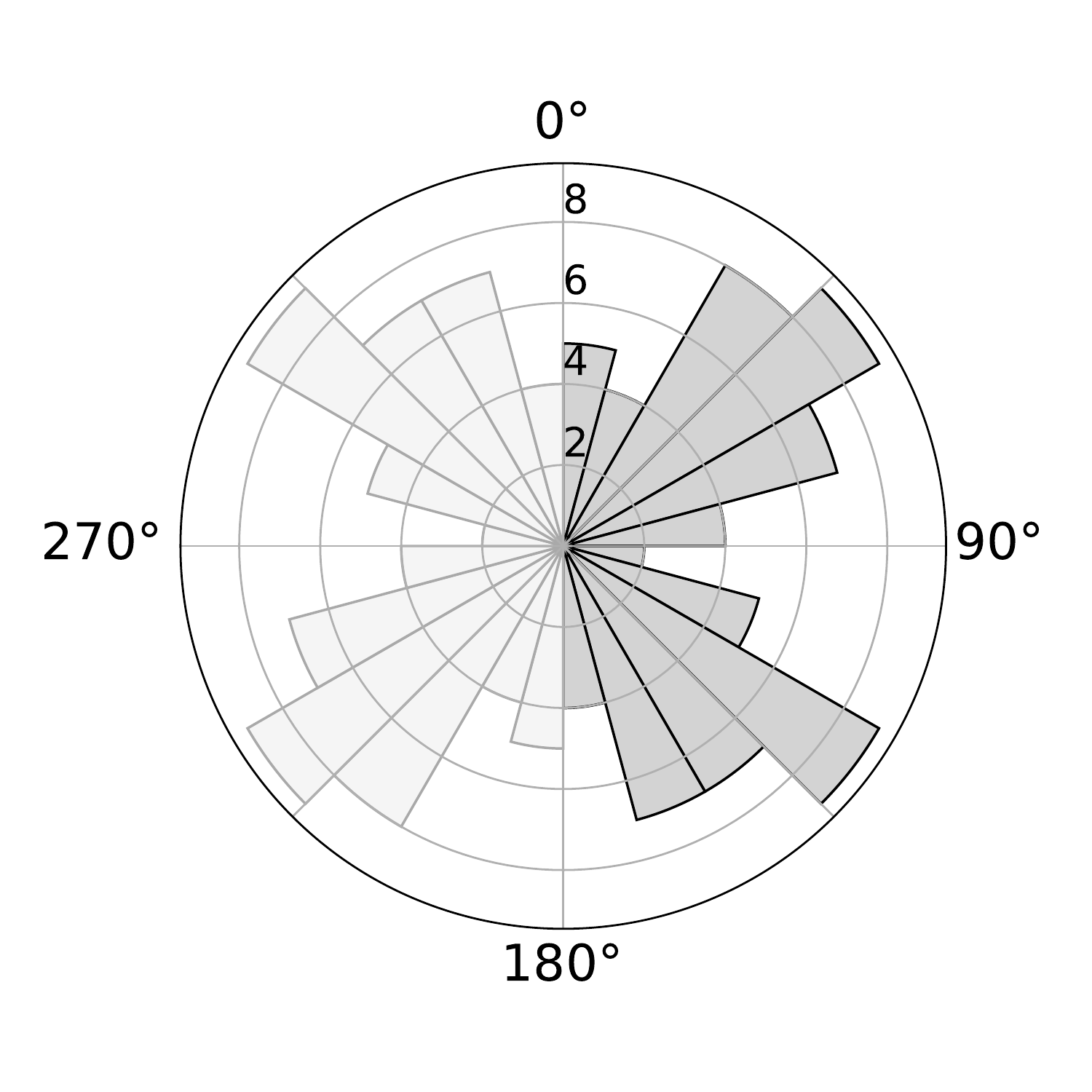} \label{subRoseUnweighted}}
	\subfigure[Weighted]{\includegraphics[width=0.48\columnwidth]{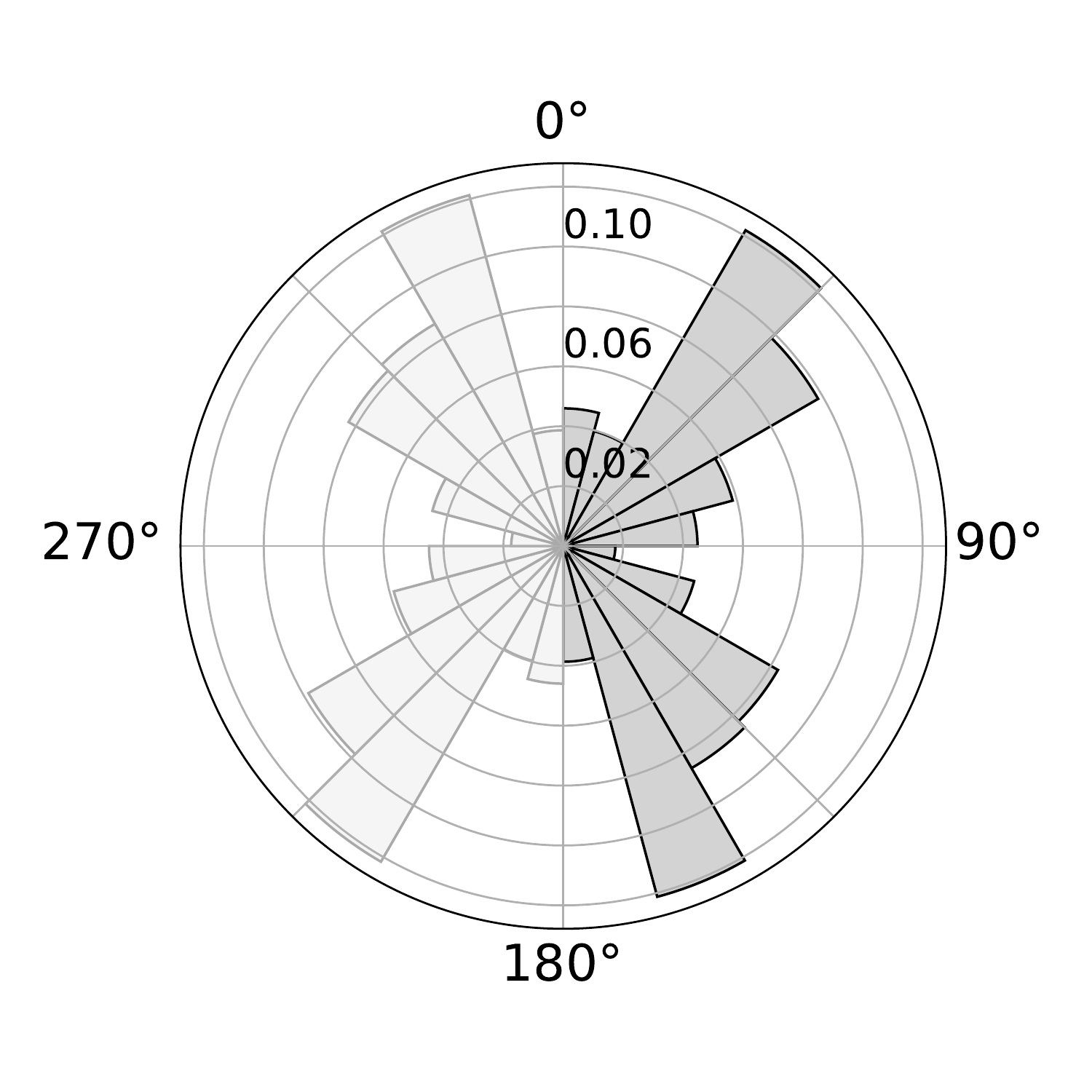} \label{subRoseWeighted}}
	\caption[Position angles with and without weighting (rose diagram)]{As Fig.~\ref{figPAHist} but represented as a rose diagram; again \protect\subref{subRoseUnweighted} is unweighted and \protect\subref{subRoseWeighted} is ODR goodness-of-fit weighted, both with $15^{\circ}$ bins. As is conventional for axial data the [$0^{\circ}$, $180^{\circ}$) data are duplicated on the opposite side of the rose diagram [$180^{\circ}$, $360^{\circ}$); duplicates shown in a lighter shade.}
	\label{figPARose}
\end{figure}

Figs.~\ref{figPAHist} (histograms) and \ref{figPARose} (rose diagrams) show LQG PAs after this parallel transport, both unweighted and weighted by orthogonal distance regression goodness-of-fit $w = \ell/\sigma^2$ (Eq.~\ref{eqWeight}). For axial data [$0^{\circ}$, $180^{\circ}$), where $0^{\circ}$ and $180^{\circ}$ are equivalent, a conventional histogram (Fig.~\ref{figPAHist}) can be misleading, since it represents data that are close together (e.g.\ $1^{\circ}$ and $179^{\circ}$) at opposite extremes of the distribution.

An alternative representation is the rose diagram (Fig.~\ref{figPARose}), where wedge length is proportional to the count and spanning angle denotes the bins. Note this is not a true histogram; the area displayed for a single count increases with increasing length. For undirected (i.e.\ non-directional, axial) data it is conventional to duplicate the [$0^{\circ}$, $180^{\circ}$) wedges on the opposite side of the rose diagram [$180^{\circ}$, $360^{\circ}$). For clarity, we reiterate that the PA data are axial [$0^{\circ}$, $180^{\circ}$) and not circular [$0^{\circ}$, $360^{\circ}$); we do not have PAs in the range [$180^{\circ}$, $360^{\circ}$) (the lighter shade in Fig.~\ref{figPARose}).

In both Figs.~\ref{figPAHist} and \ref{figPARose} the data appear bimodal, with peaks at $\theta\sim45^{\circ}$ and $\theta\sim135^{\circ}$ (in the absence of goodness-of-fit weighting). The peaks are thus separated by $\Delta\theta\sim90^{\circ}$, indicating that some LQGs may have PAs that are preferentially parallel (i.e.\ aligned) whilst others are preferentially orthogonal (this is described as `anti-aligned' by \citet{Hutsemekers2014} and \citet{Pelgrims2016thesis}) to one another. We analyse the PA distribution in detail in section (\ref{LQGStatsResults}).

\clearpage

\clearpage

% NB changes to this table may also necessitate changes to tabLQGDataFull
\begin{table}
    \centering
    \small
    \caption[Large quasar group locations and position angles (example LQGs)]{Example large quasar groups, where $m$ is the number of members, and $\bar{\alpha}$, $\bar{\delta}$, and $\bar{z}$ are the mean right ascension, declination, and redshift of the member quasars. The normalized goodness-of-fit weight $w$ (Eq.~\ref{eqWeight}) is scaled by $w_{71} = w \times 71$ for clarity, and to distinguish those LQGs weighted higher ($w_{71} > 1$) or lower ($w_{71} < 1$) than the mean $\bar{w}$. Position angle $\theta$ and half-width confidence interval $\gamma_h$ are shown for both the 2D and 3D approaches. The ratio of 3D eigenvalues (and ellipsoid axes lengths) is given by $a:b:c$. See Appendix~\ref{appLQGData} for full sample of all 71 LQGs.}
    \label{tabLQGDataPart}
    \begin{tabular}{crrrrrrrrc}
        \hline
        & \multicolumn{2}{r}{J2000 ($^\circ$)} &&& \multicolumn{2}{r}{2D PA ($^\circ$)} & \multicolumn{2}{r}{3D PA ($^\circ$)} & \\
        $m$ & $\bar{\alpha}$ & $\bar{\delta}$ & $\bar{z}$ & $w_{71}$ & $\theta$ & $\gamma_h$ & $\theta$ & $\gamma_h$ & $a:b:c$ \\
        \hline
        20 & 121.1 & 27.9 & 1.73 & 1.13 & 119.7 & 10.4 & 115.1 & 10.7 & 0.50:0.30:0.21 \\
        20 & 151.5 & 48.6 & 1.46 & 1.21 & 144.4 & 9.2 & 144.3 & 11.5 & 0.50:0.33:0.17 \\
        20 & 155.9 & 12.8 & 1.50 & 0.32 & 120.7 & 25.2 & 117.2 & 37.0 & 0.44:0.43:0.14 \\
        20 & 163.6 & 16.9 & 1.57 & 1.00 & 0.9 & 8.8 & 5.7 & 9.2 & 0.47:0.33:0.20 \\

        &&&&&\dots&&&& \\

        23 & 209.5 & 34.3 & 1.65 & 1.99 & 152.5 & 2.8 & 152.3 & 2.8 & 0.68:0.17:0.15 \\
        23 & 214.3 & 31.8 & 1.48 & 0.27 & 18.6 & 32.6 & 87.7 & 32.5 & 0.47:0.35:0.18 \\

        &&&&&\dots&&&& \\
        
        26 & 160.3 & 53.5 & 1.18 & 0.33 & 110.6 & 23.2 & 111.5 & 25.1 & 0.38:0.34:0.28 \\
        26 & 171.7 & 24.2 & 1.10 & 0.78 & 48.5 & 8.3 & 47.3 & 8.1 & 0.46:0.29:0.24 \\
        
        &&&&&\dots&&&& \\
        
        55 & 196.5 & 27.1 & 1.59 & 0.95 & 107.5 & 3.4 & 107.0 & 3.4 & 0.58:0.24:0.18 \\
        56 & 167.0 & 33.8 & 1.11 & 0.81 & 110.2 & 3.5 & 110.4 & 3.8 & 0.50:0.29:0.21 \\
        64 & 196.4 & 39.9 & 1.14 & 0.83 & 133.6 & 3.0 & 133.9 & 3.2 & 0.48:0.36:0.17 \\
        73 & 164.1 & 14.1 & 1.27 & 0.76 & 156.6 & 4.2 & 156.3 & 4.5 & 0.55:0.28:0.16 \\
        \hline
    \end{tabular}
\end{table}

\clearpage

\subsection{LQG PAs as a function of redshift} \label{subLQGPARedshift}

Much of this work concentrates on examining the apparent bimodality of the PA distribution, and the alignment and orthogonality in large-scale structure that this may indicate. We serendipitously notice that the redshift distribution also appears bimodal. In this section we briefly examine the relationship between LQG position angle and redshift, both of LQGs and their member quasars.

\begin{figure} % PA_z_marginal_plot
	\centering
    \includegraphics[width=0.8\columnwidth]{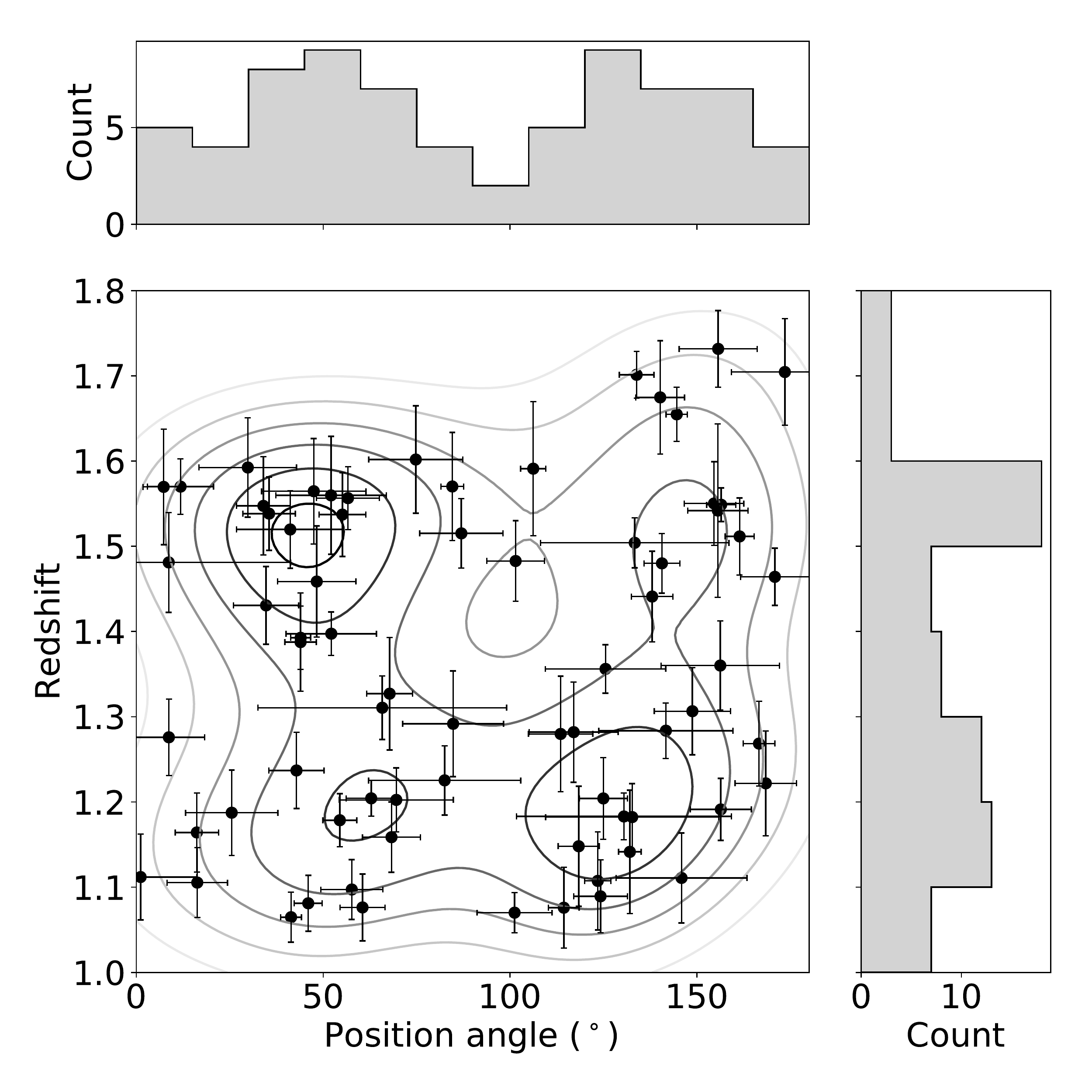}
    \caption[Marginal plot of the LQG PA $\times$ redshift plane]{Marginal plot of the LQG PA $\times$ redshift plane. Error bars are PA half-width confidence interval and redshift standard deviation. Marginal histograms show PA and redshift distributions with $\Delta \theta = 15^{\circ}$ and $\Delta z = 0.1$ bins. Contours are a Gaussian kernel density estimation of the scatter plot. Histograms and KDE are unweighted.}
    \label{figPAzMarginal}
\end{figure}

Fig.~\ref{figPAzMarginal} shows a marginal plot of the LQG PA $\times$ redshift plane; redshift here is LQG redshift, defined as the mean redshift of its member quasars. From Fig.~\ref{subHistUnweighted}, we expect the PA histogram (top margin) to be bimodal, as seen. The redshift histogram (right margin) also exhibits some bimodality. To investigate the relationship between these two variables, and whether there is any correlation between their modes, we add kernel density estimation (\gls{KDE}) contours to the scatter plot. This shows hints of three or four modes, although the correlation is weak. We note that the apparently stronger modes at $\theta \sim 45^{\circ}$ $\times$ $z \sim 1.5$ and $\theta \sim 135^{\circ}$ $\times$ $z \sim 1.2$ result from the points with the greatest uncertainty. Conversely, the weaker modes at $\theta \sim 55^{\circ}$ $\times$ $z \sim 1.2$ and $\theta \sim 150^{\circ}$ $\times$ $z \sim 1.5$ result from the points with the smallest uncertainty. Statistical analysis is left to future work.

\begin{figure} % PA_z_marginal_plot
	\centering
    \includegraphics[width=0.6\columnwidth]{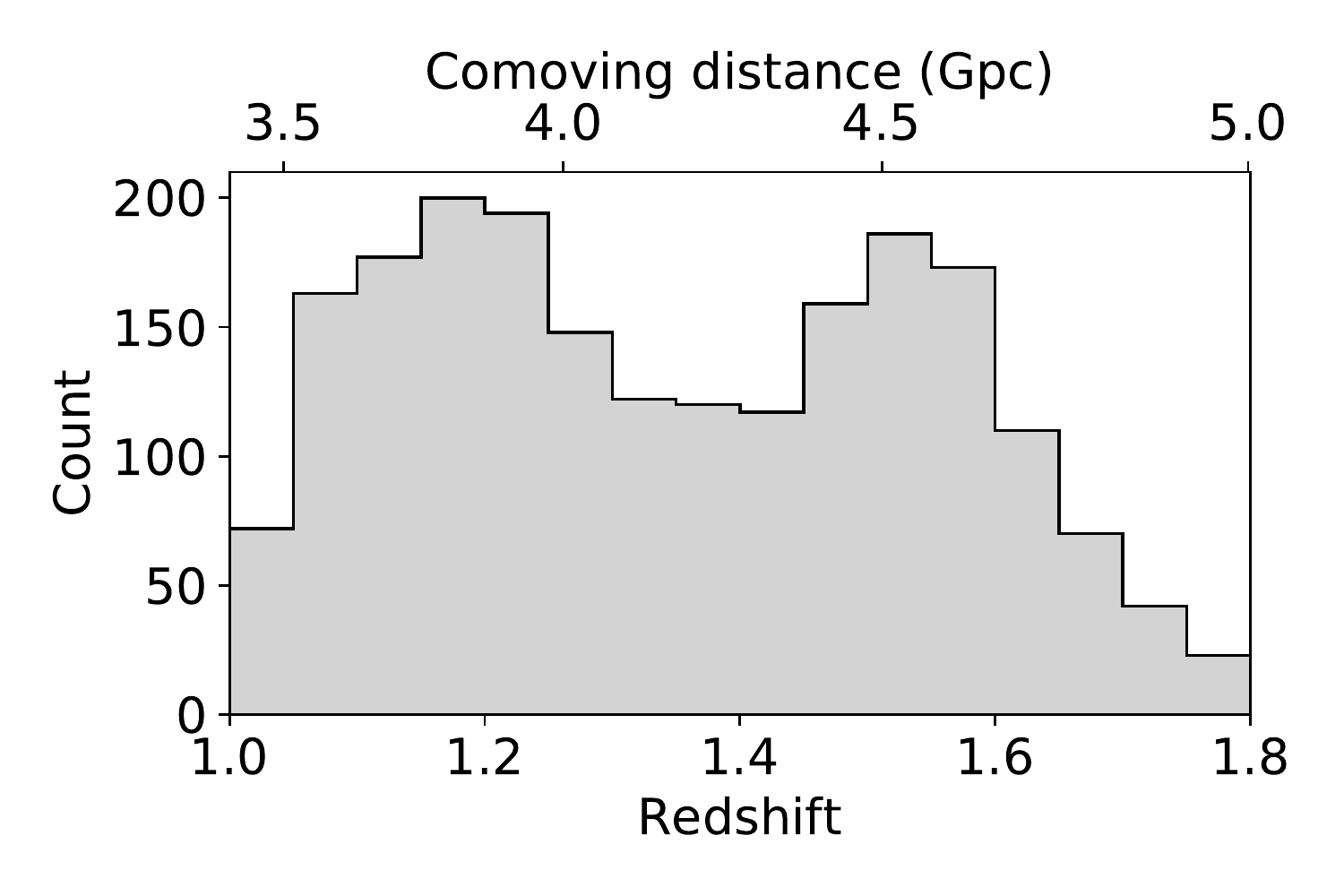}
    \caption[Redshift distribution of member quasars from all 71 LQGs]{Redshift distribution of the 2,076 member quasars comprising our sample of 71 LQGs, with $\Delta z = 0.05$ bins, unweighted. Top x-axis shows approximate comoving radial distance. Distribution is bimodal, with modes (peaks) at $z \sim 1.15$ and $z \sim 1.55$.}
    \label{figQSOz}
\end{figure}

Due to their scale, LQGs extend considerably in the radial direction, and some features may be lost when we only analyse their mean redshift. Our sample of 71 LQGs collectively comprise 2,076 member quasars. Fig.~\ref{figQSOz} shows the redshift distribution of these. The distribution appears bimodal with modes (peaks) at $z \sim 1.15$ and $z \sim 1.55$, similar to that for LQG redshifts (Fig.~\ref{figPAzMarginal}, right margin).

\begin{figure} % PA_z_marginal_plot
	\centering
    \includegraphics[width=0.6\columnwidth]{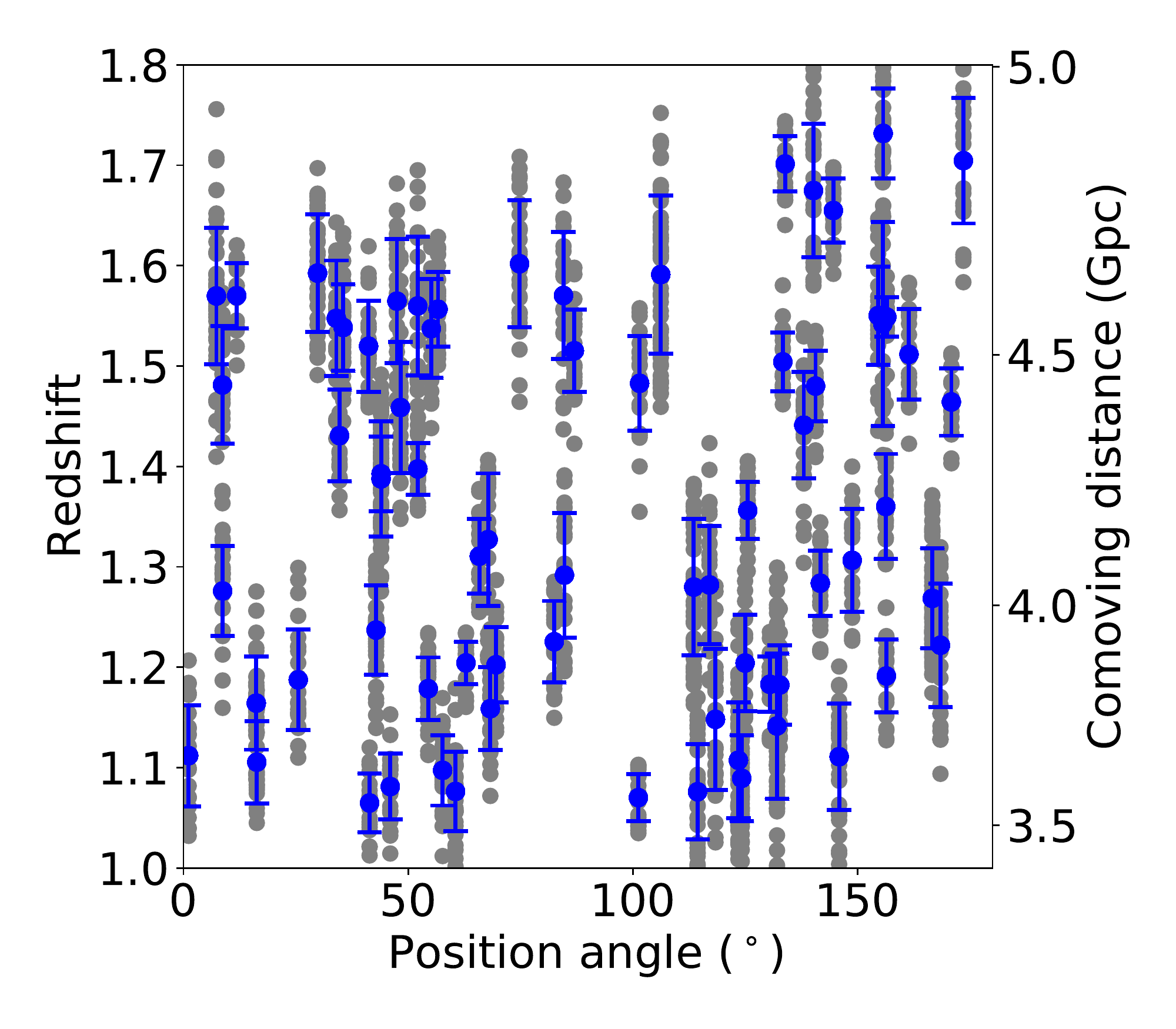}
    \caption[Scatter plot of the LQG PA $\times$ redshift plane]{Scatter plot of the LQG PA $\times$ redshift plane. Right-hand y-axis shows approximate comoving radial distance. Redshift ($z$) of member quasars are shown as grey dots. LQG redshifts ($\bar{z}$) are shown as blue dots with error bars of standard deviation in $z$. LQGs are `deep' in redshift, with several encroaching on our redshift boundaries of $1.0 \leq z \leq 1.8$.}
    \label{figPAQSOz}
\end{figure}

Finally, we present a scatter plot of the LQG PA $\times$ redshift plane in Fig.~\ref{figPAQSOz}, showing both LQG redshift (blue dots) and member quasar redshift (grey dots). This demonstrates the redshift `depth' of the LQGs, and also the number which encroach on our redshift boundaries of $1.0 \leq z \leq 1.8$, particularly at higher values of position angle, around the second PA mode.

\clearpage

\subsection{LQG PA weights} \label{subLQGPAWeights}

The two-dimensional approach of determining position angles includes orthogonal distance regression of tangent plane projected quasars. We evaluate the ODR goodness-of-fit, which may be used to weight the results, e.g.\ the weighted histogram in Fig.~\ref{subHistWeighted} and the weighted rose diagram in Fig.~\ref{subRoseWeighted}. These weights are also used for some of the statistical analysis, so are worth briefly exploring here.

\begin{figure} [h!] % LQG_meta_v2
	\centering
    \includegraphics[width=0.6\columnwidth]{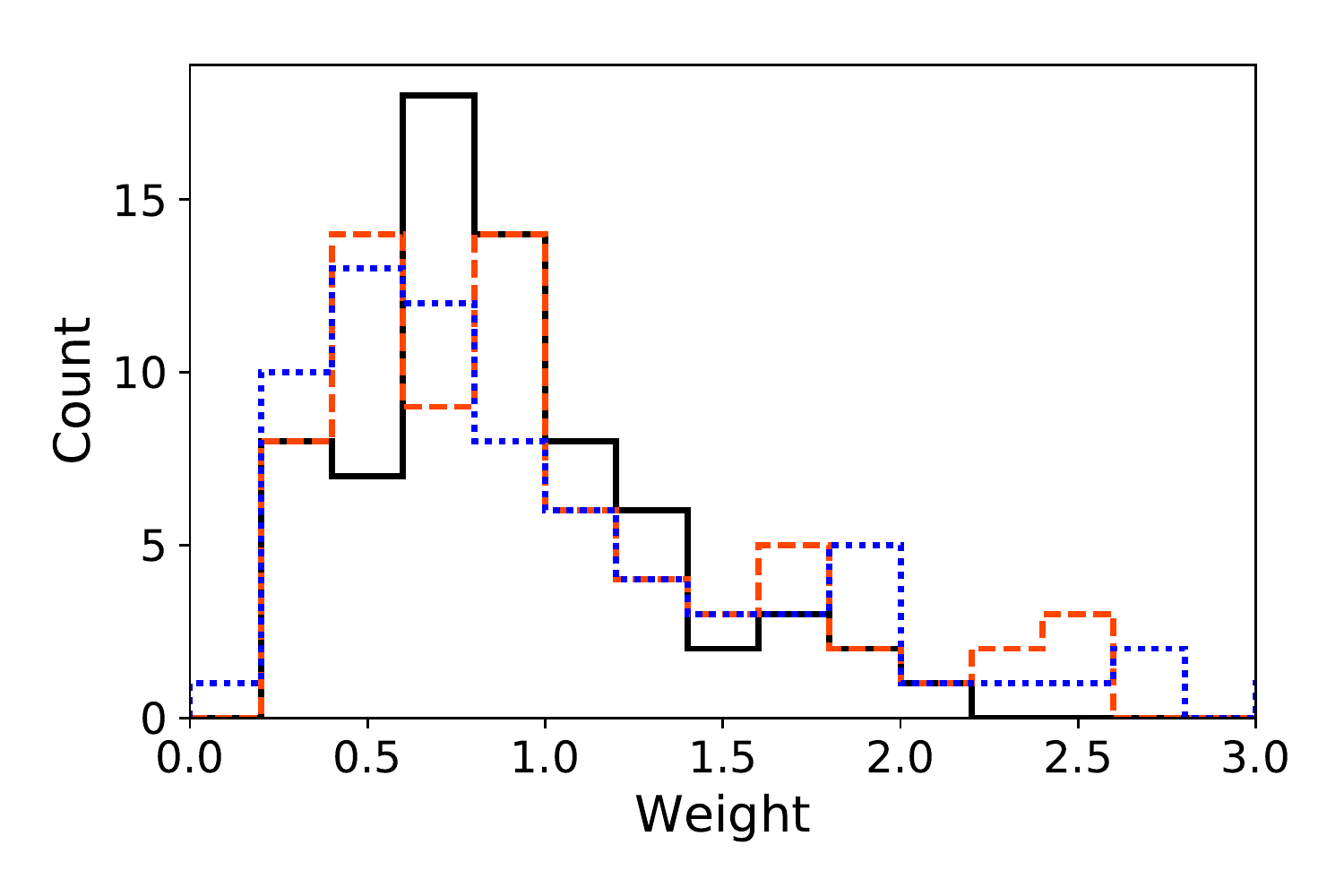}
    \caption[Distribution of goodness-of-fit weights]{Distribution of goodness-of-fit weights $w = \ell / \sigma^2$ (solid black), normalized and scaled by $w_{71} = w_i \times 71$ for clarity, and to distinguish those LQGs weighted higher ($w_{71} > 1$) or lower ($w_{71} < 1$) than mean weight $\bar{w}$. Also shown, weights $w = 1 / \gamma_h$ where $\gamma_h$ is the half-width confidence interval from bootstraps of the 2D (orange dashed) and 3D (blue dotted) PA approaches, similarly normalized and scaled.}
    \label{figHistWeights}
\end{figure}

Fig.~\ref{figHistWeights} (solid black line) shows the goodness-of-fit weights $w = \ell / \sigma^2$ of our sample of $n = 71$ LQGs, after they are normalized to $\sum_{i = 1}^{n} w_i = 1$ then scaled by $w_{71} = w_i \times 71$. This distinguishes those LQGs weighted higher ($w_{71} > 1$) or lower ($w_{71} < 1$) than the mean weight $\bar{w} = \sum_{i = 1}^{n} w_i / n$ of our sample of 71 LQGs.

An alternative weighting scheme is based on measurement uncertainties, or half-width confidence intervals (HWCIs), estimated from 10,000 bootstraps of the PA calculations. Fig.~\ref{figHistWeights} shows weights $w = 1 / \gamma_h$ from this empirical method, where $\gamma_h$ is the HWCI determined by the two-dimensional (orange dashed line) and three-dimensional (blue dotted line) PA approaches. Weights are normalized and scaled as for the goodness-of-fit weights.

\begin{figure}[t!] % LQG_meta_v2
    \centering
	\subfigure[ODR GoF weighted]{\includegraphics[width=0.48\columnwidth]{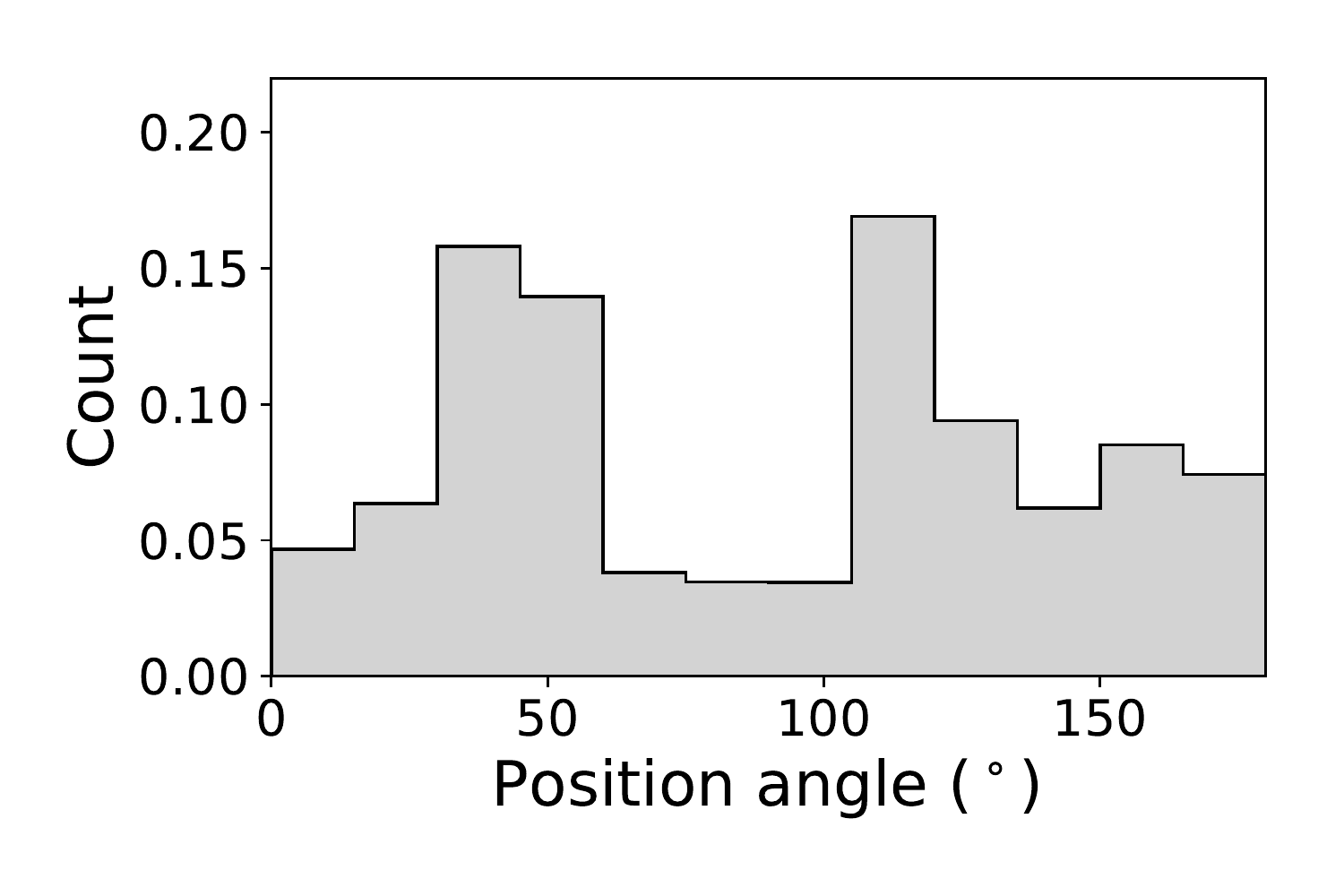} \label{subHistGoFWeighted}}
	\subfigure[HWCI weighted]{\includegraphics[width=0.48\columnwidth]{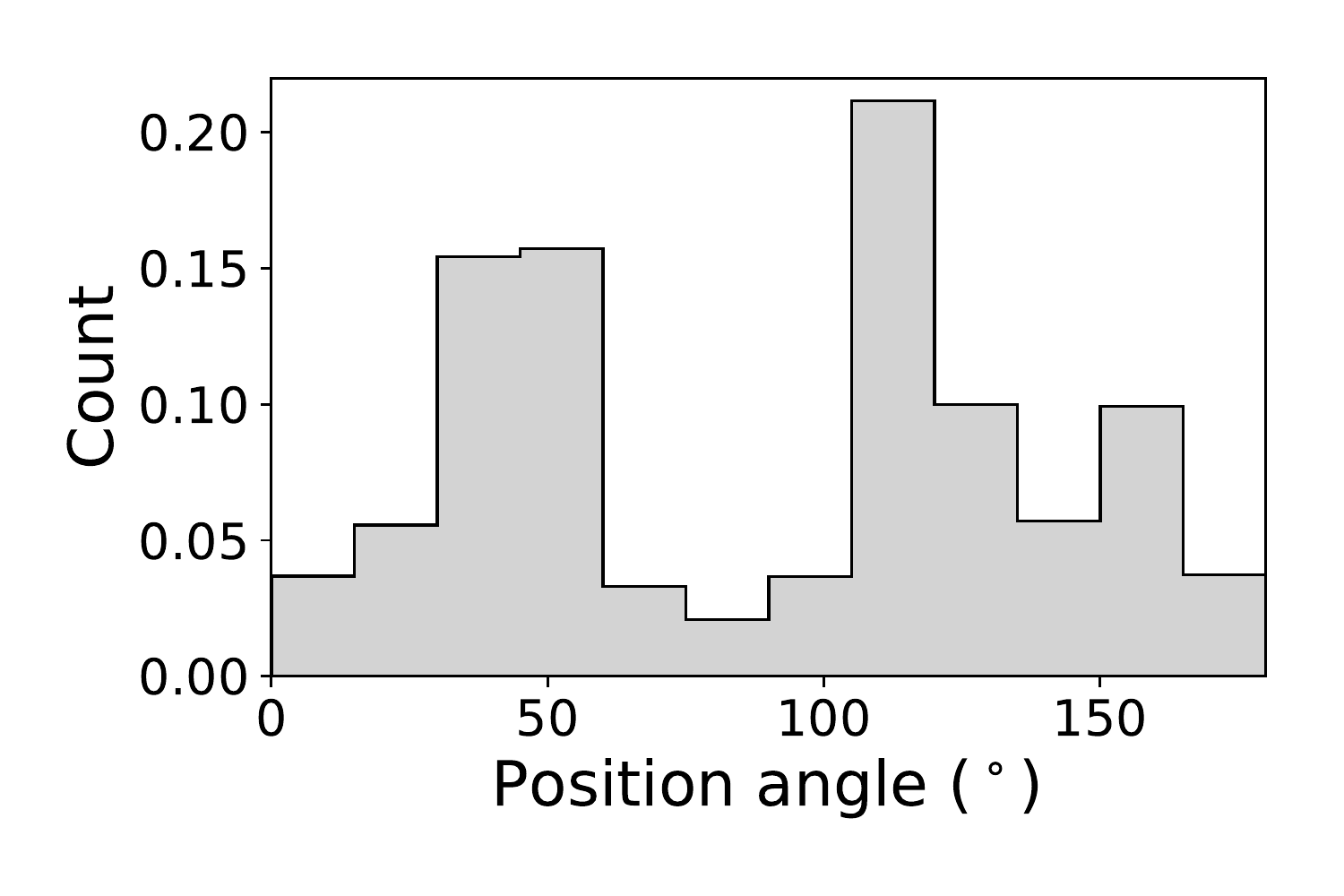} \label{subHistHWCIWeighted}}
	\caption[Position angles with alternative weighting (histogram)]{Position angles calculated by 2D approach, with no parallel transport, and $15^{\circ}$ bins. \protect\subref{subHistGoFWeighted} is goodness-of-fit weighted and \protect\subref{subHistHWCIWeighted} is half-width confidence interval weighted. The bimodal distribution is robust to which method of weighting is used.}
	\label{figWeightedHist}
\end{figure}

The distributions of goodness-of-fit, 2D HWCI, and 3D HWCI weights are similar. The means of all three distributions are, by design, $\bar{w} = 1$. The medians of the goodness-of-fit, 2D HWCI, and 3D HWCI distributions are, respectively, 0.83, 0.85, and 0.78. This is consistent with the 2D approach to calculating PAs having smaller uncertainties than the 3D approach.

We further show that the PA distribution is robust between these alternative weighting schemes. Fig.~\ref{figWeightedHist} shows PAs, determined by the 2D approach, with no parallel transport, and weighted by (a) goodness-of-fit weights $w = \ell / \sigma^2$, and (b) HWCI weights $w = 1 / \gamma_h$. The bimodal distribution is robust to which method of weighting is used; we continue to use the former in this work as it is more physically motivated and, judging by Fig.~\ref{figWeightedHist}, is also slightly more conservative.

\clearpage

\subsection{LQG morphology} \label{subLQGPAMorphology}

\begin{figure} % LQG_meta_v2
	\centering
    \includegraphics[width=0.65\columnwidth]{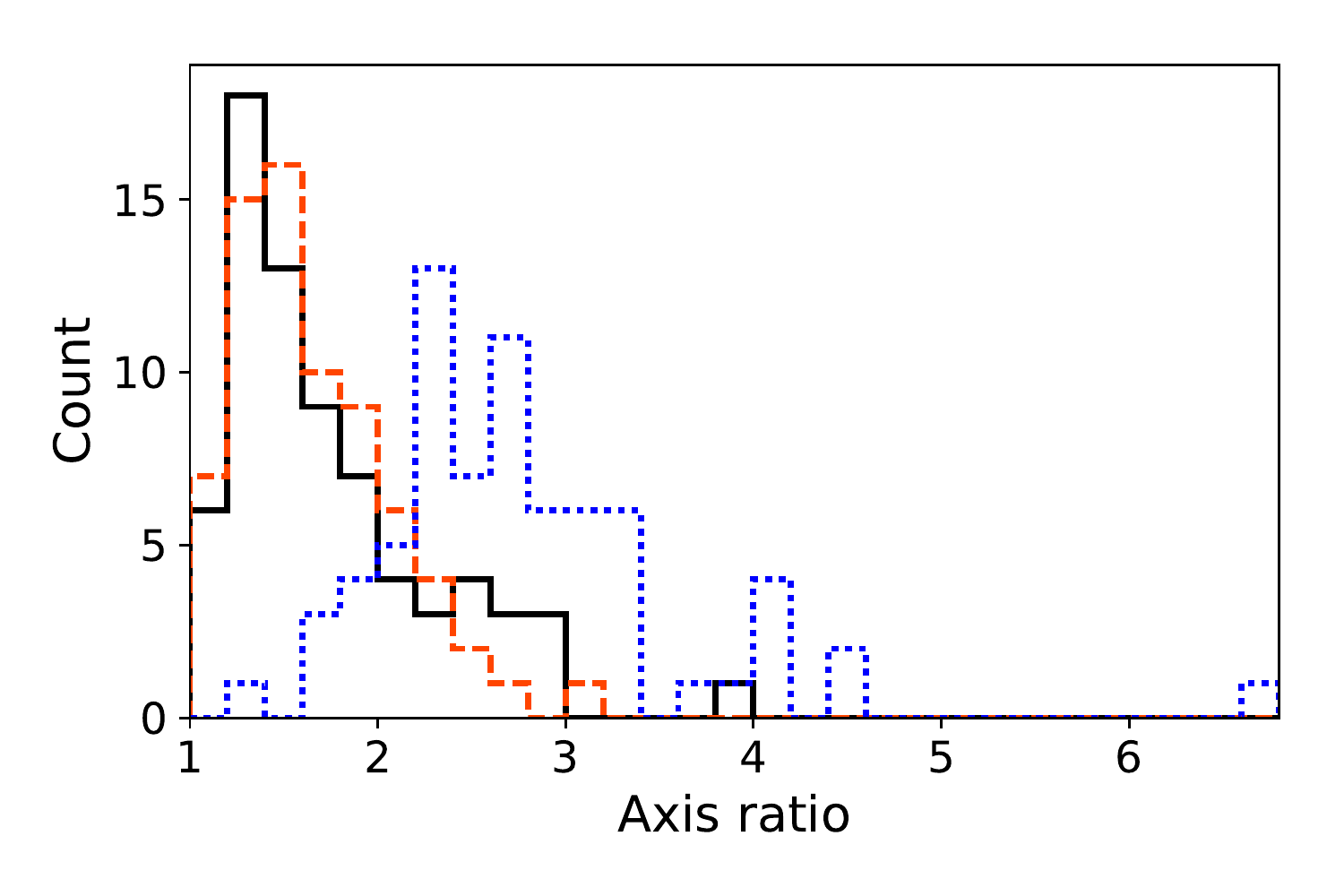}
    \caption[LQG ellipsoid axis length ratios $a:b:c$ (histogram)]{LQG ellipsoid axis length ratios $a:b:c$, represented as fractions $a/b$ (solid black), $b/c$ (dashed orange), and $a/c$ (dotted blue).}
    \label{figHistAxisRatios}
\end{figure}

The three-dimensional approach of determining position angles yields eigenvectors and eigenvalues. These are used to build a confidence ellipsoid, where the length $\ell$ of each axis is a function of its eigenvalue $\lambda$, specifically $\ell \propto \sqrt{\lambda}$. We compute the ratio of the axis lengths $a:b:c$, where $a$ is the length of the major axis and $a > b > c$. These ratios are shown in  Tables~\ref{tabLQGDataPart} (example LQGs) and \ref{tabLQGDataFull} (full LQG sample), and in Fig.~\ref{figHistAxisRatios}.

\begin{figure} % LQG_meta_v2
	\centering
    \includegraphics[width=0.65\columnwidth]{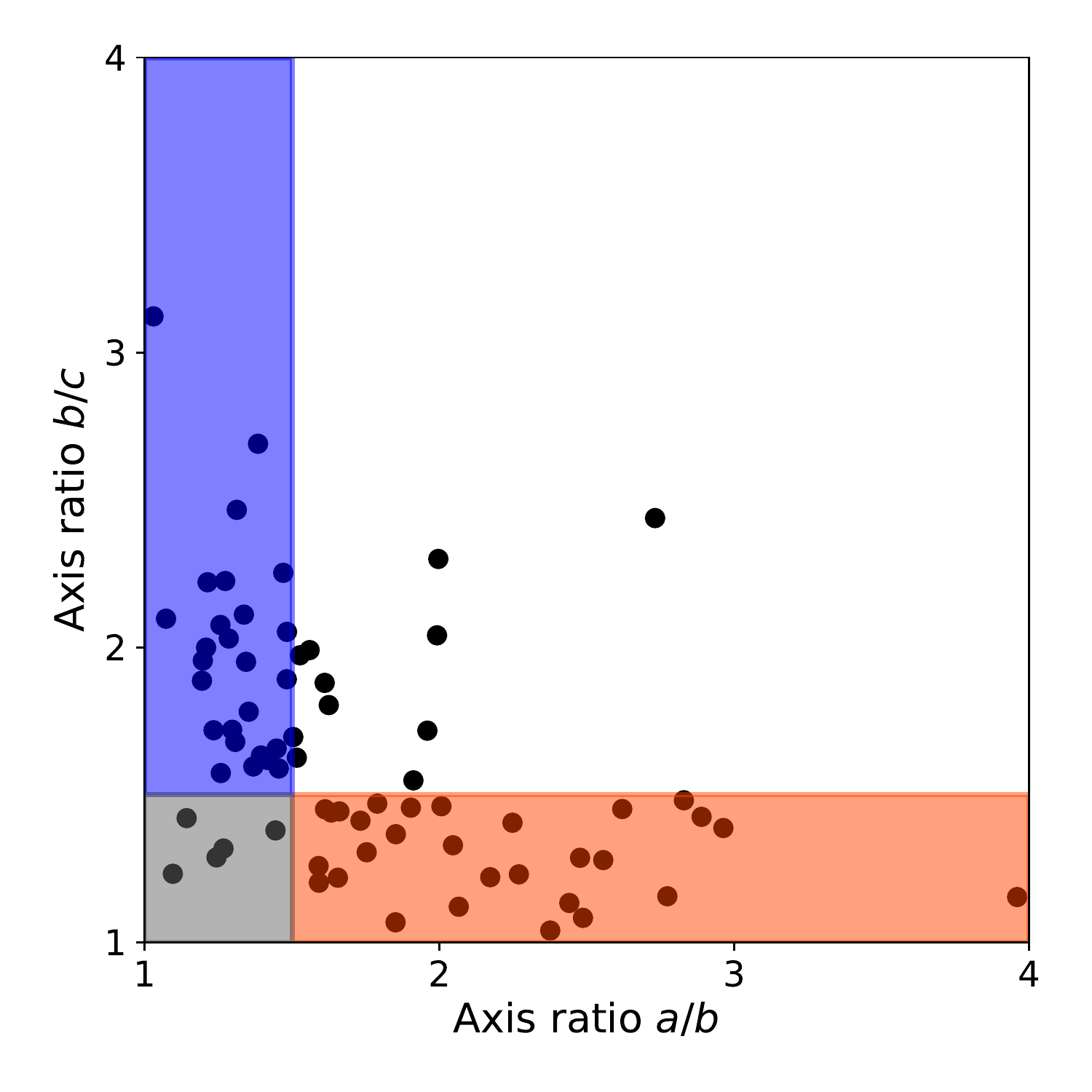}
    \caption[LQG ellipsoid axis length ratios $a/b$ versus $b/c$ (scatter plot)]{LQG ellipsoid axis length ratios $a:b:c$, represented as fractions $a/b$ and $b/c$. LQGs categorised roughly as `cigar-shaped' in the orange region, `discus-shaped' in the blue region, and spheroidal in the grey region. The boundaries of these regions are arbitrary and for illustration only.}
    \label{figScatterAxisRatios}
\end{figure}

The ellipsoid axis length ratios can roughly categorise the morphology of LQGs; if the $a$ axis is much longer than the other axes (and $b \sim c$) it is `cigar-shaped', if the $c$ axis is much shorter than the other axes (and $a \sim b$) it is `discus-shaped', and if all three axes are similar lengths ($a \sim b \sim c$) it is spheroidal. For our sample of 71 LQGs we find that 29 have $a/b \geq 1.5$ and $b/c \leq 1.5$, tending towards `cigars', 26 have $a/b \leq 1.5$ and $b/c \geq 1.5$, tending towards `discuses', and 5 have $a/b \leq 1.5$~and~$b/c \leq 1.5$, tending towards spheroids. See, respectively, orange, blue, and grey shaded regions in Fig.~\ref{figScatterAxisRatios}. Generally, LQGs are somewhere between these simple categories, and our choice of cut-off ($a/b = 1.5$ and $b/c = 1.5$) is arbitrary. 

For our sample of 71 LQGs, the mean (median) ratios are $a/b = 1.75$ ($a/b = 1.59$), $b/c = 1.65$ ($b/c = 1.57$), and $a/c = 2.77$ ($a/c = 2.62$). For determining PAs, the $a/b$ ratio is important; if $a \sim b$ then the orientation of the LQG with respect to the line-of-sight will greatly affect which axis \textit{appears} longer in the tangent plane projection of the 2D approach.

This work focuses on LQG alignment; we leave the study of LQG morphology to future work. We also suggest investigating any correlation of LQG alignment or morphology with respect to the line of sight, which may help identify or reject systematic errors due to redshift-space distortions or other line-of-sight biases (e.g.\ larger uncertainties in the position angle of LQGs orientated towards the line-of-sight) in future LQG analyses.

\section{Results: statistical analysis} \label{LQGStatsResults}

In this section we show how likely it is that the PAs are drawn from a uniform distribution, how significant their bimodality is and where the peaks are, and how likely their alignment/orthogonality is compared to randoms/simulations.

All analysis in this section is performed on the PAs of our sample of 71 LQGs, including parallel transport corrections. The uniformity and bimodality analyses are performed on PAs parallel transported to the centre of the A1 region \citep{Hutsemekers1998}. The correlation analysis is performed using the coordinate invariant version of the S test \citep{Jain2004}, which incorporates parallel transport corrections.

\subsection{Uniformity: LQG PAs are unlikely to be uniform} \label{subsecUniformityResults}

The results from applying the uniformity tests are listed in Table~\ref{tabUniformityStats}.

\textbf{Weighting.} Kuiper's and $\chi^2$ tests are evaluated for PAs with and without goodness-of-fit weighting. The Hermans-Rasson (HR) test is only evaluated for unweighted PAs. 

\textbf{2-axial data.} Kuiper's and $\chi^2$ tests are evaluated for raw PAs ($\theta$). Kuiper's test is also evaluated for PAs of the form $\Theta_{2ax} = 2\theta$; results are consistent, as expected. For the $\chi^2$ test we use $20^{\circ}$ bins. The HR test requires circular data, so is only evaluated for PAs of the form $\Theta_{2ax} = 2\theta$. The p-values for these are shown in the `2-axial' columns, with unweighted and weighted sub-columns.

\textbf{4-axial data.} Kuiper's and $\chi^2$ tests are both evaluated for PAs of the form $\theta_{4ax}$. Kuiper's test is also evaluated for PAs of the form $\Theta_{4ax} = 4\theta_{4ax}$; again results are consistent. For the $\chi^2$ test we use $10^{\circ}$ bins, half that for 2-axial data; 4-axial data would otherwise have twice the frequency of 2-axial data. The HR test is evaluated for PAs of the form $\Theta_{4ax} = 4\theta_{4ax}$ (i.e.\ circular). The p-values for these are shown in the `4-axial' columns, with unweighted and weighted sub-columns.

The $\chi^2$ test does not show evidence of non-uniformity for 2-axial PAs without weighting ($16\%$ significance level), but does show some evidence of non-uniformity with weighting ($1\%$ significance level). For 4-axial PAs it shows marginal evidence of non-uniformity both with and without weighting ($2\%$ and $7\%$ respectively).

Kuiper's test does not show evidence of non-uniformity for 2-axial PAs, with or without weighting. However, the PA distribution is bimodal, which dramatically reduces the test's discriminatory power, so the absence of a signal is unsurprising. For 4-axial PAs Kuiper's test indicates a rejection of the null hypothesis of uniformity at the $0.8\%$ and $0.9\%$ significance level (weighted and unweighted), i.e.\ $\sim 2.4\sigma$.

\begin{table} % Hermans_Rasson_test
    \centering
    \small
    \caption[Results of uniformity tests ($\chi^2$, Kuiper's, Hermans-Rasson)]{Results (p-values) of uniformity tests. For 2-axial (4-axial) PAs the $\chi^2$ test is evaluated using $20^{\circ}$ ($10^{\circ}$) bins. The $\chi^2$ and Kuiper's tests are computed both with and without goodness-of-fit weighting. For 2-axial PAs, only the $\chi^2$ test shows evidence for non-uniformity (of weighted PAs). For 4-axial PAs, all tests show evidence for non-uniformity, mostly marginal, with Kuiper's being the most significant.}
    \label{tabUniformityStats}
    \begin{tabular}{l c c c c}
        \hline
        & \multicolumn{4}{c}{p-value} \\
        & \multicolumn{2}{c}{2-axial} & \multicolumn{2}{c}{4-axial} \\
        test & unweighted & weighted & unweighted & weighted \\
        \hline
        $\chi^2$ ($20^{\circ}\ /\ 10^{\circ}$ bins) & 0.16 & 0.01 & 0.07 & 0.02 \\
        Kuiper's & 0.62 & 0.59 & 0.009 & 0.008 \\
        Hermans-Rasson & 0.07 & - & 0.04 & - \\
        \hline
    \end{tabular}
\end{table}

Finally, the HR test shows marginal evidence of non-uniformity for 2-axial and 4-axial PAs ($7\%$ and $4\%$ significance levels, respectively), both without weighting.

Based on the results from these three uniformity tests we cannot confidently reject the null hypothesis that the observed PAs are drawn from a uniform distribution. The most appropriate test for the axial and bimodal nature of the PA data is the HR test, which shows marginal evidence of non-uniformity.

However, if we \emph{a priori} expect \emph{f}-fold symmetry (specifically 2-fold, in case of a bimodal distribution) then the conversion of PAs to 4-axial becomes physically well motivated as well as statistically legitimate. In this case, based on the 4-axial results of Kuiper's test, we could confidently reject the null hypothesis and conclude that the PAs are non-uniform.

\subsubsection{Uniformity of mock LQG catalogues}

The results from applying uniformity tests to the mock LQGs of section~\ref{secMockLQGs} are listed in Table~\ref{tabMockChi2}. For tests of multiple samples we calculate the mean p-value. In these cases we also examine the p-value distribution, and find that $\sim 4\%$ ($\sim 5\%$) of the 110 individual 2-axial (4-axial) mocks have a $\chi^2$ p-value of $< 0.05$. However, comparisons across multiple samples, even if those samples are entirely random, occasionally yields a `significant' result due to the introduction of additional degrees of freedom. This practice, now known as $p$-hacking, is highlighted by \citet{Simmons2011} who note ``undisclosed flexibility...allows presenting anything as significant''. To compensate for this we must correct the alpha-level. Using the \citet{Sidak1967} correction, to obtain an overall alpha-level of $\alpha = 0.05$ for all 110 mocks combined requires an individual alpha-level of $\alpha = 0.0005$ per mock. None has a $\chi^2$ p-value of $< 0.0005$, and we conclude the few low p-values of $< 0.05$ are not significant. None of the 10 stacks of 11 sub-volumes have a $\chi^2$ p-value of $< 0.05$.

The $\chi^2$ and Hermans-Rasson tests show no evidence for non-uniformity of the mock LQGs. This result is consistent between 2 and 4-axial PAs, individual mocks, stacks of sub-volumes, and the stack of all 3,296 mock LQGs, notwithstanding our concerns about the independence of the latter (section~\ref{secMockLQGs}). Kuiper's test also shows no evidence for non-uniformity, except for the stack of all mock LQGs evaluated as 4-axial data (and then only marginally). This could be an artefact of the realizations not being truly independent, but if so it is unclear why this would reveal itself only in one of the six tests on this sample. Further, it is an anomaly amongst otherwise consistent results, and we urged caution interpreting any results manifest only in 4-axial data. We therefore conclude that all three tests indicate statistical uniformity of the mock LQGs.

\begin{table}[]
    \centering
    \small
    \caption[Results of uniformity tests applied to mock LQGs]{Results (p-values) of uniformity tests applied to mock LQGs. For 2-axial (4-axial) PAs the $\chi^2$ test is evaluated using $15^{\circ}$ ($7.5^{\circ}$) bins, except for individual mocks where it is evaluated using $30^{\circ}$ ($15^{\circ}$) bins. For multiple samples the number of LQGs and p-values are means plus the standard error on the mean. All PAs are unweighted and are parallel transported to, and measured at, the centre of the Al region \citep{Hutsemekers1998}. Most samples and tests show no evidence for non-uniformity.}
    \label{tabMockChi2}
    \begin{tabular}{l l l l l}
         \hline
         & No. of & & \multicolumn{2}{c}{p-value} \\
         sample(s) & LQGs & test & \multicolumn{1}{c}{2-axial} & \multicolumn{1}{c}{4-axial} \\
         \hline
         \multirow{3}{*}{110 individual mocks} & \multirow{3}{*}{$30\pm 0.4$} & $\chi^2$ ($30^{\circ}$/$15^{\circ}$ bins) & $0.53 \pm 0.03$ & $0.51 \pm 0.03$ \\
         & & Kuiper's & $0.53 \pm 0.03$ & $0.50 \pm 0.03$ \\
         & & Hermans-Rasson & $0.48 \pm 0.03$ & $0.50 \pm 0.03$ \\
         \arrayrulecolor{lightgray}\hline
         \multirow{3}{*}{10 stacks of 11 sub-volumes} & \multirow{3}{*}{$330\pm 3$} & $\chi^2$ ($15^{\circ}$/$7.5^{\circ}$ bins) & $0.47 \pm 0.10$ & $0.41 \pm 0.08$ \\
         & & Kuiper's & $0.56 \pm 0.10$ & $0.44 \pm 0.09$ \\
         & & Hermans-Rasson & $0.46 \pm 0.11$ & $0.62 \pm 0.09$ \\
         \arrayrulecolor{lightgray}\hline
         \multirow{3}{*}{1 stack of 110 mocks} & \multirow{3}{*}{3,296} & $\chi^2$ ($15^{\circ}$/$7.5^{\circ}$ bins) & 0.20 & 0.31 \\
         & & Kuiper's & 0.27 & 0.03 \\
         & & Hermans-Rasson & 0.16 & 0.15 \\
         \arrayrulecolor{black}\hline
    \end{tabular}
\end{table}

\clearpage

\subsection{Bimodality: LQG PA distribution is bimodal} \label{subsecBimodalityResults}

\subsubsection{Bimodality coefficient}

\begin{figure} % plot_PAs
	\centering
	\subfigure[Unweighted; $10^{\circ}$ bins]{\includegraphics[width=0.48\columnwidth]{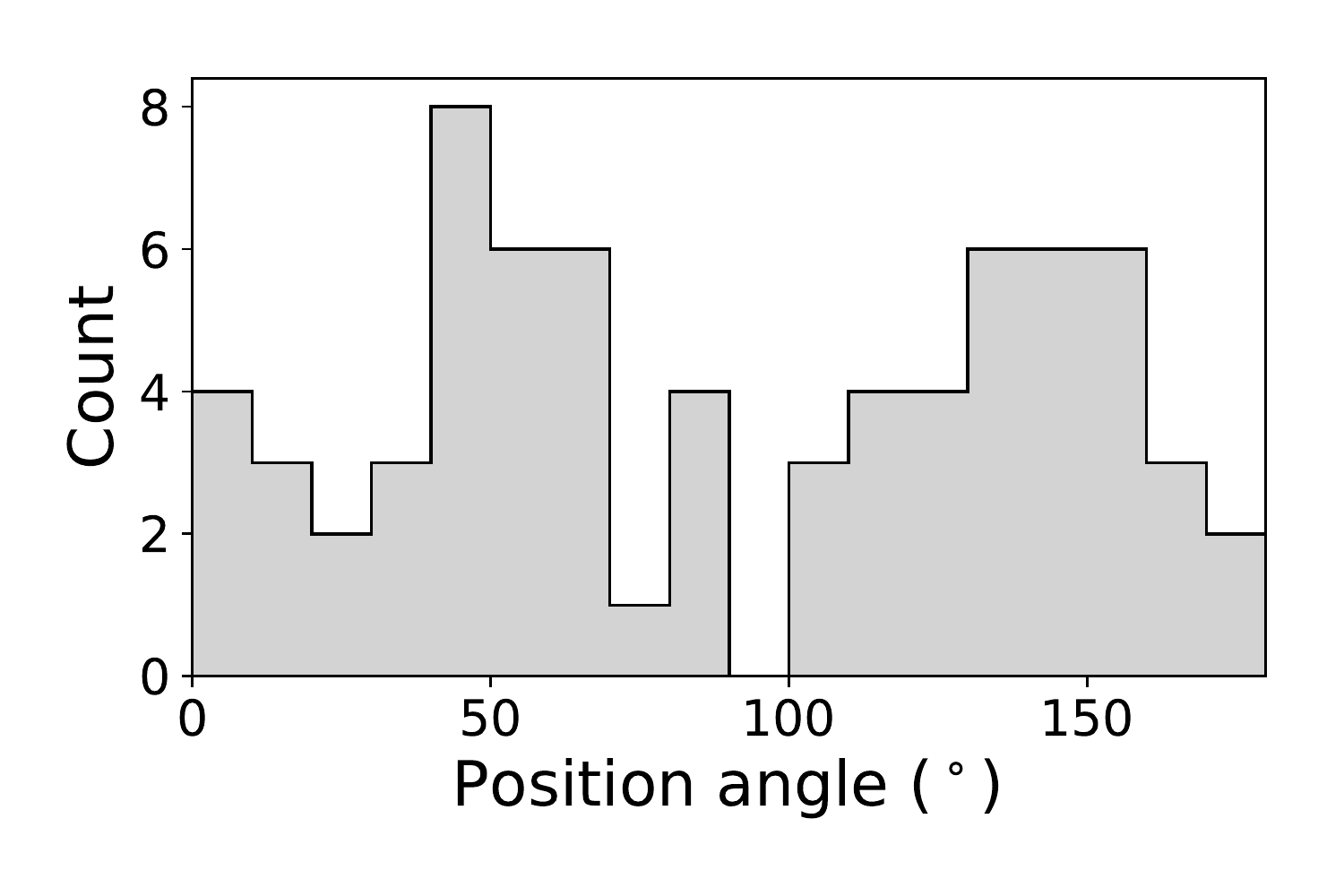} \label{sub10}}
	\subfigure[Weighted; $10^{\circ}$ bins]{\includegraphics[width=0.48\columnwidth]{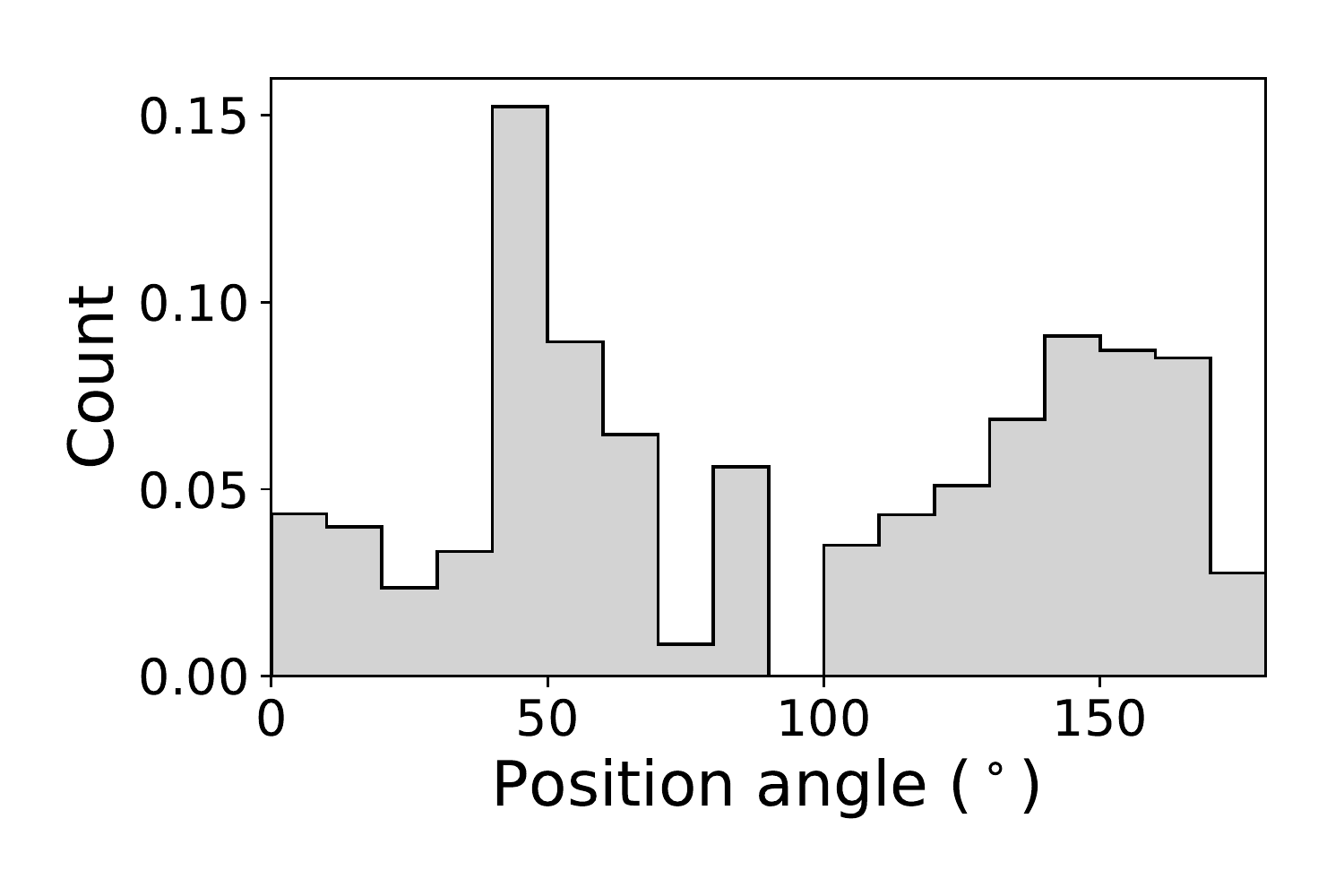} \label{sub10w}}
	\\
	\subfigure[Unweighted; $20^{\circ}$ bins]{\includegraphics[width=0.48\columnwidth]{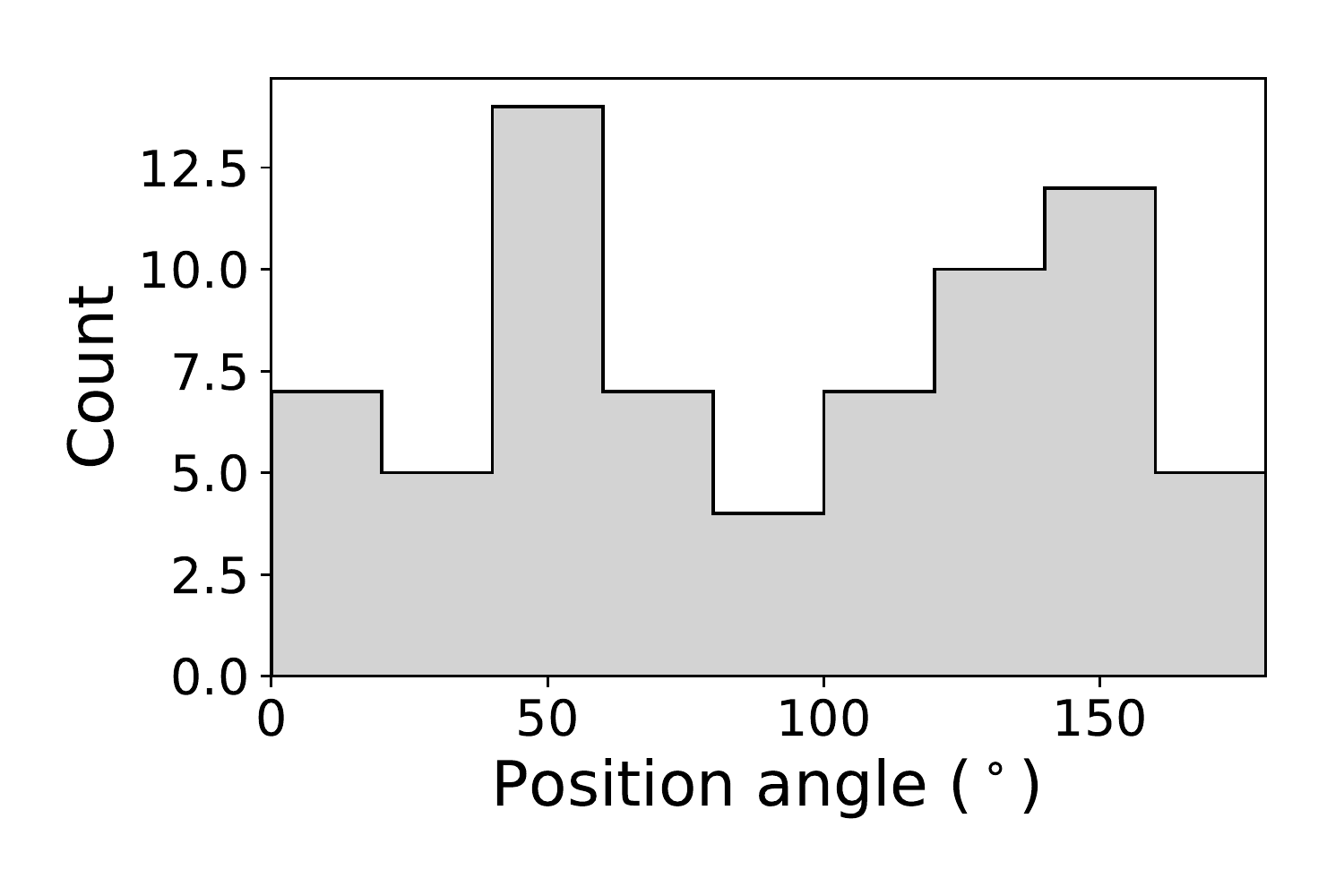} \label{sub20}}
	\subfigure[Weighted; $20^{\circ}$ bins]{\includegraphics[width=0.48\columnwidth]{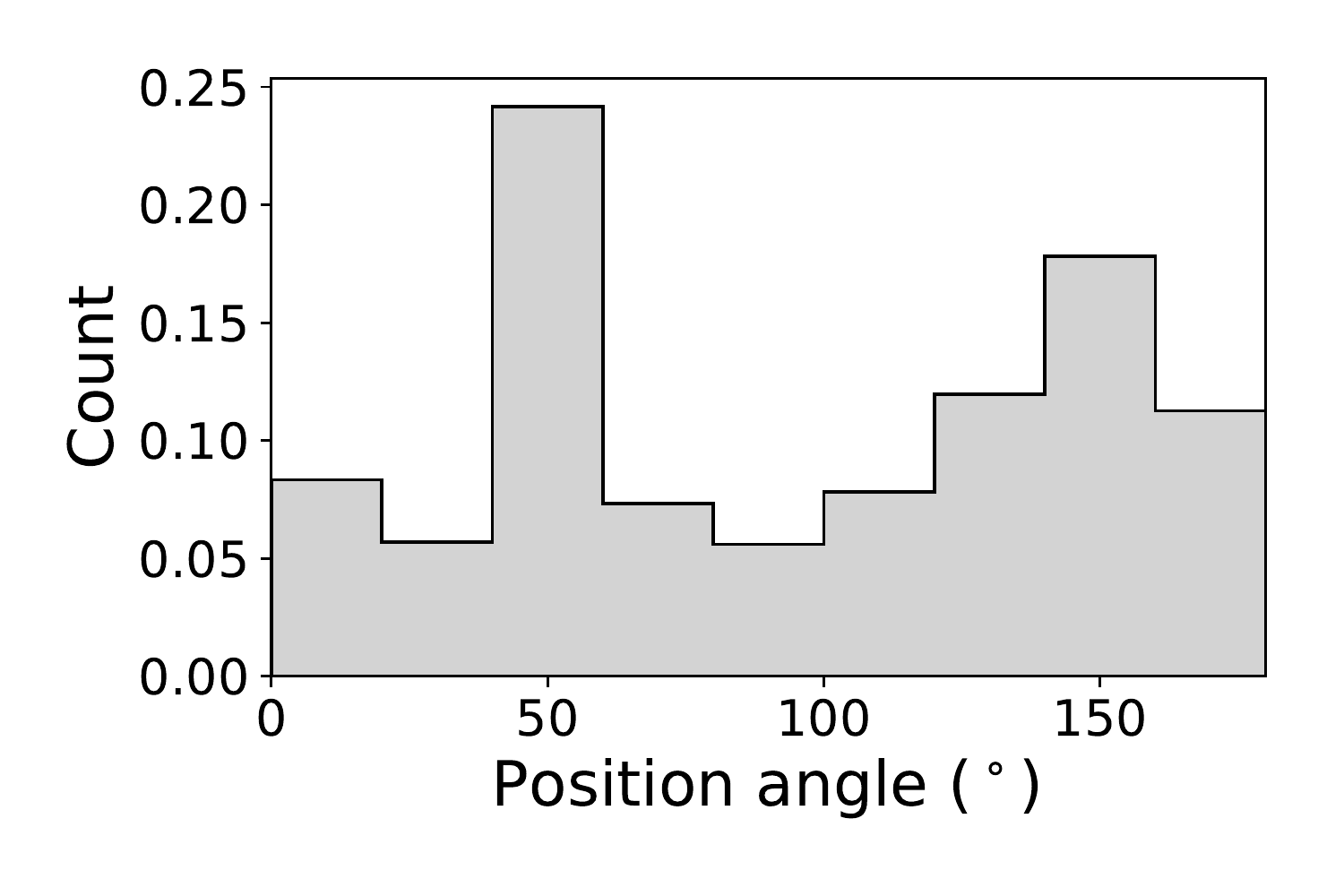} \label{sub20w}}
	\\
	\subfigure[Unweighted; $30^{\circ}$ bins]{\includegraphics[width=0.48\columnwidth]{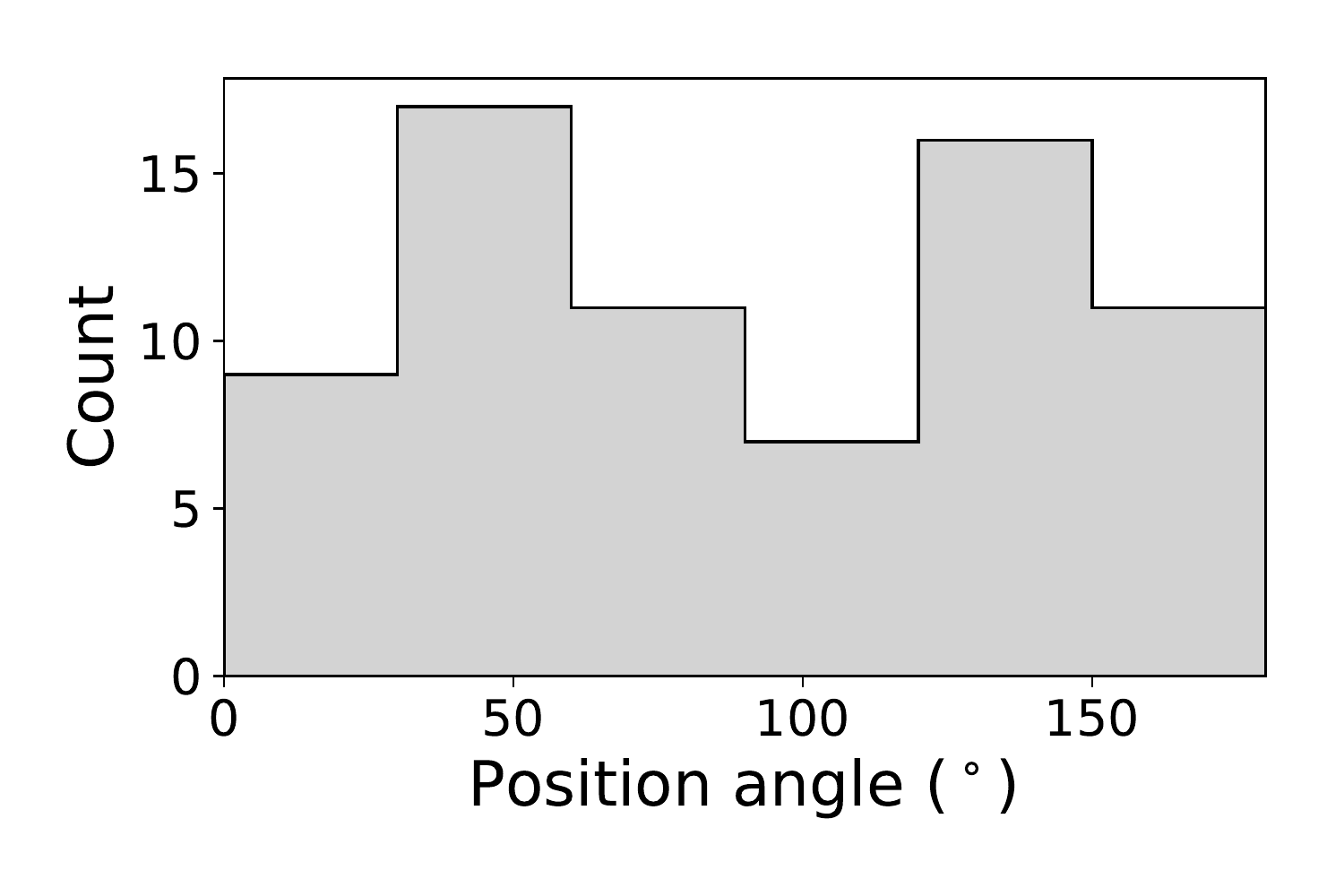} \label{sub30}}
	\subfigure[Weighted; $30^{\circ}$ bins]{\includegraphics[width=0.48\columnwidth]{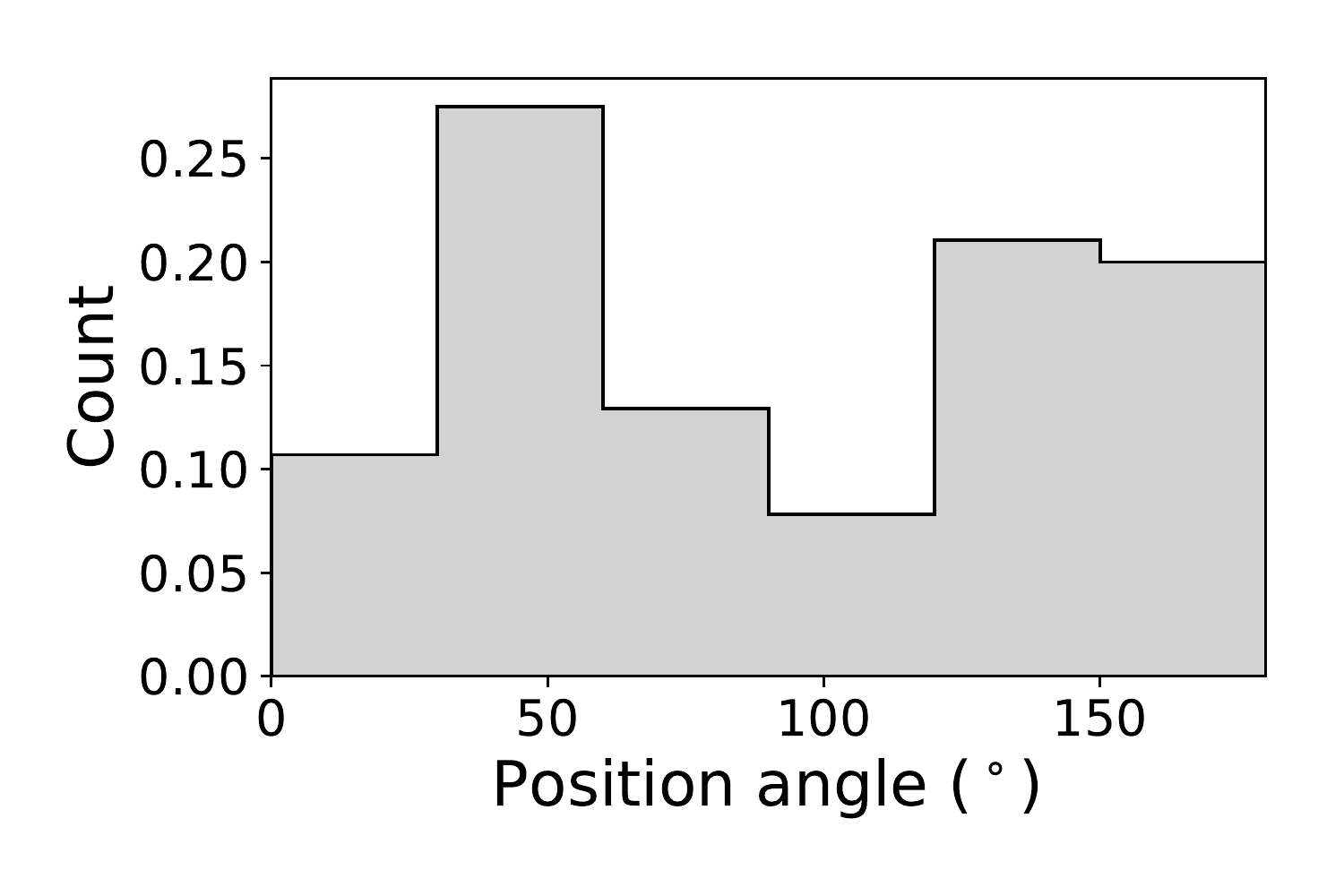} \label{sub30w}}
	\caption[As Fig.~\ref{figPAHist} (LQG PA histogram) but for different bin widths]{LQG position angles as in Fig.~\ref{figPAHist} but with $10^{\circ}$, $20^{\circ}$, and $30^{\circ}$ bins (respectively, rows 1, 2, and 3). Column 1 is unweighted and column 2 is ODR goodness-of-fit weighted. The BC is calculated (Table~\ref{tabBC}) from the skewness and kurtosis of these histograms (Table~\ref{tabSkewKurtosis}).}
	\label{figPA102030}
\end{figure}

The results from calculating the bimodality coefficient (BC) are listed in Table~\ref{tabBC}. As BC is calculated using categorical data, we calculate it for three different bin widths ($10^{\circ}$, $20^{\circ}$, and $30^{\circ}$). We also calculate BC before and after goodness-of-fit weighting. Histograms of each combination of bin width and weighting are shown in Fig.~\ref{figPA102030}.

The BC is evaluated for 2-axial PAs, using both raw PAs ($\theta$) and PAs of the form $\Theta_{2ax} = 2\theta$; results are not consistent, probably due to sensitivity to bin size. The BC is not calculated for 4-axial PAs, since this conversion yields unimodal data, for which the BC is not meaningful.

The BC generally indicates unimodality with $10^{\circ}$ bins, and bimodality with $20^{\circ}$ and $30^{\circ}$ bins (remember $BC_{crit} \approx 0.555$), with a few exceptions. This trend is true for both 2-axial and circular PAs, again with a few exceptions, although the value of the BC varies significantly between the two. This dependence of BC on bin width, and which form of PA is used, is troubling.

Table~\ref{tabSkewKurtosis} lists the skewness and kurtosis of the histograms in Fig.~\ref{figPA102030}, from which the BC is derived. All the histograms are, to varying degrees, skewed to the right (positive $m_3$). High skewness, in whichever direction, can lead to a bogus identification of bimodality \citep{Pfister2013}. Furthermore, for unweighted data especially, the excess kurtosis becomes increasingly negative (light tailed) with larger bins, which will also tend to increase the BC. For these reasons, and the limitations identified in section~\ref{subsecBimodalityMethods}, we do not consider the BC further.

\begin{table}
    \centering
    \small
    \caption[Bimodality coefficient for different bin widths (see also Fig.~\ref{figPA102030})]{Bimodality coefficient calculated using $10^{\circ}$, $20^{\circ}$, and $30^{\circ}$ bins (histograms in Fig.~\ref{figPA102030}). The BC is calculated for 2-axial ($\theta$) and circular ($\Theta_{2ax} = 2\theta$) PAs, both before and after goodness-of-fit weighting. The results are not always consistent and are not robust to choice of bin width.}
    \label{tabBC}
    \begin{tabular}{c c c c c}
        \hline
        & \multicolumn{4}{c}{bimodality coefficient ($BC_{crit} \approx 0.555$)} \\
        & \multicolumn{2}{c}{2-axial PAs ($\theta$)} & \multicolumn{2}{c}{circular PAs ($\Theta=2\theta$)} \\
        bin width & unweighted & weighted & unweighted & weighted \\
        \hline
        $10^{\circ}$ & 0.40 & 0.44 & 0.42 & 0.65 \\
        $20^{\circ}$ & 0.63 & 0.71 & 0.49 & 0.65 \\
        $30^{\circ}$ & 0.60 & 0.57 & 0.67 & 0.78 \\
        \hline
    \end{tabular}
\end{table}

\begin{table}
    \centering
    \small
    \caption[Skewness and kurtosis of Fig.~\ref{figPA102030} used to calculate the BC]{Skewness ($m_3$) and kurtosis ($m_4$) of the categorical PA data (histograms in Fig.~\ref{figPA102030}) used to calculate the BC (Table~\ref{tabBC}), for 2-axial ($\theta$) and circular ($\Theta_{2ax} = 2\theta$) PAs, both before and after goodness-of-fit weighting. Values of high skewness and negative kurtosis may lead to the instability we see in the BC.}
    \label{tabSkewKurtosis}
    \begin{tabular}{c r r r r r r r r}
        \hline
        & \multicolumn{8}{c}{skewness ($m_3$) and kurtosis ($m_4$)} \\
        & \multicolumn{4}{c}{2-axial PAs ($\theta$)} & \multicolumn{4}{c}{circular PAs ($\Theta=2\theta$)} \\
        bin width & \multicolumn{2}{c}{unweighted} & \multicolumn{2}{c}{weighted} & \multicolumn{2}{c}{unweighted} & \multicolumn{2}{c}{weighted} \\
        & $m_3$ & $m_4$ & $m_3$ & $m_4$ & $m_3$ & $m_4$ & $m_3$ & $m_4$ \\
        \hline
        $10^{\circ}$ & $0.04$ & $-0.60$ & $0.87$ & $0.89$ & $0.34$ & $-0.46$ & $2.05$ & $4.80$ \\
        $20^{\circ}$ & $0.66$ & $-0.84$ & $1.16$ & $0.14$ & $0.37$ & $-0.79$ & $1.24$ & $0.81$ \\
        $30^{\circ}$ & $0.27$ & $-1.35$ & $0.23$ & $-1.27$ & $0.17$ & $-1.59$ & $0.76$ & $-1.12$ \\
        \hline
    \end{tabular}
\end{table}

\clearpage

\subsubsection{Hartigans' dip statistic} \label{secHDSResults}

We calculate Hartigans' dip statistic (HDS) for continuous unweighted PA data, using sensitivity $\alpha = 0.03$ and merge distance $=5^{\circ}$. This recovers two peaks between $\sim36^{\circ} - 83^{\circ}$ and $\sim114^{\circ} - 156^{\circ}$, with 97\% confidence. We do not perform HDS on weighted data.

The unweighted (weighted) means of PAs in these two ranges are $\sim54\pm2^{\circ}$ ($\sim52\pm2^{\circ}$) and $\sim136\pm3^{\circ}$ ($\sim137\pm3^{\circ}$), consistent with the peaks we see in the distribution of categorical PA data (e.g.\ Fig.~\ref{figPAHist}). The bimodality observed in the categorical data is therefore unlikely to be an artefact of binning.

The bimodal peaks are identified by HDS after all PAs are parallel transported to the centre of the A1 region. Recalling section~\ref{secParallelTransport} \citep[see also][]{Jain2004}, the process of parallel transport rotates PAs; the amount of rotation depending on the path taken (direction and distance). Therefore it is reasonable to check whether parallel transporting to a different location would affect the position of the PA peaks.

\begin{figure} % fit_gaussians_to_pas
    \centering
	\includegraphics[width=0.98\columnwidth]{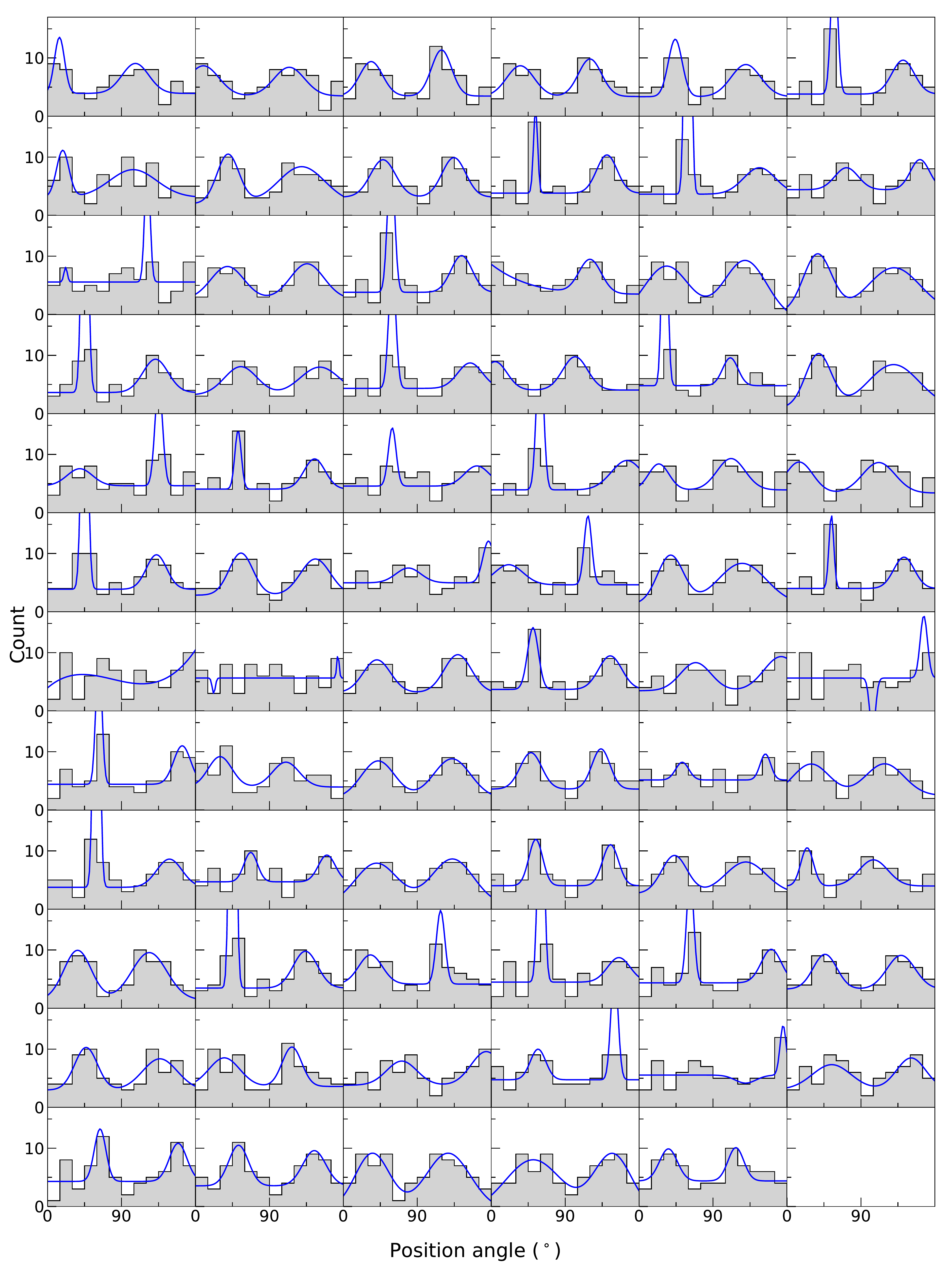}
	\caption[Position angles parallel transported to each LQG position]{LQG position angles after parallel transport to the location of each LQG; $15^{\circ}$ bins. Solid blue lines are double Gaussian fit. The bimodal distribution is generally robust to parallel transport destination.}
	\label{figGaussians}
\end{figure}

We parallel transport all 71 PAs to the location of each of the 71 LQGs, and at each one fit a double Gaussian to the unweighted histogram of these PAs. The height, width, and location of each Gaussian are fitted using Python's \texttt{scipy.optimize. leastsq}. The results are shown in Fig.~\ref{figGaussians}; at most ($\gtrsim 80\%$) LQG locations the PAs still show a similar bimodal distribution.

\begin{figure} % fit_gaussians_to_pas
    \centering
	\includegraphics[width=0.6\columnwidth]{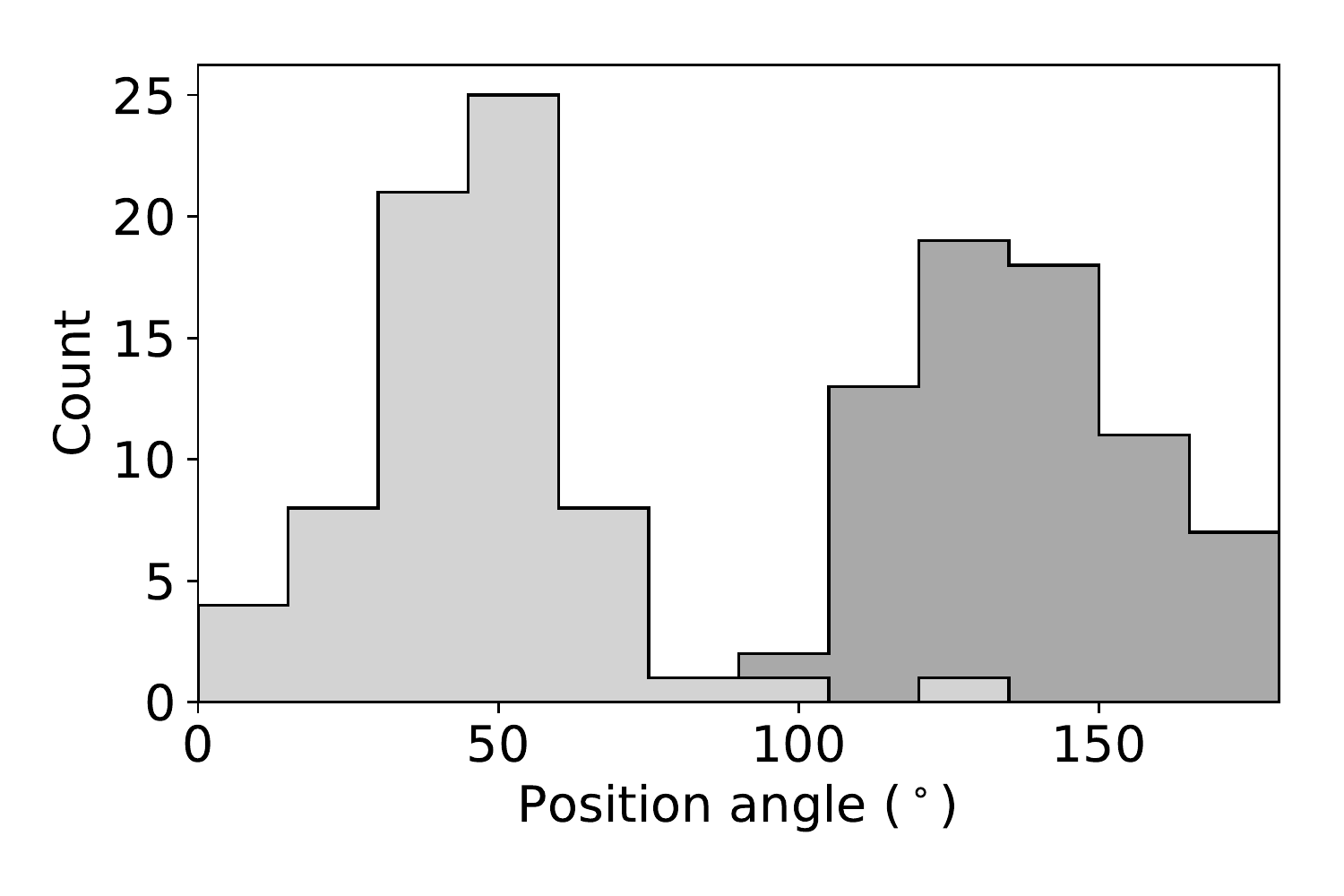}
	\caption[Location of double Gaussian peaks from Fig.~\ref{figGaussians}]{Location of the centres of the double Gaussian peaks in Fig.~\ref{figGaussians}; $15^{\circ}$ bins. The locations of the two peaks are consistent, regardless of parallel transport destination. The mean locations are $\sim47\pm 2^{\circ}$ and $\sim138\pm 2^{\circ}$, consistent (within $\lesssim10^{\circ}$) with the HDS result.}
	\label{figPeaks}
\end{figure}

The location of the centres of the double Gaussian peaks, in each of the 71 LQG locations, are shown in Fig.~\ref{figPeaks}. The median (mean) location of the first peak is $\sim45\pm2^{\circ}$ ($\sim47\pm2^{\circ}$), and the median (mean) location of the second peak is $\sim136\pm2^{\circ}$ ($\sim138\pm2^{\circ}$). These are consistent (within the mean PA half-width confidence interval of $\sim10^{\circ}$) with the peaks identified by HDS at the centre of the A1 region ($\sim54\pm2^{\circ}$ and $\sim136\pm3^{\circ}$ for unweighted PAs).

\subsection{Correlation: LQG PAs are aligned and orthogonal} \label{subsecCorrelationResults}

\begin{figure} % S_test_LQGs
    \centering
    \includegraphics[width=0.98\columnwidth]{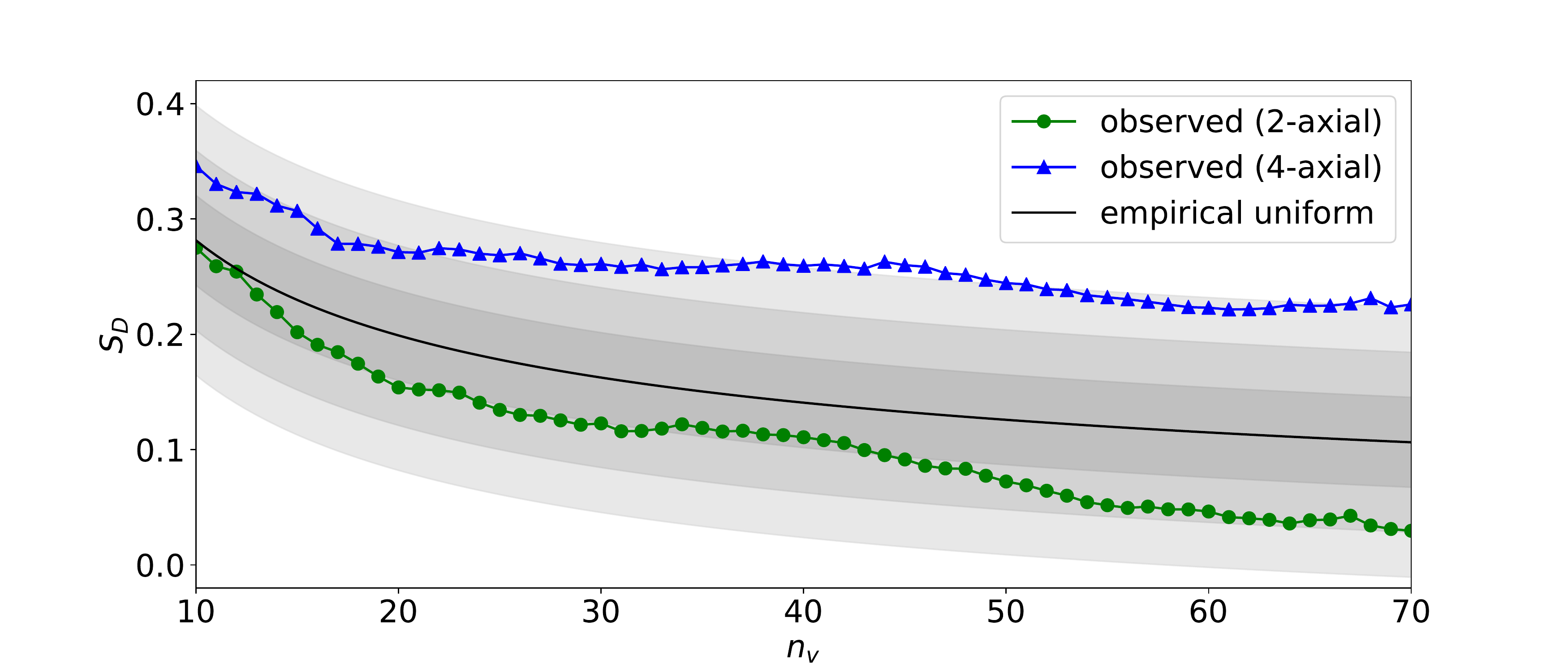}
    \caption[S test statistic for 2-axial and 4-axial position angles]{The S test statistic $S_D$ calculated for 2-axial (green circles) and 4-axial (blue triangles) PAs as a function of nearest neighbours $n_v$ determined in 3D. Also shown, empirical values estimated for a uniform distribution (black) and their approximate $\pm 1\sigma,\ 2\sigma,\ 3\sigma$ confidence intervals (grey). 4-axial PAs show more alignment than uniforms, whilst 2-axial PAs show less.}
    \label{figSD}
\end{figure}

We compute the S statistic $S_D$ using the \citet{Jain2004} coordinate invariant version of the S test for samples of $n_v$ nearest neighbours, where $10 \leq n_v \leq 70$. In Fig.~\ref{figSD} we show the values of $S_D$ calculated for observed LQG PAs, represented as both 2-axial (green circles) and 4-axial (blue triangles) data, as a function of nearest neighbours $n_v$. Nearest neighbours are determined in 3D by taking into account radial distance. Results are entirely consistent with neighbours determined in 2D (i.e.\ by angular separation).

Larger values of $S_D$ indicate stronger alignment. So Fig.~\ref{figSD} illustrates that 4-axial LQG PAs are more aligned than expected if they are randomly drawn from a uniform distribution. Conversely, it also suggests that 2-axial LQG PAs are less aligned (more orthogonal) than expected if they are randomly drawn from a uniform distribution. The physical interpretation of the latter is not straightforward, but likely to be due to the contribution from orthogonal PAs leading to large dispersion, and hence a small value of $S_D$.

\begin{figure} % S_test_LQGs
    \centering
    \includegraphics[width=0.98\columnwidth]{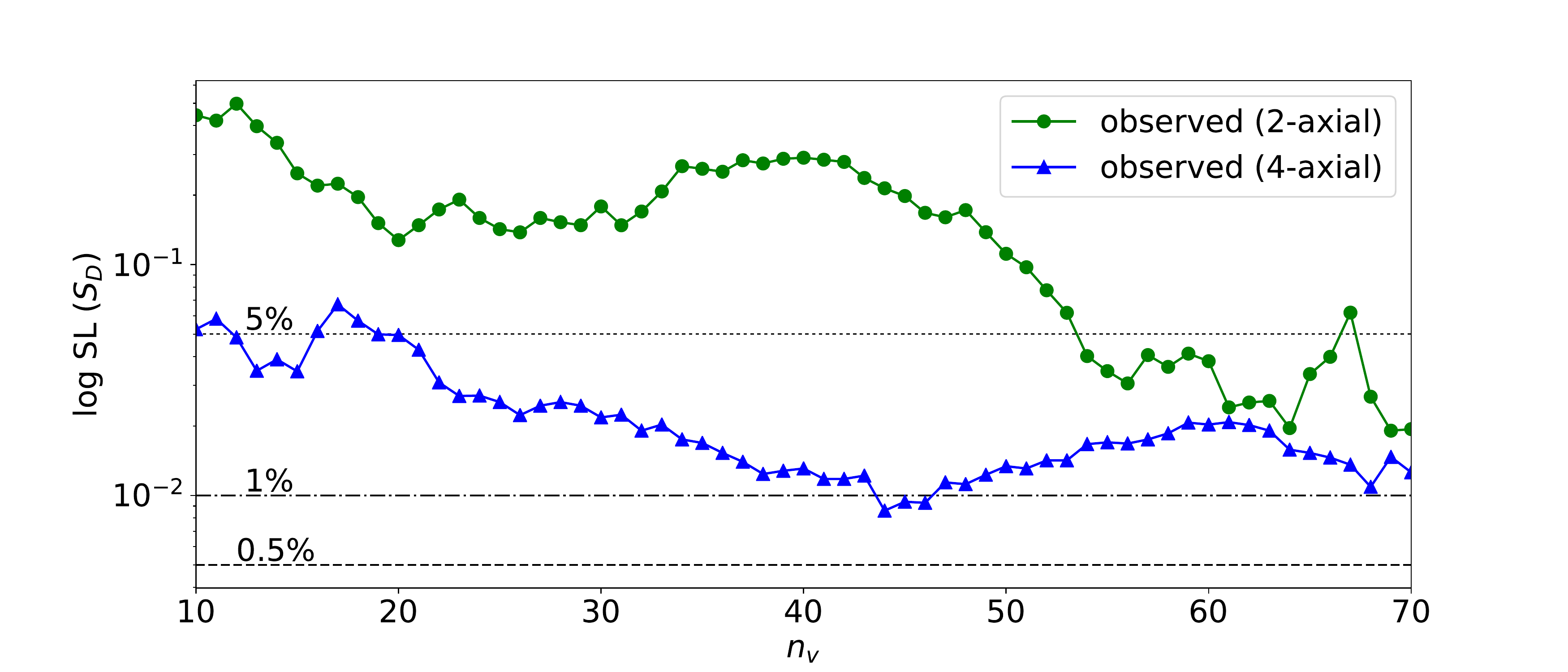}
    \caption[S test significance level for 2-axial and 4-axial position angles]{The logarithmic significance level (SL) of the S test calculated for 2-axial (green circles) and 4-axial (blue triangles) PAs as a function of nearest neighbours $n_v$ determined in 3D. The dotted, dash-dotted and dashed horizontal lines indicate SL = 0.05, 0.01 and 0.005 respectively. 2-axial LQG PAs show correlation ($\mathrm{\overline{SL}}\sim 3.3\%$) above $n_v \ge 54$. 4-axial PAs show correlation ($\mathrm{\overline{SL}}\sim 1.5\%$) above $n_v \ge 30$.}
    \label{figSL}
\end{figure}

The approximate empirical standard deviation, and hence the $\pm 1\sigma, 2\sigma$, and $3\sigma$ confidence intervals shown in Fig.~\ref{figSD}, are only valid for large $n$ while $n_v \ll n$ \citep{Jain2004}. These are not valid assumptions for much of our range of $S_D$. Therefore, due to this and the mutual dependence of groups of nearest neighbours, we calculate the significance level of the S test using numerical simulations. In Fig.~\ref{figSL} we show the significance level of the S test calculated using 10,000 numerical simulations, for values of $S_D$ determined using both 2-axial (green circles) and 4-axial (blue triangles) PAs. Again, this is shown as a function of nearest neighbours $n_v$.

For 4-axial LQG PAs we are testing for alignment and orthogonality by combining the modes, resulting in alignment only (right-hand $S_D$ tail), so the significance level (SL) is defined as the proportion of numerical simulations with an S test statistic $S_D$ higher than that observed (Eq.~\ref{eqSL4ax}). We find significance levels generally between 1\% and 5\% for most numbers of nearest neighbours $n_v$. For most of the $n_v$ range ($30 \le n_v \le 70$) the significance level is $0.9\% \le \mathrm{SL} \le 2.2\%$, with a mean (median) of 1.5\% (1.5\%). It is most significant for $n_v \sim 45$, with $\mathrm{SL} \simeq 0.8\%$.

For 2-axial LQG PAs the orthogonal mode dominates (left-hand $S_D$ tail), so the SL is defined as the proportion of simulations with an S test statistic $S_D$ lower than that observed (Eq~\ref{eqSL2ax}). We find significance levels generally above 5\% until $n_v \ge 54$. For the remainder of the range of $n_v$ ($54 \le n_v \le 70$) the SL is $1.9\% \le \mathrm{SL} \le 6.2\%$, with a mean (median) of 3.3\% (3.4\%).

\subsubsection{Typical angular and comoving separations}

\begin{figure} % nv_to_sep
    \centering
    \subfigure[Median angular separation as a function of $n_v$]{\includegraphics[width=0.98\columnwidth]{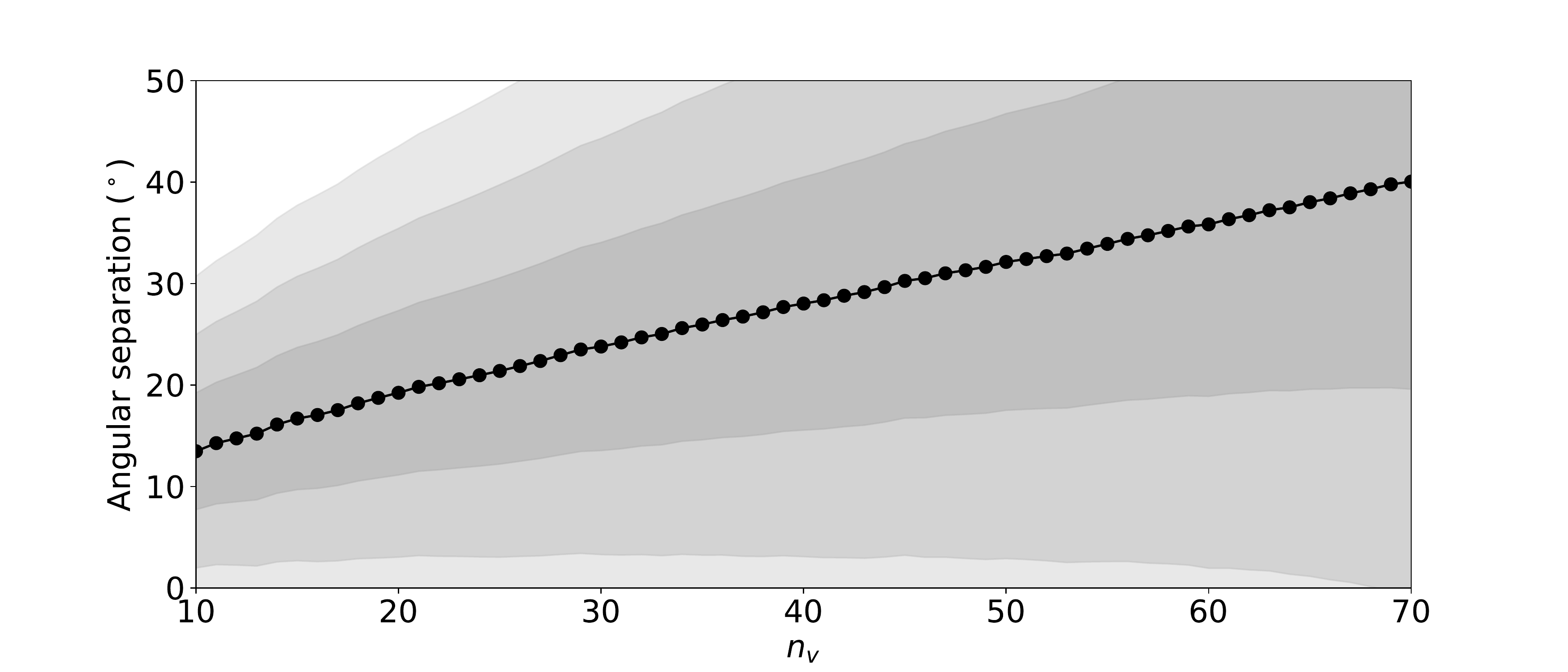} \label{subAngSep}}\\
    \subfigure[Median comoving separation as a function of $n_v$]{\includegraphics[width=0.98\columnwidth]{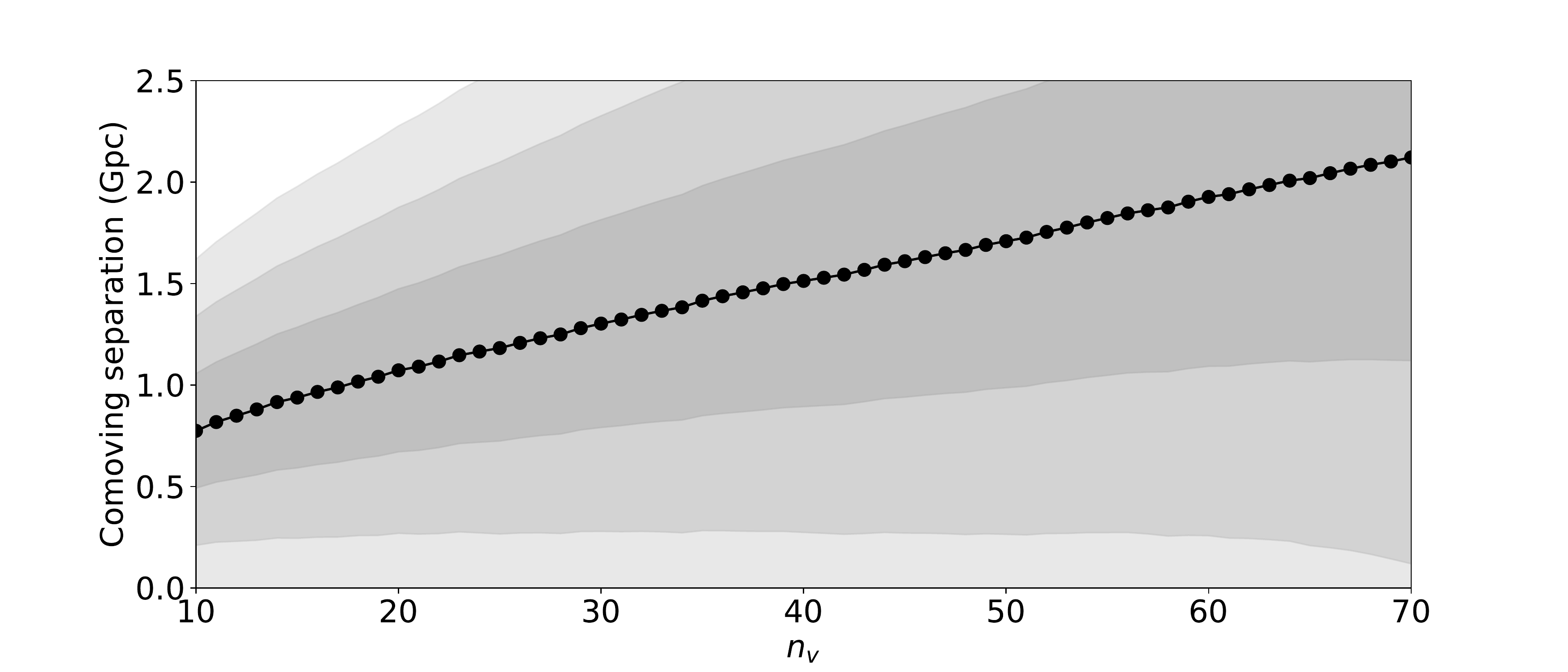} \label{subCartSep}}
    \caption[Median angular and comoving LQG separation as a function of $n_v$]{The median \protect\subref{subAngSep} angular separation and \protect\subref{subCartSep} comoving separation of LQGs as a function of their number of nearest neighbours $n_v$. Grey bands illustrate the distribution dispersions as $\pm 1\sigma,\ 2\sigma,\ 3\sigma$ standard deviations.}
    \label{figNvFnSep}
\end{figure}

The number of LQG nearest neighbours $n_v$ is a free parameter explored by the S test. In this work we identify these neighbours using three-dimensional comoving positions of each LQG centroid. The parameter $n_v$ is related to the scale of the nearest neighbour groups, but because LQGs are not homogeneously distributed we cannot directly interpret it as corresponding to a particular scale. In Fig.~\ref{figNvFnSep} we show the relationship between the parameter $n_v$ and the typical scale of the nearest neighbour groups, defined as the median angular (comoving) separation between each LQG and its $n_v$ nearest neighbours. The relationships are nearly linear when $n_v \gtrsim 20$, and the distributions naturally become wider with increasing $n_v$.

\begin{figure} % nv_to_sep
    \centering
    \subfigure[S test significance level as a function of angular separation]{\includegraphics[width=0.98\columnwidth]{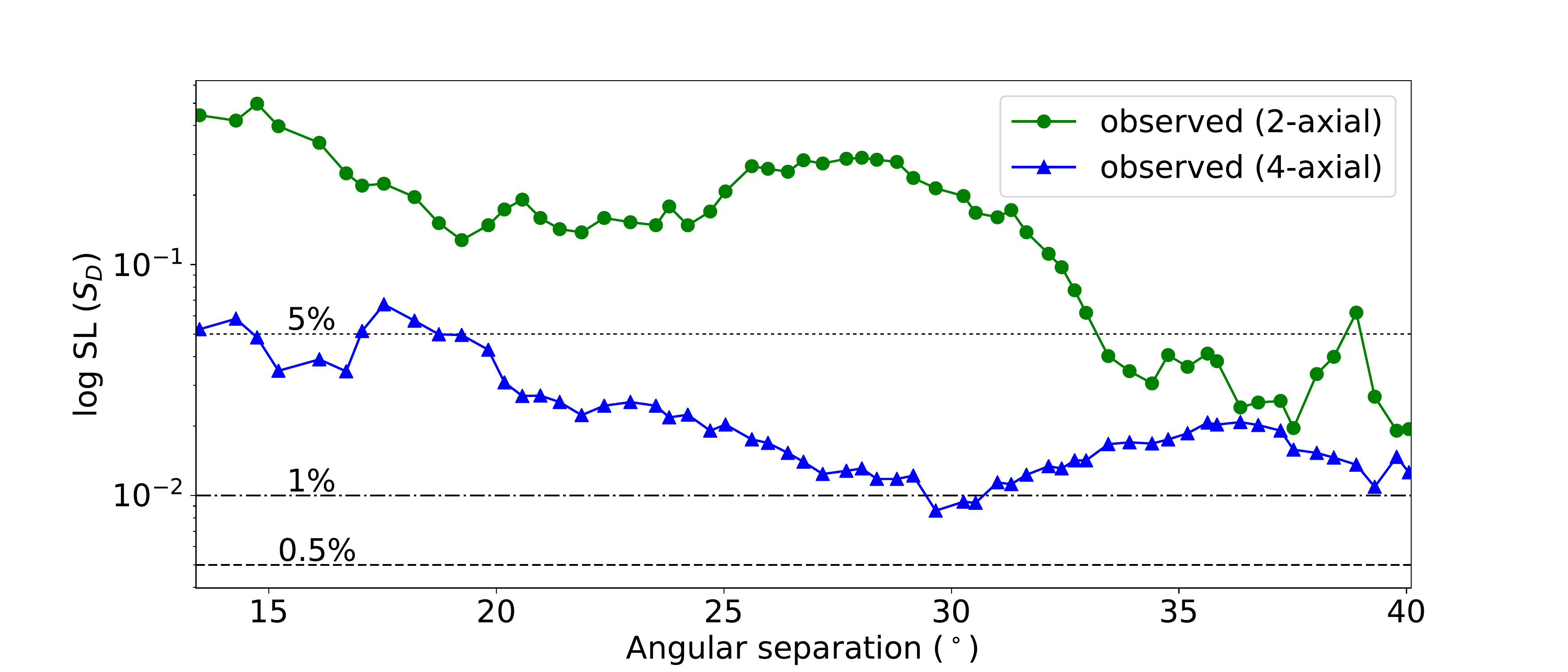} \label{subSLAngSep}}\\
    \subfigure[S test significance level as a function of comoving separation]{\includegraphics[width=0.98\columnwidth]{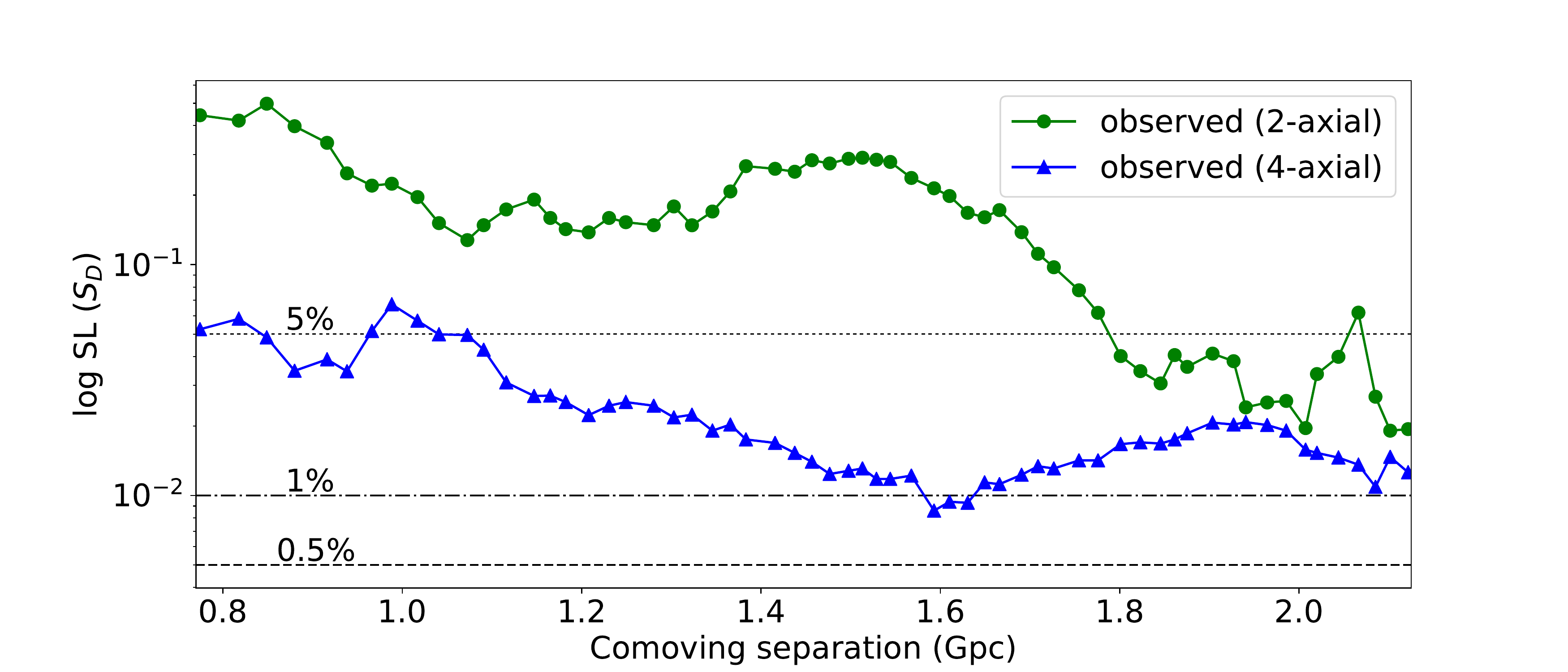} \label{subSLCartSep}}
    \caption[S test significance level as a function of LQG separation]{The logarithmic significance level (SL) of the S test calculated for 2-axial (green circles) and 4-axial (blue triangles) PAs as in Fig.~\ref{figSL} but as a function of typical \protect\subref{subSLAngSep} angular separation and \protect\subref{subSLCartSep} comoving separation of LQGs. The dotted, dash-dotted and dashed horizontal lines indicate SL = 0.05, 0.01 and 0.005 respectively.}
    \label{figSLFnSep}
\end{figure}

Using these relationships, in Fig.~\ref{figSLFnSep} we show the S test significance levels as a function of typical angular (comoving) separation instead of nearest neighbours $n_v$. The functions do not differ significantly in shape, because the relationships are generally linear, but they show that the correlation is most significant for typical angular (comoving) separations of $\sim 30^{\circ}$ (1.6 Gpc).

\clearpage

\section{Discussion and conclusions} \label{secLQGOrientationsConclusions}

The orientations of our sample of 71 large quasars groups are expressed as position angles, which are intrinsically 2-axial [$0^{\circ}$, $180^{\circ}$). We note that multimodality in the PA distribution reduces the power of many statistical tests. A statistically legitimate approach to overcome this in cases of 2-fold symmetry \citep{Fisher1993} is to convert PAs to 4-axial [$0^{\circ}$, $90^{\circ}$) for analysis. We argue this is also physically legitimate; we have \textit{a priori} motivation to suspect this type of distribution from \cite{Pelgrims2015}.

We demonstrate that the 4-axial PAs are unlikely to be drawn from a uniform distribution. Using a variety of tests, with and without goodness-of-fit weighting, the probability of uniformity is 0.8\% - 7\%. However, mock LQG catalogues constructed from cosmological simulations \citep{Marinello2016} show no evidence for non-uniformity of either 2-axial or 4-axial PAs. We further conclude that the 2-axial PA distribution is bimodal, with 97\% confidence. Hartigans' dip statistic for weighted 2-axial data indicates modes (peaks) at $\bar{\theta}\sim52\pm2^{\circ}$ and $\bar{\theta}\sim137\pm3^{\circ}$, remarkably close to the mean angles of radio quasar polarization, $\bar{\theta}\simeq42^{\circ}$ and $\bar{\theta}\simeq131^{\circ}$, reported by \citet{Pelgrims2015}\footnote{Mean angles derived by the \citet{Hawley1975} method; errors were not reported}.

In addition to a bimodal PA distribution, we also observe a potentially bimodal redshift distribution with modes at $z \sim 1.2$ and $z \sim 1.5$. However, we cannot identify a correlation between the modes of the PA and redshift distributions. In other words, we cannot attribute the $\bar{\theta}\sim52^{\circ}$ PA mode to LQGs from the $z \sim 1.2$ redshift mode etc. It is possible that both PA modes ($\bar{\theta}\sim52^{\circ}$ and $\bar{\theta}\sim137^{\circ}$) exist at both redshift modes ($z \sim 1.2$ and $z \sim 1.5$), although a plausible physical mechanism for this is far from obvious.

The uniformity and bimodality tests are applied to PAs parallel transported to the centre of the A1 region \citep{Hutsemekers1998}. To check that results are not dependent on the parallel transport destination, we further parallel transport all PAs to the location of each of the 71 LQGs. Analysis of the resultant histograms indicates the median locations of the modes are $\bar{\theta}\sim45\pm2^{\circ}$ and $\bar{\theta}\sim136\pm2^{\circ}$, consistent with results at the centre of the A1 region. The bimodality is therefore unlikely to be an artefact of our choice of parallel transport destination.

Finally, we assess PA correlation using the S test. We find 4-axial PAs are more aligned than numerical simulations, with mean significance level $\overline{\mathrm{SL}} \simeq 1.5\%$ for number of nearest neighbours $n_v \geq 30$. This is most significant ($\mathrm{SL} \simeq 0.8\%$) for $n_v \sim 45$. Note that based on this test alone we cannot distinguish between `alignment only' and `alignment plus orthogonality' signals. We find 2-axial PAs are more orthogonal than simulations, with $\overline{\mathrm{SL}} \simeq 3.3\%$ for $n_v \geq 54$. We conclude that PAs are \textit{both} aligned and orthogonal, consistent with two modes separated by $\Delta\theta \sim 90^{\circ}$, and that the orthogonal mode somewhat dominates, particularly on the largest scales.

The two regions of radio quasar polarization alignment identified by \citet{Pelgrims2015} are coincident to our LQG sample, and the preferred angles they report ($\bar{\theta}\simeq42^{\circ}$ and $\bar{\theta}\simeq131^{\circ}$) are remarkably close to our LQG PA modes. They note that these values are close to $45^{\circ}$ and $135^{\circ}$, and these particular values (multiples of $45^{\circ}$) lead them to consider systematic bias in the dataset \citep{Battye2008}. They conclude this is unlikely. Furthermore, \citet{Hutsemekers2014} and \citet{Pelgrims2016} report that quasar polarization vectors are preferentially aligned and orthogonal to LQG axes.

LQG axes being preferentially aligned at $\bar{\theta}\sim45\pm2^{\circ}$ and $136\pm2^{\circ}$ (this work, median modes), together with quasar polarization vectors being preferentially parallel and orthogonal to LQG axes \citep{Hutsemekers2014,Pelgrims2016}, would result in polarization vectors with preferred angles of $\sim42^{\circ}$ and $\sim131^{\circ}$ \citep{Pelgrims2015}.

% END

%% file: chapterDiscussion.tex
\chapter{Discussion} \label{chaptDiscussion}

This work comprised two complementary projects which investigated correlations between three different tracers of the cosmic web, namely the cosmic microwave background (CMB), supernovae (SNe), and large quasar groups (LQGs).

\section{General data and methods contributions}

Much of this work was devoted to thorough understanding of the data and its careful analysis using appropriate statistical methods, specifically to examine the apparent correlations between CMB temperature and SNe redshifts, and LQG orientations. Before discussing our specific projects in the remainder of this chapter, in this section we discuss some general contributions which are equally applicable to other projects and fields.

We talked in detail about the challenges of statistical analysis of axial and/or bimodal data in chapters~\ref{chaptLQGSample} and \ref{chaptLQGStatsMethods}. We will not reprise that discussion here, except to note that these characteristics reduce the discriminatory power many statistical methods, as discussed previously. This means there is a higher chance we would incorrectly fail to reject the null hypothesis (such as uniformity), i.e.\ an increased chance of a type II error, or a false negative.

\begin{figure}[t!] % A1_SNe_all_fields_remainder
    \centering
	\subfigure[Binned]{\includegraphics[width=0.42\columnwidth]{binned_scatter_fields_1to7_all_square.pdf} \label{subBinned}}
	\subfigure[Unbinned]{\includegraphics[width=0.55\columnwidth]{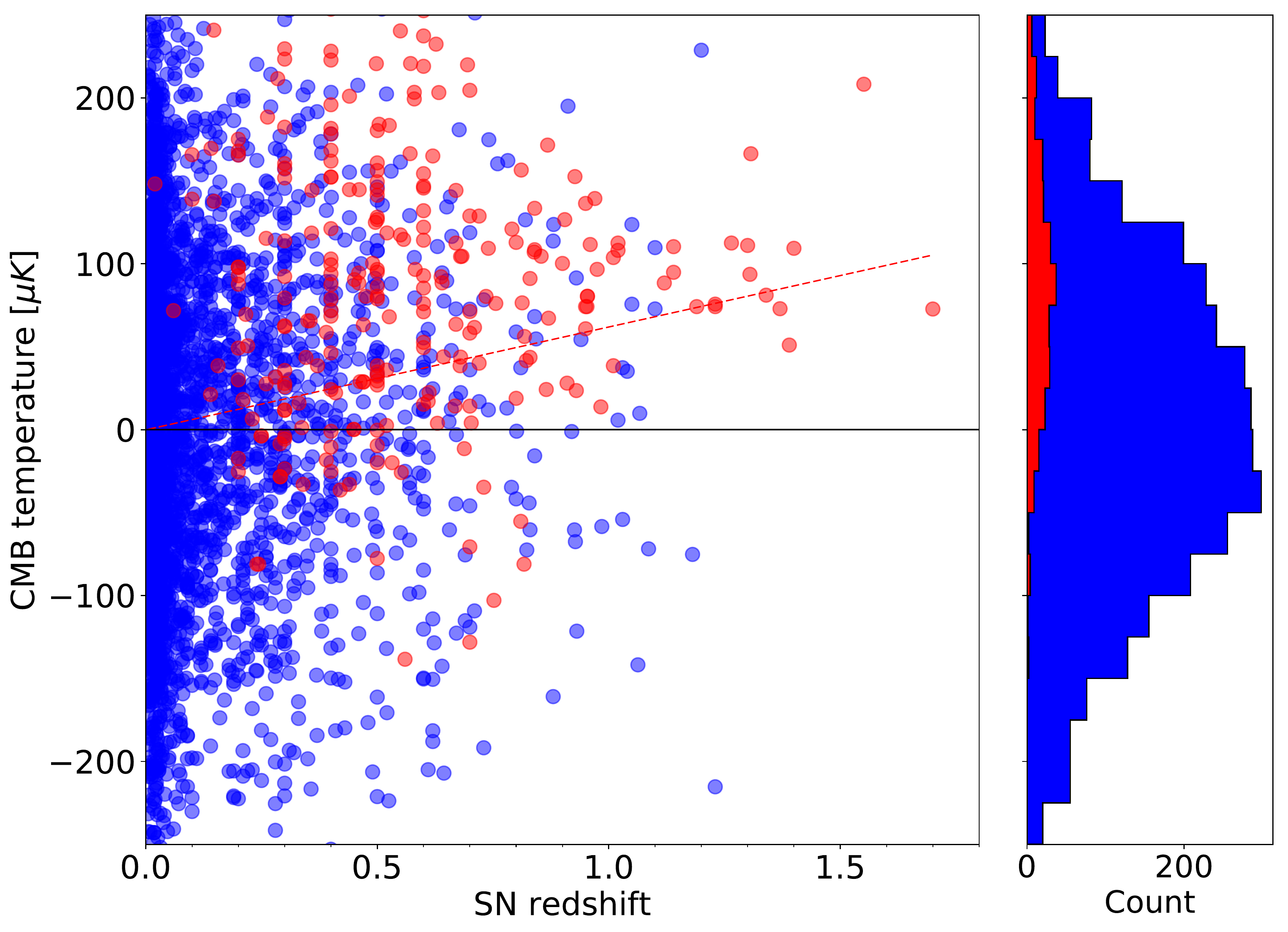} \label{subUnbinned}}
	\caption[CMB $T$ as a function of SNe $z$ (binned and unbinned)]{Plot of CMB temperature at SNe locations versus SNe redshift. (a) is as Fig.~\ref{figBinnedScatter}, showing data binned with bin size $\Delta z=0.01$ and error bars of the standard error on the bin mean. (b) is the same data but without binning, showing SNe from the seven deep survey fields in red and the remainder in blue; the marginal histogram shows their different CMB temperature distributions with bin size $\Delta T=25\mu K$. Dashed lines indicate ordinary least squares linear regression of (a) binned and (b) unbinned data.}
	\label{figBinning}
\end{figure}

In the first project, analysing the correlation between CMB temperature and SNe redshifts, we revisited the data and methods used by \citet{Yershov2012,Yershov2014} and found them generally appropriate. Indeed, we found similar results. However, \citet{Yershov2012,Yershov2014} present and analyse the apparent correlation using binned data, masking its heteroscedasticity. Specifically, the variance of CMB temperature at SNe locations is unequal across the range of redshifts. This is illustrated in Fig.~\ref{figBinning}, which shows (a) the binned data with its more compelling correlation and (b) the raw (unbinned) data with wider variance at low redshifts.

High-redshift SNe tend to be in deep survey fields which, due to chance alignments, generally have hot \textit{Planck} CMB pixel temperatures. Low-redshift SNe are more uniformly scattered across the sky and thus have far wider variance of hot and cold pixels. Identifying this led us to a different interpretation of the data to \citet{Yershov2012,Yershov2014}. We conclude that low-redshift SNe are not necessarily associated with pixels of average CMB temperature ($\Delta T \sim 0$), in fact many are on particularly hot or cold pixels (Fig.~\ref{subUnbinned}). Rather, low-redshift bins \textit{average} to a CMB temperature approaching zero ($\overline{\Delta T} \sim 0$). Further, whilst high-redshift bins do average to a hotter mean, remember SNe pixels are generally no hotter than other pixels in their deep field.

We also found that binning was problematic in the second project, examining the LQG position angle distribution. The $\chi^2$ test was somewhat sensitive to bin size, although not significantly enough to affect our conclusions. The bimodality coefficient, however, varied so wildly with bin size that its results were wholly untrustworthy. This could have led to a higher chance of us incorrectly rejecting the null hypothesis (in this case unimodality), i.e.\ an increased chance of a type I error, or a false positive. We only employed binning after first examining the raw data, and then we explored a range of bin sizes to confirm our results were robust.

We used parallel transport to ensure our analysis was coordinate-invariant. The size of parallel transport corrections becomes larger the more widely the data are distributed across the celestial sphere. We note that without ensuring coordinate invariance, the choice of coordinate frame could dramatically affect the results of orientation analyses, potentially leading to type I or type II errors.

\section{CMB $T$ and SNe $z$ correlation is due to bias}

In chapter~\ref{chaptSNeCMBT} we investigated the cross-correlation of SNe position and CMB temperature, reported by \citet{Yershov2012,Yershov2014} and attributed by them to the integrated Sachs-Wolfe effect and/or to some unrelated distant foreground emission. We were motivated to re-examine their result after noticing the potentially misleading effect of binning (Fig.~\ref{figBinning}) and that despite calculating errors (Fig.~\ref{subBinned} error bars) they did not perform weighted regression. In this work, we analysed the apparent correlation independently, and our results (including weighted regression) supported the prima facie existence of the correlation. Our interpretation, however, differs.

We identified seven regions with a high surface density of SNe, all of which are coincident with deep survey fields. There were 3 fields from the Supernova Legacy Survey \citep[SNLS,][]{Astier2006}, 2 from the ESSENCE supernova survey \citep{Miknaitis2007}, the Hubble Deep Field North \citep[HDF-N,][]{Williams1996}, and the Chandra Deep Field South \citep[CDF-S,][]{Giacconi2001}.

We show that all seven fields are aligned with CMB hotspots. The CMB temperatures of the ESSENCE wdd, SNLS D4, and CDF-S fields are particularly extreme, with means at SNe locations of $\overline{T} = 130.8 \pm 9.2 \,\mu K$, $\overline{T} = 199.3 \pm 13.6 \,\mu K$, and $\overline{T} = 120.5 \pm 12.0 \,\mu K$ respectively. As deep survey fields, unsurprisingly all seven fields are also associated with high-redshift SNe. The HDF-N, SNLS D4, and CDF-S fields are particularly extreme, with means of $\overline{z} = 0.94 \pm 0.06$, $\overline{z} = 0.67 \pm 0.04$ and $\overline{z} = 0.89 \pm 0.11$ respectively.

We also identified a high surface density of SNe in SDSS Stripe 82. However, although Stripe 82 is slightly deeper it is not hotter, and does not significantly contribute to the correlation.

We concluded \citep{Friday2018} that, rather than having an astrophysical origin, the apparent correlation between CMB temperature and SNe redshifts is a composite selection bias caused by a combination of

\begin{enumerate}[(a)]
\item high-redshift SNe being preferentially located in deep survey fields, plus
\item the chance alignment of seven of those fields with CMB hotspots.
\end{enumerate}

\noindent Using Monte Carlo simulations and random CMB realisations we estimated the likelihood of this alignment occurring by chance is $6.8\%-8.9\%$, so not exceptionally unlikely.

\citet{Yershov2020} subsequently claimed the distant foreground explanation of the apparent correlation between CMB temperature and SNe redshifts could also resolve the much debated discrepancy between local measurements of the Hubble parameter $H_0$ and the \textit{Planck}-derived value \citep[see review by][]{Riess2019}. They argue that contamination from cold ($\sim 5$K) extragalactic dust would increase the dispersion of CMB temperature fluctuations, reducing the \textit{Planck}-derived value of $H_0$. This may be plausible, but still does not explain their claim that high-redshift SNe are preferentially associated with hotter CMB temperatures, nor have they acknowledged that this could simply be due to chance alignment.

The apparent correlation between CMB temperature and SNe redshifts is one example of deep survey fields biasing CMB cross-correlation. Deep fields could, potentially, bias any cross-correlation analyses between astronomical objects (e.g., SNe, galaxies, GRBs, quasars) and the CMB (temperature or lensing). It is conceivable that deep fields could, by chance, also be aligned with distant large-scale structures, voids, cosmic bulk flows, or directions of anisotropic cosmic expansion (should they exist), which may similarly bias other cross-correlation analyses.

Furthermore, SNe datasets are spatially non-uniform, with a wide swathe of low-redshift SNe together with narrow pencil beams of high-redshift SNe, such as our seven deep survey fields. Perhaps this non-uniformity, together with chance alignments, could help explain some of the tensions that have been reported between and within SNe datasets \citep[e.g.,][]{Choudhury2005, Nesseris2007, BuenoSanchez2009, Karpenka2015}.

\subsection{Potential CMB and LQG correlations}

Following the CMB $T$ and SNe $z$ correlation work, we briefly explored potential cross-correlation between our LQG sample and the CMB temperature and lensing fields. This was initially motivated by the reported coincidence of the Huge-LQG with a CMB temperature anomaly \citep{Romano2015}.

For most LQGs we did not identify any CMB temperature anomaly, although 9 of the largest 10 LQGs were associated with hotter than average CMB temperatures. A `filament' of 4 LQGs arranged end-to-end was coincident with a large, linear, hot feature in the CMB ($\Delta T = 71.0 \pm 0.3\; \mu\mathrm{K}$). It was this that prompted our investigation into LQG orientations. Very preliminary analysis using random CMB realisations indicated that the likelihood of the linear hotspot occurring by chance was $\lesssim 0.5\%$. Further, this feature was also associated with a well defined branch of a minimal spanning tree constructed from peaks in the CMB lensing field.

We conclude that it is possible these 4 LQGs are tracing a linear matter over-density, which leaves an imprint on both CMB temperature and lensing. But we leave the investigation of this potential feature, and the statistical analysis of its potential correlations, to future work.

\section{Large-scale correlation of LQG orientations}

In the remainder of this work (chapters \ref{chaptLQGSample}, \ref{chaptLQGStatsMethods}, and \ref{chaptLQGOrientation}) we investigated for the first time the intriguing potential correlation of LQG orientations. Such correlations, if real, would inform the debate about the `realness', origin, and significance of LQGs, and whether or not they may challenge the cosmological principle \citep[e.g.][]{Nadathur2013,Pilipenko2013,Park2015}. Additionally, coherent orientation of their host LQGs could offer a plausible origin for the alignment of quasar polarization, first reported by \citet{Hutsemekers1998} and most recently by \cite{Pelgrims2019}. It would provide corroborating evidence for, and enhancement of, the intrinsic alignment interpretation of the results from these and many other quasar polarization studies \citep[e.g.][]{Hutsemekers2001,Jain2004,Cabanac2005,Tiwari2013,Pelgrims2014,Pelgrims2015}.

In chapter~\ref{chaptLQGSample} we presented a sample of 71 LQGs with members $20 \leq m \leq 73$ at redshifts $1.0 \leq z \leq 1.8$. We described two and three-dimensional methods for determining LQG orientation, and showed that position angles determined by each are largely consistent. The two-dimensional approach has slightly smaller uncertainties, with a mean half-width confidence interval $\sim10^{\circ}$ estimated from 10,000 bootstraps. This, together with concerns about redshift errors and possible line-of-sight biases in the three-dimensional method, motivated us to primarily use the two-dimensional approach in this work. To account for the variation in PA uncertainty between different LQGs we analysed PAs both with and without goodness-of-fit weighting, and found most results robust to whether or not data were weighted.

We introduced the methods necessary to analyse axial [$0^{\circ}$, $180^{\circ}$) PA data, and to test for alignment and, simultaneously, orthogonality. The latter motivated, in part, by reports that quasar polarization is preferentially either aligned or orthogonal to LQG axes \citep{Hutsemekers2014,Pelgrims2015}. We described the need for coordinate-invariance and explained parallel transport \citep{Jain2004}, including a variant applying it for the first time to PAs transported to a single common destination before analysis.

We found that the PA distribution is bimodal, which together with the axial nature of PAs, and their wide distribution on the celestial sphere, made statistical analysis challenging. We devoted chapter~\ref{chaptLQGStatsMethods} to describing and evaluating in detail several statistical methods to analyse the uniformity, bimodality, and correlation of these data. Then in chapter~\ref{chaptLQGOrientation} we presented the results of our statistical analyses, which indicate that LQG position angles are unlikely to be drawn from a uniform distribution, their distribution is bimodal, and they are correlated such that the modes ($\bar{\theta}\sim52\pm2^{\circ}, 137\pm3^{\circ}$ for weighted PAs) are roughly aligned and orthogonal ($\Delta\theta \sim 90^{\circ}$).

\subsection{Summary of methods evaluation and results}

In this section we recap our evaluation of certain methods to analyse the uniformity, bimodality, and correlation of LQG position angles, and we summarise the results of these tests.

%\vspace{5mm}
\noindent \textbf{Uniformity tests}

\textbf{Hermans-Rasson test}. We found the most generally applicable test of uniformity for our data is the Fourier-type Hermans-Rasson test \citep[HR,][]{Hermans1985}, which to the best of our knowledge we bring to astrophysical data for the first time. The HR test hinted at non-uniformity, with p-values of 0.07 (0.04) for 2-axial (4-axial) data. For mock LQGs constructed from cosmological simulations \citep{Marinello2016} the HR test showed no evidence for non-uniformity using either 2-axial or 4-axial PAs (both unweighted).

\textbf{Kuiper's test}. Kuiper's test \citep{Kuiper1960} of uniformity suffers dramatically reduced discriminatory power for multimodal distributions. This makes it inappropriate (low power) for our 2-axial PAs, which are bimodal, and as expected Kuiper's test showed no evidence of non-uniformity for these data. We found Kuiper's test powerful for our 4-axial PAs, where combining the two modes results in a unimodal distribution, and for these data it indicated non-uniformity with p-values of 0.009 (0.008) using unweighted (weighted) PAs. For mock LQGs Kuiper's test generally showed no evidence for non-uniformity using either 2-axial or 4-axial PAs (both unweighted). The one exception was for the stack of all mock LQGs when evaluated as 4-axial data, which indicated marginal evidence of non-uniformity with a p-value of 0.03. Our concerns about the statistical analysis of this particular stack, the otherwise highly consistent results, and our caution about interpreting results manifest only in 4-axial data, led to this anomalous result being discredited. We concluded that Kuiper's test indicated statistical uniformity of the mock LQGs.

\clearpage

\textbf{$\chi^2$ test}. We included the $\chi^2$ test due to its simplicity and familiarity, but noted it has lower discriminatory power than tests applied to continuous data, is somewhat sensitive to bin size, and is not well suited to axial data. The $\chi^2$ test indicated marginal evidence of non-uniformity, particularly for weighted PAs with p-values of 0.01 (0.02) for 2-axial (4-axial) data. For mock LQGs the $\chi^2$ test showed no evidence for non-uniformity using either 2-axial or 4-axial PAs (both unweighted).

%\vspace{5mm}
\noindent \textbf{Bimodality tests}

\textbf{Bimodality coefficient}. We investigated the bimodality coefficient, but noted it was unsuitable for hypothesis significance testing, has an undesirable sensitivity to skew, and found it inconsistent with different bin sizes. We concluded the bimodality coefficient was too capricious for us to draw any conclusions from its results.

\textbf{Hartigans' dip statistic}. To examine PA bimodality we found Hartigans' dip statistic \citep[HDS,][]{Hartigan1985} to be robust and informative. HDS indicated that the PA distribution is bimodal with 97\% confidence. It identified modes (peaks) at $\bar{\theta}\sim52\pm2^{\circ}$ and $\bar{\theta}\sim137\pm3^{\circ}$. By choosing alternative parallel transport destinations, we demonstrated that the PA bimodality was robust to choice of destination, with the median location of the peaks at $\bar{\theta}\sim45\pm2^{\circ}$ and $\bar{\theta}\sim136\pm2^{\circ}$. These are remarkably close to the mean angles of radio quasar polarization reported by \cite{Pelgrims2015}\footnote{Errors on these means were not reported}, $\bar{\theta}\simeq42^{\circ}$ and $\bar{\theta}\simeq131^{\circ}$, in two regions coincident with our LQG sample. They note that these angles are close to $45^{\circ}$ and $135^{\circ}$, and these particular values lead them to consider systematic bias in the dataset \citep{Battye2008}, which they concluded is unlikely.

In addition to a bimodal PA distribution, we also observed hints of a bimodal redshift distribution with modes at at $z \sim 1.2$ and $z \sim 1.5$. We did not identify a correlation between the modes of the PA and redshift distributions; further investigation would comprise a follow-up study of the redshift distribution.

\vspace{5mm}
\noindent \textbf{Correlation tests}

\textbf{Z test}. We considered the Z test since \citet{Hutsemekers2014} report it is better suited to small samples than the S test. However, Pelgrims (2019, private communication) noted that the Z test uses a measure of the correlation between PA and position, so is unable to detect `global' alignments (i.e.\ correlations present throughout the survey volume). As expected, we do not find a signal with the Z test, probably because the PA correlations are global rather than local.

\textbf{S test}. We found the \citet{Jain2004} version of the S test was appropriate for our data. We dedicated several sections to explaining the application and interpretation of the S test, particularly in relation to simultaneously testing for alignment and orthogonality by combining the modes, which was not initially intuitive. We demonstrated the effects on the S test statistic $S_D$ distribution of (1) changing the number of nearest neighbours, (2) converting from 2-axial to 4-axial data (the S test had not previously been used with 4-axial data), and (3) applying parallel transport corrections. We explained how all these factors are incorporated into numerical simulations to estimate the significance of the PA correlation.

Using the S test we examined the correlation in LQG PAs as a function of nearest neighbours $n_v$ and found that 4-axial PAs are more aligned ($\overline{\mathrm{SL}} \simeq 1.5\%, 2.2\sigma$, for $n_v \geq 30$) and 2-axial PAs are more orthogonal ($\overline{\mathrm{SL}} \simeq 3.3\%, 1.8\sigma$, for $n_v \geq 54$) than numerical simulations, where $\overline{\mathrm{SL}}$ is the mean significance level. This is most significant ($\mathrm{SL} \simeq 0.8\%, 2.4\sigma$) for 4-axial PAs with $n_v \sim 45$, which corresponds to a typical angular (comoving) separation of $\sim 30^{\circ}$ (1.6 Gpc). We conclude that PAs are both aligned and orthogonal, consistent with two modes separated by $\Delta\theta \sim 90^{\circ}$, and that the orthogonal mode somewhat dominates, particularly on the largest scales.

\vspace{5mm}
\noindent \textbf{Summary of results}

To summarise, LQG PAs are unlikely to be drawn from a uniform distribution (p-values $0.008 \lesssim p \lesssim 0.07$). However, similar non-uniformity is not found in mock LQG catalogues, indicating the LQG correlation is not found in cosmological simulations. Further, the LQG PA distribution is bimodal, with modes for weighted PAs at $\bar{\theta}\sim52\pm2^{\circ}, 137\pm3^{\circ}$ (97\% confidence). This bimodality is robust to parallel transport destination, with the median location of the peaks at all 71 LQG locations of $\bar{\theta}\sim45\pm2^{\circ}, 136\pm2^{\circ}$. LQGs are aligned and orthogonal across very large scales, with a maximum significance of $\simeq 0.8\%, 2.4\sigma$ at separations of $\sim 30^{\circ}$ (1.6 Gpc).

\subsection{Origin of LQG orientation correlation}

The two LQG position angle modes we report are roughly orthogonal, separated by $\Delta\theta \sim 90^{\circ}$. Comparing the 2D and 3D PA approaches (section~\ref{subsec2D3DEval}), we found that the 2D approach sometimes fits a regression comparable to the first minor axis of the 3D approach, if the major axis is orientated towards the line-of-sight. These axes are orthogonal, so potentially there could be a geometric explanation for the orthogonal modes in 2D, if the LQG major axes were aligned in 3D. However, preliminary inspection indicates that two modes are also present in three dimensions (Fig.~\ref{figLQGs3D} and Appendix~\ref{appLQGs3DPerspectives}), disfavouring this explanation. We leave thorough three-dimensional analysis of LQG orientations to future work.

We discuss next what astrophysical mechanisms could potentially cause large-scale correlations in LQG orientations. \citet{vandeWeygaert1994} argues that Voronoi tessellation \citep{Voronoi1908} is a reasonable approximation of the cosmic web. This cellular structure (described in Appendix~\ref{appVoronoi}) is caused by gravitational collapse, and consists of voids surrounded by walls of matter. Voronoi tessellations are non-regular, yet exhibit well defined preferred angles between adjoining walls: $\sim129^{\circ}$ and $\sim120^{\circ}$ in two and three-dimensional analysis \citep{Icke1987,vandeWeygaert1994}. These angles are consistent, regardless of the scale, orientation, or spatial position of individual Voronoi cells.

\citet{vandeWeygaert2007} uses Voronoi tessellation to describe the cosmic web on $\gtrsim 100$ Mpc scales, and demonstrates that the observed clustering of galaxy clusters can be explained by a cellular structure with a typical cell size $\sim 70\:h^{-1}$ Mpc. The scales over which we detect coherent alignment (and orthogonality) of LQG position angles are at least an order of magnitude larger than this. We do not suggest Gpc-scale Voronoi cells are a plausible explanation for our results; voids are generally $\lesssim 100$ Mpc \citep{Carroll2006}. Perhaps, though, a superposition of non-regular Voronoi cells throughout our survey volume could translate into our observed position angle distribution. Or maybe a more regular (crystalline) cellular structure leading to a repeating pattern of cell walls (or edges) could induce coherent large-scale structure orientations.

Alternatively, perhaps anisotropies in the early Universe could account for the large-scale correlations in LQG orientations. \citet{Poltis2010} proposed primordial cosmic strings as an explanation for the quasar polarization alignments. They suggest that the decay of these would seed correlated primordial magnetic fields, which could affect the polarization of electromagnetic radiation during propagation. Further, they suggest the torque from such magnetic fields could align the spins of protogalaxies, also supporting intrinsic alignment explanation. This mechanism could potentially explain LQG correlations, if we accept that it can be extrapolated up to Gpc scales. However, investigation of the potential imprint of primordial magnetic fields on the CMB constrains their amplitude to less than a few nanoGauss \citep{Planck2016h}, so this explanation seems unlikely.

\citet{Hutsemekers2005} report the mean optical quasar polarization angle appears to rotate with redshift at the rate of $\sim 30^{\circ}$ per Gpc. They suggest this may be caused by a global rotation of the Universe, such as that invoked by \citet{Jaffe2005} to explain large-scale anisotropies in the CMB data. It is plausible that cosmological rotation or torsion, particularly in the early Universe, may lead to global anisotropy of the Universe, but there currently appear to be no accepted observables to empirically test this theory. Further, from CMB temperature and polarization analysis \citet{Saadeh2016} conclude that Universe is neither rotating nor anisotropically stretched, disfavouring this explanation.

The correlation in LQG orientation is intriguing, and its origin remains unexplained.

\afterpage{\null\newpage} % because this ended on an odd page so the next chapter started on an even page

%% file: chapterConclusion.tex
\chapter{Conclusions and future work}
\label{chaptConclusion}

The large-scale structure of the Universe leaves a detectable imprint on the CMB, offering a probe of the cosmic web along the line-of-sight. In this work we examined claims that SNe tracing the cosmic web leave an imprint on the CMB temperature. This was reported by \citet{Yershov2012,Yershov2014} as an apparent cross-correlation between SNe redshift and CMB temperature at SNe position.

We concluded \citep{Friday2018} that, rather than having an astrophysical origin, the apparent correlation between CMB temperature and SNe redshifts is a composite selection bias, caused by high-redshift SNe being preferentially located in deep survey fields which are by chance aligned with CMB hotspots. The likelihood of this occurring by chance is $6.8\%-8.9\%$, so it cannot reasonably be rejected as a hypothesis. It has so far been ignored in those authors' extended work claiming a solution to the Hubble tension \citep{Yershov2020}. We suggest that it may be prudent to again contribute our hypothesis to the literature here.

We briefly discussed another potential correlation between large-scale structure and the CMB, in this case between our LQG sample and CMB temperature and lensing. Certain LQGs appear to be associated with CMB hotspots and CMB lensing peaks. We believe that further investigation of these potential cross-correlations is likely to be productive, but we leave their analysis to future work.

In the remainder of this work we examined for the first time the unexpected large-scale correlation of LQG orientations. Analyses of the LQG morphological position angles indicate they are unlikely to be drawn from a uniform distribution, their distribution is bimodal, and they are correlated such that the two modes ($\bar{\theta} \sim52\pm2^{\circ}, 137\pm3^{\circ}$ for weighted PAs) are aligned and orthogonal ($\Delta\theta \sim 90^{\circ}$) across very large scales. The median location of the modes at all 71 LQG locations is $\bar{\theta}\sim45\pm2^{\circ}, 136\pm2^{\circ}$. We found that the correlation is most significant for groups of $\sim45$ nearest neighbour LQGs, which corresponds to typical angular (comoving) separations of $\sim 30^{\circ}$ (1.6 Gpc).

The statistical significance of this correlation is marginal, from $\sim 0.8\%$ ($2.4 \sigma$) with Kuiper's test to $\sim 3.3\%$ ($1.8 \sigma$) with the S test (maximum significance $\sim 0.8\%$). We therefore cannot exclude it being a chance statistical anomaly. However, its coincidence with regions of quasar polarization alignment \citep[e.g.][]{Hutsemekers1998,Pelgrims2015}, the link between quasar polarization and LQG axes \citep{Hutsemekers2014,Pelgrims2016}, and the similarity between LQG position angles and the preferred angles of quasar radio polarization alignment \citep{Pelgrims2015}, suggest an interesting result.

A plausible physical mechanism for the correlation of LQG orientations on such large scales is not obvious. We considered the geometry of the cosmic web, and whether a cellular structure such as Voronoi tessellation or a more regular crystalline structure could cause such an effect. We also contemplated primordial anisotropies induced by scenarios such as cosmic strings or rotating cosmologies. For both types of explanation it was unclear how they could translate into our observed position angle distribution. Unfortunately, the exploration of such mechanisms was outside the scope of this work.

We found no evidence of $\Lambda$CDM cosmological simulations predicting correlations between objects on Gpc scales, but this had not been specifically examined for LQGs. Using mock LQG catalogues \citep{Marinello2016} we found no evidence of LQG correlation in the Horizon Run 2 simulation \citep{Kim2011}. This suggests the cosmic web of the observed Universe differs on the largest scales to this dark matter only $N$-body simulation. It hints that there could, given the caveats associated with the simulations, be aspects of the large-scale structure that are not captured by the power spectrum. It is clear that running the LQG finder on other cosmological simulations would be informative. If the correlation in LQG orientation is found then perhaps it is an unexpected feature of known physics. If it is not seen then maybe something is missing from the simulations (e.g.\ primordial anisotropies) or it is, after all, a statistical fluke which coincidentally gives rise to the aligned quasar polarizations.

The LQG orientation correlation we found offers a plausible explanation for the quasar polarization alignments reported by many studies \citep[e.g.][]{Hutsemekers1998,Hutsemekers2001,Jain2004,Cabanac2005,Tiwari2013,Pelgrims2014,Pelgrims2015,Pelgrims2019}. If LQG axes are preferentially aligned at $\bar{\theta}\sim45\pm2^{\circ}, 136\pm2^{\circ}$ (this work, median modes), and if quasar polarization vectors are preferentially parallel and orthogonal to LQG axes \citep{Hutsemekers2014,Pelgrims2016}, this could result in polarization vectors with preferred angles of $\sim42^{\circ}$ and $\sim131^{\circ}$ \citep{Pelgrims2015}. Our results therefore offer corroborating evidence for, and enhancement of, the intrinsic alignment interpretation of these studies.

Quasar polarization alignment is also detected in the south Galactic cap \citep[SGC, e.g.][]{Hutsemekers1998,Pelgrims2015}, which is not coincident with our LQG sample in the north Galactic cap (NGC). The forthcoming 4-metre Multi-Object Spectroscopic Telescope (4MOST) Active Galactic Nuclei survey \citep{Merloni2019} will survey a million $z \lesssim 2.5$ quasars over $\sim 10,000 \ \mathrm{deg}^2$, with first light expected in 2022. This could deliver an LQG sample in the \gls{SGC} for similar evaluation to our work in the NGC. Of particular interest would be whether LQG orientation again corresponds to the preferred angle of quasar radio polarization alignment, which differ between NCG and SCG \citep[e.g.][]{Pelgrims2016thesis}.

Our results are based on a sample of 71 LQGs at redshifts $1.0 \leq z \leq 1.8$, which were detected using the SDSS DR7QSO catalogue \citep{Schneider2010} of $\sim 105$k quasars across $\sim 7,600 \ \mathrm{deg}^2$. Forthcoming spectroscopic surveys will deliver a far larger sample of quasars, e.g.\ the Dark Energy Spectroscopic Instrument \citep[DESI,][]{DESI2016} 5-year survey aims to target 1.7 million $z < 2.1$ quasars covering $\sim 14,000 \ \mathrm{deg}^2$, beginning summer 2020. This has the potential to deliver a larger sample of LQGs for a better assessment of their correlation.

If the LQG orientation correlation is real, it represents large-scale structure alignment over $\gtrsim$ Gpc scales, larger than those predicted by cosmological simulations and at least an order of magnitude larger than any so far observed, with the exception of quasar polarization alignment. However, careful statistical analysis is required before making inferences about whether such a large-scale correlation challenges the assumption of large-scale statistical isotropy and homogeneity of the Universe.

To conclude, we started this work investigating the apparent cross-correlation of SNe redshift and CMB temperature, which we found to be a chance selection bias. We then unexpectedly found the large-scale correlation of LQG orientations, which we report here for the first time. This helps explain a substantial body of work on quasar polarization alignment, but at the expense of raising potentially even more challenging questions about the origin of the LQG correlation. Forthcoming surveys and the other future work we suggest here, particularly cross-correlations with the CMB, will illuminate LQGs and their intriguing correlation further.

%% file: appendixISW.tex
\chapter{Integrated Sachs-Wolfe effect} \label{appISW}

The Sachs-Wolfe effect \citep{SachsWolfe1967, Rees1968} occurs when cosmic microwave background (CMB) photons pass through the time-evolving gravitational potentials of large-scale structure in an expanding Universe. In the case of an over-density, the potential well decays as the photons traverse it, so the blueshift from `falling in' is greater than the redshift from `climbing out'. This leads to a residual blueshift and a slight increase in temperature of the CMB (`hotspots'). For under-densities (voids), the potential hill decays, leaving a residual redshift and a slight decrease in temperature (`coldspots').

The effect is known as the \textit{integrated} Sachs-Wolfe (\gls{ISW}) effect, because the overall amplitude of the effect is dependent on the evolution of gravitational potential integrated along the line-of-sight, between the last scattering surface and today. An extension to the late-time ISW incorporating clustering is the Rees-Sciama (\gls{RS}) effect. The ISW effect introduces secondary anisotropies expected to be most prominent in the CMB spectrum at low multipoles ($\ell \lesssim 40$), or large angular scales $\gtrsim 10^\circ$ \citep{Kofman1985}. See \citet{Nishizawa2014} for a review of the physics of the ISW effect, and a summary of observational results and their interpretation.

The first detection of the ISW signal was achieved by cross-correlating NRAO VLA Sky Survey (NVSS) radio galaxy data with Wilkinson Microwave Anisotropy Probe (WMAP) first year CMB temperature data \citep{Crittenden1996}. Several other detections of the ISW signal have since been made in radio data \citep[e.g.][]{Nolta2004, Boughn2004a, Boughn2004b, Raccanelli2008}. The ISW signal has also been detected by cross-correlating the WMAP CMB temperature data with Sloan Digital Sky Survey (SDSS) galaxy data \citep{Fosalba2003, Padmanabhan2005, Granett2009, Papai2011}.

The significance of ISW detections can be increased by combining measurements from multiple catalogues. \citet{Ho2008} combined the Two Micron All-Sky Survey (2MASS), SDSS LRGs, NVSS, and SDSS QSOs catalogues and reported a $3.7 \sigma$ detection. They do not find significant evidence for an ISW signal at $z$ \textgreater $1$, which they suggest is in agreement with $\Lambda$CDM theoretical predictions. \citet{Giannantonio2008} combined the 2MASS, SDSS galaxies, SDSS LRGs, NVSS, High Energy Astrophysical Observatory (HEAO), and SDSS QSO catalogues and reported a $\sim 4.5 \sigma$ detection. The ISW signal is also detected at $4 \sigma$ in the \textit{Planck} 2015 data release, where the CMB temperature was cross-correlated with the NVSS, SDSS, and Wide-field Infrared Survey Explorer (WISE) catalogues to trace LSS, plus the \textit{Planck} 2015 lensing data \citep{Planck2016f}.

Detections of the ISW signal typically span LSS redshifts $0.1 \lesssim z \lesssim 1.5$, but usually $z \lesssim 1$, and most appear fully compatible with $\Lambda$CDM theoretical predictions. However, some measurements do appear slightly above $\Lambda$CDM expectations, e.g.\ \citet[][$\sim 1 \sigma$ above]{Giannantonio2008}, \citet[][$\gtrsim 3 \sigma$ above]{Granett2008}, \citet[][$\sim 2.2 \sigma$ above]{Goto2012}, and \citet[][$\sim 1.2 \sigma$ above]{Kovacs2017}. At present there may be a slight tension between ISW measurement and $\Lambda$CDM that is not fully understood, although this is still the subject of some debate \citep[e.g.][$\sim 1.2 \sigma$ above]{Nadathur2016}.

%% file: appendixSZ.tex
\chapter{Sunyaev-Zel\textquotesingle dovich effect} \label{appSZ}

The Sunyaev-Zel\textquotesingle dovich (\gls{SZ}) effect \citep{SunyaevZeldovich1970, SunyaevZeldovich1972} is caused by inverse Compton scattering of low energy CMB photons by much higher energy free electrons, such as found in the hot plasma of the intra-cluster medium (\gls{ICM}). This results in a distortion of the CMB spectrum in the direction of a cluster, which has been used to detect and characterize the properties of galaxy clusters \citep[e.g.][]{Halverson2009, Planck2014c, Bleem2015, Bender2016}. As a scattering effect it is independent of both redshift and the expansion of the Universe, unlike the ISW effect.

There are two components to the SZ effect. The thermal SZ (tSZ) effect is due to scattering of CMB photons by thermal electrons and is dependent on the thermal energy contained in the ICM. The kinetic SZ (kSZ) effect is due to scattering of CMB photons by a population of electrons moving with a line-of-sight peculiar velocity to the CMB rest frame. See \citet{Birkinshaw1999} and \citet{Carlstrom2002} for reviews of the physics of the SZ effect.

The tSZ effect results in a CMB spectral distortion where some lower energy photons are boosted in energy. This leads to a decrease in the observed CMB intensity below the `cross-over' frequency of $\sim 218$ GHz, and an increase at higher frequencies. The precise shape of the tSZ effect spectrum is slightly affected by electron temperature. Accurate measurements of the tSZ spectrum can thus be used to measure the average ICM temperature of galaxy clusters \citep[e.g.][]{Pointecouteau1998, Hurier2017}. Because the amplitude of tSZ signal is proportional to the integrated line-of-sight ICM pressure, it can also be used as a proxy for mass in cosmological studies \citep[e.g.][]{Bleem2015, Planck2016g}.

The wide range of frequency channels of the \textit{Planck} satellite, from 30 to 857 GHz, either side of the tSZ `cross-over' frequency, make it sensitive to the tSZ effect. At the angular resolution of \textit{Planck} ($\sim 5'$) it can predominantly detect local ($z \lesssim 1$) galaxy clusters \citep[e.g.][]{Planck2016e}. Note that catalogues of galaxy clusters identified by the tSZ effect have also been identified by the Atacama Cosmology Telescope \citep[ACT, ][]{Hasselfield2013} and the South Pole Telescope \citep[SPT, ][]{Reichardt2013}.

The kSZ effect is caused by the Doppler shift of CMB photons induced by the bulk motion of ICM electrons. It is generally at least an order of magnitude smaller than the tSZ effect for a typical galaxy cluster, and only becomes significant when the velocity of the electrons approaches 1000 km/s \citep{Birkinshaw1999}. The maximum kSZ signal is observed at $\sim 217$ GHz (if relativistic effects are negligible), very close to the `cross-over' frequency of the tSZ signal \citep{Adam2017}. The amplitude of the kSZ signal is dependent on the integrated line-of-sight ICM density and velocity with respect to the CMB rest frame.

%% file: appendixFields1to7.tex
\chapter{\textit{Planck} CMB temperature for SNe deep fields 1-7} \label{appFields1to7}

\begin{figure} % SNe_plot_field_v2
	\centering
	\subfigure[Field 1, centred on\newline SNLS D3] {\includegraphics[width=0.27\textwidth]{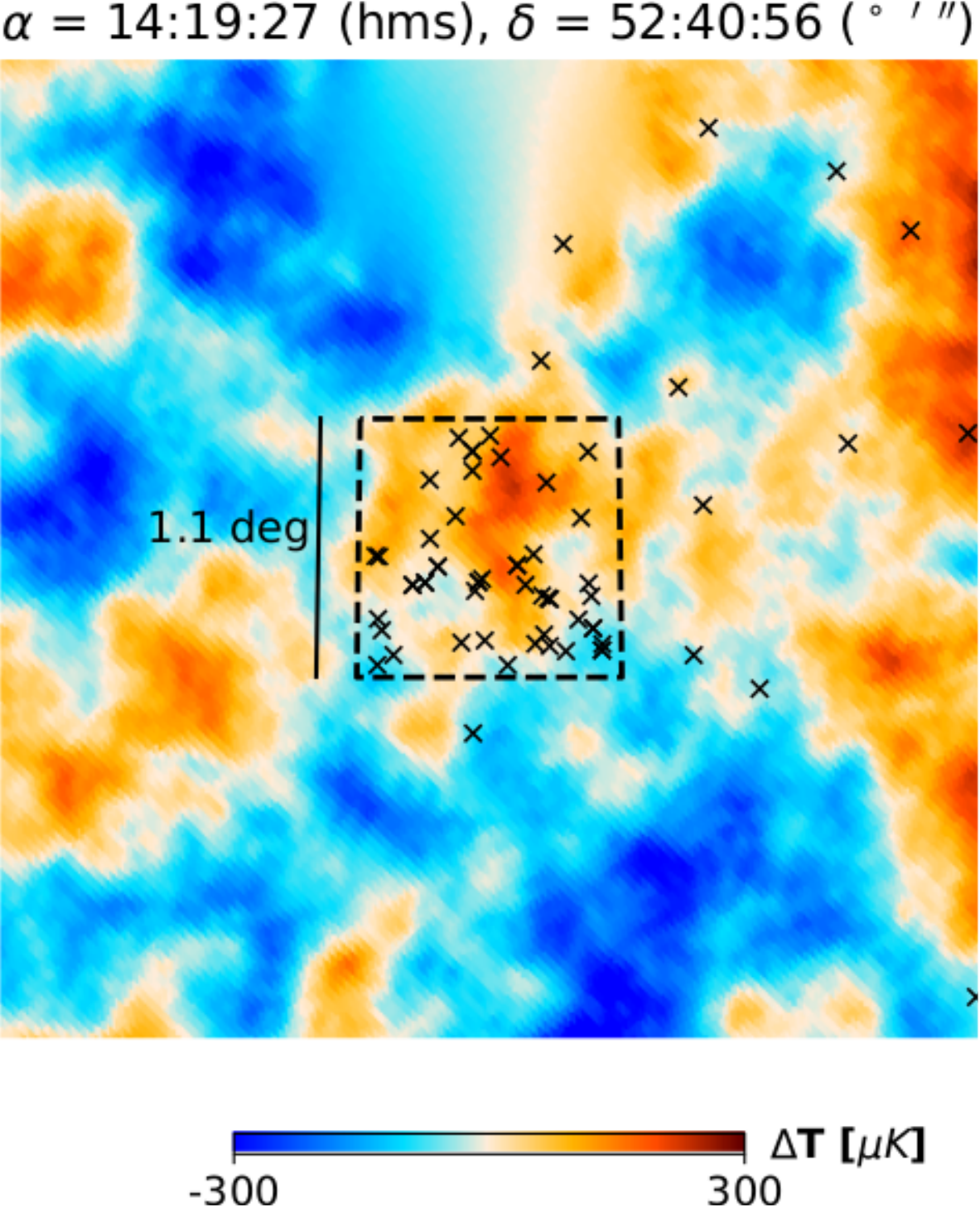} \label{subField1}}
	\subfigure[Field 2, centred on\newline HDF-N] {\includegraphics[width=0.27\textwidth]{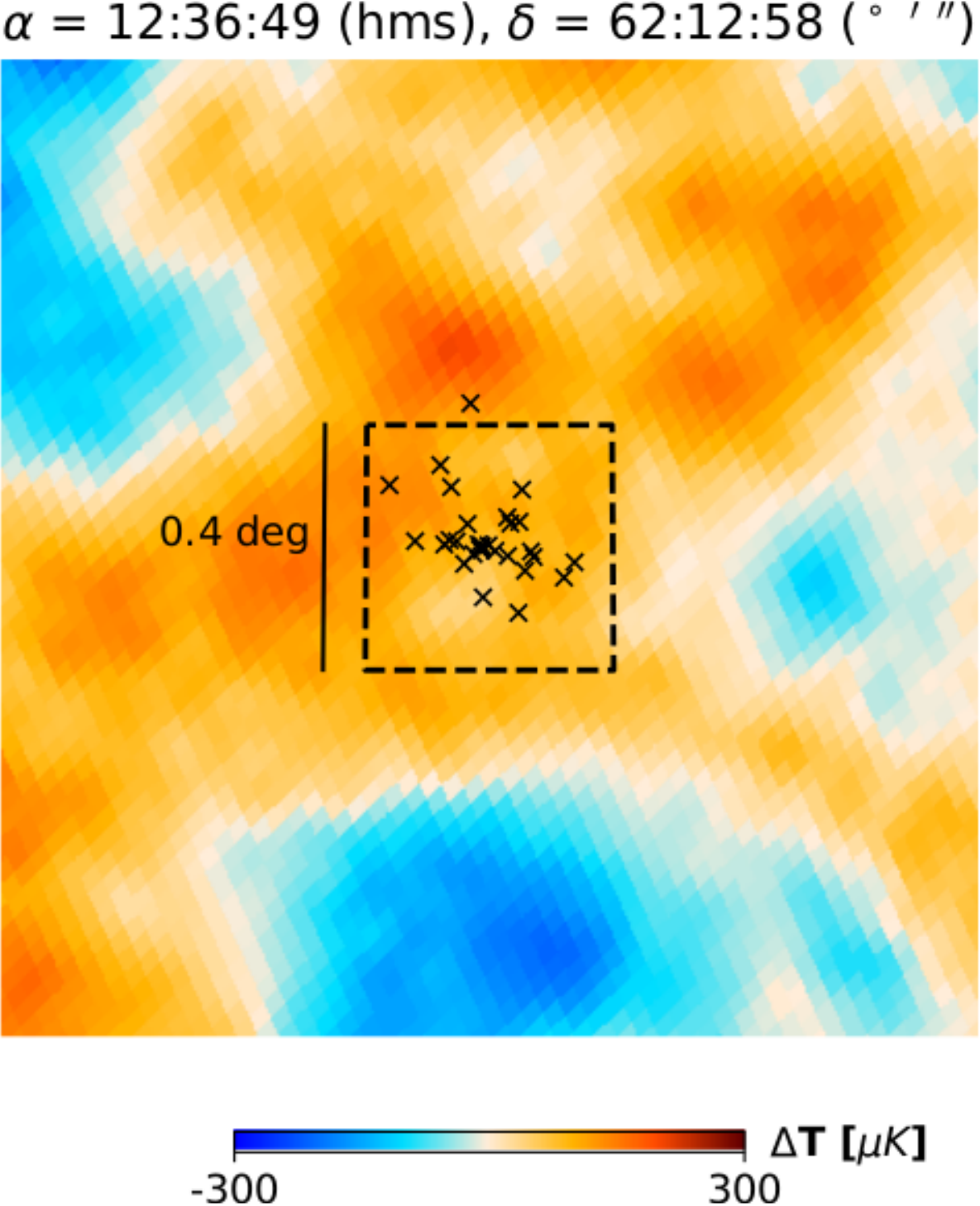} \label{subField2}}
	\subfigure[Field 3, centred on\newline ESSENCE wdd] {\includegraphics[width=0.27\textwidth]{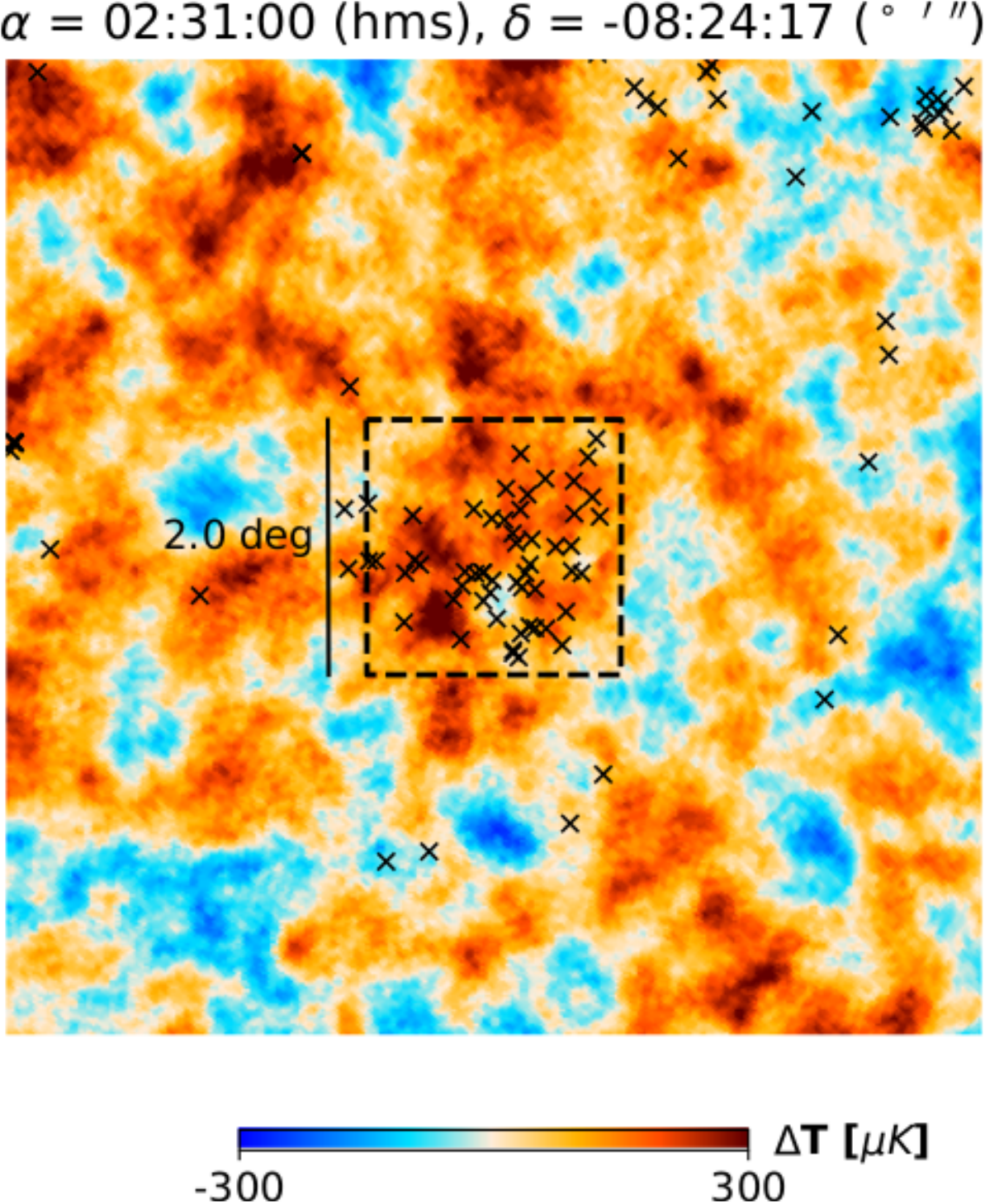} \label{subField3}}
	\\
	\subfigure[Field 4, centred on\newline SNLS D1] {\includegraphics[width=0.27\textwidth]{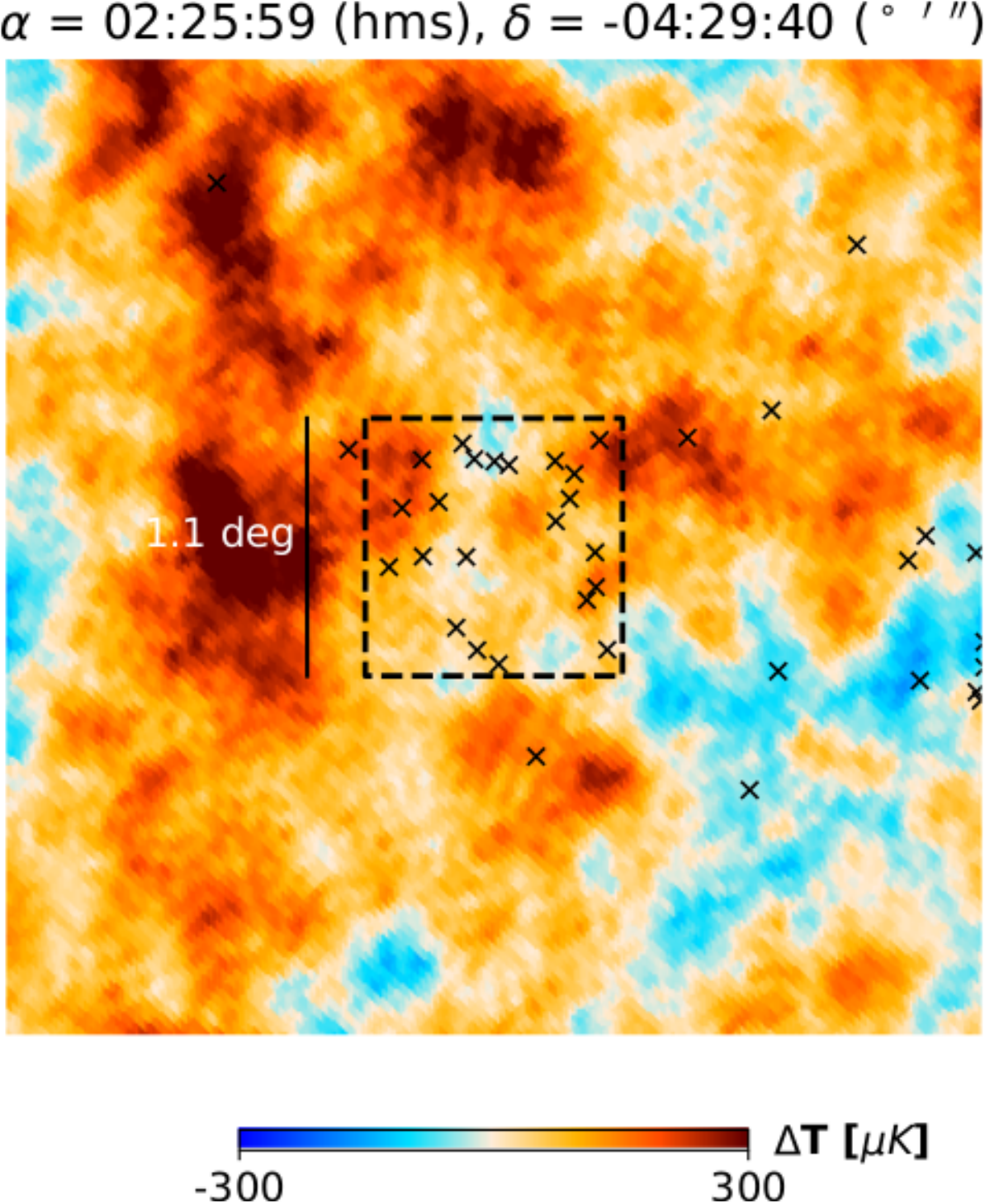} \label{subField4}}
	\subfigure[Field 5, centred on\newline ESSENCE wcc] {\includegraphics[width=0.27\textwidth]{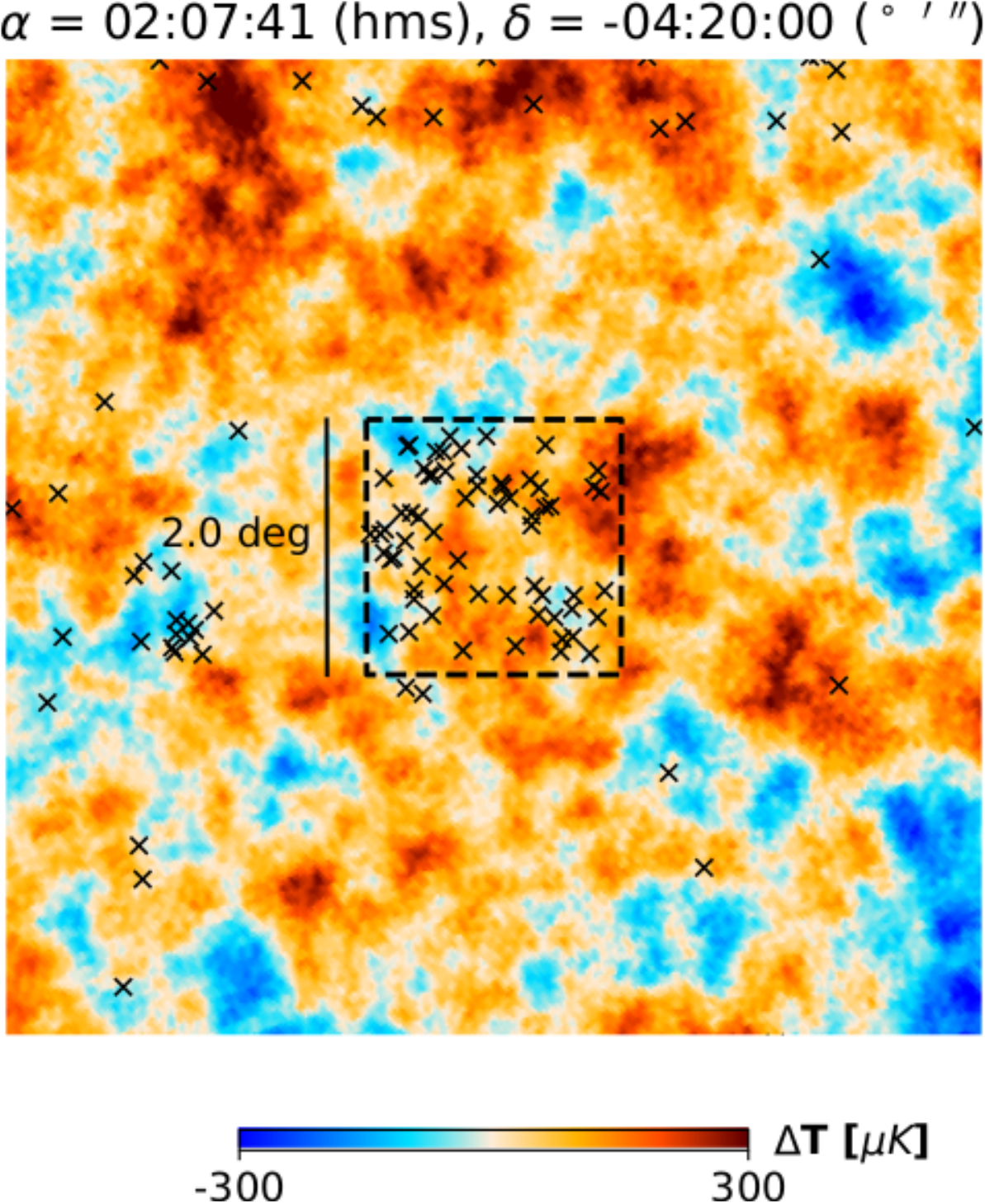} \label{subField5}}
	\subfigure[Field 6, centred on\newline SNLS D4] {\includegraphics[width=0.27\textwidth]{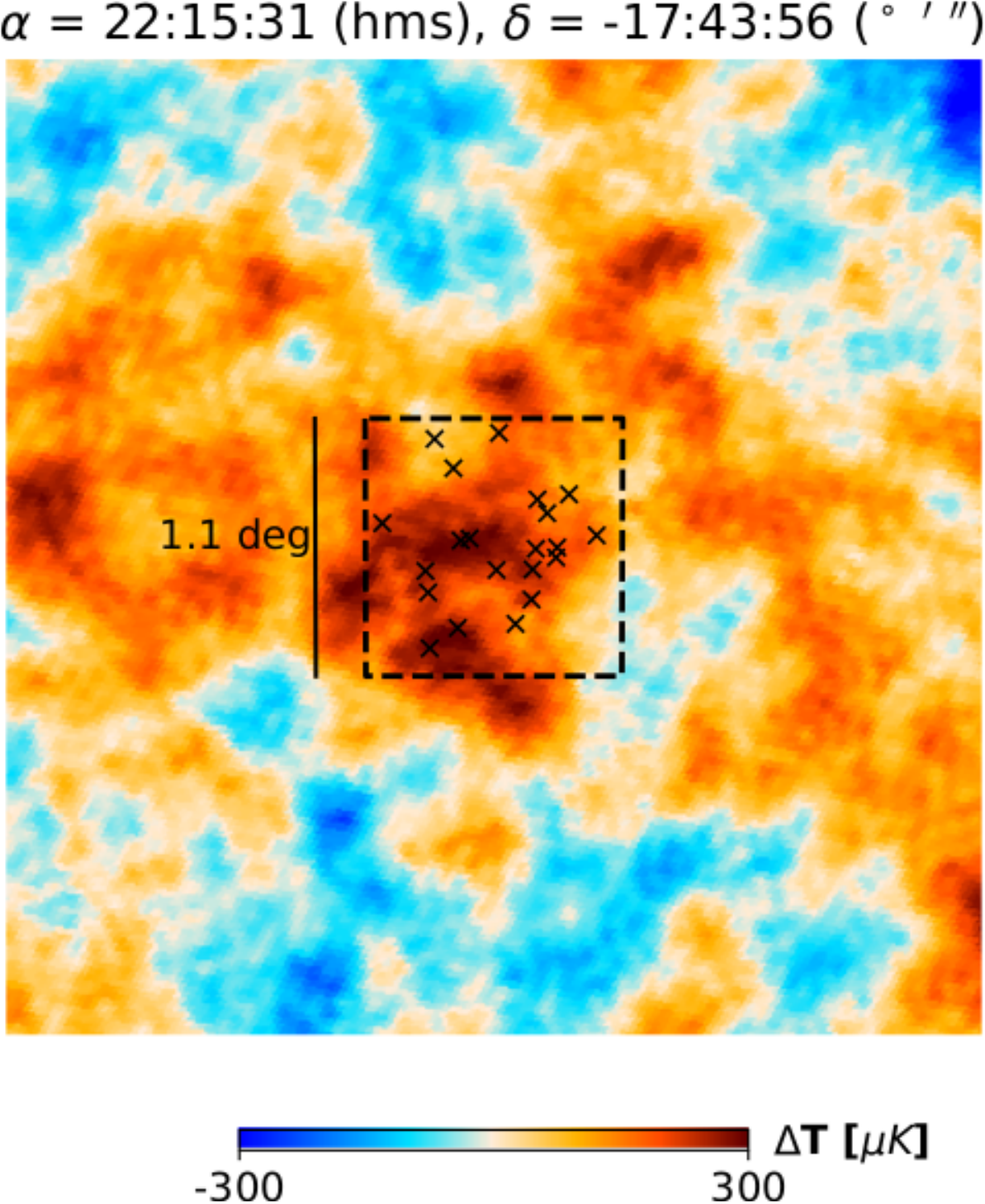} \label{subField6}}
	\\
	\subfigure[Field 7, centred on\newline CDF-S] {\includegraphics[width=0.27\textwidth]{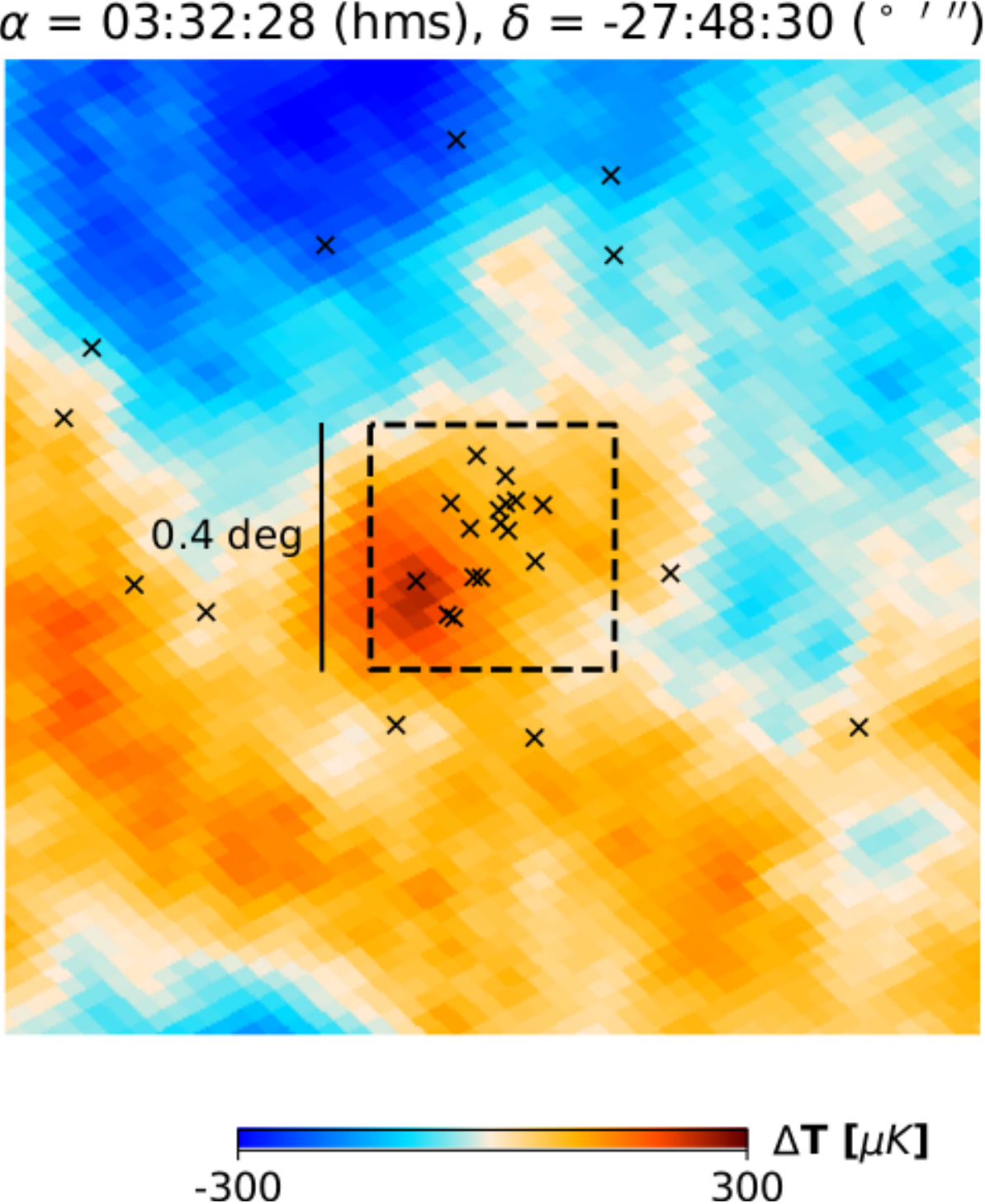} \label{subField7}}
	\caption[CMB $T$ in the vicinity of SNe fields 1-7]{Gnomic projections of the \textit{Planck} 2015 SMICA CMB map in the vicinity of fields 1-7. The location of SAI sample SNe are plotted with crosses ($\times$). The dashed squares are the boundaries of fields 1-7, centred on the corresponding deep survey fields at the specified coordinates ($\alpha$ and $\delta$, J2000).}
	\label{figFields1to7}
\end{figure}

%% file: appendixVariances.tex
\chapter{CMB $T$ and SNe $z$ correlation: variances and smoothing} \label{appVariances}

\begin{figure} % CMB_plot_noise_map
\begin{minipage}[t]{0.5\textwidth}
    \centering
    \subfigure[\label{subHMHDunweighted}Unweighted $\overline{T}$]{
    \makebox[\columnwidth][c]{
    \includegraphics[width=0.4\columnwidth]{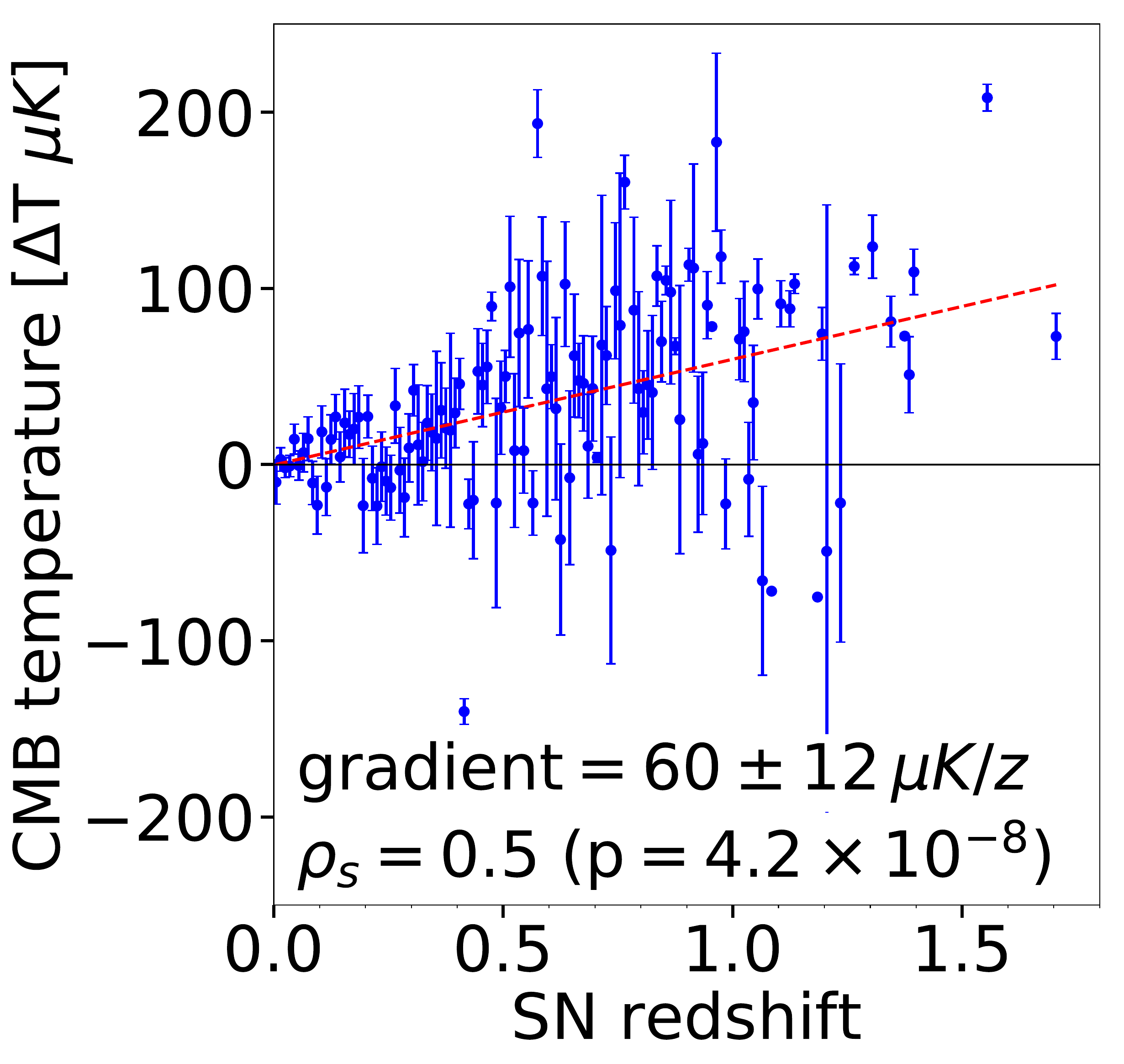}}}
	\\
	\subfigure[\label{subHMHDweightedUnsmoothed}Weighted $\overline{T}$, HMHD $\sigma_{\overline{T}}$, unsmoothed]{
    \makebox[\columnwidth][c]{
    \includegraphics[width=0.4\columnwidth]{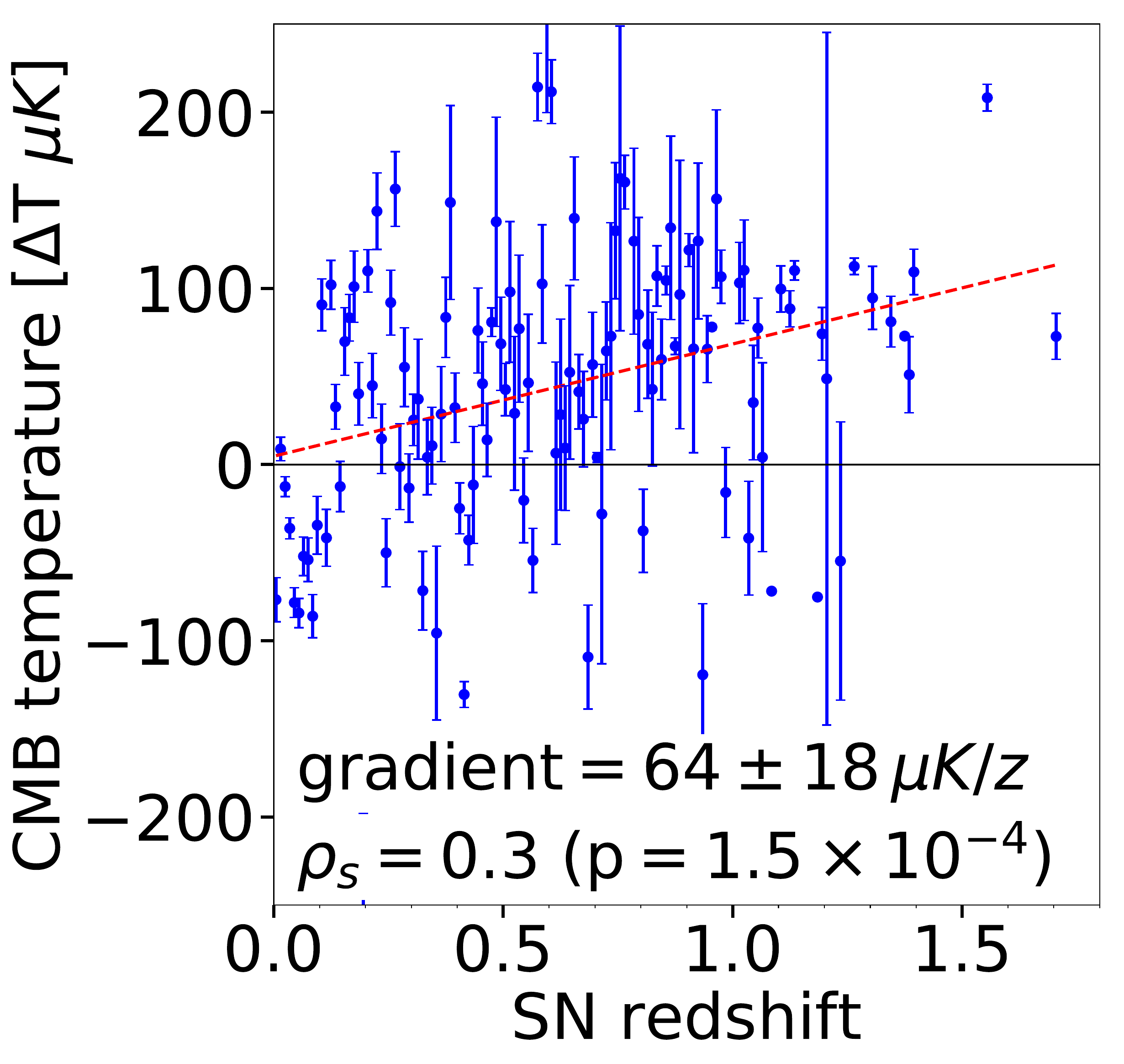}}}
	\\
	\subfigure[\label{subHMHDweightedSmoothed5arcmin}Weighted $\overline{T}$, HMHD $\sigma_{\overline{T}}$, $5'$ smoothed]{
    \makebox[\columnwidth][c]{
    \includegraphics[width=0.4\columnwidth]{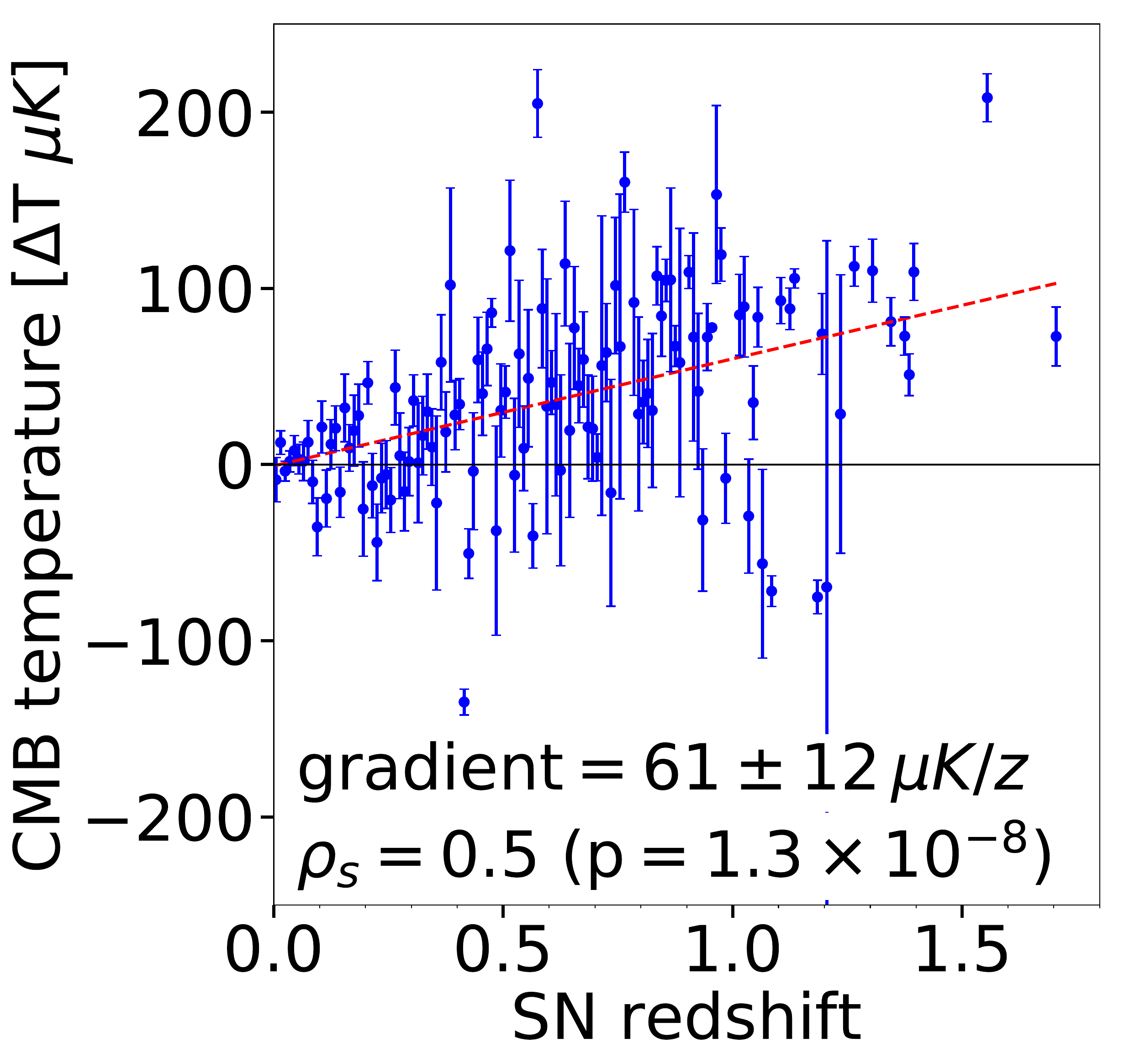}}}
	\\
	\subfigure[\label{subHMHDweightedSmoothedHalfDeg}Weighted $\overline{T}$, HMHD $\sigma_{\overline{T}}$, 0.5$^\circ$ smoothed]{
    \makebox[\columnwidth][c]{
    \includegraphics[width=0.4\columnwidth]{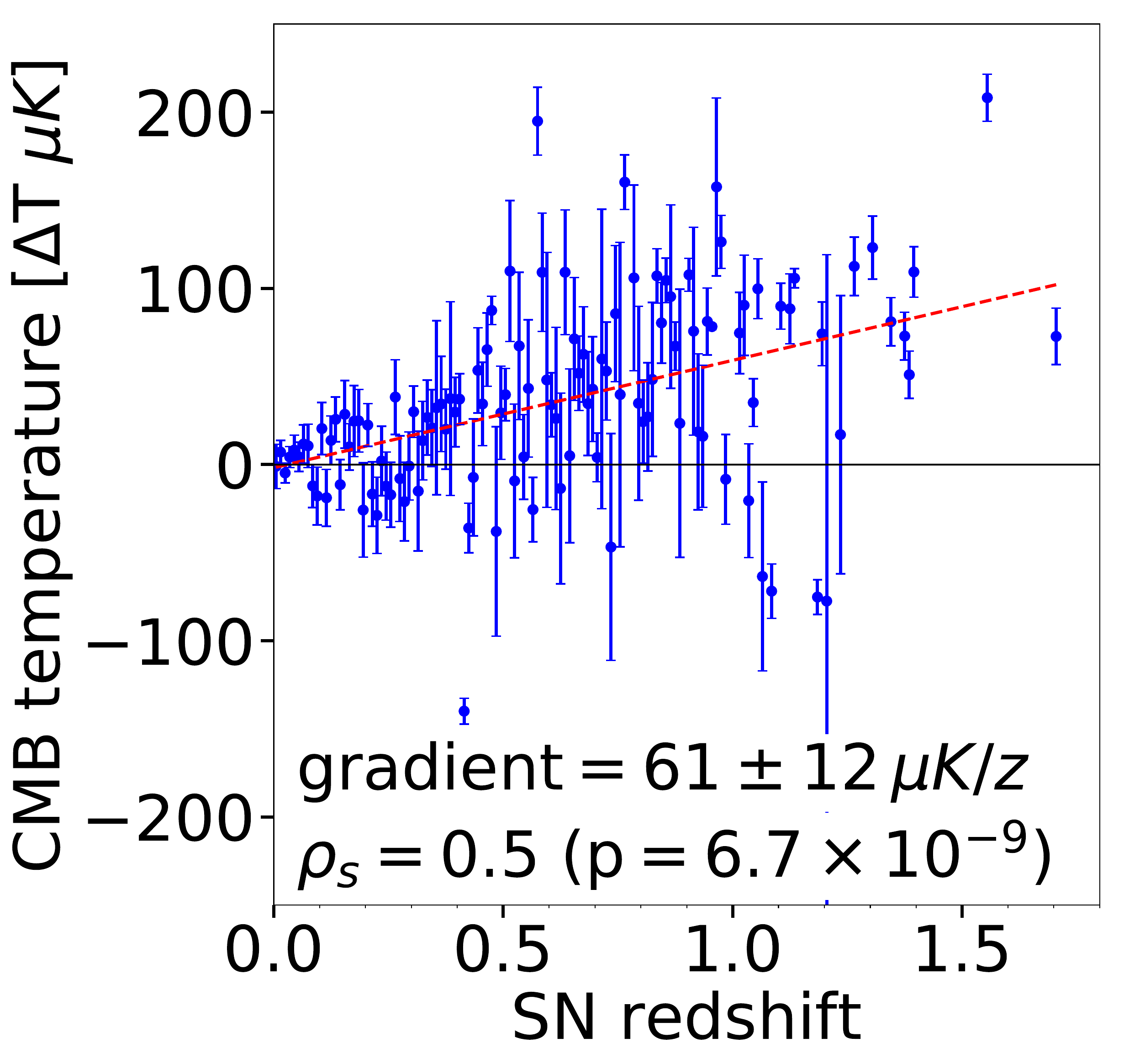}}}
	\\
	\subfigure[\label{subHMHDweightedSmoothed5deg}Weighted $\overline{T}$, HMHD $\sigma_{\overline{T}}$, 5$^\circ$ smoothed]{
    \makebox[\columnwidth][c]{
    \includegraphics[width=0.4\columnwidth]{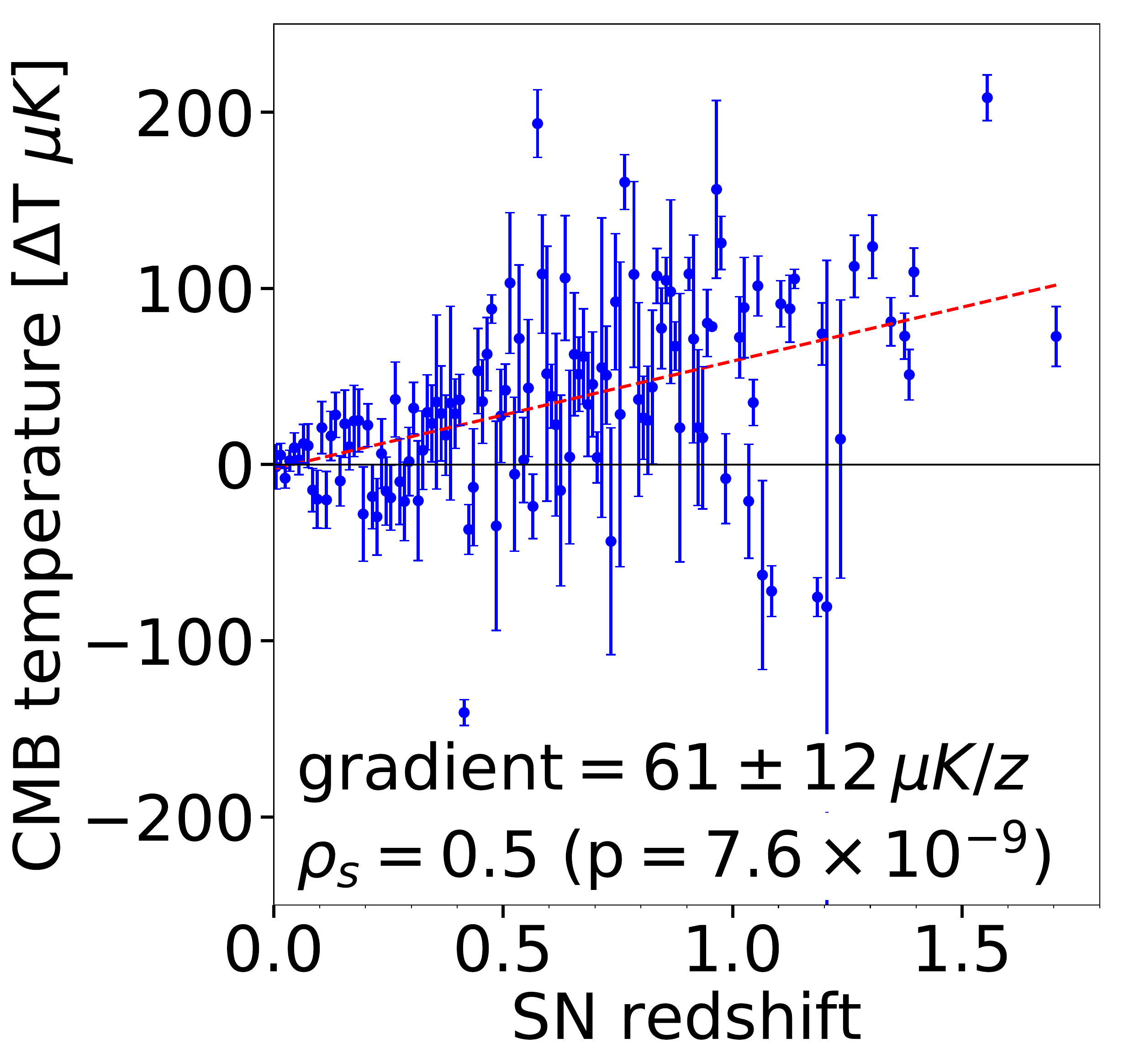}}}
	\caption[As Fig.~\ref{figBinnedScatter} but for variance estimated from HMHD maps]{Same as Fig.~\ref{figBinnedScatter} but for variance estimated using \textit{Planck} 2015 R2.01 SMICA half-mission half-difference (HMHD) maps with specified smoothing.}
	\label{figHMDH}
\end{minipage}%\hfill
\begin{minipage}[t]{0.5\textwidth}
    \centering
    \subfigure[\label{subHRHDunweighted}Unweighted $\overline{T}$]{
    \makebox[\columnwidth][c]{
    \includegraphics[width=0.4\columnwidth]{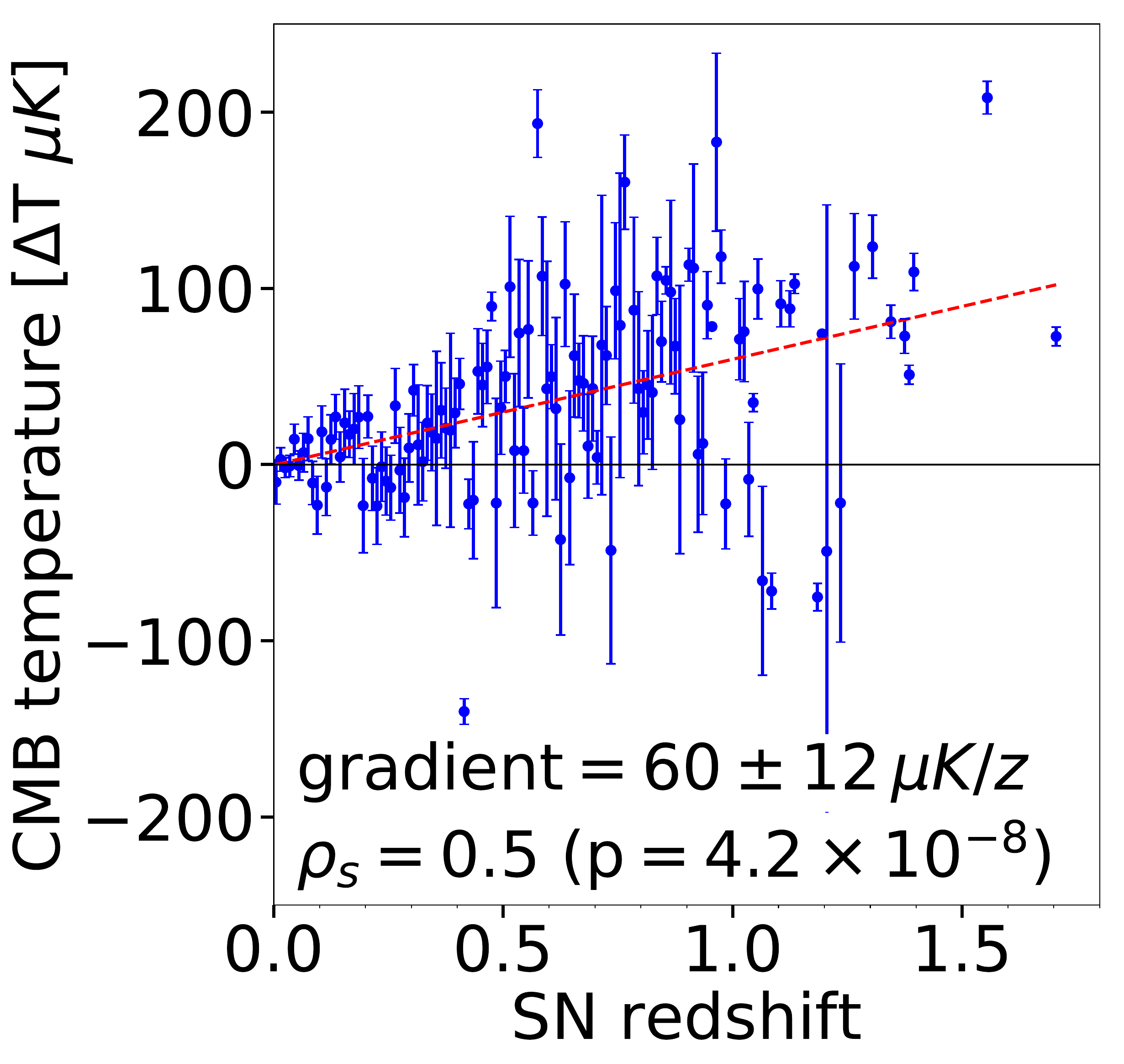}}}
	\\
	\subfigure[\label{subHRHDweightedUnsmoothed}Weighted $\overline{T}$, HRHD $\sigma_{\overline{T}}$, unsmoothed]{
    \makebox[\columnwidth][c]{
    \includegraphics[width=0.4\columnwidth]{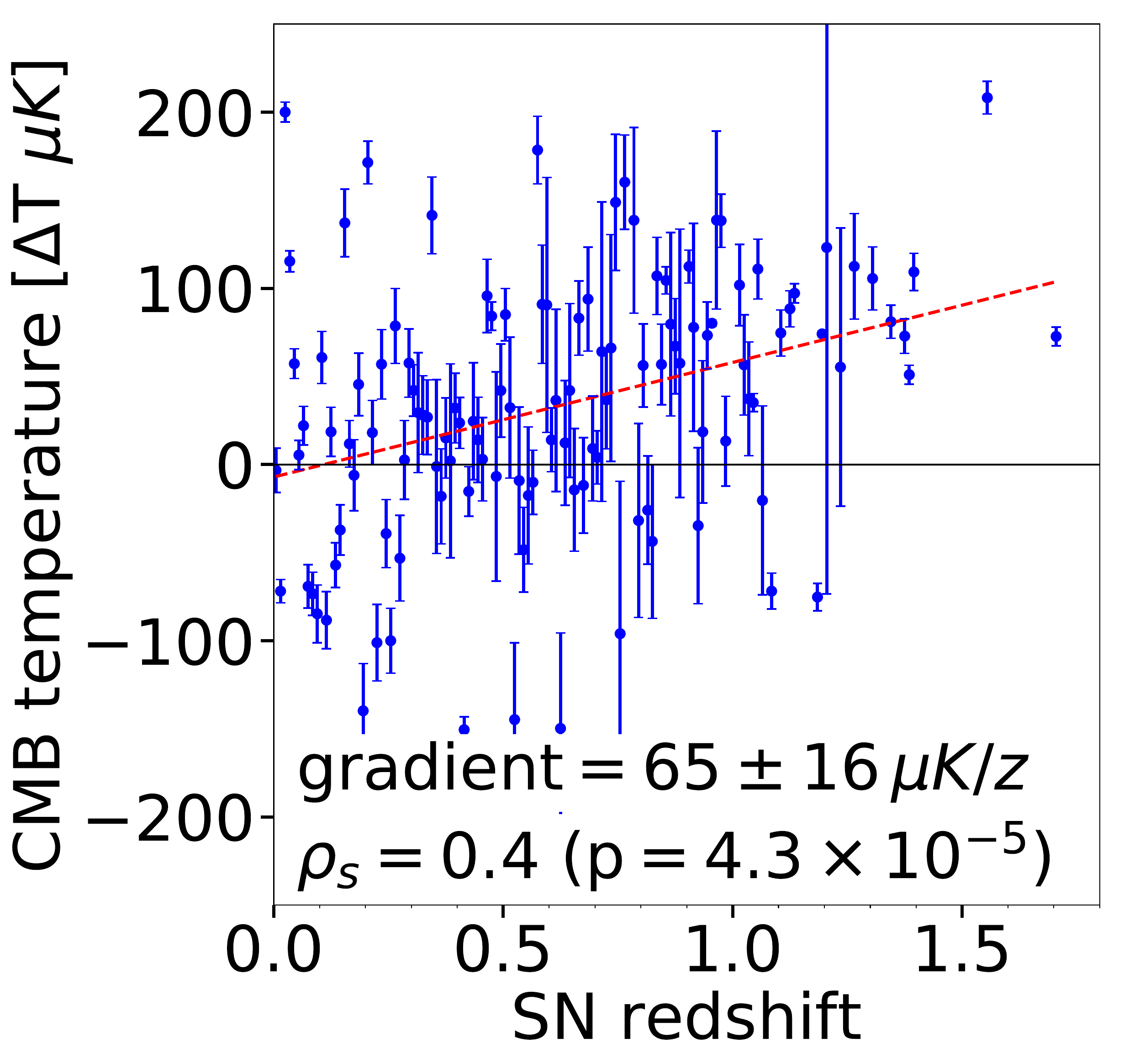}}}
	\\
	\subfigure[\label{subHRHDweightedSmoothed5arcmin}Weighted $\overline{T}$, HRHD $\sigma_{\overline{T}}$, $5'$ smoothed]{
    \makebox[\columnwidth][c]{
    \includegraphics[width=0.4\columnwidth]{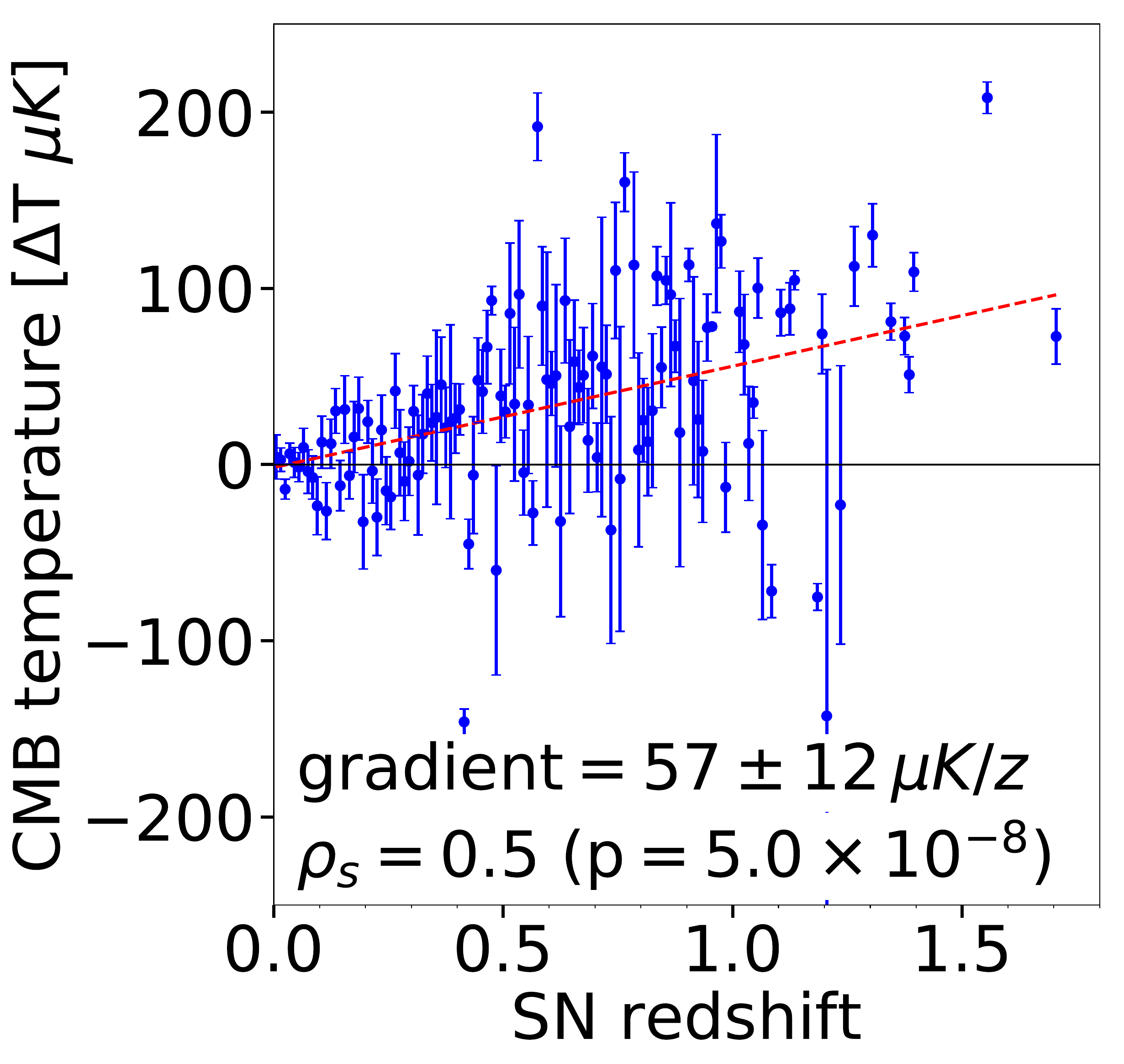}}}
	\\
	\subfigure[\label{subHRHDweightedSmoothedHalfDeg}Weighted $\overline{T}$, HRHD $\sigma_{\overline{T}}$, 0.5$^\circ$ smoothed]{
    \makebox[\columnwidth][c]{
    \includegraphics[width=0.4\columnwidth]{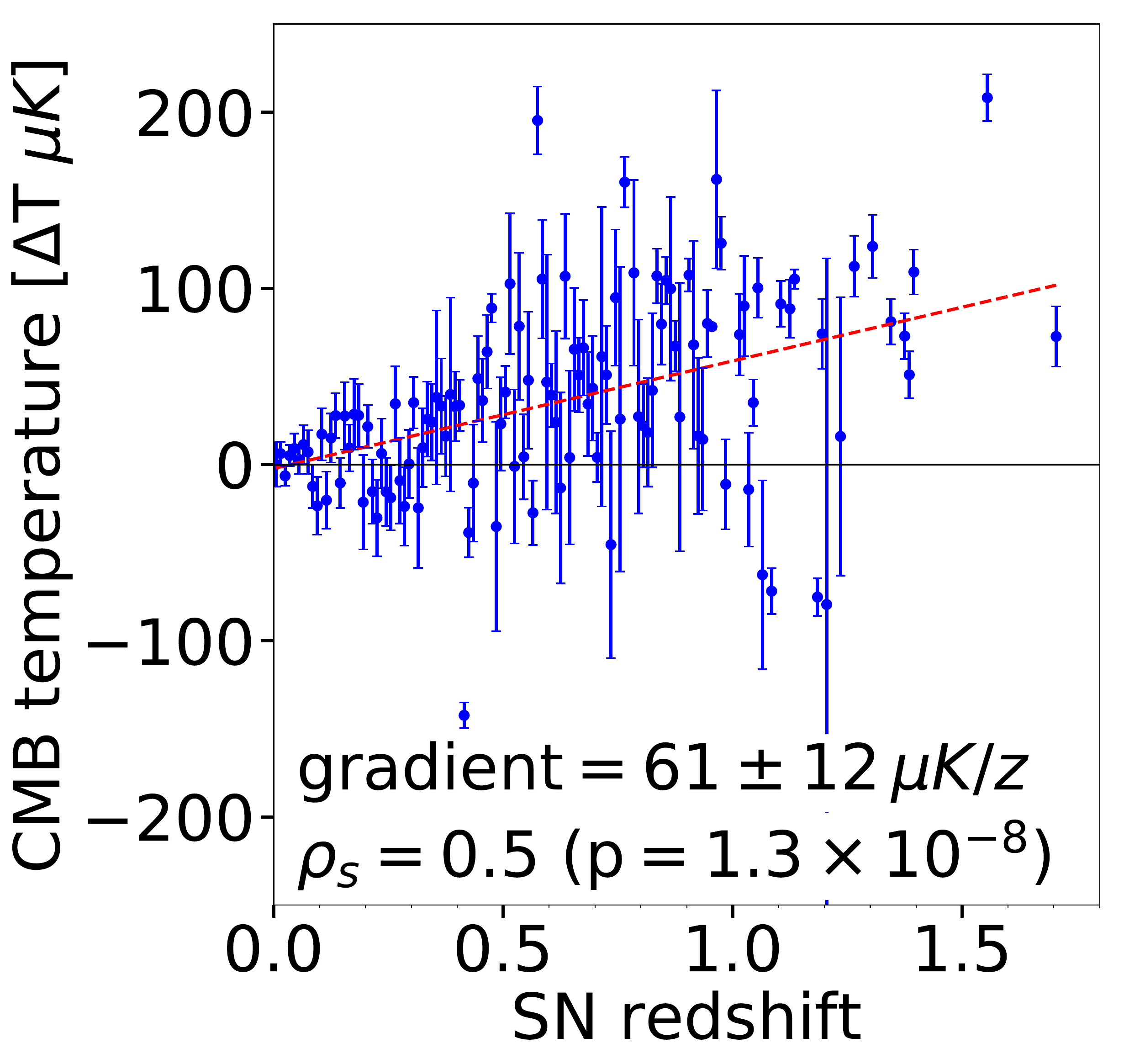}}}
	\\
	\subfigure[\label{subHRHDweightedSmoothed5deg}Weighted $\overline{T}$, HRHD $\sigma_{\overline{T}}$, 5$^\circ$ smoothed]{
    \makebox[\columnwidth][c]{
    \includegraphics[width=0.4\columnwidth]{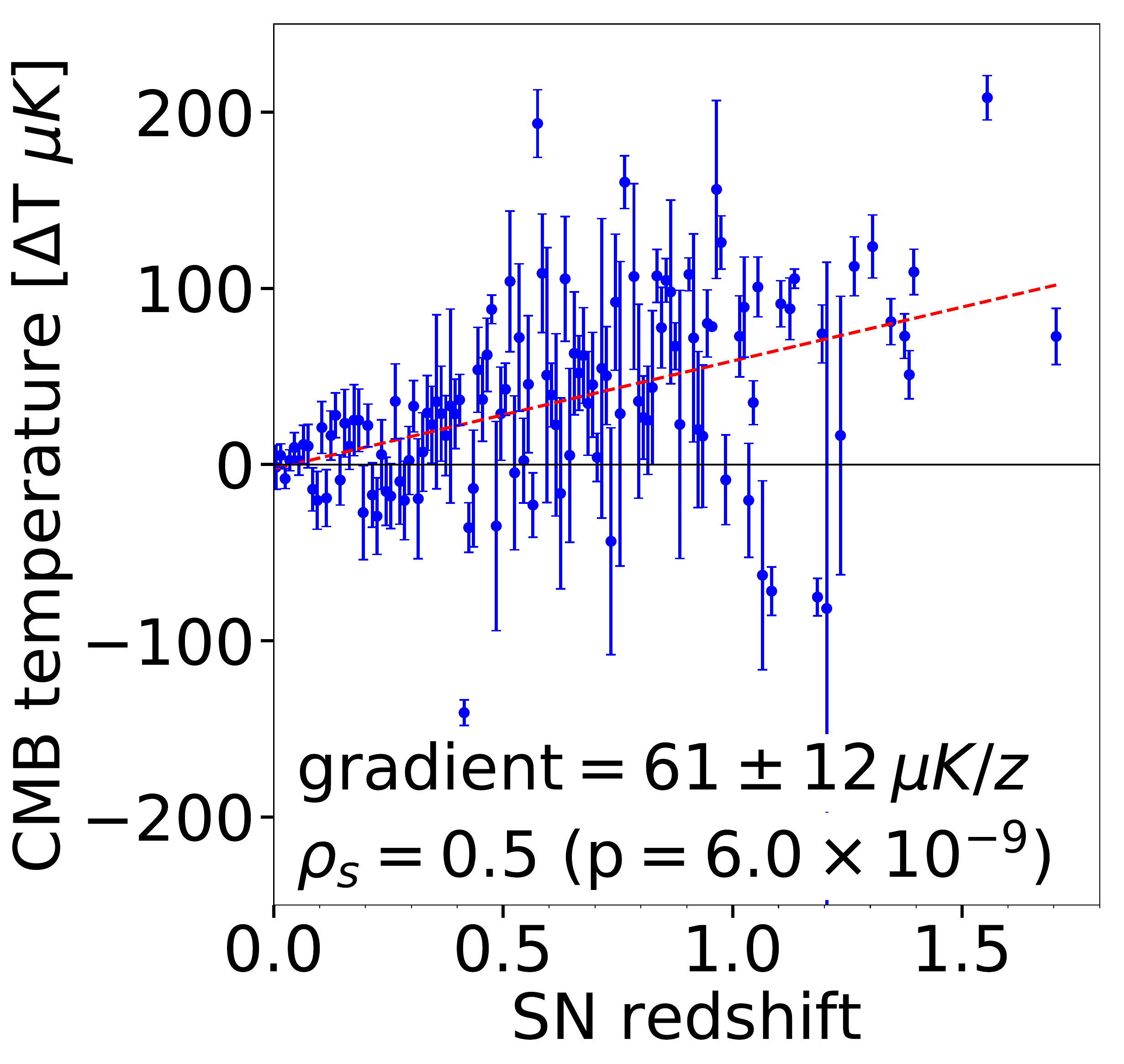}}}
	\caption[As Fig.~\ref{figBinnedScatter} but for variance estimated from HRHD maps]{Same as Fig. \ref{figHMDH} but for variance estimated using \textit{Planck} 2015 R2.01 SMICA half-ring half-difference (HRHD) maps.}
	\label{figHRHD}
\end{minipage}
\end{figure}

\begin{figure} % CMB_plot_noise_map
\begin{minipage}[t]{0.5\textwidth}
	\centering
	\subfigure[\label{sub143weightedUnsmoothed}Weighted $\overline{T}$, 143GHz covariance, unsmoothed]{
    \makebox[\columnwidth][c]{
    \includegraphics[width=0.4\columnwidth]{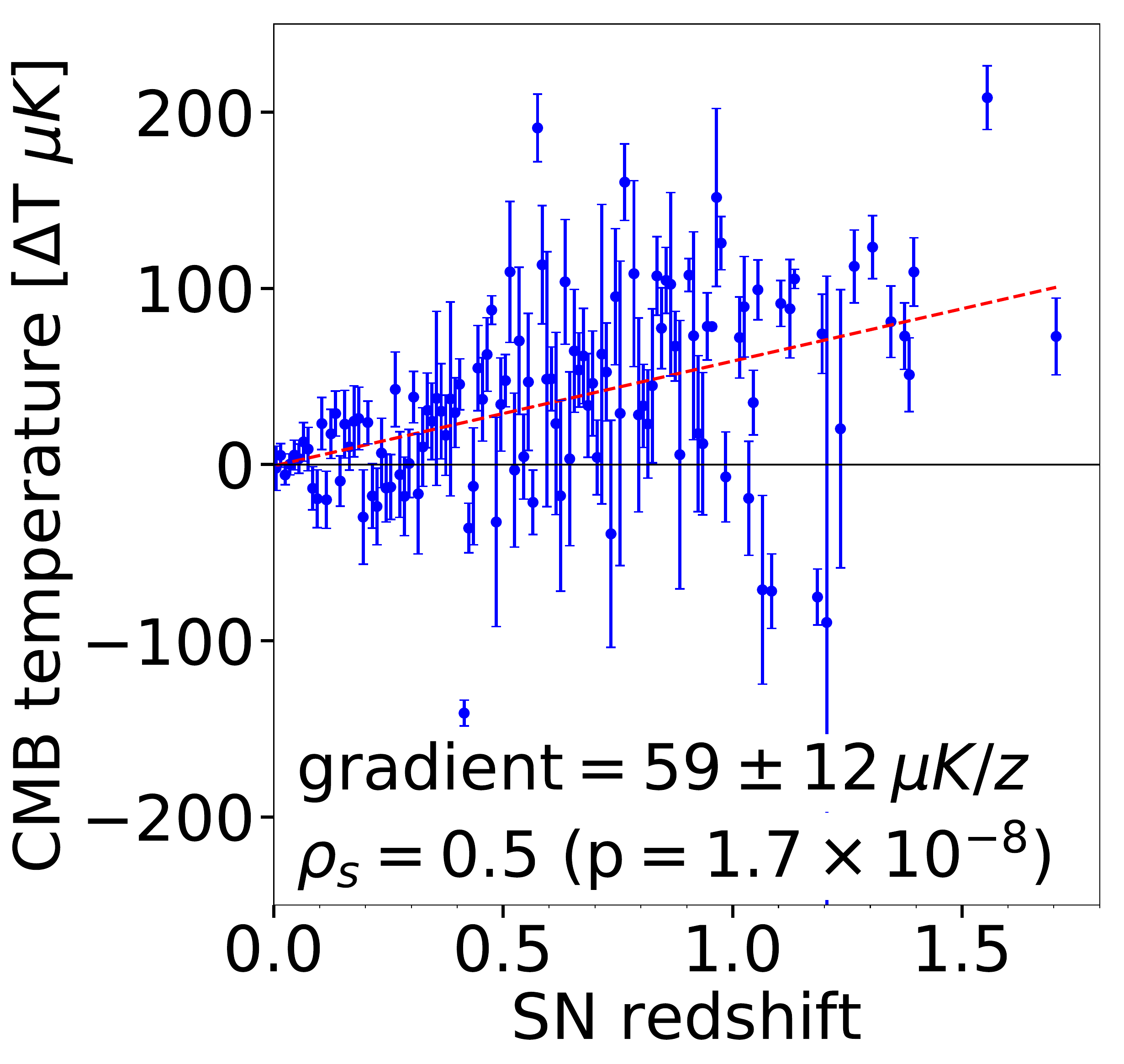}}}
	\\
	\subfigure[\label{sub217weightedUnsmoothed}Weighted $\overline{T}$, 217GHz covariance, unsmoothed]{
    \makebox[\columnwidth][c]{
    \includegraphics[width=0.4\columnwidth]{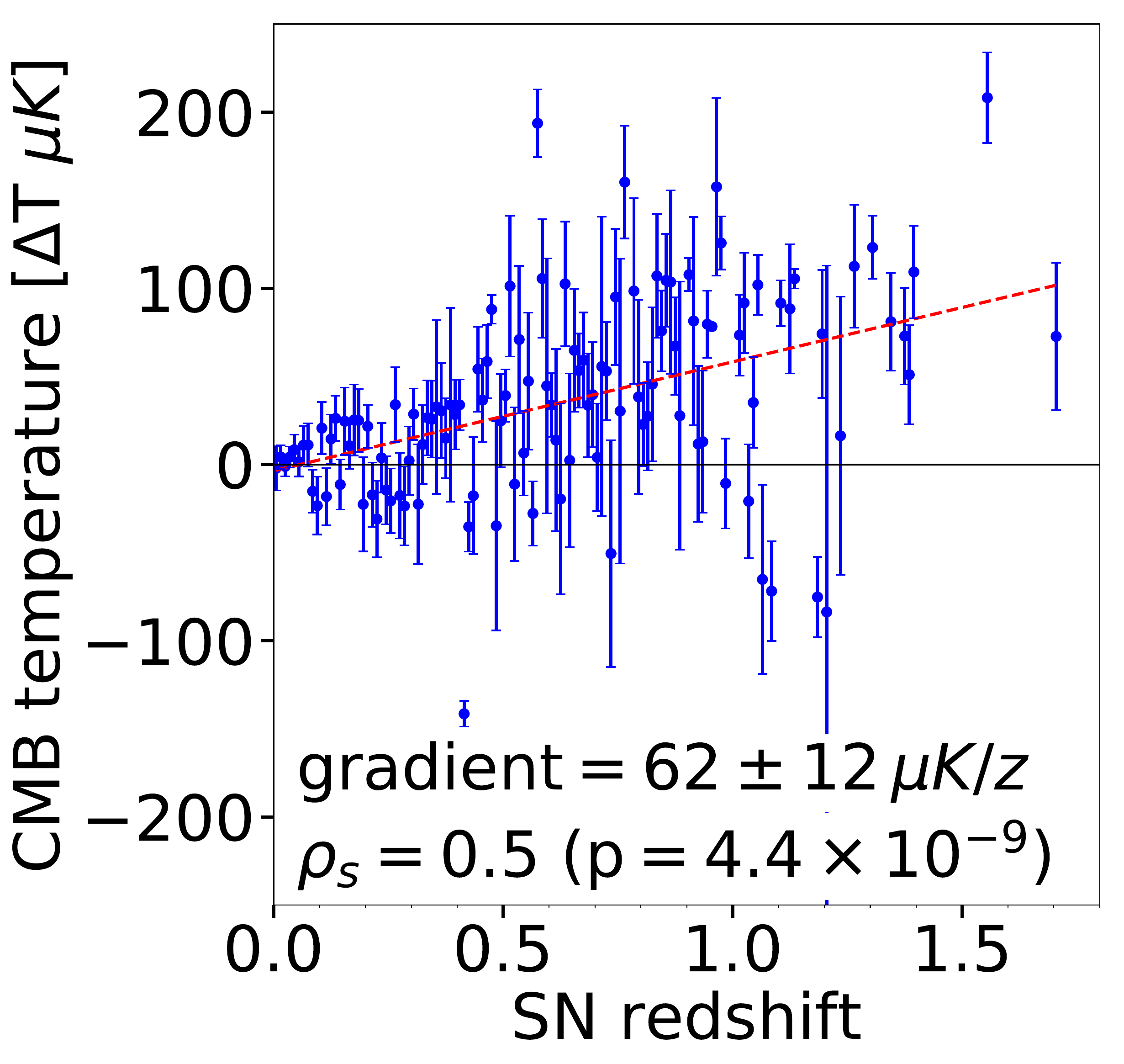}}}
	\caption[As Fig.~\ref{figBinnedScatter} but for variance estimated from 143GHz \& 217GHz maps]{Same as Fig. \ref{figHMDH} \protect\subref{subHMHDweightedUnsmoothed} but for variance estimated using intensity covariance in \textit{Planck} 2015 R2.02 143GHz and 217GHz frequency maps.}
	\label{fig143217}
\end{minipage}%\hfill
\begin{minipage}[t]{0.5\textwidth}
	\centering
	\subfigure[\label{sub2013unweighted}Unweighted $\overline{T}$]{
    \makebox[\columnwidth][c]{
    \includegraphics[width=0.4\columnwidth]{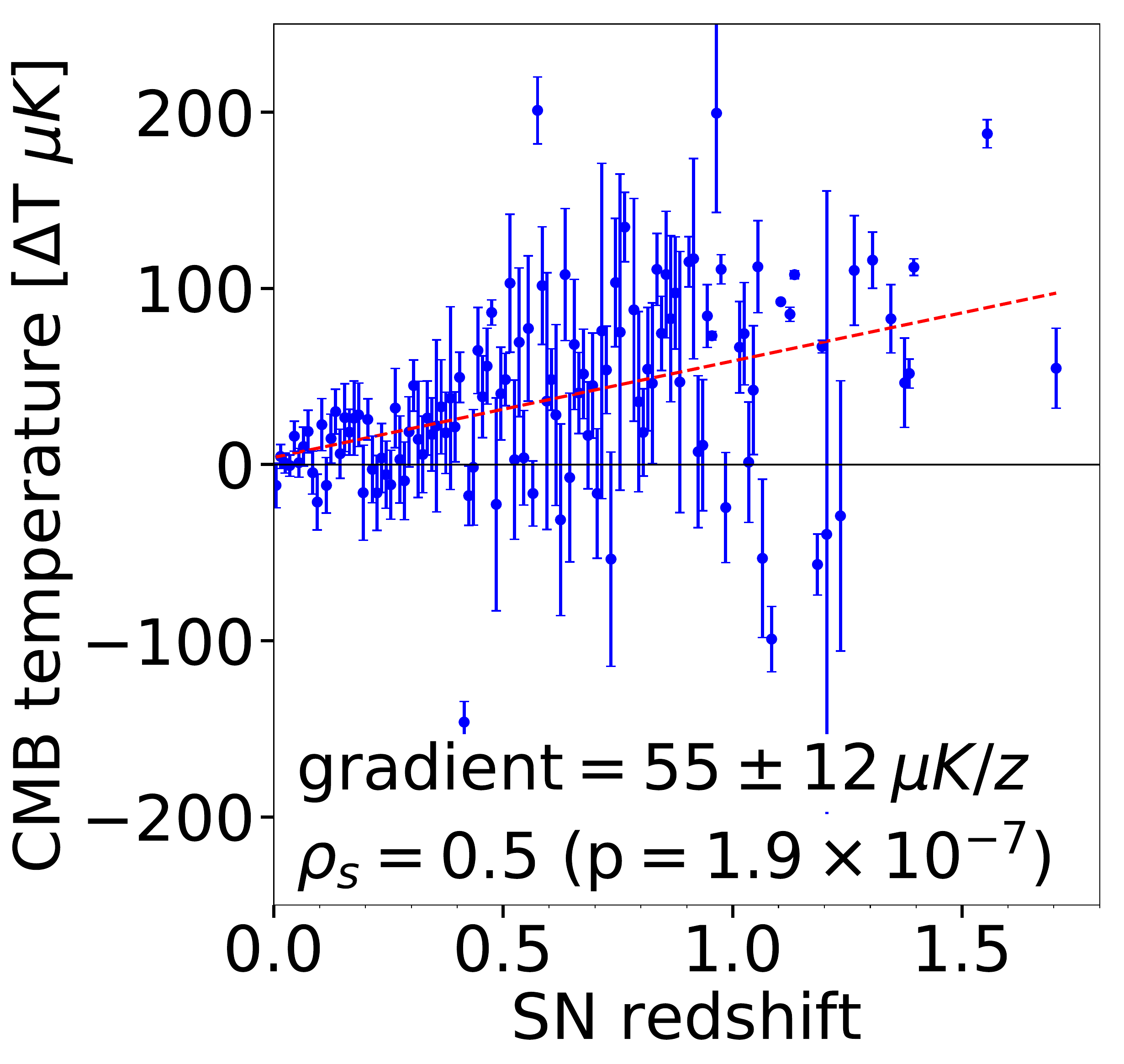}}}
	\\
	\subfigure[\label{sub2013weightedUnsmoothed}Weighted $\overline{T}$, R1.20 $\sigma_{\overline{T}}$, unsmoothed]{
    \makebox[\columnwidth][c]{
    \includegraphics[width=0.4\columnwidth]{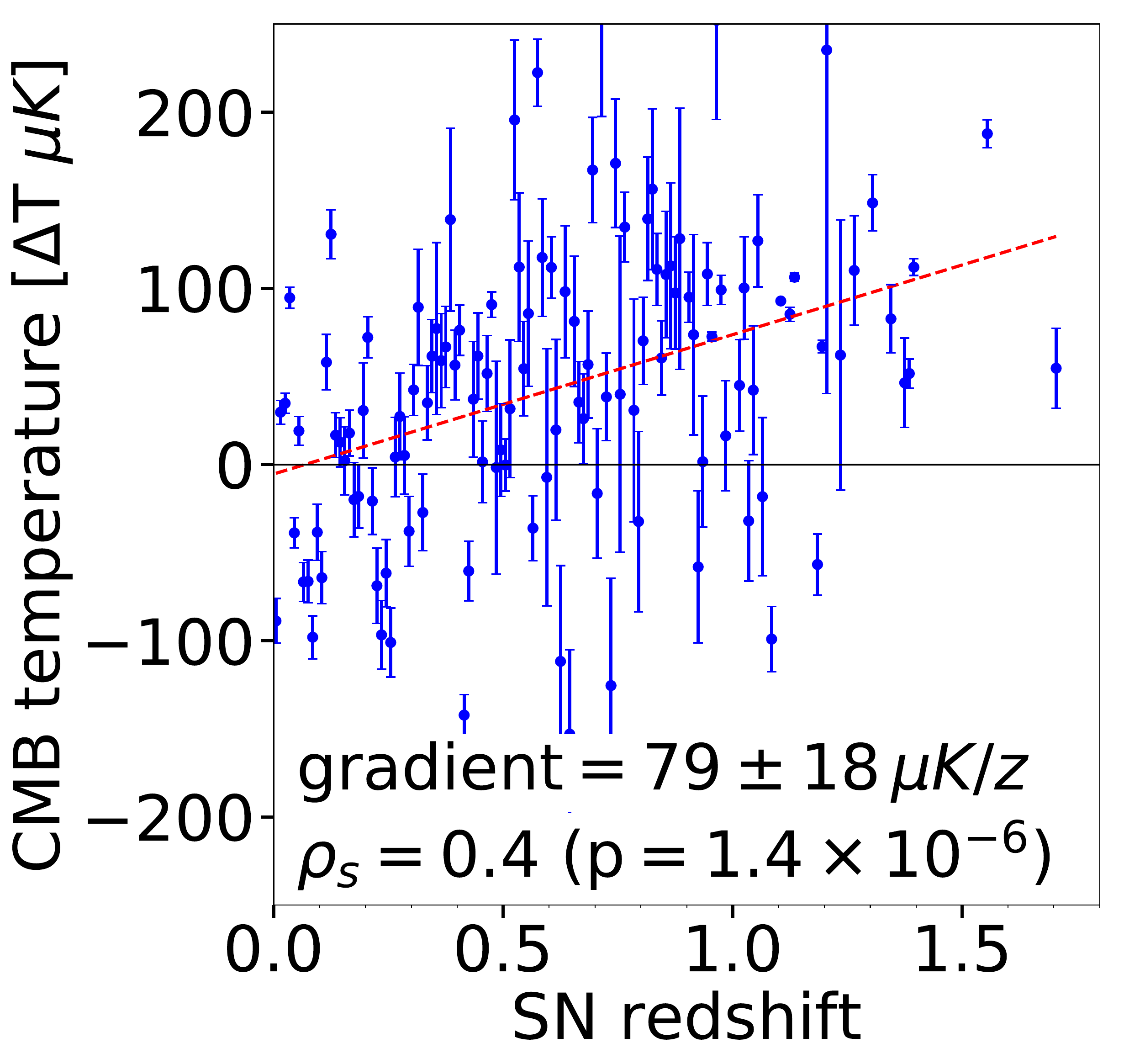}}}
	\\
	\subfigure[\label{sub2013weightedSmoothed5arcmin}Weighted $\overline{T}$, R1.20 $\sigma_{\overline{T}}$, $5'$ smoothed]{
    \makebox[\columnwidth][c]{
    \includegraphics[width=0.4\columnwidth]{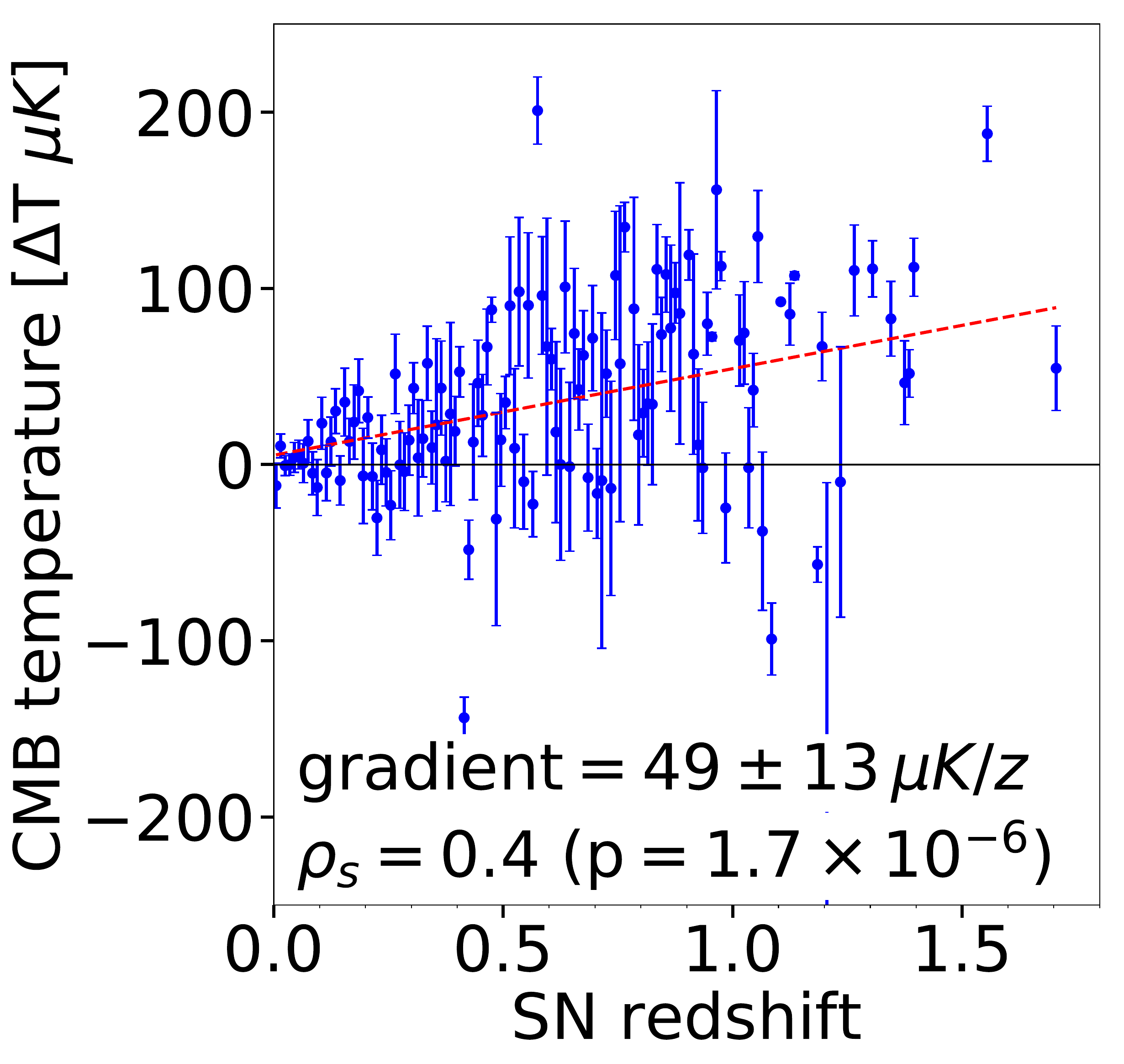}}}
	\\
	\subfigure[\label{sub2013weightedSmoothedHalfDeg}Weighted $\overline{T}$, R1.20 $\sigma_{\overline{T}}$, 0.5$^\circ$ smoothed]{
    \makebox[\columnwidth][c]{
    \includegraphics[width=0.4\columnwidth]{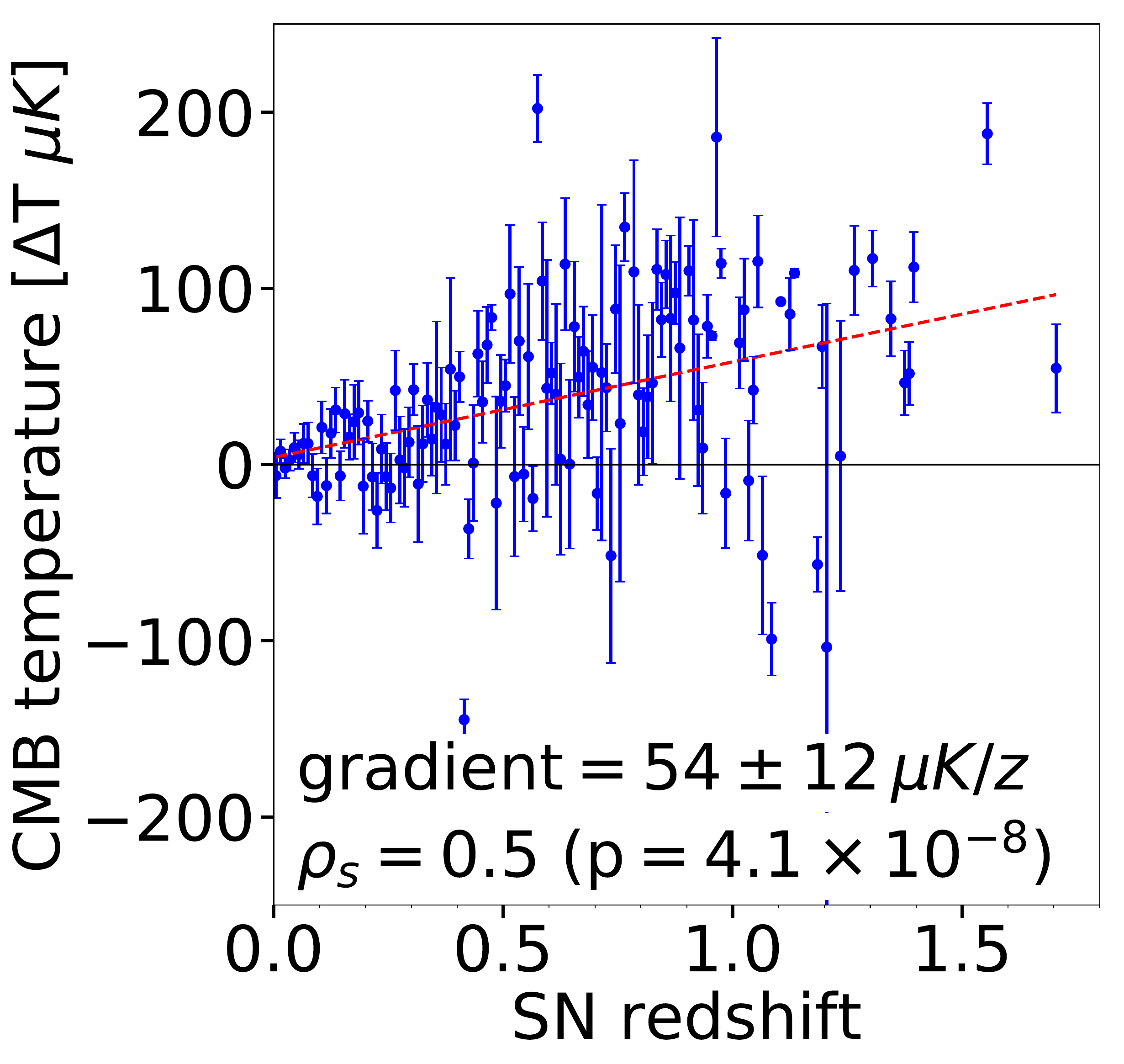}}}
	\\
	\subfigure[\label{sub2013weightedSmoothed5deg}Weighted $\overline{T}$, R1.20 $\sigma_{\overline{T}}$, 5$^\circ$ smoothed]{
    \makebox[\columnwidth][c]{
    \includegraphics[width=0.4\columnwidth]{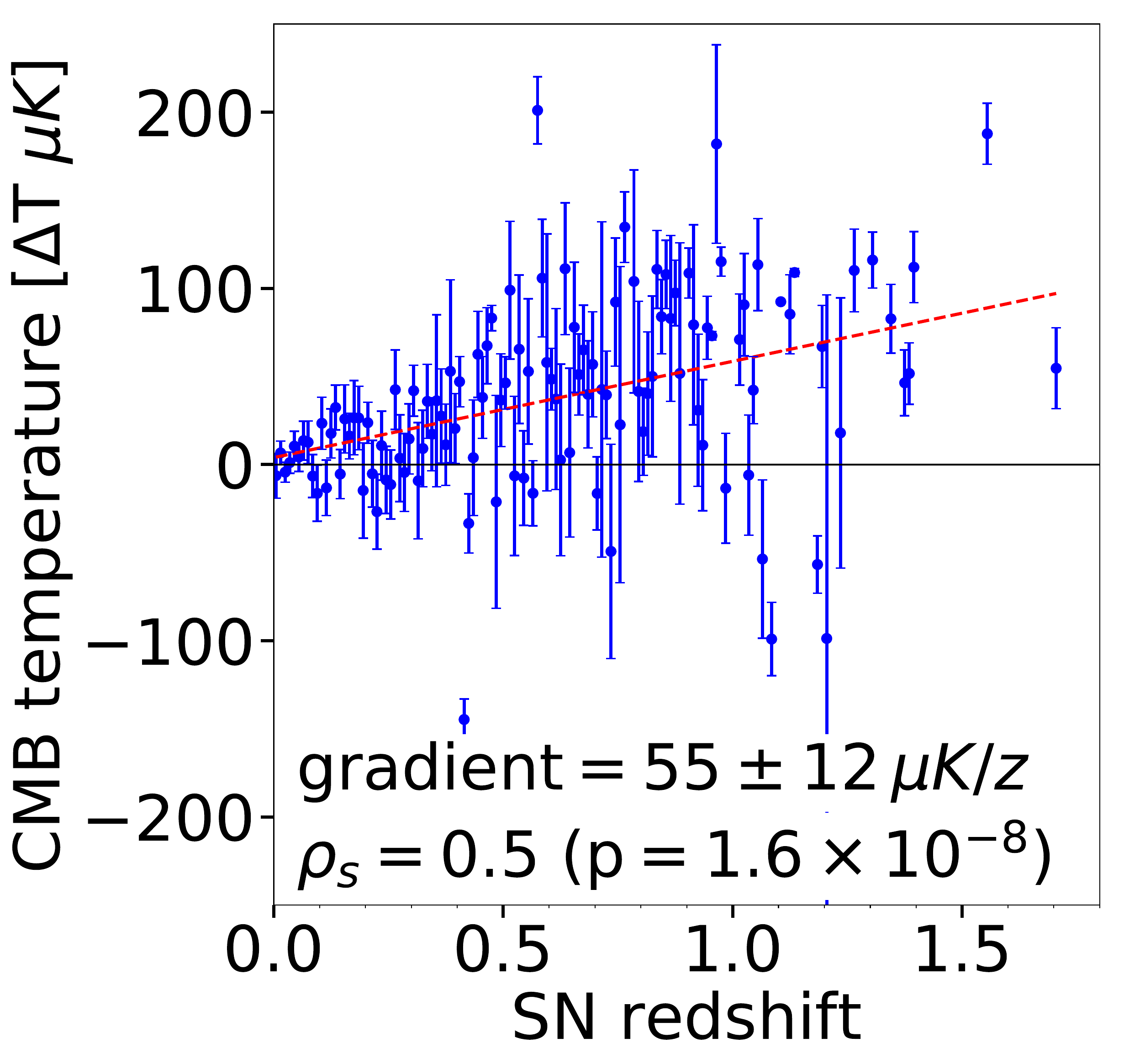}}}
	\caption[As Fig.~\ref{figBinnedScatter} but for variance estimated from SMICA noise]{Same as Fig. \ref{figHMDH} but for variance estimated using \textit{Planck} 2013 R1.20 SMICA noise.}
	\label{fig2013}
\end{minipage}
\end{figure}

\afterpage{\null\newpage} % because this ended on an odd page so appendix E started on an even page

%% file: appendixAltPlanckMaps.tex
\chapter{CMB $T$ and SNe $z$ correlation: alternative \textit{Planck} maps} \label{appAltPlanckMaps}

\setlength{\tabcolsep}{1pt} % default is 6pt
\begin{sidewaystable} % A4_SNe_find_field_z_and_T & A5_SNe_find_field_pixels
	\centering
	\small
	\caption[As Table \ref{tabHistograms} but for Commander, NILC, SEVEM, and SMICA maps]{Same as Table \ref{tabHistograms} CMB temperature columns but for temperature measured, and variance estimated, using each of the \textit{Planck} 2015 Commander, NILC, SEVEM, and SMICA maps.}
	\label{tabHistogramsAltPlanckMaps}
	\begin{tabular}{lrclcrclcrclcrclcrclcrclcrclcrclc}
		\hline
        \multirow{3}{*}{Field} & \multicolumn{31}{c}{CMB temperature ($\mu K$)} \\
        & \multicolumn{7}{c}{Commander} && \multicolumn{7}{c}{NILC} && \multicolumn{7}{c}{SEVEM} && \multicolumn{7}{c}{SMICA} \\
		& \multicolumn{3}{c}{SNe} && \multicolumn{3}{c}{pixels} && \multicolumn{3}{c}{SNe} && \multicolumn{3}{c}{pixels} && \multicolumn{3}{c}{SNe} && \multicolumn{3}{c}{pixels} && \multicolumn{3}{c}{SNe} && \multicolumn{3}{c}{pixels}\\
		\hline
		Field 1 & 32.4 & $\pm$ & 9.4 &~~~& 52.3 & $\pm$ & 1.7 &~~~& 30.5 & $\pm$ & 9.1 &~~~& 49.0 & $\pm$ & 1.7 &~~~& 35.7 & $\pm$ & 8.9 &~~~& 51.9 & $\pm$ & 1.7 &~~~& 29.1 & $\pm$ & 9.1 &~~~& 47.5 & $\pm$ & 1.7 \\
        Field 2 & 94.8 & $\pm$ & 2.8 &~~~& 95.2 & $\pm$ & 1.6 &~~~& 90.7 & $\pm$ & 3.1 &~~~& 92.9 & $\pm$ & 1.6 &~~~& 97.0 & $\pm$ & 3.1 &~~~& 96.1 & $\pm$ & 1.7 &~~~& 92.9 & $\pm$ & 3.3 && 94.3 & $\pm$ & 1.7 \\
        Field 3 & 133.0 & $\pm$ & 9.4 &~~~& 138.9 & $\pm$ & 1.0 &~~~& 132.1 & $\pm$ & 9.2 &~~~& 139.0 & $\pm$ & 1.0 &~~~& 132.2 & $\pm$ & 9.3 &~~~& 140.5 & $\pm$ & 1.0 &~~~& 130.8 & $\pm$ & 9.2 && 137.5 & $\pm$ & 1.0 \\
        Field 4 & 91.0 & $\pm$ & 14.7 &~~~& 85.7 & $\pm$ & 1.7 &~~~& 94.2 & $\pm$ & 14.5 &~~~& 88.2 & $\pm$ & 1.7 &~~~& 87.4 & $\pm$ & 13.8 &~~~& 85.6 & $\pm$ & 1.7 &~~~& 85.3 & $\pm$ & 14.3 && 81.7 & $\pm$ & 1.7 \\
        Field 5 & 51.8 & $\pm$ & 8.9 &~~~& 70.2 & $\pm$ & 1.2 &~~~& 52.1 & $\pm$ & 8.6 &~~~& 71.5 & $\pm$ & 1.2 &~~~& 51.8 & $\pm$ & 8.7 &~~~& 70.2 & $\pm$ & 1.2 &~~~& 49.7 & $\pm$ & 8.5 && 68.2 & $\pm$ & 1.2 \\
        Field 6 & 197.3 & $\pm$ & 14.5 &~~~& 169.5 & $\pm$ & 1.9 &~~~& 195.7 & $\pm$ & 14.1 &~~~& 167.4 & $\pm$ & 1.9 &~~~& 197.0 & $\pm$ & 13.8 &~~~& 167.1 & $\pm$ & 1.9 &~~~& 199.3 & $\pm$ &13.6 && 168.9 & $\pm$ & 1.9 \\
        Field 7 & 117.5 & $\pm$ & 12.3 &~~~& 100.5 & $\pm$ & 3.9 &~~~& 121.0 & $\pm$ & 12.4 &~~~& 100.8 & $\pm$ & 4.1 &~~~& 121.9 & $\pm$ & 11.9 &~~~& 103.9 & $\pm$ & 4.0 &~~~& 120.5 & $\pm$ & 12.0 && 100.8 & $\pm$ & 4.0 \\
        Fields 1-7 & 89.4 & $\pm$ & 5.0 &~~~& 103.6 & $\pm$ & 0.7 &~~~& 88.4 & $\pm$ & 5.0 &~~~& 103.7 & $\pm$ & 0.7 &~~~& 90.1 & $\pm$ & 4.9 &~~~& 103.8 & $\pm$ & 4.0 &~~~& 87.4 & $\pm$ & 5.0 && 101.5 & $\pm$ & 0.7 \\
        Stripe 82 & 4.8 & $\pm$ & 4.0 &~~~& 17.9 & $\pm$ & 0.2 &~~~& 6.1 & $\pm$ & 4.0 &~~~& 19.5 & $\pm$ & 0.2 &~~~& 5.8 & $\pm$ & 4.0 &~~~& 18.6 & $\pm$ & 0.2 &~~~& 3.8 & $\pm$ & 4.0 && 17.2 & $\pm$ & 0.2 \\
        Whole sample & 10.8 & $\pm$ & 2.0 &~~~& 3.2 & $\pm$ & 0.0 &~~~& 10.4 & $\pm$ & 2.0 &~~~& 2.2 & $\pm$ & 0.0 &~~~& 11.8 & $\pm$ & 2.0 &~~~& 3.0 & $\pm$ & 0.0 &~~~& 10.3 & $\pm$ & 2.0 && 3.1 & $\pm$ & 0.0 \\
		\hline
	\end{tabular}
\end{sidewaystable}
\setlength{\tabcolsep}{6pt}

\begin{figure} % A1_SNe_all_fields_remainder
\begin{minipage}[t]{0.5\textwidth}
	\centering
	\subfigure[Whole SNe sample]{\makebox[\columnwidth][c]{\includegraphics[width=0.7\columnwidth]{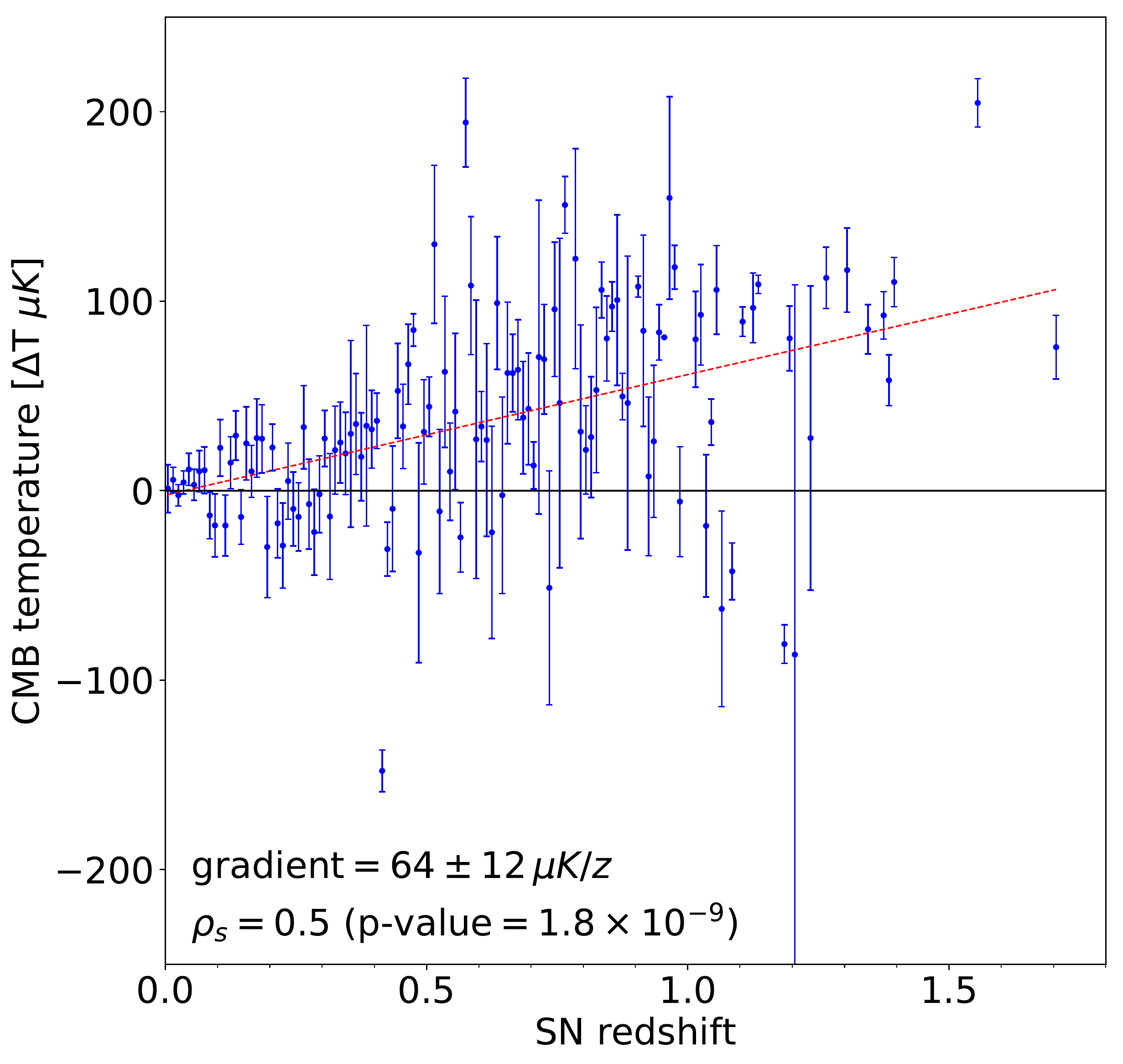}} \label{subBinnedScatterFields1to7AllCommander}}
	\\
	\subfigure[Fields 1-7 SNe sample]{\makebox[\columnwidth][c]{\includegraphics[width=0.7\columnwidth]{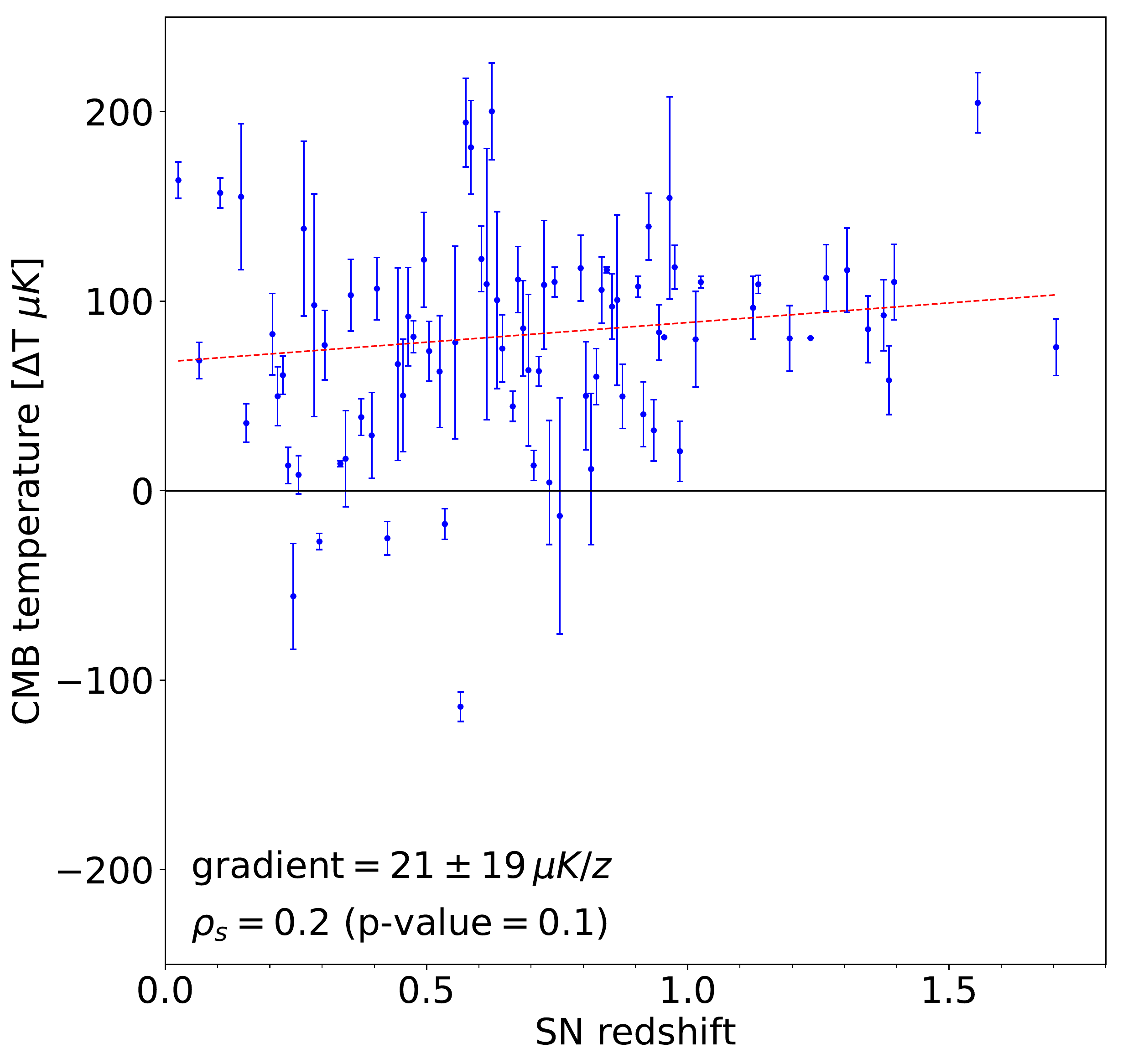}} \label{subBinnedScatterFields1to7FieldsCommander}}
	\\
	\subfigure[Remainder SNe sample]{\makebox[\columnwidth][c]{\includegraphics[width=0.7\columnwidth]{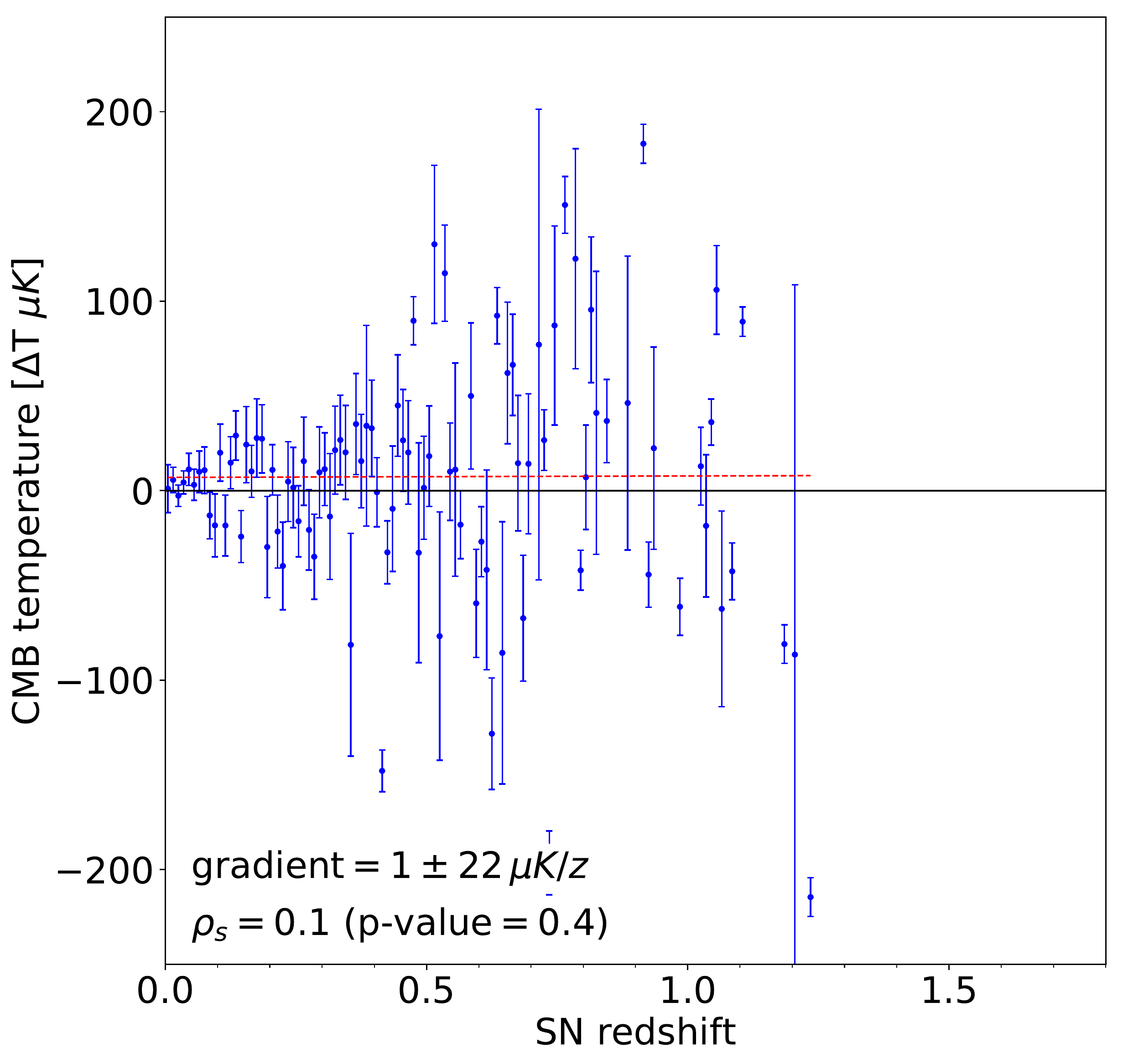}} \label{subBinnedScatterFields1to7RemainderCommander}}
	\caption[As Fig. \ref{figBinnedScatterFields1to7} but using \textit{Planck} 2015 Commander]{Same as Fig. \ref{figBinnedScatterFields1to7} but for temperature measured, and variance estimated, using \textit{Planck} 2015 Commander instead of SMICA.}
	\label{figBinnedScatterFields1to7Commander}
\end{minipage}%\hfill
\begin{minipage}[t]{0.5\textwidth}
	\centering
	\subfigure[Whole SNe sample]{\makebox[\columnwidth][c]{\includegraphics[width=0.7\columnwidth]{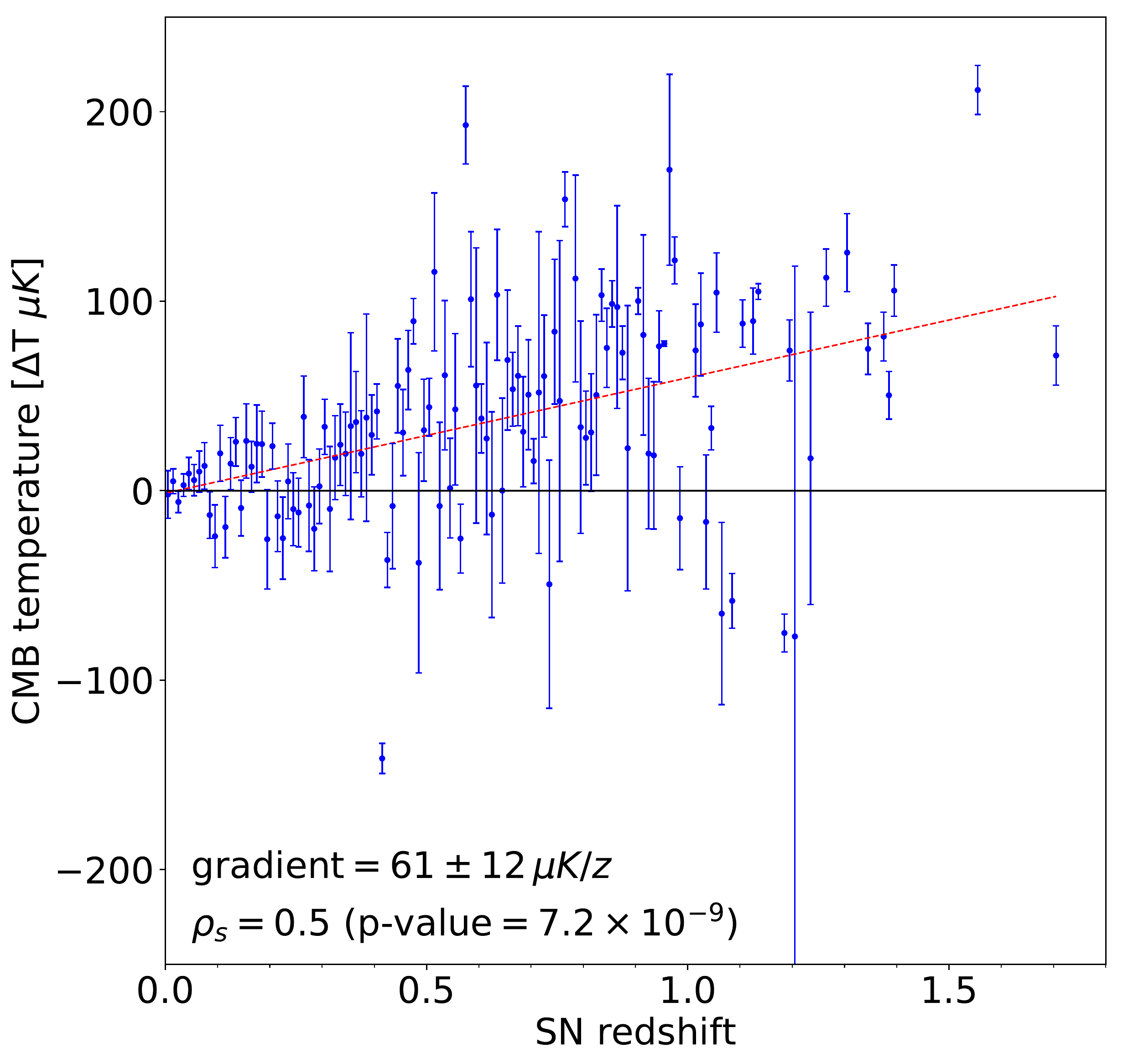}} \label{subBinnedScatterFields1to7Allnilc}}
	\\
	\subfigure[Fields 1-7 SNe sample]{\makebox[\columnwidth][c]{\includegraphics[width=0.7\columnwidth]{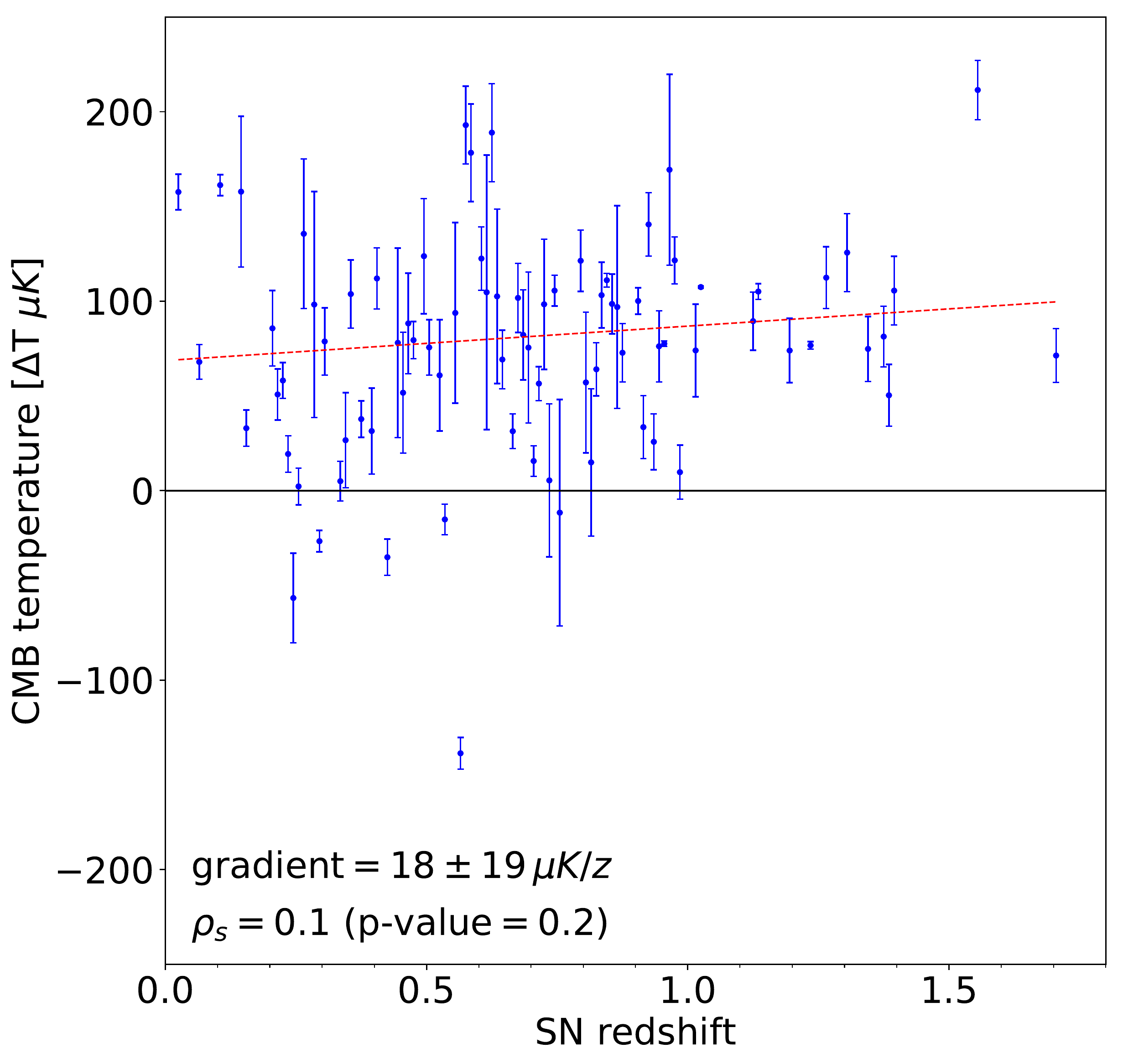}} \label{subBinnedScatterFields1to7Fieldsnilc}}
	\\
	\subfigure[Remainder SNe sample]{\makebox[\columnwidth][c]{\includegraphics[width=0.7\columnwidth]{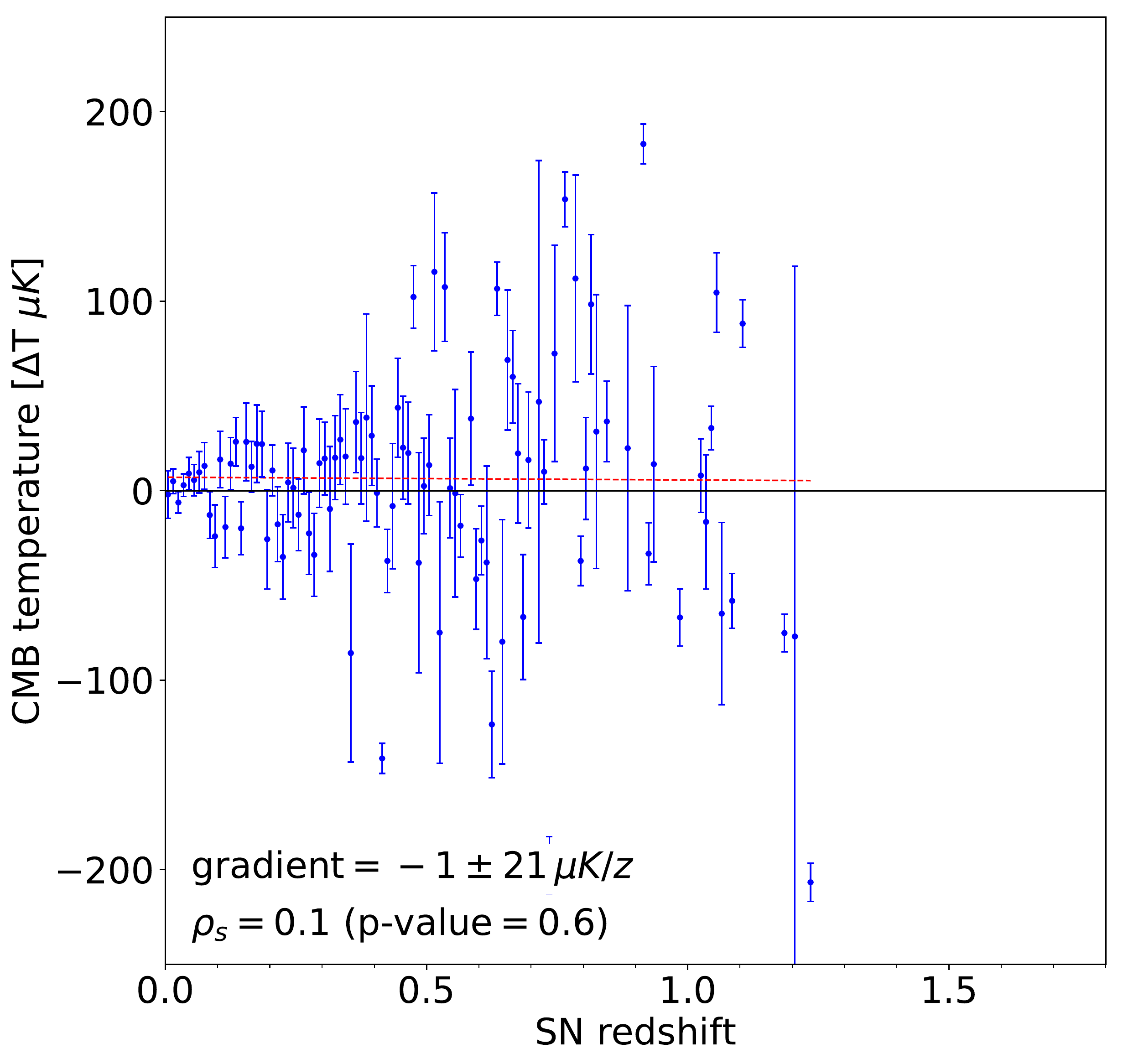}} \label{subBinnedScatterFields1to7Remaindernilc}}
	\caption[As Fig. \ref{figBinnedScatterFields1to7} but using \textit{Planck} 2015 NILC]{Same as Fig. \ref{figBinnedScatterFields1to7} but for temperature measured, and variance estimated, using \textit{Planck} 2015 NILC instead of SMICA.}
	\label{figBinnedScatterFields1to7nilc}
\end{minipage}
\end{figure}

\begin{figure} % A1_SNe_all_fields_remainder
	\centering
	\subfigure[Whole SNe sample]{\makebox[\columnwidth][c]{\includegraphics[width=0.35\columnwidth]{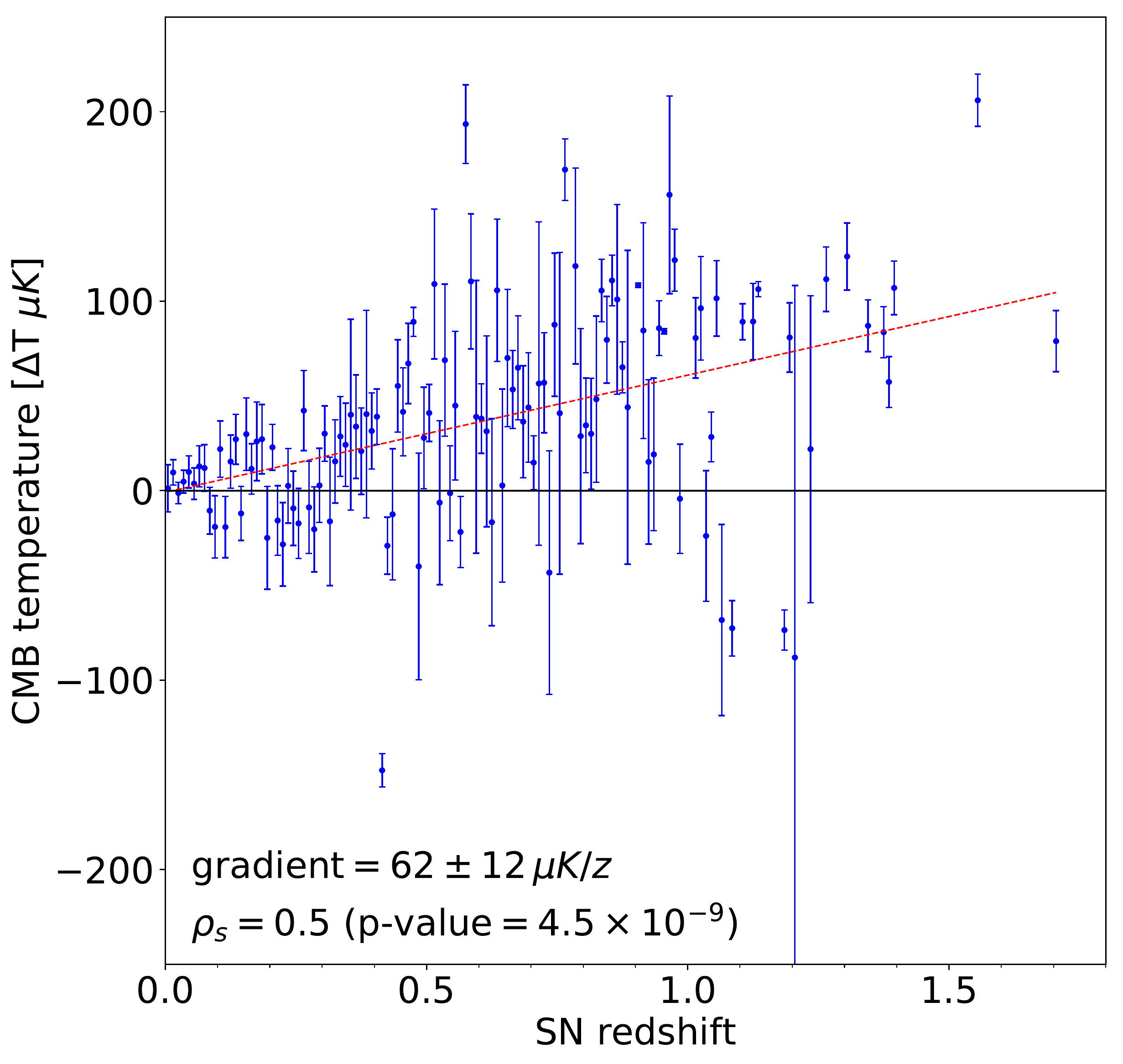}} \label{subBinnedScatterFields1to7Allsevem}}
	\\
	\subfigure[Fields 1-7 SNe sample]{\makebox[\columnwidth][c]{\includegraphics[width=0.35\columnwidth]{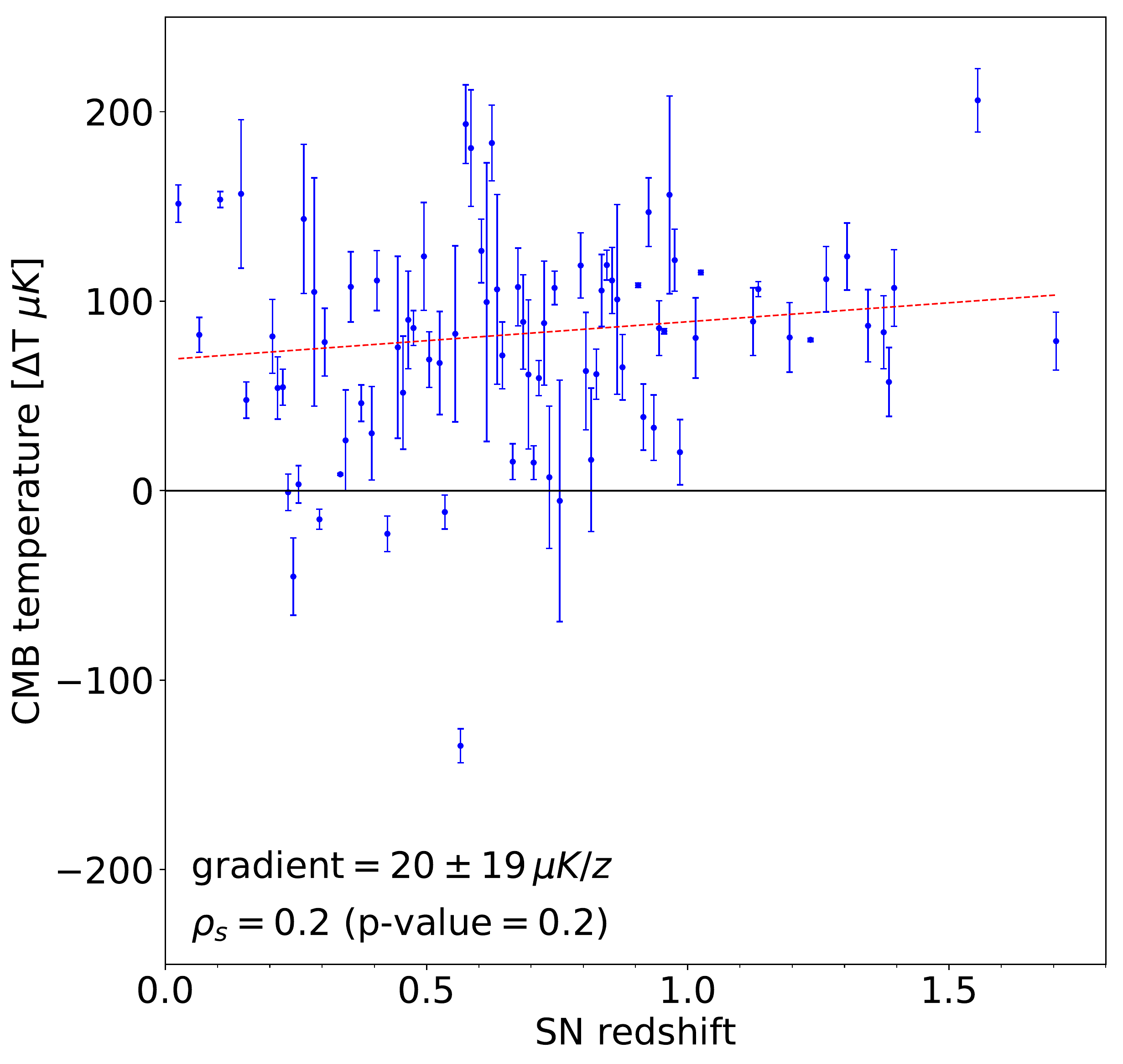}} \label{subBinnedScatterFields1to7Fieldssevem}}
	\\
	\subfigure[Remainder SNe sample]{\makebox[\columnwidth][c]{\includegraphics[width=0.35\columnwidth]{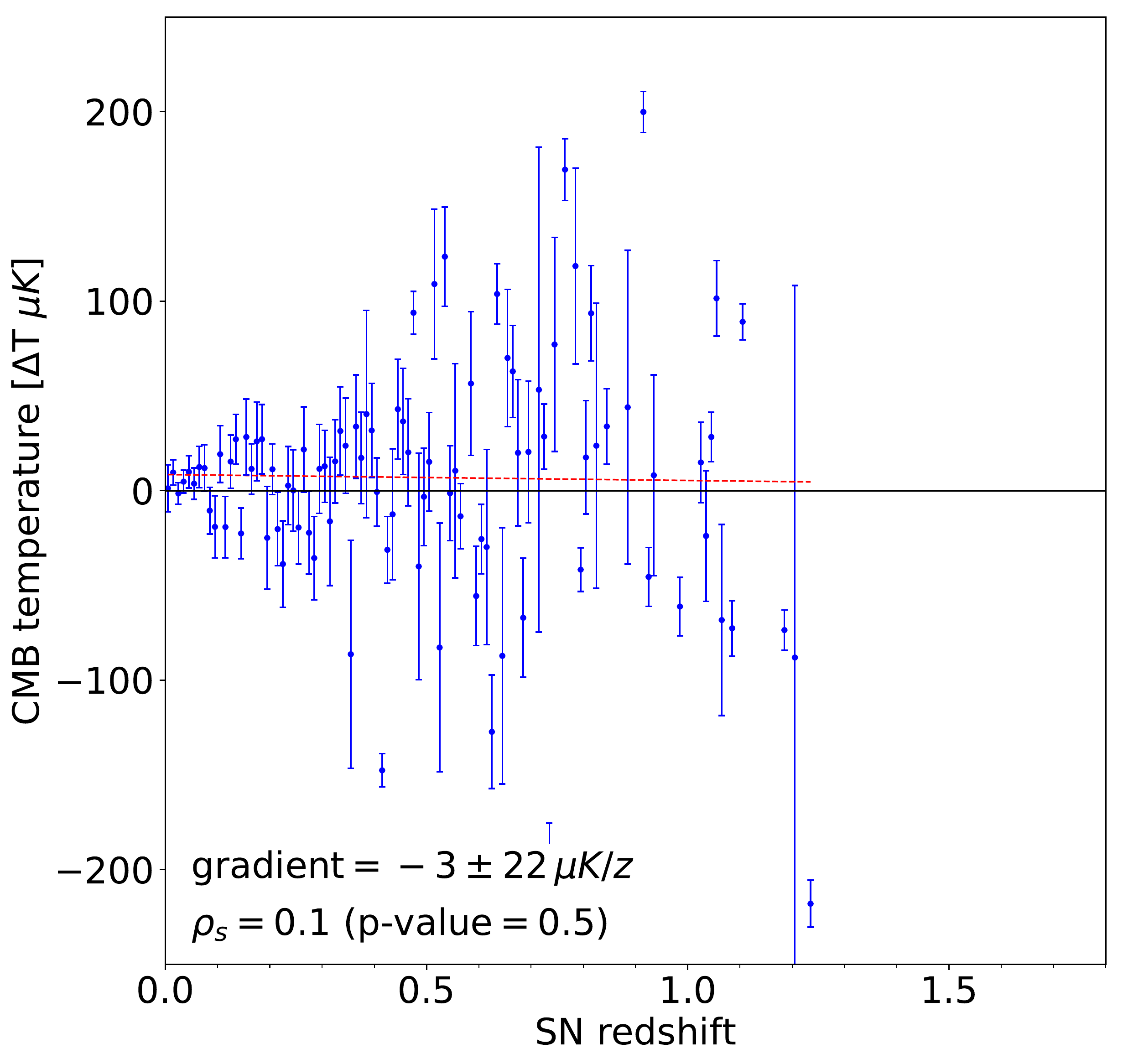}} \label{subBinnedScatterFields1to7Remaindersevem}}
	\caption[As Fig. \ref{figBinnedScatterFields1to7} but using \textit{Planck} 2015 SEVEM]{Same as Fig. \ref{figBinnedScatterFields1to7} but for temperature measured, and variance estimated, using \textit{Planck} 2015 SEVEM instead of SMICA.}
	\label{figBinnedScatterFields1to7sevem}
\end{figure}

%% file: appendixComovingDistance.tex
\chapter{Calculating comoving distance} \label{appComovingDistance}

To calculate the comoving radial distance $r(z)$ of an object at redshift $z$, we use the Friedmann equation for the expansion of an isotropic, homogeneous Universe

\begin{equation} \label{eqFriedmann}
    \left(\frac{\dot{a}}{a}\right)^2 = \frac{8 \pi G}{3}(\rho_r + \rho_m + \rho_{\Lambda}) - \frac{k c^2}{a^2}\ ,
\end{equation}

\noindent where $G$ is the gravitational constant, $\rho_r$, $\rho_m$, and $\rho_{\Lambda}$ are the radiation, matter, and dark energy densities of the Universe, $k$ is the curvature parameter, and $c$ is the speed of light. The cosmological constant $\Lambda$ has been included in the form $\rho_{\Lambda} \equiv \Lambda/8 \pi G$. The scale factor of the Universe is $a(t) = 1 / (1 + z)$, with the scale factor at the present epoch $t = t_0$ defined as $a(t_0) \equiv 1$. The critical density ($\rho_c = \rho_r + \rho_m + \rho_{\Lambda}$) for a flat Universe ($k = 0$) is

\begin{equation}
    \rho_c = \frac{3 H^2}{8 \pi G}\ ,
\end{equation}

\noindent where $H$ is the Hubble parameter, $H \equiv \dot{a} / a$. So for a flat Universe, Eq.~\ref{eqFriedmann} can be rewritten

\begin{equation} \label{eqFriedmann2}
    H^2 = \frac{H_0^2}{\rho_{c,0}}(\rho_r + \rho_m + \rho_{\Lambda})\ ,
\end{equation}

\noindent where $H_0$ is the Hubble parameter at $t_0$, $\rho_{c,0}$ is the critical density at $t_0$, and all other terms are a function of time.

\clearpage

\noindent In an expanding Universe, the energy densities scale with the scale factor $a$ according to their equations of state as follows

\begin{equation} \label{eqDensities}
\begin{aligned}
    \rho_r &= \frac{\rho_{r,0}}{a^4}\ , \\
    \rho_m &= \frac{\rho_{m,0}}{a^3}\ , \\
    \rho_{\Lambda} &= \rho_{\Lambda,0}\ ,
\end{aligned}
\end{equation}

\noindent where $\rho_{i,0}$ indicates energy density at the present epoch $t = t_0$, and $\rho_i$ and $a$ (the scale factor) are both functions of time. Substituting Eqs.~\ref{eqDensities} into Eq.~\ref{eqFriedmann2} gives

\begin{equation} \label{eqFriedmann3}
    H^2 = \frac{H_0^2}{\rho_{c,0}}\left(\frac{\rho_{r,0}}{a^4} + \frac{\rho_{m,0}}{a^3} + \rho_{\Lambda,0}\right)\ .
\end{equation}

\noindent Energy densities may be expressed as dimensionless fractions of the critical density

\begin{equation} \label{eqDensityParams}
\begin{aligned}
    \Omega_r &= \frac{\rho_r}{\rho_c}\ , \\
    \Omega_m &= \frac{\rho_m}{\rho_c}\ , \\
    \Omega_{\Lambda} &= \frac{\rho_{\Lambda}}{\rho_c}\ ,
\end{aligned}
\end{equation}

\noindent where all terms are functions of time. Substituting these values at the present epoch $t = t_0$ into Eq.~\ref{eqFriedmann3} gives

\begin{equation} \label{eqFriedmann4}
    H^2 = H_0^2\left(\frac{\Omega_{r,0}}{a^4} + \frac{\Omega_{m,0}}{a^3} + \Omega_{\Lambda,0}\right)\ ,
\end{equation}

\noindent and further substituting $a(t) = 1 / (1 + z)$ into Eq.~\ref{eqFriedmann4} gives

\begin{equation} \label{eqFriedmann5}
    H^2 = H_0^2(\Omega_{r,0}(1 + z)^4 + \Omega_{m,0}(1 + z)^3 + \Omega_{\Lambda,0})\ .
\end{equation}

\noindent For the concordance cosmological model, $\rho_r = \Omega_r = 0$, so $\rho_c = \rho_m + \rho_{\Lambda}$ and $\Omega_m + \Omega_{\Lambda} = 1$. So, finally rearranging Eq.~\ref{eqFriedmann5} gives

\begin{equation} \label{eqFriedmann6}
    \frac{H}{H_0} = E(z) = \sqrt{1 + \Omega_m((1 + z)^3 -1)}\ ,
\end{equation}

\noindent where $H_0$ is the Hubble parameter and $\Omega_m$ is the matter density, both at $t_0$. Following \citet{Hogg1999}, who in turn follows \citet{Peebles1993}, the comoving radial distance $r(z)$ of an object at redshift $z$ is therefore

\begin{equation} \label{eqRadialDistApp}
    r(z) = \frac{c}{H_0}\int_{0}^{z}\frac{dz}{E(z)}\ .
\end{equation}

%% file: appendixLQGData.tex
\chapter{Large quasar group position angles} \label{appLQGData}
 
% NB changes to this table may also necessitate changes to tabLQGDataPart
\begin{longtable}{crrrrrrrrc}
\captionsetup{font={small}, labelfont=bf, margin=1cm}

\caption[Large quasar group locations and position angles (full LQG sample)]{The 71 large quasar groups, where $m$ is the number of members, and $\bar{\alpha}$, $\bar{\delta}$, and $\bar{z}$ are the mean right ascension, declination, and redshift of the member quasars. The normalized goodness-of-fit weight $w$ (Eq.~\ref{eqWeight}) is scaled by $w_{71} = w \times 71$ for clarity, and to distinguish those LQGs weighted higher ($w_{71} > 1$) or lower ($w_{71} < 1$) than the mean $\bar{w}$. Position angle $\theta$ and half-width confidence interval $\gamma_h$ are shown for both the 2D and 3D approaches. The ratio of 3D eigenvalues (and ellipsoid axes lengths) is given by $a:b:c$. This list was summarized in Table~\ref{tabLQGDataPart}.\label{tabLQGDataFull}}\\

\hline
& \multicolumn{2}{r}{J2000 ($^\circ$)} &&& \multicolumn{2}{r}{2D PA ($^\circ$)} & \multicolumn{2}{r}{3D PA ($^\circ$)} & \\
$m$ & $\bar{\alpha}$ & $\bar{\delta}$ & $\bar{z}$ & $w_{71}$ & $\theta$ & $\gamma_h$ & $\theta$ & $\gamma_h$ & $a:b:c$ \\
\hline
\endfirsthead

\multicolumn{10}{l}{\small\textbf{Table~\ref{tabLQGDataFull}} continued...}\\
\hline
& \multicolumn{2}{r}{J2000 ($^\circ$)} &&& \multicolumn{2}{r}{2D PA ($^\circ$)} & \multicolumn{2}{r}{3D PA ($^\circ$)} & \\
$m$ & $\bar{\alpha}$ & $\bar{\delta}$ & $\bar{z}$ & $w_{71}$ & $\theta$ & $\gamma_h$ & $\theta$ & $\gamma_h$ & $a:b:c$ \\
\hline
\endhead

\hline
\endfoot

\hline\hline
\endlastfoot

20 & 121.1 & 27.9 & 1.73 & 1.13 & 119.7 & 10.4 & 115.1 & 10.7 & 0.50:0.30:0.21 \\
20 & 151.5 & 48.6 & 1.46 & 1.21 & 144.4 & 9.2 & 144.3 & 11.5 & 0.50:0.33:0.17 \\
20 & 155.9 & 12.8 & 1.50 & 0.32 & 120.7 & 25.2 & 117.2 & 37.0 & 0.44:0.43:0.14 \\
20 & 163.6 & 16.9 & 1.57 & 1.00 & 0.9 & 8.8 & 5.7 & 9.2 & 0.47:0.33:0.20 \\
20 & 178.0 & 1.2 & 1.23 & 0.37 & 78.9 & 20.3 & 77.0 & 14.7 & 0.51:0.31:0.17 \\
20 & 216.4 & 1.4 & 1.11 & 0.86 & 21.6 & 8.1 & 28.4 & 11.3 & 0.49:0.30:0.21 \\
21 & 133.5 & 41.2 & 1.40 & 0.72 & 16.9 & 12.1 & 18.5 & 10.6 & 0.49:0.33:0.18 \\
21 & 170.6 & 16.8 & 1.07 & 0.90 & 93.0 & 10.0 & 94.1 & 15.4 & 0.47:0.37:0.17 \\
21 & 191.8 & 11.0 & 1.06 & 4.43 & 40.9 & 2.8 & 41.0 & 2.9 & 0.66:0.24:0.10 \\
21 & 209.1 & 3.2 & 1.56 & 0.77 & 55.9 & 14.8 & 12.3 & 25.9 & 0.49:0.30:0.21 \\
21 & 212.9 & 12.6 & 1.55 & 1.72 & 162.7 & 3.9 & 162.8 & 3.9 & 0.63:0.22:0.15 \\
21 & 231.2 & 25.2 & 1.51 & 4.57 & 177.7 & 3.9 & 178.8 & 2.4 & 0.57:0.29:0.14 \\
22 & 136.8 & 49.5 & 1.19 & 1.03 & 119.4 & 8.2 & 126.4 & 13.2 & 0.46:0.38:0.17 \\
22 & 182.0 & 55.5 & 1.70 & 1.88 & 126.1 & 4.6 & 126.3 & 4.7 & 0.56:0.23:0.20 \\
22 & 217.8 & -0.8 & 1.31 & 0.42 & 71.0 & 33.3 & 69.2 & 27.0 & 0.42:0.33:0.25 \\
23 & 155.9 & 53.5 & 1.48 & 1.79 & 115.6 & 4.8 & 115.8 & 4.8 & 0.60:0.22:0.19 \\
23 & 166.4 & 37.1 & 1.31 & 0.97 & 134.5 & 10.2 & 126.3 & 15.9 & 0.44:0.37:0.19 \\
23 & 171.3 & 14.0 & 1.20 & 1.57 & 117.5 & 6.4 & 101.2 & 15.7 & 0.48:0.37:0.15 \\
23 & 180.5 & 6.0 & 1.29 & 1.26 & 81.2 & 13.5 & 72.8 & 11.3 & 0.57:0.25:0.18 \\
23 & 209.5 & 34.3 & 1.65 & 1.99 & 152.5 & 2.8 & 152.3 & 2.8 & 0.68:0.17:0.15 \\
23 & 214.3 & 31.8 & 1.48 & 0.27 & 18.6 & 32.6 & 87.7 & 32.5 & 0.47:0.35:0.18 \\
24 & 119.1 & 18.6 & 1.28 & 1.06 & 157.3 & 9.6 & 162.8 & 11.1 & 0.46:0.32:0.23 \\
24 & 139.6 & 2.5 & 1.20 & 0.61 & 48.9 & 6.7 & 49.0 & 6.7 & 0.58:0.29:0.13 \\
24 & 171.1 & 17.6 & 1.52 & 0.83 & 78.7 & 11.2 & 77.0 & 10.3 & 0.50:0.28:0.22 \\
24 & 179.6 & 65.0 & 1.08 & 0.71 & 35.4 & 3.7 & 35.2 & 3.9 & 0.55:0.23:0.22 \\
24 & 205.0 & 12.0 & 1.36 & 0.61 & 129.2 & 16.1 & 129.2 & 16.6 & 0.46:0.33:0.20 \\
24 & 217.1 & 33.8 & 1.11 & 0.99 & 12.8 & 15.0 & 34.9 & 21.8 & 0.54:0.28:0.18 \\
24 & 217.1 & 57.5 & 1.70 & 0.75 & 9.8 & 14.3 & 11.6 & 10.3 & 0.52:0.28:0.20 \\
25 & 142.5 & 31.3 & 1.33 & 1.70 & 42.4 & 6.1 & 41.1 & 7.2 & 0.50:0.34:0.15 \\
25 & 149.5 & 43.7 & 1.16 & 0.81 & 42.3 & 7.8 & 38.9 & 9.9 & 0.45:0.35:0.20 \\
25 & 186.0 & 3.6 & 1.19 & 0.61 & 23.6 & 12.3 & 47.3 & 15.3 & 0.48:0.32:0.20 \\
25 & 196.7 & 36.5 & 1.48 & 0.64 & 103.1 & 7.7 & 102.9 & 8.6 & 0.47:0.33:0.20 \\
25 & 231.9 & 43.6 & 1.20 & 0.30 & 91.9 & 15.3 & 75.1 & 16.9 & 0.44:0.36:0.21 \\
26 & 160.3 & 53.5 & 1.18 & 0.33 & 110.6 & 23.2 & 111.5 & 25.1 & 0.38:0.34:0.28 \\
26 & 171.7 & 24.2 & 1.10 & 0.78 & 48.5 & 8.3 & 47.3 & 8.1 & 0.46:0.29:0.24 \\
26 & 206.0 & 6.5 & 1.43 & 0.75 & 38.1 & 8.7 & 46.7 & 12.0 & 0.46:0.36:0.18 \\
26 & 224.2 & 61.6 & 1.57 & 1.52 & 106.8 & 3.1 & 106.8 & 4.3 & 0.51:0.33:0.16 \\
26 & 245.6 & 40.3 & 1.55 & 0.58 & 4.2 & 8.0 & 3.5 & 4.7 & 0.49:0.26:0.25 \\
27 & 184.1 & 52.7 & 1.18 & 0.37 & 124.4 & 28.7 & 125.7 & 28.7 & 0.40:0.35:0.25 \\
27 & 204.0 & 13.7 & 1.16 & 0.98 & 19.6 & 5.8 & 19.9 & 8.2 & 0.50:0.36:0.13 \\
27 & 221.8 & 52.4 & 1.55 & 1.12 & 52.3 & 7.2 & 54.5 & 6.6 & 0.54:0.26:0.20 \\
27 & 225.5 & 57.3 & 1.52 & 0.54 & 63.3 & 14.4 & 45.6 & 14.6 & 0.45:0.37:0.18 \\
27 & 231.2 & 15.4 & 1.56 & 1.01 & 60.8 & 14.0 & 53.2 & 18.7 & 0.43:0.35:0.22 \\
28 & 138.6 & 14.8 & 1.54 & 1.22 & 135.6 & 8.1 & 139.3 & 7.1 & 0.63:0.21:0.15 \\
28 & 177.0 & 43.1 & 1.54 & 1.10 & 45.7 & 6.2 & 45.2 & 6.1 & 0.52:0.25:0.23 \\
28 & 192.3 & 10.7 & 1.36 & 0.51 & 155.8 & 15.8 & 137.9 & 12.9 & 0.49:0.32:0.19 \\
28 & 212.0 & 61.7 & 1.15 & 1.33 & 131.7 & 5.5 & 135.9 & 7.6 & 0.61:0.23:0.16 \\
30 & 138.1 & 7.9 & 1.56 & 0.85 & 39.7 & 8.4 & 41.0 & 9.5 & 0.51:0.32:0.17 \\
30 & 217.2 & 10.5 & 1.67 & 0.99 & 147.5 & 6.5 & 146.9 & 6.8 & 0.56:0.23:0.21 \\
30 & 234.0 & 21.8 & 1.60 & 0.61 & 91.3 & 12.6 & 107.0 & 16.9 & 0.46:0.33:0.21 \\
31 & 176.5 & 38.3 & 1.28 & 0.37 & 132.6 & 17.9 & 134.4 & 16.6 & 0.46:0.34:0.19 \\
31 & 202.7 & 21.3 & 1.08 & 0.74 & 64.1 & 6.0 & 65.3 & 5.3 & 0.56:0.24:0.20 \\
32 & 177.4 & 32.7 & 1.18 & 2.11 & 46.6 & 4.5 & 46.1 & 4.6 & 0.59:0.23:0.18 \\
33 & 126.6 & 20.3 & 1.44 & 1.15 & 109.6 & 5.6 & 112.8 & 5.7 & 0.52:0.29:0.20 \\
33 & 157.8 & 20.7 & 1.59 & 1.07 & 15.6 & 13.0 & 12.7 & 11.9 & 0.41:0.33:0.26 \\
33 & 183.6 & 11.0 & 1.08 & 0.69 & 111.3 & 4.1 & 111.2 & 3.7 & 0.54:0.25:0.21 \\
34 & 162.3 & 5.3 & 1.28 & 0.60 & 108.6 & 11.9 & 153.9 & 25.5 & 0.44:0.37:0.19 \\
34 & 228.5 & 8.1 & 1.22 & 0.71 & 178.7 & 8.2 & 156.5 & 17.2 & 0.45:0.34:0.21 \\
34 & 234.5 & 10.7 & 1.24 & 0.88 & 55.9 & 7.4 & 53.5 & 8.3 & 0.48:0.29:0.24 \\
36 & 189.0 & 44.1 & 1.39 & 1.22 & 41.3 & 4.2 & 38.9 & 4.1 & 0.53:0.28:0.19 \\
37 & 189.5 & 20.3 & 1.46 & 0.66 & 46.7 & 10.5 & 62.4 & 23.4 & 0.42:0.39:0.19 \\
38 & 161.6 & 3.5 & 1.11 & 0.35 & 137.7 & 17.5 & 119.2 & 19.3 & 0.46:0.37:0.18 \\
38 & 227.6 & 41.4 & 1.54 & 0.50 & 54.7 & 7.0 & 51.2 & 7.7 & 0.50:0.34:0.16 \\
41 & 205.3 & 50.4 & 1.39 & 1.36 & 51.3 & 2.7 & 51.0 & 2.7 & 0.63:0.22:0.15 \\
43 & 231.0 & 47.8 & 1.57 & 0.76 & 30.4 & 5.5 & 35.0 & 6.0 & 0.47:0.32:0.20 \\
44 & 208.7 & 25.8 & 1.28 & 0.45 & 120.0 & 8.6 & 131.2 & 6.8 & 0.54:0.27:0.19 \\
46 & 226.7 & 16.7 & 1.09 & 0.63 & 136.2 & 7.2 & 133.9 & 7.5 & 0.47:0.30:0.23 \\
55 & 196.5 & 27.1 & 1.59 & 0.95 & 107.5 & 3.4 & 107.0 & 3.4 & 0.58:0.24:0.18 \\
56 & 167.0 & 33.8 & 1.11 & 0.81 & 110.2 & 3.5 & 110.4 & 3.8 & 0.50:0.29:0.21 \\
64 & 196.4 & 39.9 & 1.14 & 0.83 & 133.6 & 3.0 & 133.9 & 3.2 & 0.48:0.36:0.17 \\
73 & 164.1 & 14.1 & 1.27 & 0.76 & 156.6 & 4.2 & 156.3 & 4.5 & 0.55:0.28:0.16 \\
\hline
\end{longtable}

%% file: appendixLQGs3DPerspectives.tex
\chapter{Large quasar groups in three-dimensional comoving space} \label{appLQGs3DPerspectives}
    
\begin{figure} % LQG_plot_3D
    \centering
    \includegraphics[width=0.9\columnwidth]{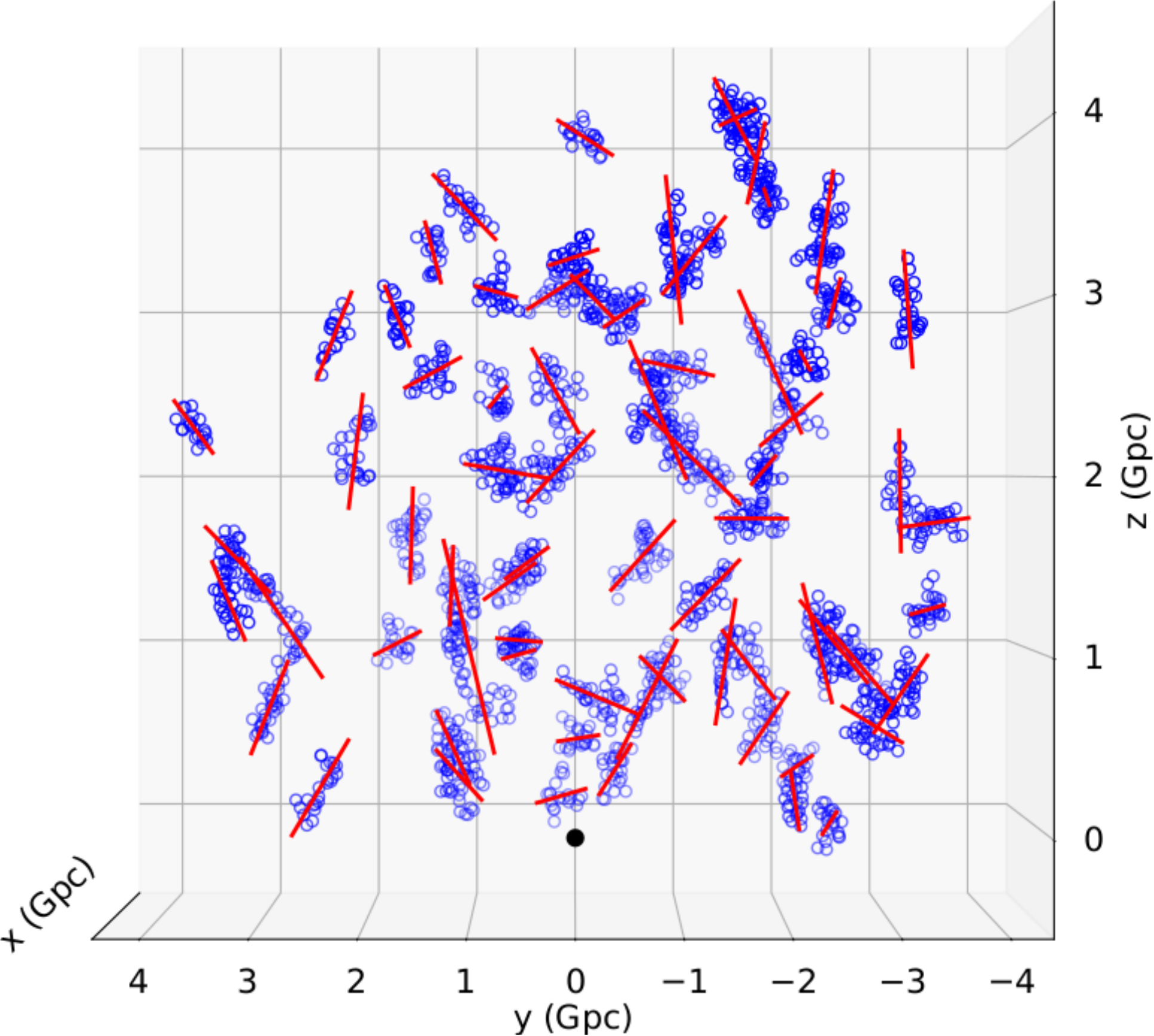}
    \caption[As Fig.~\ref{figLQGs3D} but viewed from three orthogonal perspectives]{(a) LQGs in three-dimensional comoving space, viewed with the $x$-axis orthogonal to the page, showing LQG quasars (blue circles) and 3D LQG major axes (red lines). Quasar markers are shaded to give the appearance of depth, with lighter shades representing more distant quasars. Our location at ($x, y, z$) coordinates (0, 0, 0) is indicated by a black dot.}
    \label{figLQGs3DAspectsPt1}
\end{figure}

\begin{figure}
    \ContinuedFloat
    \centering
    \includegraphics[width=0.9\columnwidth]{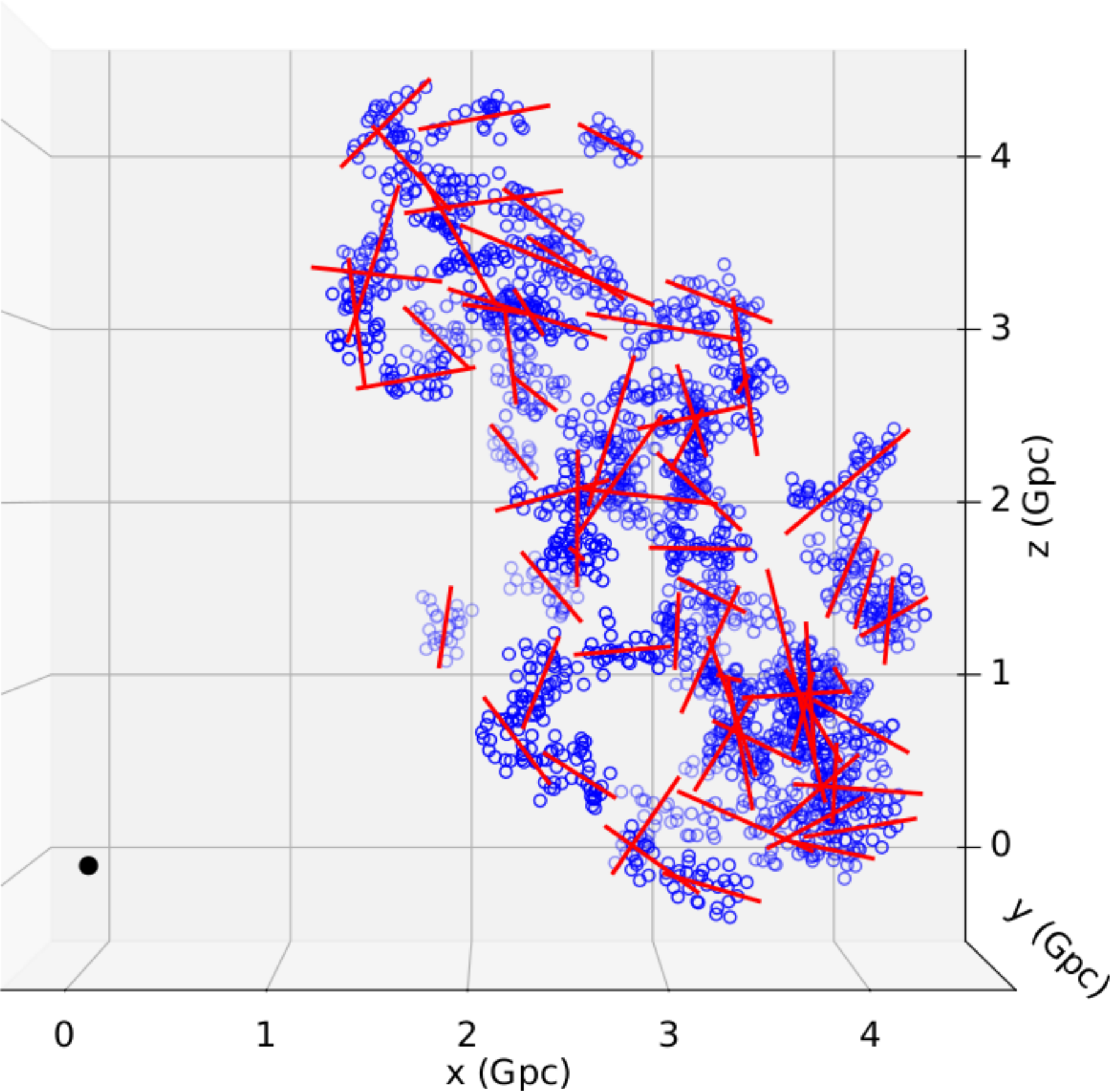}
    \caption[]{(b) LQGs in three-dimensional comoving space, viewed with the $y$-axis orthogonal to the page, showing LQG quasars (blue circles) and 3D LQG major axes (red lines). Quasar markers are shaded to give the appearance of depth, with lighter shades representing more distant quasars. Our location at ($x, y, z$) coordinates (0, 0, 0) is indicated by a black dot.}
    \label{figLQGs3DAspectsPt2}
\end{figure}

\begin{figure}
    \ContinuedFloat
    \centering
    \includegraphics[width=0.9\columnwidth]{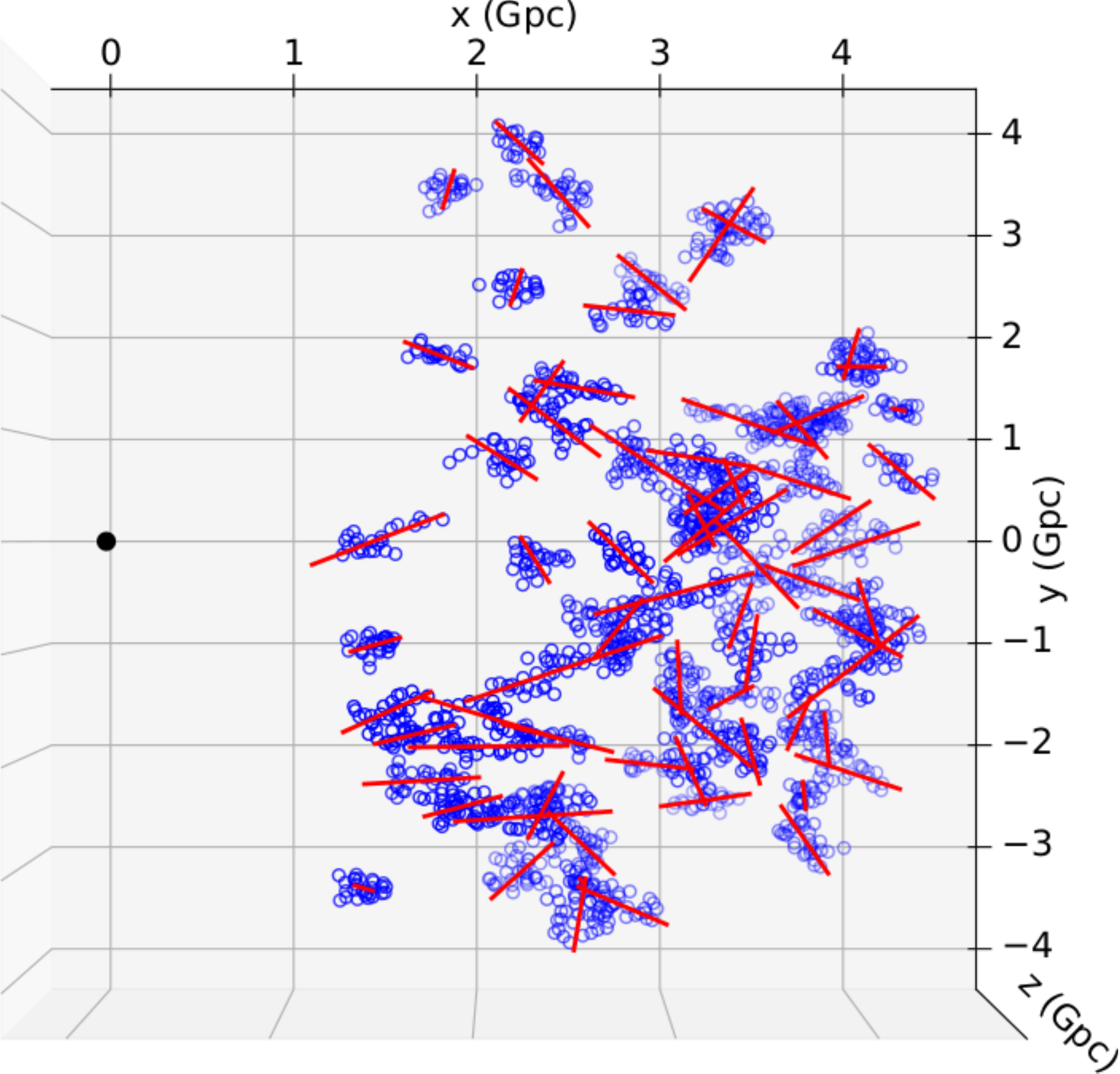}
    \caption[]{(c) LQGs in three-dimensional comoving space, viewed with the $z$-axis orthogonal to the page, showing LQG quasars (blue circles) and 3D LQG major axes (red lines). Quasar markers are shaded to give the appearance of depth, with lighter shades representing more distant quasars. Our location at ($x, y, z$) coordinates (0, 0, 0) is indicated by a black dot.}
    \label{figLQGs3DAspectsPt3}
\end{figure}

%% file: appendixVoronoi.tex
\chapter{A Voronoi tessellated Universe} \label{appVoronoi}

\begin{figure}
    \centering
    \includegraphics[width=0.35\columnwidth]{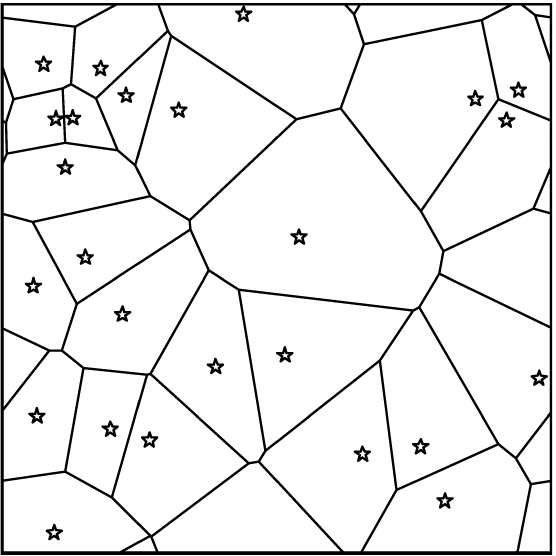}
    \caption[Two-dimensional Voronoi tessellation]{Two-dimensional Voronoi tessellation of 25 cell nuclei (stars) assuming periodic boundary conditions. Image credit: \protect\citet{vandeWeygaert2001}.}
    \label{figVoronoi}
\end{figure}

The cosmic web is the result of anisotropic gravitational collapse \citep[e.g.][]{ZelDovich1970,Shen2006}. \citet{Cautun2014} show that matter tends to flow first from voids into walls, and then via filaments into clusters. Walls are formed when matter flowing from a void encounters material flowing from an adjacent void, and then collects on planes between the voids. These planes partition space like non-regular tessellated Voronoi cells \citep{Voronoi1908}, where each cell is a polyhedron enclosing the region of space closer to its cell nucleus (void expansion centre) than any other cell nuclei. Fig.~\ref{figVoronoi} illustrates a two-dimensional example. \citet{vandeWeygaert1994} argues that because under-densities (voids) dominate the cosmic web, Voronoi tessellation is a reasonable approximation of large-scale matter distribution.

\begin{figure}
    \centering
    \includegraphics[width=0.6\columnwidth]{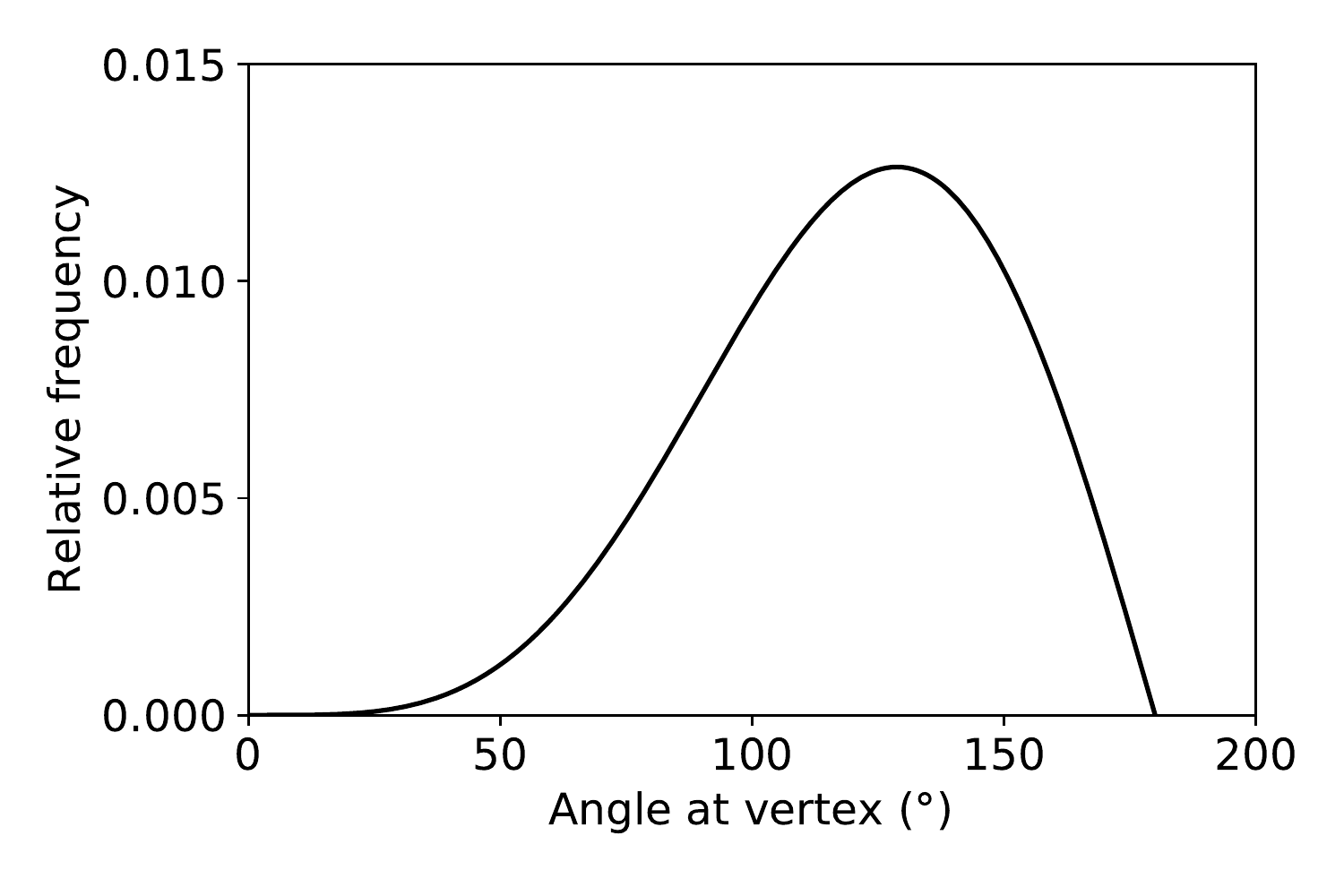}
    \caption[Angles at vertices in a two-dimensional Voronoi tessellation]{The distribution of angles at vertices in a two-dimensional Voronoi tessellation with Poissonian distribution of cell nuclei. Image credit: \protect\citet{Icke1987}}
    \label{figVoronoiAngles}
\end{figure}

In statistical analysis of two-dimensional Voronoi tessellations \citet{Icke1987} find analytically that the most likely angle between cell walls at vertices is $\sim129^{\circ}$ (Fig.~\ref{figVoronoiAngles}). \citet{vandeWeygaert1994} extends this to three dimensions using Monte Carlo simulations, and finds the mean angles between pairs of cell edges and pairs of cell walls are $\sim111^{\circ}$ and $\sim120^{\circ}$ respectively. These angles are consistent regardless of whether the distribution of cell nuclei is Poissonian, clustered (correlated), or somewhat regular (which they call `anti-correlated').

\citet{vandeWeygaert2007} uses Voronoi tessellation to describe the spatial distribution of matter in the Universe on $\gtrsim 100$ Mpc scales: (1) the interior of the cells correspond to voids, (2) cell walls represent sheets of galaxies, (3) wall edges (where two or more walls meet) correspond to filaments, and (4) vertices represent the most dense nodes in the cosmic web, namely galaxy clusters and superclusters. They demonstrate that the observed clustering of galaxy clusters can be explained by a cellular structure with a basic scale $\sim 70\:h^{-1}$ Mpc, although the 5\% richest nodes are coherent on scales at least $2-3$ times larger.

Voronoi tessellations are non-regular, yet exhibit well defined preferred angles between adjoining walls ($\sim129^{\circ}$ and $\sim120^{\circ}$ in two and three-dimensional analysis). These angles are consistent, regardless of the scale, orientation, or spatial position of individual Voronoi cells. If Voronoi tessellation is a legitimate approximation of the cosmic web, these preferred angles will be reflected in its structure. Further, \citet{Neyrinck2018} note that if the three-dimensional cosmic web is a Voronoi tessellation, so will be a two-dimensional slice through it. Therefore, depending on survey depth and LSS scale, it is plausible that either angle may be manifest in the relative orientations of structures residing within Voronoi cell walls.

%% file: thesis.bbl
\providecommand{\noopsort}[1]{}
\begin{thebibliography}{266}
\expandafter\ifx\csname natexlab\endcsname\relax\def\natexlab#1{#1}\fi

\bibitem[{{Adam} {et~al.}(2017){Adam}, {Bartalucci}, {Pratt}, {Ade},
  {Andr{\'e}}, {Arnaud}, {Beelen}, {Beno{\^i}t}, {Bideaud}, {Billot},
  {Bourdin}, {Bourrion}, {Calvo}, {Catalano}, {Coiffard}, {Comis}, {D'Addabbo},
  {De Petris}, {D{\'e}mocl{\`e}s}, {D{\'e}sert}, {Doyle}, {Egami}, {Ferrari},
  {Goupy}, {Kramer}, {Lagache}, {Leclercq}, {Mac{\'{\i}}as-P{\'e}rez},
  {Maurogordato}, {Mauskopf}, {Mayet}, {Monfardini}, {Mroczkowski}, {Pajot},
  {Pascale}, {Perotto}, {Pisano}, {Pointecouteau}, {Ponthieu}, {Rev{\'e}ret},
  {Ritacco}, {Rodriguez}, {Romero}, {Ruppin}, {Schuster}, {Sievers},
  {Triqueneaux}, {Tucker}, {Zemcov}, \& {Zylka}}]{Adam2017}
{Adam}, R., {Bartalucci}, I., {Pratt}, G.~W., {et~al.} 2017, \aap, 598, A115

\bibitem[{Akaike(1974)}]{Akaike1974}
Akaike, H. 1974, IEEE Transactions on Automatic Control, 19, 716

\bibitem[{{Alpher} {et~al.}(1948){Alpher}, {Bethe}, \& {Gamow}}]{Alpher1948}
{Alpher}, R.~A., {Bethe}, H., \& {Gamow}, G. 1948, Physical Review, 73, 803

\bibitem[{{Angulo} {et~al.}(2012){Angulo}, {Springel}, {White}, {Cole},
  {Jenkins}, {Baugh}, \& {Frenk}}]{Angulo2012}
{Angulo}, R.~E., {Springel}, V., {White}, S.~D.~M., {et~al.} 2012, \mnras, 425,
  2722

\bibitem[{{Ansarinejad} \& {Shanks}(2018)}]{Ansarinejad2018}
{Ansarinejad}, B. \& {Shanks}, T. 2018, \mnras, 479, 4091

\bibitem[{{Astier} {et~al.}(2006){Astier}, {Guy}, {Regnault}, {Pain},
  {Aubourg}, {Balam}, {Basa}, {Carlberg}, {Fabbro}, {Fouchez}, {Hook},
  {Howell}, {Lafoux}, {Neill}, {Palanque-Delabrouille}, {Perrett}, {Pritchet},
  {Rich}, {Sullivan}, {Taillet}, {Aldering}, {Antilogus}, {Arsenijevic},
  {Balland}, {Baumont}, {Bronder}, {Courtois}, {Ellis}, {Filiol}, {Gon{\c
  c}alves}, {Goobar}, {Guide}, {Hardin}, {Lusset}, {Lidman}, {McMahon},
  {Mouchet}, {Mourao}, {Perlmutter}, {Ripoche}, {Tao}, \&
  {Walton}}]{Astier2006}
{Astier}, P., {Guy}, J., {Regnault}, N., {et~al.} 2006, \aap, 447, 31

\bibitem[{{Astropy Collaboration} {et~al.}(2018){Astropy Collaboration},
  {Price-Whelan}, {Sip{\H{o}}cz}, {G{\"u}nther}, {Lim}, {Crawford}, {Conseil},
  {Shupe}, {Craig}, {Dencheva}, {Ginsburg}, {Vand erPlas}, {Bradley},
  {P{\'e}rez-Su{\'a}rez}, {de Val-Borro}, {Aldcroft}, {Cruz}, {Robitaille},
  {Tollerud}, {Ardelean}, {Babej}, {Bach}, {Bachetti}, {Bakanov}, {Bamford},
  {Barentsen}, {Barmby}, {Baumbach}, {Berry}, {Biscani}, {Boquien}, {Bostroem},
  {Bouma}, {Brammer}, {Bray}, {Breytenbach}, {Buddelmeijer}, {Burke},
  {Calderone}, {Cano Rodr{\'\i}guez}, {Cara}, {Cardoso}, {Cheedella}, {Copin},
  {Corrales}, {Crichton}, {D'Avella}, {Deil}, {Depagne}, {Dietrich}, {Donath},
  {Droettboom}, {Earl}, {Erben}, {Fabbro}, {Ferreira}, {Finethy}, {Fox},
  {Garrison}, {Gibbons}, {Goldstein}, {Gommers}, {Greco}, {Greenfield},
  {Groener}, {Grollier}, {Hagen}, {Hirst}, {Homeier}, {Horton}, {Hosseinzadeh},
  {Hu}, {Hunkeler}, {Ivezi{\'c}}, {Jain}, {Jenness}, {Kanarek}, {Kendrew},
  {Kern}, {Kerzendorf}, {Khvalko}, {King}, {Kirkby}, {Kulkarni}, {Kumar},
  {Lee}, {Lenz}, {Littlefair}, {Ma}, {Macleod}, {Mastropietro}, {McCully},
  {Montagnac}, {Morris}, {Mueller}, {Mumford}, {Muna}, {Murphy}, {Nelson},
  {Nguyen}, {Ninan}, {N{\"o}the}, {Ogaz}, {Oh}, {Parejko}, {Parley}, {Pascual},
  {Patil}, {Patil}, {Plunkett}, {Prochaska}, {Rastogi}, {Reddy Janga},
  {Sabater}, {Sakurikar}, {Seifert}, {Sherbert}, {Sherwood-Taylor}, {Shih},
  {Sick}, {Silbiger}, {Singanamalla}, {Singer}, {Sladen}, {Sooley},
  {Sornarajah}, {Streicher}, {Teuben}, {Thomas}, {Tremblay}, {Turner},
  {Terr{\'o}n}, {van Kerkwijk}, {de la Vega}, {Watkins}, {Weaver}, {Whitmore},
  {Woillez}, {Zabalza}, \& {Astropy Contributors}}]{Astropy2018}
{Astropy Collaboration}, {Price-Whelan}, A.~M., {Sip{\H{o}}cz}, B.~M., {et~al.}
  2018, \aj, 156, 123

\bibitem[{{Astropy Collaboration} {et~al.}(2013){Astropy Collaboration},
  {Robitaille}, {Tollerud}, {Greenfield}, {Droettboom}, {Bray}, {Aldcroft},
  {Davis}, {Ginsburg}, {Price-Whelan}, {Kerzendorf}, {Conley}, {Crighton},
  {Barbary}, {Muna}, {Ferguson}, {Grollier}, {Parikh}, {Nair}, {Unther},
  {Deil}, {Woillez}, {Conseil}, {Kramer}, {Turner}, {Singer}, {Fox}, {Weaver},
  {Zabalza}, {Edwards}, {Azalee Bostroem}, {Burke}, {Casey}, {Crawford},
  {Dencheva}, {Ely}, {Jenness}, {Labrie}, {Lim}, {Pierfederici}, {Pontzen},
  {Ptak}, {Refsdal}, {Servillat}, \& {Streicher}}]{Astropy2013}
{Astropy Collaboration}, {Robitaille}, T.~P., {Tollerud}, E.~J., {et~al.} 2013,
  \aap, 558, A33

\bibitem[{{Barrow} {et~al.}(1985){Barrow}, {Bhavsar}, \& {Sonoda}}]{Barrow1985}
{Barrow}, J.~D., {Bhavsar}, S.~P., \& {Sonoda}, D.~H. 1985, \mnras, 216, 17

\bibitem[{{Bartunov} {et~al.}(2007){Bartunov}, {Tsvetkov}, \&
  {Pavlyuk}}]{Bartunov2007}
{Bartunov}, O.~S., {Tsvetkov}, D.~Y., \& {Pavlyuk}, N.~N. 2007, Highlights
  Astron., 14, 316

\bibitem[{{Bate} {et~al.}(2020){Bate}, {Chisari}, {Codis}, {Martin}, {Dubois},
  {Devriendt}, {Pichon}, \& {Slyz}}]{Bate2020}
{Bate}, J., {Chisari}, N.~E., {Codis}, S., {et~al.} 2020, \mnras, 491, 4057

\bibitem[{{Battye} {et~al.}(2008){Battye}, {Browne}, \& {Jackson}}]{Battye2008}
{Battye}, R.~A., {Browne}, I.~W.~A., \& {Jackson}, N. 2008, \mnras, 385, 274

\bibitem[{{Baugh} {et~al.}(2019){Baugh}, {Gonzalez-Perez}, {Lagos}, {Lacey},
  {Helly}, {Jenkins}, {Frenk}, {Benson}, {Bower}, \& {Cole}}]{Baugh2019}
{Baugh}, C.~M., {Gonzalez-Perez}, V., {Lagos}, C. d.~P., {et~al.} 2019, \mnras,
  483, 4922

\bibitem[{{Bender} {et~al.}(2016){Bender}, {Kennedy}, {Ade}, {Basu},
  {Bertoldi}, {Burkutean}, {Clarke}, {Dahlin}, {Dobbs}, {Ferrusca}, {Flanigan},
  {Halverson}, {Holzapfel}, {Horellou}, {Johnson}, {Kermish}, {Klein},
  {Kneissl}, {Lanting}, {Lee}, {Mehl}, {Menten}, {Muders}, {Nagarajan},
  {Pacaud}, {Reichardt}, {Richards}, {Schaaf}, {Schwan}, {Sommer}, {Spieler},
  {Tucker}, \& {Westbrook}}]{Bender2016}
{Bender}, A.~N., {Kennedy}, J., {Ade}, P.~A.~R., {et~al.} 2016, \mnras, 460,
  3432

\bibitem[{{Bennett} {et~al.}(2013){Bennett}, {Larson}, {Weiland}, {Jarosik},
  {Hinshaw}, {Odegard}, {Smith}, {Hill}, {Gold}, {Halpern}, {Komatsu}, {Nolta},
  {Page}, {Spergel}, {Wollack}, {Dunkley}, {Kogut}, {Limon}, {Meyer}, {Tucker},
  \& {Wright}}]{Bennett2013}
{Bennett}, C.~L., {Larson}, D., {Weiland}, J.~L., {et~al.} 2013, \apjs, 208, 20

\bibitem[{{Beno{\^i}t} {et~al.}(2003){Beno{\^i}t}, {Ade}, {Amblard}, {Ansari},
  {Aubourg}, {Bargot}, {Bartlett}, {Bernard}, {Bhatia}, {Blanchard}, {Bock},
  {Boscaleri}, {Bouchet}, {Bourrachot}, {Camus}, {Couchot}, {de Bernardis},
  {Delabrouille}, {D{\'e}sert}, {Dor{\'e}}, {Douspis}, {Dumoulin}, {Dupac},
  {Filliatre}, {Fosalba}, {Ganga}, {Gannaway}, {Gautier}, {Giard},
  {Giraud-H{\'e}raud}, {Gispert}, {Guglielmi}, {Hamilton}, {Hanany},
  {Henrot-Versill{\'e}}, {Kaplan}, {Lagache}, {Lamarre}, {Lange},
  {Mac{\'{\i}}as-P{\'e}rez}, {Madet}, {Maffei}, {Magneville}, {Marrone},
  {Masi}, {Mayet}, {Murphy}, {Naraghi}, {Nati}, {Patanchon}, {Perrin}, {Piat},
  {Ponthieu}, {Prunet}, {Puget}, {Renault}, {Rosset}, {Santos}, {Starobinsky},
  {Strukov}, {Sudiwala}, {Teyssier}, {Tristram}, {Tucker}, {Vanel}, {Vibert},
  {Wakui}, \& {Yvon}}]{Benoit2003}
{Beno{\^i}t}, A., {Ade}, P., {Amblard}, A., {et~al.} 2003, \aap, 399, L19

\bibitem[{{\noopsort{Bernardis}}{de Bernardis}
  {et~al.}(2000){\noopsort{Bernardis}}{de Bernardis}, {Ade}, {Bock}, {Bond},
  {Borrill}, {Boscaleri}, {Coble}, {Crill}, {De Gasperis}, {Farese},
  {Ferreira}, {Ganga}, {Giacometti}, {Hivon}, {Hristov}, {Iacoangeli}, {Jaffe},
  {Lange}, {Martinis}, {Masi}, {Mason}, {Mauskopf}, {Melchiorri}, {Miglio},
  {Montroy}, {Netterfield}, {Pascale}, {Piacentini}, {Pogosyan}, {Prunet},
  {Rao}, {Romeo}, {Ruhl}, {Scaramuzzi}, {Sforna}, \&
  {Vittorio}}]{deBernardis2000}
{\noopsort{Bernardis}}{de Bernardis}, P., {Ade}, P.~A.~R., {Bock}, J.~J.,
  {et~al.} 2000, \nat, 404, 955

\bibitem[{{Beutler} {et~al.}(2011){Beutler}, {Blake}, {Colless}, {Jones},
  {Staveley-Smith}, {Campbell}, {Parker}, {Saunders}, \&
  {Watson}}]{Beutler2011}
{Beutler}, F., {Blake}, C., {Colless}, M., {et~al.} 2011, \mnras, 416, 3017

\bibitem[{{Bharadwaj} {et~al.}(2004){Bharadwaj}, {Bhavsar}, \&
  {Sheth}}]{Bharadwaj2004}
{Bharadwaj}, S., {Bhavsar}, S.~P., \& {Sheth}, J.~V. 2004, \apj, 606, 25

\bibitem[{{Bietenholz}(1986)}]{Bietenholz1986}
{Bietenholz}, M.~F. 1986, \aj, 91, 1249

\bibitem[{{Binggeli}(1982)}]{Binggeli1982}
{Binggeli}, B. 1982, \aap, 107, 338

\bibitem[{{Birkinshaw}(1999)}]{Birkinshaw1999}
{Birkinshaw}, M. 1999, \physrep, 310, 97

\bibitem[{{Bleem} {et~al.}(2015){Bleem}, {Stalder}, {de Haan}, {Aird}, {Allen},
  {Applegate}, {Ashby}, {Bautz}, {Bayliss}, {Benson}, {Bocquet}, {Brodwin},
  {Carlstrom}, {Chang}, {Chiu}, {Cho}, {Clocchiatti}, {Crawford}, {Crites},
  {Desai}, {Dietrich}, {Dobbs}, {Foley}, {Forman}, {George}, {Gladders},
  {Gonzalez}, {Halverson}, {Hennig}, {Hoekstra}, {Holder}, {Holzapfel},
  {Hrubes}, {Jones}, {Keisler}, {Knox}, {Lee}, {Leitch}, {Liu}, {Lueker},
  {Luong-Van}, {Mantz}, {Marrone}, {McDonald}, {McMahon}, {Meyer}, {Mocanu},
  {Mohr}, {Murray}, {Padin}, {Pryke}, {Reichardt}, {Rest}, {Ruel}, {Ruhl},
  {Saliwanchik}, {Saro}, {Sayre}, {Schaffer}, {Schrabback}, {Shirokoff},
  {Song}, {Spieler}, {Stanford}, {Staniszewski}, {Stark}, {Story}, {Stubbs},
  {Vanderlinde}, {Vieira}, {Vikhlinin}, {Williamson}, {Zahn}, \&
  {Zenteno}}]{Bleem2015}
{Bleem}, L.~E., {Stalder}, B., {de Haan}, T., {et~al.} 2015, \apjs, 216, 27

\bibitem[{{Blinov} {et~al.}(2020){Blinov}, {Casadio}, {Mandarakas}, \&
  {Angelakis}}]{Blinov2020}
{Blinov}, D., {Casadio}, C., {Mandarakas}, N., \& {Angelakis}, E. 2020, arXiv
  e-prints, arXiv:2002.00982

\bibitem[{{Bond} {et~al.}(1996){Bond}, {Kofman}, \& {Pogosyan}}]{Bond1996}
{Bond}, J.~R., {Kofman}, L., \& {Pogosyan}, D. 1996, \nat, 380, 603

\bibitem[{{Bond} {et~al.}(2010){Bond}, {Strauss}, \& {Cen}}]{Bond2010}
{Bond}, N.~A., {Strauss}, M.~A., \& {Cen}, R. 2010, \mnras, 406, 1609

\bibitem[{{Boughn} \& {Crittenden}(2004{\natexlab{a}})}]{Boughn2004a}
{Boughn}, S. \& {Crittenden}, R. 2004{\natexlab{a}}, \nat, 427, 45

\bibitem[{{Boughn} \& {Crittenden}(2004{\natexlab{b}})}]{Boughn2004b}
{Boughn}, S.~P. \& {Crittenden}, R.~G. 2004{\natexlab{b}}, \apj, 612, 647

\bibitem[{{Bourdin} {et~al.}(2015){Bourdin}, {Mazzotta}, \&
  {Rasia}}]{Bourdin2015}
{Bourdin}, H., {Mazzotta}, P., \& {Rasia}, E. 2015, \apj, 815, 92

\bibitem[{{Bremer} {et~al.}(2010){Bremer}, {Silk}, {Davies}, \&
  {Lehnert}}]{Bremer2010}
{Bremer}, M.~N., {Silk}, J., {Davies}, L.~J.~M., \& {Lehnert}, M.~D. 2010,
  \mnras, 404, L69

\bibitem[{{Bueno Sanchez} {et~al.}(2009){Bueno Sanchez}, {Nesseris}, \&
  {Perivolaropoulos}}]{BuenoSanchez2009}
{Bueno Sanchez}, J.~C., {Nesseris}, S., \& {Perivolaropoulos}, L. 2009, \jcap,
  11, 029

\bibitem[{{Cabanac} {et~al.}(2005){Cabanac}, {Hutsem{\'e}kers}, {Sluse}, \&
  {Lamy}}]{Cabanac2005}
{Cabanac}, R.~A., {Hutsem{\'e}kers}, D., {Sluse}, D., \& {Lamy}, H. 2005, in
  Astronomical Polarimetry: Current Status and Future Directions, Vol. 343, 498

\bibitem[{{Cai} {et~al.}(2014){Cai}, {Neyrinck}, {Szapudi}, {Cole}, \&
  {Frenk}}]{Cai2014}
{Cai}, Y.-C., {Neyrinck}, M.~C., {Szapudi}, I., {Cole}, S., \& {Frenk}, C.~S.
  2014, \apj, 786, 110

\bibitem[{{Cardoso} {et~al.}(2008){Cardoso}, {Le Jeune}, {Delabrouille},
  {Betoule}, \& {Patanchon}}]{Cardoso2008}
{Cardoso}, J.-F., {Le Jeune}, M., {Delabrouille}, J., {Betoule}, M., \&
  {Patanchon}, G. 2008, IEEE Journal of Selected Topics in Signal Processing,
  2, 735

\bibitem[{{Carlstrom} {et~al.}(2002){Carlstrom}, {Holder}, \&
  {Reese}}]{Carlstrom2002}
{Carlstrom}, J.~E., {Holder}, G.~P., \& {Reese}, E.~D. 2002, \araa, 40, 643

\bibitem[{{Carroll} \& {Ostlie}(2006)}]{Carroll2006}
{Carroll}, B.~W. \& {Ostlie}, D.~A. 2006, {An introduction to modern
  astrophysics and cosmology} (Pearson)

\bibitem[{{Carter} \& {Metcalfe}(1980)}]{Carter1980}
{Carter}, D. \& {Metcalfe}, N. 1980, \mnras, 191, 325

\bibitem[{{Cautun} {et~al.}(2014){Cautun}, {van de Weygaert}, {Jones}, \&
  {Frenk}}]{Cautun2014}
{Cautun}, M., {van de Weygaert}, R., {Jones}, B. J.~T., \& {Frenk}, C.~S. 2014,
  \mnras, 441, 2923

\bibitem[{{Challinor} \& {Chon}(2002)}]{Challinor2002}
{Challinor}, A. \& {Chon}, G. 2002, \prd, 66, 127301

\bibitem[{{Choi} {et~al.}(2010){Choi}, {Bond}, {Strauss}, {Coil}, {Davis}, \&
  {Willmer}}]{Choi2010}
{Choi}, E., {Bond}, N.~A., {Strauss}, M.~A., {et~al.} 2010, \mnras, 406, 320

\bibitem[{{Choudhury} \& {Padmanabhan}(2005)}]{Choudhury2005}
{Choudhury}, T.~R. \& {Padmanabhan}, T. 2005, \aap, 429, 807

\bibitem[{{Clampitt} {et~al.}(2016){Clampitt}, {Miyatake}, {Jain}, \&
  {Takada}}]{Clampitt2016}
{Clampitt}, J., {Miyatake}, H., {Jain}, B., \& {Takada}, M. 2016, \mnras, 457,
  2391

\bibitem[{{Clowes} \& {Campusano}(1991)}]{Clowes1991}
{Clowes}, R.~G. \& {Campusano}, L.~E. 1991, \mnras, 249, 218

\bibitem[{{Clowes} {et~al.}(2012){Clowes}, {Campusano}, {Graham}, \&
  {S{\"o}chting}}]{Clowes2012}
{Clowes}, R.~G., {Campusano}, L.~E., {Graham}, M.~J., \& {S{\"o}chting}, I.~K.
  2012, \mnras, 419, 556

\bibitem[{{Clowes} {et~al.}(2013){Clowes}, {Harris}, {Raghunathan},
  {Campusano}, {S{\"o}chting}, \& {Graham}}]{Clowes2013}
{Clowes}, R.~G., {Harris}, K.~A., {Raghunathan}, S., {et~al.} 2013, \mnras,
  429, 2910

\bibitem[{{Codis} {et~al.}(2018){Codis}, {Jindal}, {Chisari}, {Vibert},
  {Dubois}, {Pichon}, \& {Devriendt}}]{Codis2018}
{Codis}, S., {Jindal}, A., {Chisari}, N.~E., {et~al.} 2018, \mnras, 481, 4753

\bibitem[{{Colless} {et~al.}(2001){Colless}, {Dalton}, {Maddox}, {Sutherland },
  {Norberg}, {Cole}, {Bland -Hawthorn}, {Bridges}, {Cannon}, {Collins},
  {Couch}, {Cross}, {Deeley}, {De Propris}, {Driver}, {Efstathiou}, {Ellis},
  {Frenk}, {Glazebrook}, {Jackson}, {Lahav}, {Lewis}, {Lumsden}, {Madgwick},
  {Peacock}, {Peterson}, {Price}, {Seaborne}, \& {Taylor}}]{Colless2001}
{Colless}, M., {Dalton}, G., {Maddox}, S., {et~al.} 2001, \mnras, 328, 1039

\bibitem[{{Contigiani} {et~al.}(2017){Contigiani}, {de Gasperin}, {Miley},
  {Rudnick}, {Andernach}, {Banfield}, {Kapi{\'n}ska}, {Shabala}, \&
  {Wong}}]{Contigiani2017}
{Contigiani}, O., {de Gasperin}, F., {Miley}, G.~K., {et~al.} 2017, \mnras,
  472, 636

\bibitem[{{Crittenden} \& {Turok}(1996)}]{Crittenden1996}
{Crittenden}, R.~G. \& {Turok}, N. 1996, Physical Review Letters, 76, 575

\bibitem[{{Croton}(2013)}]{Croton2013}
{Croton}, D.~J. 2013, \pasa, 30, e052

\bibitem[{{Das} {et~al.}(2014){Das}, {Louis}, {Nolta}, {Addison},
  {Battistelli}, {Bond}, {Calabrese}, {Crichton}, {Devlin}, {Dicker},
  {Dunkley}, {D{\"u}nner}, {Fowler}, {Gralla}, {Hajian}, {Halpern},
  {Hasselfield}, {Hilton}, {Hincks}, {Hlozek}, {Huffenberger}, {Hughes},
  {Irwin}, {Kosowsky}, {Lupton}, {Marriage}, {Marsden}, {Menanteau}, {Moodley},
  {Niemack}, {Page}, {Partridge}, {Reese}, {Schmitt}, {Sehgal}, {Sherwin},
  {Sievers}, {Spergel}, {Staggs}, {Swetz}, {Switzer}, {Thornton}, {Trac}, \&
  {Wollack}}]{Das2014}
{Das}, S., {Louis}, T., {Nolta}, M.~R., {et~al.} 2014, \jcap, 4, 014

\bibitem[{{Dav{\'e}} {et~al.}(2019){Dav{\'e}}, {Angl{\'e}s-Alc{\'a}zar},
  {Narayanan}, {Li}, {Rafieferantsoa}, \& {Appleby}}]{Dave2019}
{Dav{\'e}}, R., {Angl{\'e}s-Alc{\'a}zar}, D., {Narayanan}, D., {et~al.} 2019,
  \mnras, 486, 2827

\bibitem[{{Dawson} {et~al.}(2009){Dawson}, {Aldering}, {Amanullah}, {Barbary},
  {Barrientos}, {Brodwin}, {Connolly}, {Dey}, {Doi}, {Donahue}, {Eisenhardt},
  {Ellingson}, {Faccioli}, {Fadeyev}, {Fakhouri}, {Fruchter}, {Gilbank},
  {Gladders}, {Goldhaber}, {Gonzalez}, {Goobar}, {Gude}, {Hattori}, {Hoekstra},
  {Huang}, {Ihara}, {Jannuzi}, {Johnston}, {Kashikawa}, {Koester}, {Konishi},
  {Kowalski}, {Lidman}, {Linder}, {Lubin}, {Meyers}, {Morokuma}, {Munshi},
  {Mullis}, {Oda}, {Panagia}, {Perlmutter}, {Postman}, {Pritchard}, {Rhodes},
  {Rosati}, {Rubin}, {Schlegel}, {Spadafora}, {Stanford}, {Stanishev}, {Stern},
  {Strovink}, {Suzuki}, {Takanashi}, {Tokita}, {Wagner}, {Wang}, {Yasuda},
  {Yee}, \& {Supernova Cosmology Project}}]{Dawson2009}
{Dawson}, K.~S., {Aldering}, G., {Amanullah}, R., {et~al.} 2009, \aj, 138, 1271

\bibitem[{{Delabrouille} {et~al.}(2009){Delabrouille}, {Cardoso}, {Le Jeune},
  {Betoule}, {Fay}, \& {Guilloux}}]{Delabrouille2009}
{Delabrouille}, J., {Cardoso}, J.-F., {Le Jeune}, M., {et~al.} 2009, \aap, 493,
  835

\bibitem[{{Delubac} {et~al.}(2015){Delubac}, {Bautista}, {Busca}, {Rich},
  {Kirkby}, {Bailey}, {Font-Ribera}, {Slosar}, {Lee}, {Pieri}, {Hamilton},
  {Aubourg}, {Blomqvist}, {Bovy}, {Brinkmann}, {Carithers}, {Dawson},
  {Eisenstein}, {Gontcho}, {Kneib}, {Le Goff}, {Margala}, {Miralda-Escud{\'e}},
  {Myers}, {Nichol}, {Noterdaeme}, {O'Connell}, {Olmstead},
  {Palanque-Delabrouille}, {P{\^a}ris}, {Petitjean}, {Ross}, {Rossi},
  {Schlegel}, {Schneider}, {Weinberg}, {Y{\`e}che}, \& {York}}]{Delubac2015}
{Delubac}, T., {Bautista}, J.~E., {Busca}, N.~G., {et~al.} 2015, \aap, 574, A59

\bibitem[{{DES Collaboration} {et~al.}(2016){DES Collaboration}, {Abbott},
  {Abdalla}, {Aleksi{\'c}}, {Allam}, {Amara}, {Bacon}, {Balbinot}, {Banerji},
  {Bechtol}, {Benoit-L{\'e}vy}, {Bernstein}, {Bertin}, {Blazek}, {Bonnett},
  {Bridle}, {Brooks}, {Brunner}, {Buckley-Geer}, {Burke}, {Caminha}, {Capozzi},
  {Carlsen}, {Carnero-Rosell}, {Carollo}, {Carrasco-Kind}, {Carretero},
  {Castander}, {Clerkin}, {Collett}, {Conselice}, {Crocce}, {Cunha},
  {D'Andrea}, {da Costa}, {Davis}, {Desai}, {Diehl}, {Dietrich}, {Dodelson},
  {Doel}, {Drlica-Wagner}, {Estrada}, {Etherington}, {Evrard}, {Fabbri},
  {Finley}, {Flaugher}, {Foley}, {Fosalba}, {Frieman}, {Garc{\'{\i}}a-Bellido},
  {Gaztanaga}, {Gerdes}, {Giannantonio}, {Goldstein}, {Gruen}, {Gruendl},
  {Guarnieri}, {Gutierrez}, {Hartley}, {Honscheid}, {Jain}, {James}, {Jeltema},
  {Jouvel}, {Kessler}, {King}, {Kirk}, {Kron}, {Kuehn}, {Kuropatkin}, {Lahav},
  {Li}, {Lima}, {Lin}, {Maia}, {Makler}, {Manera}, {Maraston}, {Marshall},
  {Martini}, {McMahon}, {Melchior}, {Merson}, {Miller}, {Miquel}, {Mohr},
  {Morice-Atkinson}, {Naidoo}, {Neilsen}, {Nichol}, {Nord}, {Ogando},
  {Ostrovski}, {Palmese}, {Papadopoulos}, {Peiris}, {Peoples}, {Percival},
  {Plazas}, {Reed}, {Refregier}, {Romer}, {Roodman}, {Ross}, {Rozo}, {Rykoff},
  {Sadeh}, {Sako}, {S{\'a}nchez}, {Sanchez}, {Santiago}, {Scarpine},
  {Schubnell}, {Sevilla-Noarbe}, {Sheldon}, {Smith}, {Smith}, {Soares-Santos},
  {Sobreira}, {Soumagnac}, {Suchyta}, {Sullivan}, {Swanson}, {Tarle}, {Thaler},
  {Thomas}, {Thomas}, {Tucker}, {Vieira}, {Vikram}, {Walker}, {Wechsler},
  {Weller}, {Wester}, {Whiteway}, {Wilcox}, {Yanny}, {Zhang}, \&
  {Zuntz}}]{DES2016}
{DES Collaboration}, {Abbott}, T., {Abdalla}, F.~B., {et~al.} 2016, \mnras,
  460, 1270

\bibitem[{{DES Collaboration} {et~al.}(2018){DES Collaboration}, {Abbott},
  {Abdalla}, {Alarcon}, {Aleksi{\'c}}, {Allam}, {Allen}, {Amara}, {Annis},
  {Asorey}, {Avila}, {Bacon}, {Balbinot}, {Banerji}, {Banik}, {Barkhouse},
  {Baumer}, {Baxter}, {Bechtol}, {Becker}, {Benoit-L{\'e}vy}, {Benson},
  {Bernstein}, {Bertin}, {Blazek}, {Bridle}, {Brooks}, {Brout}, {Buckley-Geer},
  {Burke}, {Busha}, {Campos}, {Capozzi}, {Carnero Rosell}, {Carrasco Kind},
  {Carretero}, {Castander}, {Cawthon}, {Chang}, {Chen}, {Childress}, {Choi},
  {Conselice}, {Crittenden}, {Crocce}, {Cunha}, {D'Andrea}, {da Costa}, {Das},
  {Davis}, {Davis}, {De Vicente}, {DePoy}, {DeRose}, {Desai}, {Diehl},
  {Dietrich}, {Dodelson}, {Doel}, {Drlica-Wagner}, {Eifler}, {Elliott},
  {Elsner}, {Elvin-Poole}, {Estrada}, {Evrard}, {Fang}, {Fernandez},
  {Fert{\'e}}, {Finley}, {Flaugher}, {Fosalba}, {Friedrich}, {Frieman},
  {Garc{\'\i}a-Bellido}, {Garcia-Fernandez}, {Gatti}, {Gaztanaga}, {Gerdes},
  {Giannantonio}, {Gill}, {Glazebrook}, {Goldstein}, {Gruen}, {Gruendl},
  {Gschwend}, {Gutierrez}, {Hamilton}, {Hartley}, {Hinton}, {Honscheid},
  {Hoyle}, {Huterer}, {Jain}, {James}, {Jarvis}, {Jeltema}, {Johnson},
  {Johnson}, {Kacprzak}, {Kent}, {Kim}, {King}, {Kirk}, {Kokron}, {Kovacs},
  {Krause}, {Krawiec}, {Kremin}, {Kuehn}, {Kuhlmann}, {Kuropatkin}, {Lacasa},
  {Lahav}, {Li}, {Liddle}, {Lidman}, {Lima}, {Lin}, {MacCrann}, {Maia},
  {Makler}, {Manera}, {March}, {Marshall}, {Martini}, {McMahon}, {Melchior},
  {Menanteau}, {Miquel}, {Miranda}, {Mudd}, {Muir}, {M{\"o}ller}, {Neilsen},
  {Nichol}, {Nord}, {Nugent}, {Ogando}, {Palmese}, {Peacock}, {Peiris},
  {Peoples}, {Percival}, {Petravick}, {Plazas}, {Porredon}, {Prat}, {Pujol},
  {Rau}, {Refregier}, {Ricker}, {Roe}, {Rollins}, {Romer}, {Roodman},
  {Rosenfeld}, {Ross}, {Rozo}, {Rykoff}, {Sako}, {Salvador}, {Samuroff},
  {S{\'a}nchez}, {Sanchez}, {Santiago}, {Scarpine}, {Schindler}, {Scolnic},
  {Secco}, {Serrano}, {Sevilla-Noarbe}, {Sheldon}, {Smith}, {Smith}, {Smith},
  {Soares-Santos}, {Sobreira}, {Suchyta}, {Tarle}, {Thomas}, {Troxel},
  {Tucker}, {Tucker}, {Uddin}, {Varga}, {Vielzeuf}, {Vikram}, {Vivas},
  {Walker}, {Wang}, {Wechsler}, {Weller}, {Wester}, {Wolf}, {Yanny}, {Yuan},
  {Zenteno}, {Zhang}, {Zhang}, {Zuntz}, \& {Dark Energy Survey
  Collaboration}}]{DES2017}
{DES Collaboration}, {Abbott}, T.~M.~C., {Abdalla}, F.~B., {et~al.} 2018, \prd,
  98, 043526

\bibitem[{{DESI Collaboration} {et~al.}(2016){DESI Collaboration}, {Aghamousa},
  {Aguilar}, {Ahlen}, {Alam}, {Allen}, {Allende Prieto}, {Annis}, {Bailey},
  {Balland}, {Ballester}, {Baltay}, {Beaufore}, {Bebek}, {Beers}, {Bell},
  {Bernal}, {Besuner}, {Beutler}, {Blake}, {Bleuler}, {Blomqvist}, {Blum},
  {Bolton}, {Briceno}, {Brooks}, {Brownstein}, {Buckley-Geer}, {Burden},
  {Burtin}, {Busca}, {Cahn}, {Cai}, {Cardiel-Sas}, {Carlberg}, {Carton},
  {Casas}, {Castand er}, {Cervantes-Cota}, {Claybaugh}, {Close}, {Coker},
  {Cole}, {Comparat}, {Cooper}, {Cousinou}, {Crocce}, {Cuby}, {Cunningham},
  {Davis}, {Dawson}, {de la Macorra}, {De Vicente}, {Delubac}, {Derwent},
  {Dey}, {Dhungana}, {Ding}, {Doel}, {Duan}, {Ealet}, {Edelstein},
  {Eftekharzadeh}, {Eisenstein}, {Elliott}, {Escoffier}, {Evatt}, {Fagrelius},
  {Fan}, {Fanning}, {Farahi}, {Farihi}, {Favole}, {Feng}, {Fernandez},
  {Findlay}, {Finkbeiner}, {Fitzpatrick}, {Flaugher}, {Flender}, {Font-Ribera},
  {Forero-Romero}, {Fosalba}, {Frenk}, {Fumagalli}, {Gaensicke}, {Gallo},
  {Garcia-Bellido}, {Gaztanaga}, {Pietro Gentile Fusillo}, {Gerard},
  {Gershkovich}, {Giannantonio}, {Gillet}, {Gonzalez-de-Rivera},
  {Gonzalez-Perez}, {Gott}, {Graur}, {Gutierrez}, {Guy}, {Habib}, {Heetderks},
  {Heetderks}, {Heitmann}, {Hellwing}, {Herrera}, {Ho}, {Holland}, {Honscheid},
  {Huff}, {Hutchinson}, {Huterer}, {Hwang}, {Illa Laguna}, {Ishikawa},
  {Jacobs}, {Jeffrey}, {Jelinsky}, {Jennings}, {Jiang}, {Jimenez}, {Johnson},
  {Joyce}, {Jullo}, {Juneau}, {Kama}, {Karcher}, {Karkar}, {Kehoe}, {Kennamer},
  {Kent}, {Kilbinger}, {Kim}, {Kirkby}, {Kisner}, {Kitanidis}, {Kneib},
  {Koposov}, {Kovacs}, {Koyama}, {Kremin}, {Kron}, {Kronig}, {Kueter-Young},
  {Lacey}, {Lafever}, {Lahav}, {Lambert}, {Lampton}, {Land riau}, {Lang},
  {Lauer}, {Le Goff}, {Le Guillou}, {Le Van Suu}, {Lee}, {Lee}, {Leitner},
  {Lesser}, {Levi}, {L'Huillier}, {Li}, {Liang}, {Lin}, {Linder}, {Loebman},
  {Luki{\'c}}, {Ma}, {MacCrann}, {Magneville}, {Makarem}, {Manera}, {Manser},
  {Marshall}, {Martini}, {Massey}, {Matheson}, {McCauley}, {McDonald},
  {McGreer}, {Meisner}, {Metcalfe}, {Miller}, {Miquel}, {Moustakas}, {Myers},
  {Naik}, {Newman}, {Nichol}, {Nicola}, {Nicolati da Costa}, {Nie}, {Niz},
  {Norberg}, {Nord}, {Norman}, {Nugent}, {O'Brien}, {Oh}, {Olsen}, {Padilla},
  {Padmanabhan}, {Padmanabhan}, {Palanque-Delabrouille}, {Palmese},
  {Pappalardo}, {P{\^a}ris}, {Park}, {Patej}, {Peacock}, {Peiris}, {Peng},
  {Percival}, {Perruchot}, {Pieri}, {Pogge}, {Pollack}, {Poppett}, {Prada},
  {Prakash}, {Probst}, {Rabinowitz}, {Raichoor}, {Ree}, {Refregier}, {Regal},
  {Reid}, {Reil}, {Rezaie}, {Rockosi}, {Roe}, {Ronayette}, {Roodman}, {Ross},
  {Ross}, {Rossi}, {Rozo}, {Ruhlmann-Kleider}, {Rykoff}, {Sabiu}, {Samushia},
  {Sanchez}, {Sanchez}, {Schlegel}, {Schneider}, {Schubnell}, {Secroun},
  {Seljak}, {Seo}, {Serrano}, {Shafieloo}, {Shan}, {Sharples}, {Sholl},
  {Shourt}, {Silber}, {Silva}, {Sirk}, {Slosar}, {Smith}, {Smoot}, {Som},
  {Song}, {Sprayberry}, {Staten}, {Stefanik}, {Tarle}, {Sien Tie}, {Tinker},
  {Tojeiro}, {Valdes}, {Valenzuela}, {Valluri}, {Vargas-Magana}, {Verde},
  {Walker}, {Wang}, {Wang}, {Weaver}, {Weaverdyck}, {Wechsler}, {Weinberg},
  {White}, {Yang}, {Yeche}, {Zhang}, {Zhao}, {Zheng}, {Zhou}, {Zhou}, {Zhu},
  {Zou}, \& {Zu}}]{DESI2016}
{DESI Collaboration}, {Aghamousa}, A., {Aguilar}, J., {et~al.} 2016, arXiv
  e-prints, arXiv:1611.00036

\bibitem[{{Di Matteo} {et~al.}(2008){Di Matteo}, {Colberg}, {Springel},
  {Hernquist}, \& {Sijacki}}]{DiMatteo2008}
{Di Matteo}, T., {Colberg}, J., {Springel}, V., {Hernquist}, L., \& {Sijacki},
  D. 2008, \apj, 676, 33

\bibitem[{{Dicke} {et~al.}(1965){Dicke}, {Peebles}, {Roll}, \&
  {Wilkinson}}]{Dicke1965}
{Dicke}, R.~H., {Peebles}, P.~J.~E., {Roll}, P.~G., \& {Wilkinson}, D.~T. 1965,
  \apj, 142, 414

\bibitem[{{Driver} {et~al.}(2009){Driver}, {Norberg}, {Baldry}, {Bamford},
  {Hopkins}, {Liske}, {Loveday}, {Peacock}, {Hill}, {Kelvin}, {Robotham},
  {Cross}, {Parkinson}, {Prescott}, {Conselice}, {Dunne}, {Brough}, {Jones},
  {Sharp}, {van Kampen}, {Oliver}, {Roseboom}, {Bland -Hawthorn}, {Croom},
  {Ellis}, {Cameron}, {Cole}, {Frenk}, {Couch}, {Graham}, {Proctor}, {De
  Propris}, {Doyle}, {Edmondson}, {Nichol}, {Thomas}, {Eales}, {Jarvis},
  {Kuijken}, {Lahav}, {Madore}, {Seibert}, {Meyer}, {Staveley-Smith},
  {Phillipps}, {Popescu}, {Sansom}, {Sutherland}, {Tuffs}, \&
  {Warren}}]{Driver2009}
{Driver}, S.~P., {Norberg}, P., {Baldry}, I.~K., {et~al.} 2009, Astronomy and
  Geophysics, 50, 5.12

\bibitem[{{Dubois} {et~al.}(2014){Dubois}, {Pichon}, {Welker}, {Le Borgne},
  {Devriendt}, {Laigle}, {Codis}, {Pogosyan}, {Arnouts}, {Benabed}, {Bertin},
  {Blaizot}, {Bouchet}, {Cardoso}, {Colombi}, {de Lapparent}, {Desjacques},
  {Gavazzi}, {Kassin}, {Kimm}, {McCracken}, {Milliard}, {Peirani}, {Prunet},
  {Rouberol}, {Silk}, {Slyz}, {Sousbie}, {Teyssier}, {Tresse}, {Treyer},
  {Vibert}, \& {Volonteri}}]{Dubois2014}
{Dubois}, Y., {Pichon}, C., {Welker}, C., {et~al.} 2014, \mnras, 444, 1453

\bibitem[{Efron(1979)}]{Efron1979}
Efron, B. 1979, Ann. Statist., 7, 1

\bibitem[{{Efstathiou} {et~al.}(2002){Efstathiou}, {Moody}, {Peacock},
  {Percival}, {Baugh}, {Bland-Hawthorn}, {Bridges}, {Cannon}, {Cole},
  {Colless}, {Collins}, {Couch}, {Dalton}, {de Propris}, {Driver}, {Ellis},
  {Frenk}, {Glazebrook}, {Jackson}, {Lahav}, {Lewis}, {Lumsden}, {Maddox},
  {Norberg}, {Peterson}, {Sutherland}, \& {Taylor}}]{Efstathiou2002}
{Efstathiou}, G., {Moody}, S., {Peacock}, J.~A., {et~al.} 2002, \mnras, 330,
  L29

\bibitem[{{Efstathiou} \& {Rees}(1988)}]{Efstathiou1988}
{Efstathiou}, G. \& {Rees}, M.~J. 1988, \mnras, 230, 5p

\bibitem[{{Einasto} {et~al.}(1997){Einasto}, {Tago}, {Jaaniste}, {Einasto}, \&
  {Andernach}}]{Einasto1997}
{Einasto}, M., {Tago}, E., {Jaaniste}, J., {Einasto}, J., \& {Andernach}, H.
  1997, \aaps, 123, 119

\bibitem[{{Einasto} {et~al.}(2014){Einasto}, {Tago}, {Lietzen}, {Park},
  {Hein{\"a}m{\"a}ki}, {Saar}, {Song}, {Liivam{\"a}gi}, \&
  {Einasto}}]{Einasto2014}
{Einasto}, M., {Tago}, E., {Lietzen}, H., {et~al.} 2014, \aap, 568, A46

\bibitem[{{Eisenstein} {et~al.}(2005){Eisenstein}, {Zehavi}, {Hogg},
  {Scoccimarro}, {Blanton}, {Nichol}, {Scranton}, {Seo}, {Tegmark}, {Zheng},
  {Anderson}, {Annis}, {Bahcall}, {Brinkmann}, {Burles}, {Castander},
  {Connolly}, {Csabai}, {Doi}, {Fukugita}, {Frieman}, {Glazebrook}, {Gunn},
  {Hendry}, {Hennessy}, {Ivezi{\'c}}, {Kent}, {Knapp}, {Lin}, {Loh}, {Lupton},
  {Margon}, {McKay}, {Meiksin}, {Munn}, {Pope}, {Richmond}, {Schlegel},
  {Schneider}, {Shimasaku}, {Stoughton}, {Strauss}, {SubbaRao}, {Szalay},
  {Szapudi}, {Tucker}, {Yanny}, \& {York}}]{Eisenstein2005}
{Eisenstein}, D.~J., {Zehavi}, I., {Hogg}, D.~W., {et~al.} 2005, \apj, 633, 560

\bibitem[{{Enea Romano} {et~al.}(2015){Enea Romano}, {Cornejo}, \&
  {Campusano}}]{Romano2015}
{Enea Romano}, A., {Cornejo}, D., \& {Campusano}, L.~E. 2015, arXiv e-prints
  [arXiv:1509.01879]

\bibitem[{{Epps} \& {Hudson}(2017)}]{Epps2017}
{Epps}, S.~D. \& {Hudson}, M.~J. 2017, \mnras, 468, 2605

\bibitem[{{Eriksen} {et~al.}(2008){Eriksen}, {Jewell}, {Dickinson}, {Banday},
  {G{\'o}rski}, \& {Lawrence}}]{Eriksen2008}
{Eriksen}, H.~K., {Jewell}, J.~B., {Dickinson}, C., {et~al.} 2008, \apj, 676,
  10

\bibitem[{{Feigelson} \& {Babu}(2012)}]{Feigelson2012}
{Feigelson}, E.~D. \& {Babu}, G.~J. 2012, {Modern Statistical Methods for
  Astronomy} (Cambridge University Press)

\bibitem[{{Fern{\'a}ndez-Cobos} {et~al.}(2012){Fern{\'a}ndez-Cobos}, {Vielva},
  {Barreiro}, \& {Mart{\'{\i}}nez-Gonz{\'a}lez}}]{FernandezCobos2012}
{Fern{\'a}ndez-Cobos}, R., {Vielva}, P., {Barreiro}, R.~B., \&
  {Mart{\'{\i}}nez-Gonz{\'a}lez}, E. 2012, \mnras, 420, 2162

\bibitem[{{Filippenko} \& {Riess}(1998)}]{Filippenko1998}
{Filippenko}, A.~V. \& {Riess}, A.~G. 1998, \physrep, 307, 31

\bibitem[{Fisher(1993)}]{Fisher1993}
Fisher, N.~I. 1993, Statistical Analysis of Circular Data (Cambridge University
  Press)

\bibitem[{{Fixsen}(2009)}]{Fixsen2009}
{Fixsen}, D.~J. 2009, \apj, 707, 916

\bibitem[{{Fosalba} {et~al.}(2003){Fosalba}, {Gazta{\~n}aga}, \&
  {Castander}}]{Fosalba2003}
{Fosalba}, P., {Gazta{\~n}aga}, E., \& {Castander}, F.~J. 2003, \apjl, 597, L89

\bibitem[{Freeman \& Dale(2013)}]{Freeman2013}
Freeman, J.~B. \& Dale, R. 2013, Behavior Research Methods, 45, 83

\bibitem[{{Friday} {et~al.}(2018){Friday}, {Clowes}, {Raghunathan}, \&
  {Williger}}]{Friday2018}
{Friday}, T., {Clowes}, R.~G., {Raghunathan}, S., \& {Williger}, G.~M. 2018,
  \mnras, 479, 1137

\bibitem[{{Ganeshaiah Veena} {et~al.}(2018){Ganeshaiah Veena}, {Cautun}, {van
  de Weygaert}, {Tempel}, {Jones}, {Rieder}, \& {Frenk}}]{GaneshaiahVeena2018}
{Ganeshaiah Veena}, P., {Cautun}, M., {van de Weygaert}, R., {et~al.} 2018,
  \mnras, 481, 414

\bibitem[{{Geller} \& {Huchra}(1989)}]{Geller1989}
{Geller}, M.~J. \& {Huchra}, J.~P. 1989, Science, 246, 897

\bibitem[{{Giacconi} {et~al.}(2001){Giacconi}, {Zirm}, {Wang}, {Rosati},
  {Nonino}, {Tozzi}, {Gilli}, {Mainieri}, {Hasinger}, {Kewley}, {Bergeron},
  {Borgani}, {Gilmozzi}, {Grogin}, {Koekemoer}, {Schreier}, {Zheng}, \&
  {Norman}}]{Giacconi2001}
{Giacconi}, R., {Zirm}, A., {Wang}, J., {et~al.} 2001, arXiv e-prints,
  astro-ph/0112184

\bibitem[{{Giannantonio} {et~al.}(2016){Giannantonio}, {Fosalba}, {Cawthon},
  {Omori}, {Crocce}, {Elsner}, {Leistedt}, {Dodelson}, {Benoit-L{\'e}vy},
  {Gazta{\~n}aga}, {Holder}, {Peiris}, {Percival}, {Kirk}, {Bauer}, {Benson},
  {Bernstein}, {Carretero}, {Crawford}, {Crittenden}, {Huterer}, {Jain},
  {Krause}, {Reichardt}, {Ross}, {Simard}, {Soergel}, {Stark}, {Story},
  {Vieira}, {Weller}, {Abbott}, {Abdalla}, {Allam}, {Armstrong}, {Banerji},
  {Bernstein}, {Bertin}, {Brooks}, {Buckley-Geer}, {Burke}, {Capozzi},
  {Carlstrom}, {Carnero Rosell}, {Carrasco Kind}, {Castander}, {Chang},
  {Cunha}, {da Costa}, {D'Andrea}, {DePoy}, {Desai}, {Diehl}, {Dietrich},
  {Doel}, {Eifler}, {Evrard}, {Neto}, {Fernandez}, {Finley}, {Flaugher},
  {Frieman}, {Gerdes}, {Gruen}, {Gruendl}, {Gutierrez}, {Holzapfel},
  {Honscheid}, {James}, {Kuehn}, {Kuropatkin}, {Lahav}, {Li}, {Lima}, {March},
  {Marshall}, {Martini}, {Melchior}, {Miquel}, {Mohr}, {Nichol}, {Nord},
  {Ogando}, {Plazas}, {Romer}, {Roodman}, {Rykoff}, {Sako}, {Saliwanchik},
  {Sanchez}, {Schubnell}, {Sevilla-Noarbe}, {Smith}, {Soares-Santos},
  {Sobreira}, {Suchyta}, {Swanson}, {Tarle}, {Thaler}, {Thomas}, {Vikram},
  {Walker}, {Wechsler}, \& {Zuntz}}]{Giannantonio2016}
{Giannantonio}, T., {Fosalba}, P., {Cawthon}, R., {et~al.} 2016, \mnras, 456,
  3213

\bibitem[{{Giannantonio} {et~al.}(2008){Giannantonio}, {Scranton},
  {Crittenden}, {Nichol}, {Boughn}, {Myers}, \& {Richards}}]{Giannantonio2008}
{Giannantonio}, T., {Scranton}, R., {Crittenden}, R.~G., {et~al.} 2008, \prd,
  77, 123520

\bibitem[{{G{\'o}rski} {et~al.}(2005){G{\'o}rski}, {Hivon}, {Banday},
  {Wandelt}, {Hansen}, {Reinecke}, \& {Bartelmann}}]{Gorski2005}
{G{\'o}rski}, K.~M., {Hivon}, E., {Banday}, A.~J., {et~al.} 2005, \apj, 622,
  759

\bibitem[{{Goto} {et~al.}(2012){Goto}, {Szapudi}, \& {Granett}}]{Goto2012}
{Goto}, T., {Szapudi}, I., \& {Granett}, B.~R. 2012, \mnras, 422, L77

\bibitem[{{Gott} {et~al.}(2005){Gott}, {Juri{\'c}}, {Schlegel}, {Hoyle},
  {Vogeley}, {Tegmark}, {Bahcall}, \& {Brinkmann}}]{Gott2005}
{Gott}, J.~Richard, I., {Juri{\'c}}, M., {Schlegel}, D., {et~al.} 2005, \apj,
  624, 463

\bibitem[{{\noopsort{Graaf}}{de Graaff} {et~al.}(2019){\noopsort{Graaf}}{de
  Graaff}, {Cai}, {Heymans}, \& {Peacock}}]{deGraaff2019}
{\noopsort{Graaf}}{de Graaff}, A., {Cai}, Y.-C., {Heymans}, C., \& {Peacock},
  J.~A. 2019, \aap, 624, A48

\bibitem[{{Graham} {et~al.}(1995){Graham}, {Clowes}, \&
  {Campusano}}]{Graham1995}
{Graham}, M.~J., {Clowes}, R.~G., \& {Campusano}, L.~E. 1995, \mnras, 275, 790

\bibitem[{{Granett} {et~al.}(2008){Granett}, {Neyrinck}, \&
  {Szapudi}}]{Granett2008}
{Granett}, B.~R., {Neyrinck}, M.~C., \& {Szapudi}, I. 2008, \apjl, 683, L99

\bibitem[{{Granett} {et~al.}(2009){Granett}, {Neyrinck}, \&
  {Szapudi}}]{Granett2009}
{Granett}, B.~R., {Neyrinck}, M.~C., \& {Szapudi}, I. 2009, \apj, 701, 414

\bibitem[{{Guillochon} {et~al.}(2017){Guillochon}, {Parrent}, {Kelley}, \&
  {Margutti}}]{Guillochon2017}
{Guillochon}, J., {Parrent}, J., {Kelley}, L.~Z., \& {Margutti}, R. 2017, \apj,
  835, 64

\bibitem[{{Halverson} {et~al.}(2009){Halverson}, {Lanting}, {Ade}, {Basu},
  {Bender}, {Benson}, {Bertoldi}, {Cho}, {Chon}, {Clarke}, {Dobbs}, {Ferrusca},
  {G{\"u}sten}, {Holzapfel}, {Kov{\'a}cs}, {Kennedy}, {Kermish}, {Kneissl},
  {Lee}, {Lueker}, {Mehl}, {Menten}, {Muders}, {Nord}, {Pacaud}, {Plagge},
  {Reichardt}, {Richards}, {Schaaf}, {Schilke}, {Schuller}, {Schwan},
  {Spieler}, {Tucker}, {Weiss}, \& {Zahn}}]{Halverson2009}
{Halverson}, N.~W., {Lanting}, T., {Ade}, P.~A.~R., {et~al.} 2009, \apj, 701,
  42

\bibitem[{Hartigan {et~al.}(1985)Hartigan, Hartigan, {et~al.}}]{Hartigan1985}
Hartigan, J.~A., Hartigan, P.~M., {et~al.} 1985, The annals of Statistics, 13,
  70

\bibitem[{{Hasselfield} {et~al.}(2013){Hasselfield}, {Hilton}, {Marriage},
  {Addison}, {Barrientos}, {Battaglia}, {Battistelli}, {Bond}, {Crichton},
  {Das}, {Devlin}, {Dicker}, {Dunkley}, {D{\"u}nner}, {Fowler}, {Gralla},
  {Hajian}, {Halpern}, {Hincks}, {Hlozek}, {Hughes}, {Infante}, {Irwin},
  {Kosowsky}, {Marsden}, {Menanteau}, {Moodley}, {Niemack}, {Nolta}, {Page},
  {Partridge}, {Reese}, {Schmitt}, {Sehgal}, {Sherwin}, {Sievers}, {Sif{\'o}n},
  {Spergel}, {Staggs}, {Swetz}, {Switzer}, {Thornton}, {Trac}, \&
  {Wollack}}]{Hasselfield2013}
{Hasselfield}, M., {Hilton}, M., {Marriage}, T.~A., {et~al.} 2013, \jcap, 7,
  008

\bibitem[{{Hawley} \& {Peebles}(1975)}]{Hawley1975}
{Hawley}, D.~L. \& {Peebles}, P.~J.~E. 1975, \aj, 80, 477

\bibitem[{{He} {et~al.}(2018){He}, {Alam}, {Ferraro}, {Chen}, \& {Ho}}]{He2018}
{He}, S., {Alam}, S., {Ferraro}, S., {Chen}, Y.-C., \& {Ho}, S. 2018, Nature
  Astronomy, 2, 401

\bibitem[{Hermans \& Rasson(1985)}]{Hermans1985}
Hermans, M. \& Rasson, J.~P. 1985, Biometrika, 72, 698

\bibitem[{{Hewett} \& {Wild}(2010)}]{Hewett2010}
{Hewett}, P.~C. \& {Wild}, V. 2010, \mnras, 405, 2302

\bibitem[{{Hinshaw} {et~al.}(2013){Hinshaw}, {Larson}, {Komatsu}, {Spergel},
  {Bennett}, {Dunkley}, {Nolta}, {Halpern}, {Hill}, {Odegard}, {Page}, {Smith},
  {Weiland}, {Gold}, {Jarosik}, {Kogut}, {Limon}, {Meyer}, {Tucker}, {Wollack},
  \& {Wright}}]{Hinshaw2013}
{Hinshaw}, G., {Larson}, D., {Komatsu}, E., {et~al.} 2013, \apjs, 208, 19

\bibitem[{{Ho} {et~al.}(2008){Ho}, {Hirata}, {Padmanabhan}, {Seljak}, \&
  {Bahcall}}]{Ho2008}
{Ho}, S., {Hirata}, C., {Padmanabhan}, N., {Seljak}, U., \& {Bahcall}, N. 2008,
  \prd, 78, 043519

\bibitem[{{Hogg}(1999)}]{Hogg1999}
{Hogg}, D.~W. 1999, arXiv e-prints, astro-ph/9905116

\bibitem[{{Hogg} {et~al.}(2005){Hogg}, {Eisenstein}, {Blanton}, {Bahcall},
  {Brinkmann}, {Gunn}, \& {Schneider}}]{Hogg2005}
{Hogg}, D.~W., {Eisenstein}, D.~J., {Blanton}, M.~R., {et~al.} 2005, \apj, 624,
  54

\bibitem[{{Hu} \& {Dodelson}(2002)}]{Hu2002}
{Hu}, W. \& {Dodelson}, S. 2002, \araa, 40, 171

\bibitem[{{Hurier} \& {Tchernin}(2017)}]{Hurier2017}
{Hurier}, G. \& {Tchernin}, C. 2017, \aap, 604, A94

\bibitem[{{Husband} {et~al.}(2013){Husband}, {Bremer}, {Stanway}, {Davies},
  {Lehnert}, \& {Douglas}}]{Husband2013}
{Husband}, K., {Bremer}, M.~N., {Stanway}, E.~R., {et~al.} 2013, \mnras, 432,
  2869

\bibitem[{{Husband} {et~al.}(2015){Husband}, {Bremer}, {Stanway}, \&
  {Lehnert}}]{Husband2015}
{Husband}, K., {Bremer}, M.~N., {Stanway}, E.~R., \& {Lehnert}, M.~D. 2015,
  \mnras, 452, 2388

\bibitem[{{Hutsem{\'e}kers}(1998)}]{Hutsemekers1998}
{Hutsem{\'e}kers}, D. 1998, \aap, 332, 410

\bibitem[{{Hutsem{\'e}kers} {et~al.}(2010){Hutsem{\'e}kers}, {Borguet},
  {Sluse}, {Cabanac}, \& {Lamy}}]{Hutsemekers2010}
{Hutsem{\'e}kers}, D., {Borguet}, B., {Sluse}, D., {Cabanac}, R., \& {Lamy}, H.
  2010, \aap, 520, L7

\bibitem[{{Hutsem{\'e}kers} {et~al.}(2014){Hutsem{\'e}kers}, {Braibant},
  {Pelgrims}, \& {Sluse}}]{Hutsemekers2014}
{Hutsem{\'e}kers}, D., {Braibant}, L., {Pelgrims}, V., \& {Sluse}, D. 2014,
  \aap, 572, A18

\bibitem[{{Hutsem{\'e}kers} {et~al.}(2005){Hutsem{\'e}kers}, {Cabanac}, {Lamy},
  \& {Sluse}}]{Hutsemekers2005}
{Hutsem{\'e}kers}, D., {Cabanac}, R., {Lamy}, H., \& {Sluse}, D. 2005, \aap,
  441, 915

\bibitem[{{Hutsem{\'e}kers} \& {Lamy}(2001)}]{Hutsemekers2001}
{Hutsem{\'e}kers}, D. \& {Lamy}, H. 2001, \aap, 367, 381

\bibitem[{{Hutsem{\'e}kers} {et~al.}(2011){Hutsem{\'e}kers}, {Payez},
  {Cabanac}, {Lamy}, {Sluse}, {Borguet}, \& {Cudell}}]{Hutsemekers2011}
{Hutsem{\'e}kers}, D., {Payez}, A., {Cabanac}, R., {et~al.} 2011, Astronomical
  Society of the Pacific Conference Series, Vol. 449, {}, 441

\bibitem[{{Ibata} {et~al.}(2013){Ibata}, {Lewis}, {Conn}, {Irwin},
  {McConnachie}, {Chapman}, {Collins}, {Fardal}, {Ferguson}, {Ibata}, {Mackey},
  {Martin}, {Navarro}, {Rich}, {Valls-Gabaud}, \& {Widrow}}]{Ibata2013}
{Ibata}, R.~A., {Lewis}, G.~F., {Conn}, A.~R., {et~al.} 2013, \nat, 493, 62

\bibitem[{{Icke} \& {van de Weygaert}(1987)}]{Icke1987}
{Icke}, V. \& {van de Weygaert}, R. 1987, \aap, 184, 16

\bibitem[{{Inoue} \& {Silk}(2006)}]{Inoue2006}
{Inoue}, K.~T. \& {Silk}, J. 2006, \apj, 648, 23

\bibitem[{{Isobe} {et~al.}(1990){Isobe}, {Feigelson}, {Akritas}, \&
  {Babu}}]{Isobe1990}
{Isobe}, T., {Feigelson}, E.~D., {Akritas}, M.~G., \& {Babu}, G.~J. 1990, \apj,
  364, 104

\bibitem[{{Jackson}(1972)}]{Jackson1972}
{Jackson}, J.~C. 1972, \mnras, 156, 1P

\bibitem[{{Jaffe} {et~al.}(2005){Jaffe}, {Banday}, {Eriksen}, {G{\'o}rski}, \&
  {Hansen}}]{Jaffe2005}
{Jaffe}, T.~R., {Banday}, A.~J., {Eriksen}, H.~K., {G{\'o}rski}, K.~M., \&
  {Hansen}, F.~K. 2005, \apjl, 629, L1

\bibitem[{{Jain} {et~al.}(2004){Jain}, {Narain}, \& {Sarala}}]{Jain2004}
{Jain}, P., {Narain}, G., \& {Sarala}, S. 2004, \mnras, 347, 394

\bibitem[{{Jauzac} {et~al.}(2012){Jauzac}, {Jullo}, {Kneib}, {Ebeling},
  {Leauthaud}, {Ma}, {Limousin}, {Massey}, \& {Richard}}]{Jauzac2012}
{Jauzac}, M., {Jullo}, E., {Kneib}, J.-P., {et~al.} 2012, \mnras, 426, 3369

\bibitem[{{Joachimi} {et~al.}(2011){Joachimi}, {Mandelbaum}, {Abdalla}, \&
  {Bridle}}]{Joachimi2011}
{Joachimi}, B., {Mandelbaum}, R., {Abdalla}, F.~B., \& {Bridle}, S.~L. 2011,
  \aap, 527, A26

\bibitem[{{Joshi} {et~al.}(2007){Joshi}, {Battye}, {Browne}, {Jackson},
  {Muxlow}, \& {Wilkinson}}]{Joshi2007}
{Joshi}, S.~A., {Battye}, R.~A., {Browne}, I.~W.~A., {et~al.} 2007, \mnras,
  380, 162

\bibitem[{{Kaiser}(1987)}]{Kaiser1987}
{Kaiser}, N. 1987, \mnras, 227, 1

\bibitem[{{Karpenka} {et~al.}(2015){Karpenka}, {Feroz}, \&
  {Hobson}}]{Karpenka2015}
{Karpenka}, N.~V., {Feroz}, F., \& {Hobson}, M.~P. 2015, \mnras, 449, 2405

\bibitem[{{Kessler} {et~al.}(2015){Kessler}, {Marriner}, {Childress},
  {Covarrubias}, {D'Andrea}, {Finley}, {Fischer}, {Foley}, {Goldstein},
  {Gupta}, {Kuehn}, {Marcha}, {Nichol}, {Papadopoulos}, {Sako}, {Scolnic},
  {Smith}, {Sullivan}, {Wester}, {Yuan}, {Abbott}, {Abdalla}, {Allam},
  {Benoit-L{\'e}vy}, {Bernstein}, {Bertin}, {Brooks}, {Carnero Rosell},
  {Carrasco Kind}, {Castander}, {Crocce}, {da Costa}, {Desai}, {Diehl},
  {Eifler}, {Fausti Neto}, {Flaugher}, {Frieman}, {Gerdes}, {Gruen}, {Gruendl},
  {Honscheid}, {James}, {Kuropatkin}, {Li}, {Maia}, {Marshall}, {Martini},
  {Miller}, {Miquel}, {Nord}, {Ogando}, {Plazas}, {Reil}, {Romer}, {Roodman},
  {Sanchez}, {Sevilla-Noarbe}, {Smith}, {Soares-Santos}, {Sobreira}, {Tarle},
  {Thaler}, {Thomas}, {Tucker}, {Walker}, \& {DES Collaboration}}]{Kessler2015}
{Kessler}, R., {Marriner}, J., {Childress}, M., {et~al.} 2015, \aj, 150, 172

\bibitem[{{Kiessling} {et~al.}(2015){Kiessling}, {Cacciato}, {Joachimi},
  {Kirk}, {Kitching}, {Leonard}, {Mandelbaum}, {Sch{\"a}fer}, {Sif{\'o}n},
  {Brown}, \& {Rassat}}]{Kiessling2015}
{Kiessling}, A., {Cacciato}, M., {Joachimi}, B., {et~al.} 2015, \ssr, 193, 67

\bibitem[{{Kim} {et~al.}(2011){Kim}, {Park}, {Rossi}, {Lee}, \&
  {Gott}}]{Kim2011}
{Kim}, J., {Park}, C., {Rossi}, G., {Lee}, S.~M., \& {Gott}, J.~Richard, I.
  2011, Journal of Korean Astronomical Society, 44, 217

\bibitem[{Knapp(2007)}]{Knapp2007}
Knapp, T.~R. 2007, Journal of Modern Applied Statistical Methods, 6, 3

\bibitem[{{Kofman} \& {Starobinskij}(1985)}]{Kofman1985}
{Kofman}, L. \& {Starobinskij}, A.~A. 1985, Pisma v Astronomicheskii Zhurnal,
  11, 643

\bibitem[{{Komberg} {et~al.}(1996){Komberg}, {Kravtsov}, \&
  {Lukash}}]{Komberg1996}
{Komberg}, B.~V., {Kravtsov}, A.~V., \& {Lukash}, V.~N. 1996, \mnras, 282, 713

\bibitem[{{Kov{\'a}cs} {et~al.}(2017){Kov{\'a}cs}, {S{\'a}nchez},
  {Garc{\'{\i}}a-Bellido}, {Nadathur}, {Crittenden}, {Gruen}, {Huterer},
  {Bacon}, {Clampitt}, {DeRose}, {Dodelson}, {Gazta{\~n}aga}, {Jain}, {Kirk},
  {Lahav}, {Miquel}, {Naidoo}, {Peacock}, {Soergel}, {Whiteway}, {Abdalla},
  {Allam}, {Annis}, {Benoit-L{\'e}vy}, {Bertin}, {Brooks}, {Buckley-Geer},
  {Rosell}, {Carrasco Kind}, {Carretero}, {Cunha}, {D'Andrea}, {da Costa},
  {DePoy}, {Desai}, {Eifler}, {Finley}, {Flaugher}, {Fosalba}, {Frieman},
  {Giannantonio}, {Goldstein}, {Gruendl}, {Gutierrez}, {James}, {Kuehn},
  {Kuropatkin}, {Marshall}, {Melchior}, {Menanteau}, {Nord}, {Ogando},
  {Plazas}, {Romer}, {Sanchez}, {Scarpine}, {Sevilla-Noarbe}, {Sobreira},
  {Suchyta}, {Swanson}, {Tarle}, {Thomas}, {Walker}, \& {DES
  Collaboration}}]{Kovacs2017}
{Kov{\'a}cs}, A., {S{\'a}nchez}, C., {Garc{\'{\i}}a-Bellido}, J., {et~al.}
  2017, \mnras, 465, 4166

\bibitem[{{Kraljic} {et~al.}(2020){Kraljic}, {Dav{\'e}}, \&
  {Pichon}}]{Kraljic2020}
{Kraljic}, K., {Dav{\'e}}, R., \& {Pichon}, C. 2020, \mnras, 237

\bibitem[{Kuiper(1960)}]{Kuiper1960}
Kuiper, N.~H. 1960, Indagationes Mathematicae (Proceedings), 63, 38

\bibitem[{Landler {et~al.}(2018)Landler, Ruxton, \& Malkemper}]{Landler2018}
Landler, L., Ruxton, G.~D., \& Malkemper, E.~P. 2018, Behavioral Ecology and
  Sociobiology, 72, 128

\bibitem[{Landler {et~al.}(2019)Landler, Ruxton, \& Malkemper}]{Landler2019}
Landler, L., Ruxton, G.~D., \& Malkemper, E.~P. 2019, BMC Ecology, 19, 30

\bibitem[{{\noopsort{Lapparent}}{de Lapparent}
  {et~al.}(1986){\noopsort{Lapparent}}{de Lapparent}, {Geller}, \&
  {Huchra}}]{deLapparent1986}
{\noopsort{Lapparent}}{de Lapparent}, V., {Geller}, M.~J., \& {Huchra}, J.~P.
  1986, \apjl, 302, L1

\bibitem[{{Lee}(2011)}]{Lee2011}
{Lee}, J. 2011, \apj, 732, 99

\bibitem[{{Libeskind} {et~al.}(2015){Libeskind}, {Hoffman}, {Tully},
  {Courtois}, {Pomar{\`e}de}, {Gottl{\"o}ber}, \& {Steinmetz}}]{Libeskind2015}
{Libeskind}, N.~I., {Hoffman}, Y., {Tully}, R.~B., {et~al.} 2015, \mnras, 452,
  1052

\bibitem[{{Louis} {et~al.}(2017){Louis}, {Grace}, {Hasselfield}, {Lungu},
  {Maurin}, {Addison}, {Ade}, {Aiola}, {Allison}, {Amiri}, {Angile},
  {Battaglia}, {Beall}, {de Bernardis}, {Bond}, {Britton}, {Calabrese}, {Cho},
  {Choi}, {Coughlin}, {Crichton}, {Crowley}, {Datta}, {Devlin}, {Dicker},
  {Dunkley}, {D{\"u}nner}, {Ferraro}, {Fox}, {Gallardo}, {Gralla}, {Halpern},
  {Henderson}, {Hill}, {Hilton}, {Hilton}, {Hincks}, {Hlozek}, {Ho}, {Huang},
  {Hubmayr}, {Huffenberger}, {Hughes}, {Infante}, {Irwin}, {Muya Kasanda},
  {Klein}, {Koopman}, {Kosowsky}, {Li}, {Madhavacheril}, {Marriage}, {McMahon},
  {Menanteau}, {Moodley}, {Munson}, {Naess}, {Nati}, {Newburgh}, {Nibarger},
  {Niemack}, {Nolta}, {Nu{\~n}ez}, {Page}, {Pappas}, {Partridge}, {Rojas},
  {Schaan}, {Schmitt}, {Sehgal}, {Sherwin}, {Sievers}, {Simon}, {Spergel},
  {Staggs}, {Switzer}, {Thornton}, {Trac}, {Treu}, {Tucker}, {Van Engelen},
  {Ward}, \& {Wollack}}]{Louis2017}
{Louis}, T., {Grace}, E., {Hasselfield}, M., {et~al.} 2017, \jcap, 6, 031

\bibitem[{{LSST Science Collaboration} {et~al.}(2009){LSST Science
  Collaboration}, {Abell}, {Allison}, {Anderson}, {Andrew}, {Angel}, {Armus},
  {Arnett}, {Asztalos}, {Axelrod}, \& et~al.}]{LSST2009}
{LSST Science Collaboration}, {Abell}, P.~A., {Allison}, J., {et~al.} 2009,
  ArXiv e-prints, arXiv:0912.0201

\bibitem[{{Mackenzie} {et~al.}(2017){Mackenzie}, {Shanks}, {Bremer}, {Cai},
  {Gunawardhana}, {Kov{\'a}cs}, {Norberg}, \& {Szapudi}}]{Mackenzie2017}
{Mackenzie}, R., {Shanks}, T., {Bremer}, M.~N., {et~al.} 2017, \mnras, 470,
  2328

\bibitem[{Mardia \& Jupp(2000)}]{Mardia2000}
Mardia, K. \& Jupp, P. 2000, Directional statistics, Wiley series in
  probability and statistics (Wiley)

\bibitem[{{Marinello}(2015)}]{Marinello2015thesis}
{Marinello}, G.~E. 2015, PhD thesis, University of Central Lancashire

\bibitem[{{Marinello} {et~al.}(2016){Marinello}, {Clowes}, {Campusano},
  {Williger}, {S{\"o}chting}, \& {Graham}}]{Marinello2016}
{Marinello}, G.~E., {Clowes}, R.~G., {Campusano}, L.~E., {et~al.} 2016, \mnras,
  461, 2267

\bibitem[{Maurus \& Plant(2016)}]{Maurus2016}
Maurus, S. \& Plant, C. 2016, ACM SIGKDD Conference on Knowledge Discovery and
  Data Mining (Proceedings), 1055

\bibitem[{{McGreer} {et~al.}(2014){McGreer}, {Fan}, {Strauss}, {Haiman},
  {Richards}, {Jiang}, {Bian}, \& {Schneider}}]{McGreer2014}
{McGreer}, I.~D., {Fan}, X., {Strauss}, M.~A., {et~al.} 2014, \aj, 148, 73

\bibitem[{{Merloni} {et~al.}(2019){Merloni}, {Alexander}, {Banerji}, {Boller},
  {Comparat}, {Dwelly}, {Fotopoulou}, {McMahon}, {Nandra}, {Salvato}, {Croom},
  {Finoguenov}, {Krumpe}, {Lamer}, {Rosario}, {Schwope}, {Shanks}, {Steinmetz},
  {Wisotzki}, \& {Worseck}}]{Merloni2019}
{Merloni}, A., {Alexander}, D.~A., {Banerji}, M., {et~al.} 2019, The Messenger,
  175, 42

\bibitem[{{Miknaitis} {et~al.}(2007){Miknaitis}, {Pignata}, {Rest},
  {Wood-Vasey}, {Blondin}, {Challis}, {Smith}, {Stubbs}, {Suntzeff}, {Foley},
  {Matheson}, {Tonry}, {Aguilera}, {Blackman}, {Becker}, {Clocchiatti},
  {Covarrubias}, {Davis}, {Filippenko}, {Garg}, {Garnavich}, {Hicken}, {Jha},
  {Krisciunas}, {Kirshner}, {Leibundgut}, {Li}, {Miceli}, {Narayan}, {Prieto},
  {Riess}, {Salvo}, {Schmidt}, {Sollerman}, {Spyromilio}, \&
  {Zenteno}}]{Miknaitis2007}
{Miknaitis}, G., {Pignata}, G., {Rest}, A., {et~al.} 2007, \apj, 666, 674

\bibitem[{{Miller} {et~al.}(1999){Miller}, {Caldwell}, {Devlin}, {Dorwart},
  {Herbig}, {Nolta}, {Page}, {Puchalla}, {Torbet}, \& {Tran}}]{Miller1999}
{Miller}, A.~D., {Caldwell}, R., {Devlin}, M.~J., {et~al.} 1999, \apjl, 524, L1

\bibitem[{{M{\o}ller} \& {Fynbo}(2001)}]{Moller2001}
{M{\o}ller}, P. \& {Fynbo}, J.~U. 2001, \aap, 372, L57

\bibitem[{Monahan(2011)}]{Monahan2011}
Monahan, J.~F. 2011, Numerical Methods of Statistics, 2nd edn., Cambridge
  Series in Statistical and Probabilistic Mathematics (Cambridge University
  Press)

\bibitem[{{Nadathur}(2013)}]{Nadathur2013}
{Nadathur}, S. 2013, \mnras, 434, 398

\bibitem[{{Nadathur} \& {Crittenden}(2016)}]{Nadathur2016}
{Nadathur}, S. \& {Crittenden}, R. 2016, \apjl, 830, L19

\bibitem[{{Nesseris} \& {Perivolaropoulos}(2007)}]{Nesseris2007}
{Nesseris}, S. \& {Perivolaropoulos}, L. 2007, \jcap, 2, 025

\bibitem[{{Neyrinck} {et~al.}(2018){Neyrinck}, {Hidding}, {Konstantatou}, \&
  {van de Weygaert}}]{Neyrinck2018}
{Neyrinck}, M.~C., {Hidding}, J., {Konstantatou}, M., \& {van de Weygaert}, R.
  2018, Royal Society Open Science, 5, 171582

\bibitem[{{Niederste-Ostholt} {et~al.}(2010){Niederste-Ostholt}, {Strauss},
  {Dong}, {Koester}, \& {McKay}}]{NiedersteOstholt2010}
{Niederste-Ostholt}, M., {Strauss}, M.~A., {Dong}, F., {Koester}, B.~P., \&
  {McKay}, T.~A. 2010, \mnras, 405, 2023

\bibitem[{{Nishizawa}(2014)}]{Nishizawa2014}
{Nishizawa}, A.~J. 2014, Progress of Theoretical and Experimental Physics,
  2014, 06B110

\bibitem[{{Nolta} {et~al.}(2004){Nolta}, {Wright}, {Page}, {Bennett},
  {Halpern}, {Hinshaw}, {Jarosik}, {Kogut}, {Limon}, {Meyer}, {Spergel},
  {Tucker}, \& {Wollack}}]{Nolta2004}
{Nolta}, M.~R., {Wright}, E.~L., {Page}, L., {et~al.} 2004, \apj, 608, 10

\bibitem[{{Padmanabhan} {et~al.}(2005){Padmanabhan}, {Hirata}, {Seljak},
  {Schlegel}, {Brinkmann}, \& {Schneider}}]{Padmanabhan2005}
{Padmanabhan}, N., {Hirata}, C.~M., {Seljak}, U., {et~al.} 2005, \prd, 72,
  043525

\bibitem[{{Pandey} \& {Sarkar}(2015)}]{Pandey2015}
{Pandey}, B. \& {Sarkar}, S. 2015, \mnras, 454, 2647

\bibitem[{{P{\'a}pai} {et~al.}(2011){P{\'a}pai}, {Szapudi}, \&
  {Granett}}]{Papai2011}
{P{\'a}pai}, P., {Szapudi}, I., \& {Granett}, B.~R. 2011, \apj, 732, 27

\bibitem[{{Park} {et~al.}(2012){Park}, {Choi}, {Kim}, {Gott}, {Kim}, \&
  {Kim}}]{Park2012}
{Park}, C., {Choi}, Y.-Y., {Kim}, J., {et~al.} 2012, \apjl, 759, L7

\bibitem[{{Park} {et~al.}(2015){Park}, {Song}, {Einasto}, {Lietzen}, \&
  {Heinamaki}}]{Park2015}
{Park}, C., {Song}, H., {Einasto}, M., {Lietzen}, H., \& {Heinamaki}, P. 2015,
  Journal of Korean Astronomical Society, 48, 75

\bibitem[{{Park} {et~al.}(2017){Park}, {Hyun}, {Noh}, \& {Hwang}}]{Park2017}
{Park}, C.-G., {Hyun}, H., {Noh}, H., \& {Hwang}, J.-c. 2017, \mnras, 469, 1924

\bibitem[{{Pawlowski}(2018)}]{Pawlowski2018}
{Pawlowski}, M.~S. 2018, Modern Physics Letters A, 33, 1830004

\bibitem[{{Pawlowski} {et~al.}(2013){Pawlowski}, {Kroupa}, \&
  {Jerjen}}]{Pawlowski2013}
{Pawlowski}, M.~S., {Kroupa}, P., \& {Jerjen}, H. 2013, \mnras, 435, 1928

\bibitem[{{Payez} {et~al.}(2008){Payez}, {Cudell}, \&
  {Hutsem{\'e}kers}}]{Payez2008}
{Payez}, A., {Cudell}, J.~R., \& {Hutsem{\'e}kers}, D. 2008, in American
  Institute of Physics Conference Series, Vol. 1038, American Institute of
  Physics Conference Series

\bibitem[{{Payez} {et~al.}(2011){Payez}, {Cudell}, \&
  {Hutsem{\'e}kers}}]{Payez2011}
{Payez}, A., {Cudell}, J.~R., \& {Hutsem{\'e}kers}, D. 2011, \prd, 84, 085029

\bibitem[{{Peebles}(1968)}]{Peebles1968}
{Peebles}, P.~J.~E. 1968, \apj, 153, 1

\bibitem[{{Peebles}(1993)}]{Peebles1993}
{Peebles}, P.~J.~E. 1993, {Principles of Physical Cosmology} (Princeton
  University Press)

\bibitem[{{Peebles} \& {Yu}(1970)}]{Peebles1970}
{Peebles}, P.~J.~E. \& {Yu}, J.~T. 1970, \apj, 162, 815

\bibitem[{{Pelgrims}(2016)}]{Pelgrims2016thesis}
{Pelgrims}, V. 2016, PhD thesis, Universit\'e de Li\`ege, arXiv e-prints,
  arXiv:1604.05141

\bibitem[{{Pelgrims}(2019)}]{Pelgrims2019}
{Pelgrims}, V. 2019, A\&A, 622, A145

\bibitem[{{Pelgrims} \& {Cudell}(2014)}]{Pelgrims2014}
{Pelgrims}, V. \& {Cudell}, J.~R. 2014, \mnras, 442, 1239

\bibitem[{{Pelgrims} \& {Hutsem{\'e}kers}(2015)}]{Pelgrims2015}
{Pelgrims}, V. \& {Hutsem{\'e}kers}, D. 2015, \mnras, 450, 4161

\bibitem[{{Pelgrims} \& {Hutsem{\'e}kers}(2016)}]{Pelgrims2016}
{Pelgrims}, V. \& {Hutsem{\'e}kers}, D. 2016, \aap, 590, A53

\bibitem[{{Pen} {et~al.}(2000){Pen}, {Lee}, \& {Seljak}}]{Pen2000}
{Pen}, U.-L., {Lee}, J., \& {Seljak}, U. 2000, \apjl, 543, L107

\bibitem[{{Penzias} \& {Wilson}(1965)}]{Penzias1965}
{Penzias}, A.~A. \& {Wilson}, R.~W. 1965, \apj, 142, 419

\bibitem[{{Percival} {et~al.}(2010){Percival}, {Reid}, {Eisenstein}, {Bahcall},
  {Budavari}, {Frieman}, {Fukugita}, {Gunn}, {Ivezi{\'c}}, {Knapp}, {Kron},
  {Loveday}, {Lupton}, {McKay}, {Meiksin}, {Nichol}, {Pope}, {Schlegel},
  {Schneider}, {Spergel}, {Stoughton}, {Strauss}, {Szalay}, {Tegmark},
  {Vogeley}, {Weinberg}, {York}, \& {Zehavi}}]{Percival2010}
{Percival}, W.~J., {Reid}, B.~A., {Eisenstein}, D.~J., {et~al.} 2010, \mnras,
  401, 2148

\bibitem[{{Pereyra} {et~al.}(2019){Pereyra}, {Sgr{\'o}}, {Merch{\'a}n},
  {Stasyszyn}, \& {Paz}}]{Pereyra2019}
{Pereyra}, L.~A., {Sgr{\'o}}, M.~A., {Merch{\'a}n}, M.~E., {Stasyszyn}, F.~A.,
  \& {Paz}, D.~J. 2019, arXiv e-prints, arXiv:1911.06768

\bibitem[{{Perlmutter} {et~al.}(1999){Perlmutter}, {Aldering}, {Goldhaber},
  {Knop}, {Nugent}, {Castro}, {Deustua}, {Fabbro}, {Goobar}, {Groom}, {Hook},
  {Kim}, {Kim}, {Lee}, {Nunes}, {Pain}, {Pennypacker}, {Quimby}, {Lidman},
  {Ellis}, {Irwin}, {McMahon}, {Ruiz-Lapuente}, {Walton}, {Schaefer}, {Boyle},
  {Filippenko}, {Matheson}, {Fruchter}, {Panagia}, {Newberg}, {Couch}, \&
  {Project}}]{Perlmutter1999}
{Perlmutter}, S., {Aldering}, G., {Goldhaber}, G., {et~al.} 1999, \apj, 517,
  565

\bibitem[{Pfister {et~al.}(2013)Pfister, Schwarz, Janczyk, Dale, \&
  Freeman}]{Pfister2013}
Pfister, R., Schwarz, K., Janczyk, M., Dale, R., \& Freeman, J. 2013, Frontiers
  in Psychology, 4, 700

\bibitem[{Pilipenko(2007)}]{Pilipenko2007}
Pilipenko, S. 2007, Astronomy Reports, 51, 820

\bibitem[{{Pilipenko} \& {Malinovsky}(2013)}]{Pilipenko2013}
{Pilipenko}, S. \& {Malinovsky}, A. 2013, arXiv e-prints, arXiv:1306.3970

\bibitem[{{Planck Collaboration} {et~al.}(2016{\natexlab{a}}){Planck
  Collaboration}, {Adam}, {Ade}, {Aghanim}, {Akrami}, {Alves}, {Arg{\"u}eso},
  {Arnaud}, {Arroja}, {Ashdown}, \& et~al.}]{Planck2016a}
{Planck Collaboration}, {Adam}, R., {Ade}, P.~A.~R., {et~al.}
  2016{\natexlab{a}}, \aap, 594, A1

\bibitem[{{Planck Collaboration} {et~al.}(2016{\natexlab{b}}){Planck
  Collaboration}, {Adam}, {Ade}, {Aghanim}, {Arnaud}, {Ashdown}, {Aumont},
  {Baccigalupi}, {Banday}, {Barreiro}, \& et~al.}]{Planck2016c}
{Planck Collaboration}, {Adam}, R., {Ade}, P.~A.~R., {et~al.}
  2016{\natexlab{b}}, \aap, 594, A9

\bibitem[{{Planck Collaboration} {et~al.}(2015{\natexlab{a}}){Planck
  Collaboration}, {Ade}, {Aghanim}, {Alina}, {Alves}, {Armitage-Caplan},
  {Arnaud}, {Arzoumanian}, {Ashdown}, {Atrio-Barand ela}, {Aumont},
  {Baccigalupi}, {Banday}, {Barreiro}, {Battaner}, {Benabed}, \&
  et~al.}]{Planck2015a}
{Planck Collaboration}, {Ade}, P.~A.~R., {Aghanim}, N., {et~al.}
  2015{\natexlab{a}}, \aap, 576, A104

\bibitem[{{Planck Collaboration} {et~al.}(2015{\natexlab{b}}){Planck
  Collaboration}, {Ade}, {Aghanim}, {Alina}, {Aniano}, {Armitage-Caplan},
  {Arnaud}, {Ashdown}, {Atrio-Barandela}, {Aumont}, {Baccigalupi}, {Banday},
  {Barreiro}, {Battaner}, {Beichman}, \& et~al.}]{Planck2015b}
{Planck Collaboration}, {Ade}, P.~A.~R., {Aghanim}, N., {et~al.}
  2015{\natexlab{b}}, \aap, 576, A106

\bibitem[{{Planck Collaboration} {et~al.}(2014{\natexlab{a}}){Planck
  Collaboration}, {Ade}, {Aghanim}, {Alves}, {Armitage-Caplan}, {Arnaud},
  {Ashdown}, {Atrio-Barandela}, {Aumont}, {Aussel}, \& et~al.}]{Planck2014a}
{Planck Collaboration}, {Ade}, P.~A.~R., {Aghanim}, N., {et~al.}
  2014{\natexlab{a}}, \aap, 571, A1

\bibitem[{{Planck Collaboration} {et~al.}(2014{\natexlab{b}}){Planck
  Collaboration}, {Ade}, {Aghanim}, {Armitage-Caplan}, {Arnaud}, {Ashdown},
  {Atrio-Barandela}, {Aumont}, {Aussel}, {Baccigalupi}, \&
  et~al.}]{Planck2014c}
{Planck Collaboration}, {Ade}, P.~A.~R., {Aghanim}, N., {et~al.}
  2014{\natexlab{b}}, \aap, 571, A29

\bibitem[{{Planck Collaboration} {et~al.}(2014{\natexlab{c}}){Planck
  Collaboration}, {Ade}, {Aghanim}, {Armitage-Caplan}, {Arnaud}, {Ashdown},
  {Atrio-Barandela}, {Aumont}, {Baccigalupi}, {Banday}, \&
  et~al.}]{Planck2014b}
{Planck Collaboration}, {Ade}, P.~A.~R., {Aghanim}, N., {et~al.}
  2014{\natexlab{c}}, \aap, 571, A12

\bibitem[{{Planck Collaboration} {et~al.}(2016{\natexlab{c}}){Planck
  Collaboration}, {Ade}, {Aghanim}, {Arnaud}, {Arroja}, {Ashdown}, {Aumont},
  {Baccigalupi}, {Ballardini}, {Banday}, {Barreiro}, {Bartolo}, {Battaner},
  {Benabed}, {Beno{\^\i}t}, {Benoit-L{\'e}vy}, {Bernard}, {Bersanelli},
  {Bielewicz}, {Bock}, {Bonaldi}, {Bonavera}, {Bond}, {Borrill}, {Bouchet},
  {Bucher}, {Burigana}, {Butler}, {Calabrese}, {Cardoso}, {Catalano},
  {Chamballu}, {Chiang}, {Chluba}, {Christensen}, {Church}, {Clements},
  {Colombi}, {Colombo}, {Combet}, {Couchot}, {Coulais}, {Crill}, {Curto},
  {Cuttaia}, {Danese}, {Davies}, {Davis}, {de Bernardis}, {de Rosa}, {de
  Zotti}, {Delabrouille}, {D{\'e}sert}, {Diego}, {Dolag}, {Dole}, {Donzelli},
  {Dor{\'e}}, {Douspis}, {Ducout}, {Dupac}, {Efstathiou}, {Elsner},
  {En{\ss}lin}, {Eriksen}, {Fergusson}, {Finelli}, {Florido}, {Forni},
  {Frailis}, {Fraisse}, {Franceschi}, {Frejsel}, {Galeotta}, {Galli}, {Ganga},
  {Giard}, {Giraud-H{\'e}raud}, {Gjerl{\o}w}, {Gonz{\'a}lez-Nuevo},
  {G{\'o}rski}, {Gratton}, {Gregorio}, {Gruppuso}, {Gudmundsson}, {Hansen},
  {Hanson}, {Harrison}, {Helou}, {Henrot-Versill{\'e}},
  {Hern{\'a}ndez-Monteagudo}, {Herranz}, {Hildebrand t}, {Hivon}, {Hobson},
  {Holmes}, {Hornstrup}, {Hovest}, {Huffenberger}, {Hurier}, {Jaffe}, {Jaffe},
  {Jones}, {Juvela}, {Keih{\"a}nen}, {Keskitalo}, {Kim}, {Kisner}, {Knoche},
  {Kunz}, {Kurki-Suonio}, {Lagache}, {L{\"a}hteenm{\"a}ki}, {Lamarre},
  {Lasenby}, {Lattanzi}, {Lawrence}, {Leahy}, {Leonardi}, {Lesgourgues},
  {Levrier}, {Liguori}, {Lilje}, {Linden-V{\o}rnle}, {L{\'o}pez-Caniego},
  {Lubin}, {Mac{\'\i}as-P{\'e}rez}, {Maggio}, {Maino}, {Mand olesi},
  {Mangilli}, {Maris}, {Martin}, {Mart{\'\i}nez-Gonz{\'a}lez}, {Masi},
  {Matarrese}, {McGehee}, {Meinhold}, {Melchiorri}, {Mendes}, {Mennella},
  {Migliaccio}, {Mitra}, {Miville-Desch{\^e}nes}, {Molinari}, {Moneti},
  {Montier}, {Morgante}, {Mortlock}, {Moss}, {Munshi}, {Murphy}, {Naselsky},
  {Nati}, {Natoli}, {Netterfield}, {N{\o}rgaard-Nielsen}, {Noviello},
  {Novikov}, {Novikov}, {Oppermann}, {Oxborrow}, {Paci}, {Pagano}, {Pajot},
  {Paoletti}, {Pasian}, {Patanchon}, {Perdereau}, {Perotto}, {Perrotta},
  {Pettorino}, {Piacentini}, {Piat}, {Pierpaoli}, {Pietrobon}, {Plaszczynski},
  {Pointecouteau}, {Polenta}, {Popa}, {Pratt}, {Pr{\'e}zeau}, {Prunet},
  {Puget}, {Rachen}, {Rebolo}, {Reinecke}, {Remazeilles}, {Renault}, {Renzi},
  {Ristorcelli}, {Rocha}, {Rosset}, {Rossetti}, {Roudier},
  {Rubi{\~n}o-Mart{\'\i}n}, {Ruiz-Granados}, {Rusholme}, {Sandri}, {Santos},
  {Savelainen}, {Savini}, {Scott}, {Seiffert}, {Shellard}, {Shiraishi},
  {Spencer}, {Stolyarov}, {Stompor}, {Sudiwala}, {Sunyaev}, {Sutton},
  {Suur-Uski}, {Sygnet}, {Tauber}, {Terenzi}, {Toffolatti}, {Tomasi},
  {Tristram}, {Tucci}, {Tuovinen}, {Umana}, {Valenziano}, {Valiviita}, {Van
  Tent}, {Vielva}, {Villa}, {Wade}, {Wandelt}, {Wehus}, {Yvon}, {Zacchei}, \&
  {Zonca}}]{Planck2016h}
{Planck Collaboration}, {Ade}, P.~A.~R., {Aghanim}, N., {et~al.}
  2016{\natexlab{c}}, \aap, 594, A19

\bibitem[{{Planck Collaboration} {et~al.}(2016{\natexlab{d}}){Planck
  Collaboration}, {Ade}, {Aghanim}, {Arnaud}, {Ashdown}, {Aumont},
  {Baccigalupi}, {Banday}, {Barreiro}, {Barrena}, \& et~al.}]{Planck2016e}
{Planck Collaboration}, {Ade}, P.~A.~R., {Aghanim}, N., {et~al.}
  2016{\natexlab{d}}, \aap, 594, A27

\bibitem[{{Planck Collaboration} {et~al.}(2016{\natexlab{e}}){Planck
  Collaboration}, {Ade}, {Aghanim}, {Arnaud}, {Ashdown}, {Aumont},
  {Baccigalupi}, {Banday}, {Barreiro}, {Bartlett}, \& et~al.}]{Planck2016b}
{Planck Collaboration}, {Ade}, P.~A.~R., {Aghanim}, N., {et~al.}
  2016{\natexlab{e}}, \aap, 594, A13

\bibitem[{{Planck Collaboration} {et~al.}(2016{\natexlab{f}}){Planck
  Collaboration}, {Ade}, {Aghanim}, {Arnaud}, {Ashdown}, {Aumont},
  {Baccigalupi}, {Banday}, {Barreiro}, {Bartlett}, \& et~al.}]{Planck2016g}
{Planck Collaboration}, {Ade}, P.~A.~R., {Aghanim}, N., {et~al.}
  2016{\natexlab{f}}, \aap, 594, A24

\bibitem[{{Planck Collaboration} {et~al.}(2016{\natexlab{g}}){Planck
  Collaboration}, {Ade}, {Aghanim}, {Arnaud}, {Ashdown}, {Aumont},
  {Baccigalupi}, {Banday}, {Barreiro}, {Bartolo}, \& et~al.}]{Planck2016f}
{Planck Collaboration}, {Ade}, P.~A.~R., {Aghanim}, N., {et~al.}
  2016{\natexlab{g}}, \aap, 594, A21

\bibitem[{{Planck Collaboration} {et~al.}(2016{\natexlab{h}}){Planck
  Collaboration}, {Aghanim}, {Arnaud}, {Ashdown}, {Aumont}, {Baccigalupi},
  {Banday}, {Barreiro}, {Bartlett}, {Bartolo}, \& et~al.}]{Planck2016d}
{Planck Collaboration}, {Aghanim}, N., {Arnaud}, M., {et~al.}
  2016{\natexlab{h}}, \aap, 594, A11

\bibitem[{{Plionis} {et~al.}(2003){Plionis}, {Benoist}, {Maurogordato},
  {Ferrari}, \& {Basilakos}}]{Plionis2003}
{Plionis}, M., {Benoist}, C., {Maurogordato}, S., {Ferrari}, C., \&
  {Basilakos}, S. 2003, \apj, 594, 144

\bibitem[{{Pointecouteau} {et~al.}(1998){Pointecouteau}, {Giard}, \&
  {Barret}}]{Pointecouteau1998}
{Pointecouteau}, E., {Giard}, M., \& {Barret}, D. 1998, \aap, 336, 44

\bibitem[{{Poltis} \& {Stojkovic}(2010)}]{Poltis2010}
{Poltis}, R. \& {Stojkovic}, D. 2010, \prl, 105, 161301

\bibitem[{{Press} \& {Davis}(1982)}]{Press1982}
{Press}, W.~H. \& {Davis}, M. 1982, \apj, 259, 449

\bibitem[{{Pritchet} \& {SNLS Collaboration}(2005)}]{Pritchet2005}
{Pritchet}, C.~J. \& {SNLS Collaboration}. 2005, in Astronomical Society of the
  Pacific Conference Series, Vol. 339, Observing Dark Energy, 60

\bibitem[{{Raccanelli} {et~al.}(2008){Raccanelli}, {Bonaldi}, {Negrello},
  {Matarrese}, {Tormen}, \& {de Zotti}}]{Raccanelli2008}
{Raccanelli}, A., {Bonaldi}, A., {Negrello}, M., {et~al.} 2008, \mnras, 386,
  2161

\bibitem[{Rayleigh(1880)}]{Rayleigh1880}
Rayleigh, {\relax Lord}. 1880, The London, Edinburgh, and Dublin Philosophical
  Magazine and Journal of Science, 10, 73

\bibitem[{{Rees} \& {Sciama}(1968)}]{Rees1968}
{Rees}, M.~J. \& {Sciama}, D.~W. 1968, \nat, 217, 511

\bibitem[{{Reichardt} {et~al.}(2013){Reichardt}, {Stalder}, {Bleem}, {Montroy},
  {Aird}, {Andersson}, {Armstrong}, {Ashby}, {Bautz}, {Bayliss}, {Bazin},
  {Benson}, {Brodwin}, {Carlstrom}, {Chang}, {Cho}, {Clocchiatti}, {Crawford},
  {Crites}, {de Haan}, {Desai}, {Dobbs}, {Dudley}, {Foley}, {Forman}, {George},
  {Gladders}, {Gonzalez}, {Halverson}, {Harrington}, {High}, {Holder},
  {Holzapfel}, {Hoover}, {Hrubes}, {Jones}, {Joy}, {Keisler}, {Knox}, {Lee},
  {Leitch}, {Liu}, {Lueker}, {Luong-Van}, {Mantz}, {Marrone}, {McDonald},
  {McMahon}, {Mehl}, {Meyer}, {Mocanu}, {Mohr}, {Murray}, {Natoli}, {Padin},
  {Plagge}, {Pryke}, {Rest}, {Ruel}, {Ruhl}, {Saliwanchik}, {Saro}, {Sayre},
  {Schaffer}, {Shaw}, {Shirokoff}, {Song}, {Spieler}, {Staniszewski}, {Stark},
  {Story}, {Stubbs}, {{\v S}uhada}, {van Engelen}, {Vanderlinde}, {Vieira},
  {Vikhlinin}, {Williamson}, {Zahn}, \& {Zenteno}}]{Reichardt2013}
{Reichardt}, C.~L., {Stalder}, B., {Bleem}, L.~E., {et~al.} 2013, \apj, 763,
  127

\bibitem[{{Richards} {et~al.}(2006){Richards}, {Strauss}, {Fan}, {Hall},
  {Jester}, {Schneider}, {Vanden Berk}, {Stoughton}, {Anderson}, {Brunner},
  {Gray}, {Gunn}, {Ivezi{\'c}}, {Kirkland }, {Knapp}, {Loveday}, {Meiksin},
  {Pope}, {Szalay}, {Thakar}, {Yanny}, {York}, {Barentine}, {Brewington},
  {Brinkmann}, {Fukugita}, {Harvanek}, {Kent}, {Kleinman}, {Krzesi{\'n}ski},
  {Long}, {Lupton}, {Nash}, {Neilsen}, {Nitta}, {Schlegel}, \&
  {Snedden}}]{Richards2006}
{Richards}, G.~T., {Strauss}, M.~A., {Fan}, X., {et~al.} 2006, \aj, 131, 2766

\bibitem[{{Riess}(2019)}]{Riess2019}
{Riess}, A.~G. 2019, Nature Reviews Physics, 2, 10

\bibitem[{{Riess} {et~al.}(1998){Riess}, {Filippenko}, {Challis},
  {Clocchiatti}, {Diercks}, {Garnavich}, {Gilliland}, {Hogan}, {Jha},
  {Kirshner}, {Leibundgut}, {Phillips}, {Reiss}, {Schmidt}, {Schommer},
  {Smith}, {Spyromilio}, {Stubbs}, {Suntzeff}, \& {Tonry}}]{Riess1998}
{Riess}, A.~G., {Filippenko}, A.~V., {Challis}, P., {et~al.} 1998, \aj, 116,
  1009

\bibitem[{{Rost} {et~al.}(2020){Rost}, {Stasyszyn}, {Pereyra}, \&
  {Mart{\'\i}nez}}]{Rost2019}
{Rost}, A., {Stasyszyn}, F., {Pereyra}, L., \& {Mart{\'\i}nez}, H.~J. 2020,
  \mnras, 493, 1936

\bibitem[{{Saadeh} {et~al.}(2016){Saadeh}, {Feeney}, {Pontzen}, {Peiris}, \&
  {McEwen}}]{Saadeh2016}
{Saadeh}, D., {Feeney}, S.~M., {Pontzen}, A., {Peiris}, H.~V., \& {McEwen},
  J.~D. 2016, \prl, 117, 131302

\bibitem[{{Sachs} \& {Wolfe}(1967)}]{SachsWolfe1967}
{Sachs}, R.~K. \& {Wolfe}, A.~M. 1967, \apj, 147, 73

\bibitem[{{Salpeter}(1964)}]{Salpeter1964}
{Salpeter}, E.~E. 1964, \apj, 140, 796

\bibitem[{{Sastry}(1968)}]{Sastry1968}
{Sastry}, G.~N. 1968, \pasp, 80, 252

\bibitem[{{Schaffer} {et~al.}(2011){Schaffer}, {Crawford}, {Aird}, {Benson},
  {Bleem}, {Carlstrom}, {Chang}, {Cho}, {Crites}, {de Haan}, {Dobbs}, {George},
  {Halverson}, {Holder}, {Holzapfel}, {Hoover}, {Hrubes}, {Joy}, {Keisler},
  {Knox}, {Lee}, {Leitch}, {Lueker}, {Luong-Van}, {McMahon}, {Mehl}, {Meyer},
  {Mohr}, {Montroy}, {Padin}, {Plagge}, {Pryke}, {Reichardt}, {Ruhl},
  {Shirokoff}, {Spieler}, {Stalder}, {Staniszewski}, {Stark}, {Story},
  {Vanderlinde}, {Vieira}, \& {Williamson}}]{Schaffer2011}
{Schaffer}, K.~K., {Crawford}, T.~M., {Aird}, K.~A., {et~al.} 2011, \apj, 743,
  90

\bibitem[{{Schneider} {et~al.}(2010){Schneider}, {Richards}, {Hall}, {Strauss},
  {Anderson}, {Boroson}, {Ross}, {Shen}, {Brandt}, {Fan}, {Inada}, {Jester},
  {Knapp}, {Krawczyk}, {Thakar}, {Vanden Berk}, {Voges}, {Yanny}, {York},
  {Bahcall}, {Bizyaev}, {Blanton}, {Brewington}, {Brinkmann}, {Eisenstein},
  {Frieman}, {Fukugita}, {Gray}, {Gunn}, {Hibon}, {Ivezi{\'c}}, {Kent}, {Kron},
  {Lee}, {Lupton}, {Malanushenko}, {Malanushenko}, {Oravetz}, {Pan}, {Pier},
  {Price}, {Saxe}, {Schlegel}, {Simmons}, {Snedden}, {SubbaRao}, {Szalay}, \&
  {Weinberg}}]{Schneider2010}
{Schneider}, D.~P., {Richards}, G.~T., {Hall}, P.~B., {et~al.} 2010, \aj, 139,
  2360

\bibitem[{{Scodeggio} {et~al.}(2018){Scodeggio}, {Guzzo}, {Garilli}, {Granett},
  {Bolzonella}, {de la Torre}, {Abbas}, {Adami}, {Arnouts}, {Bottini}, {Cappi},
  {Coupon}, {Cucciati}, {Davidzon}, {Franzetti}, {Fritz}, {Iovino}, {Krywult},
  {Le Brun}, {Le F{\`e}vre}, {Maccagni}, {Ma{\l}ek}, {Marchetti}, {Marulli},
  {Polletta}, {Pollo}, {Tasca}, {Tojeiro}, {Vergani}, {Zanichelli}, {Bel},
  {Branchini}, {De Lucia}, {Ilbert}, {McCracken}, {Moutard}, {Peacock},
  {Zamorani}, {Burden}, {Fumana}, {Jullo}, {Marinoni}, {Mellier}, {Moscardini},
  \& {Percival}}]{Scodeggio2018}
{Scodeggio}, M., {Guzzo}, L., {Garilli}, B., {et~al.} 2018, \aap, 609, A84

\bibitem[{{Scrimgeour} {et~al.}(2012){Scrimgeour}, {Davis}, {Blake}, {James},
  {Poole}, {Staveley-Smith}, {Brough}, {Colless}, {Contreras}, {Couch},
  {Croom}, {Croton}, {Drinkwater}, {Forster}, {Gilbank}, {Gladders},
  {Glazebrook}, {Jelliffe}, {Jurek}, {Li}, {Madore}, {Martin}, {Pimbblet},
  {Pracy}, {Sharp}, {Wisnioski}, {Woods}, {Wyder}, \& {Yee}}]{Scrimgeour2012}
{Scrimgeour}, M.~I., {Davis}, T., {Blake}, C., {et~al.} 2012, \mnras, 425, 116

\bibitem[{Seabold \& Perktold(2010)}]{Seabold2010}
Seabold, S. \& Perktold, J. 2010, in 9th Python in Science Conference

\bibitem[{{Shectman} {et~al.}(1996){Shectman}, {Landy}, {Oemler}, {Tucker},
  {Lin}, {Kirshner}, \& {Schechter}}]{Shectman1996}
{Shectman}, S.~A., {Landy}, S.~D., {Oemler}, A., {et~al.} 1996, \apj, 470, 172

\bibitem[{{Shen} {et~al.}(2006){Shen}, {Abel}, {Mo}, \& {Sheth}}]{Shen2006}
{Shen}, J., {Abel}, T., {Mo}, H.~J., \& {Sheth}, R.~K. 2006, \apj, 645, 783

\bibitem[{{Shurtleff}(2014)}]{Shurtleff2014}
{Shurtleff}, R. 2014, arXiv e-prints, arXiv:1408.2514

\bibitem[{{\v{S}id\'{a}k}(1967)}]{Sidak1967}
{\v{S}id\'{a}k}, Z. 1967, Journal of the American Statistical Association, 62,
  626

\bibitem[{{Simmons} {et~al.}(2011){Simmons}, {Nelson}, \&
  {Simonsohn}}]{Simmons2011}
{Simmons}, J.~P., {Nelson}, L.~D., \& {Simonsohn}, U. 2011, Psychological
  Science, 22, 1359, pMID: 22006061

\bibitem[{{Singh} {et~al.}(2015){Singh}, {Mandelbaum}, \& {More}}]{Singh2015}
{Singh}, S., {Mandelbaum}, R., \& {More}, S. 2015, \mnras, 450, 2195

\bibitem[{{Smoot} {et~al.}(1992){Smoot}, {Bennett}, {Kogut}, {Wright}, {Aymon},
  {Boggess}, {Cheng}, {de Amici}, {Gulkis}, {Hauser}, {Hinshaw}, {Jackson},
  {Janssen}, {Kaita}, {Kelsall}, {Keegstra}, {Lineweaver}, {Loewenstein},
  {Lubin}, {Mather}, {Meyer}, {Moseley}, {Murdock}, {Rokke}, {Silverberg},
  {Tenorio}, {Weiss}, \& {Wilkinson}}]{Smoot1992}
{Smoot}, G.~F., {Bennett}, C.~L., {Kogut}, A., {et~al.} 1992, \apjl, 396, L1

\bibitem[{{S{\"o}chting} {et~al.}(2002){S{\"o}chting}, {Clowes}, \&
  {Campusano}}]{Sochting2002}
{S{\"o}chting}, I.~K., {Clowes}, R.~G., \& {Campusano}, L.~E. 2002, \mnras,
  331, 569

\bibitem[{{S{\"o}chting} {et~al.}(2004){S{\"o}chting}, {Clowes}, \&
  {Campusano}}]{Sochting2004}
{S{\"o}chting}, I.~K., {Clowes}, R.~G., \& {Campusano}, L.~E. 2004, \mnras,
  347, 1241

\bibitem[{{Spergel} {et~al.}(2003){Spergel}, {Verde}, {Peiris}, {Komatsu},
  {Nolta}, {Bennett}, {Halpern}, {Hinshaw}, {Jarosik}, {Kogut}, {Limon},
  {Meyer}, {Page}, {Tucker}, {Weiland}, {Wollack}, \& {Wright}}]{Spergel2003}
{Spergel}, D.~N., {Verde}, L., {Peiris}, H.~V., {et~al.} 2003, \apjs, 148, 175

\bibitem[{{Springel} {et~al.}(2005){Springel}, {White}, {Jenkins}, {Frenk},
  {Yoshida}, {Gao}, {Navarro}, {Thacker}, {Croton}, {Helly}, {Peacock}, {Cole},
  {Thomas}, {Couchman}, {Evrard}, {Colberg}, \& {Pearce}}]{Springel2005}
{Springel}, V., {White}, S. D.~M., {Jenkins}, A., {et~al.} 2005, \nat, 435, 629

\bibitem[{{Story} {et~al.}(2013){Story}, {Reichardt}, {Hou}, {Keisler}, {Aird},
  {Benson}, {Bleem}, {Carlstrom}, {Chang}, {Cho}, {Crawford}, {Crites}, {de
  Haan}, {Dobbs}, {Dudley}, {Follin}, {George}, {Halverson}, {Holder},
  {Holzapfel}, {Hoover}, {Hrubes}, {Joy}, {Knox}, {Lee}, {Leitch}, {Lueker},
  {Luong-Van}, {McMahon}, {Mehl}, {Meyer}, {Millea}, {Mohr}, {Montroy},
  {Padin}, {Plagge}, {Pryke}, {Ruhl}, {Sayre}, {Schaffer}, {Shaw}, {Shirokoff},
  {Spieler}, {Staniszewski}, {Stark}, {van Engelen}, {Vanderlinde}, {Vieira},
  {Williamson}, \& {Zahn}}]{Story2013}
{Story}, K.~T., {Reichardt}, C.~L., {Hou}, Z., {et~al.} 2013, \apj, 779, 86

\bibitem[{{Sunyaev} \& {Zel\textquotesingle
  dovich}(1970)}]{SunyaevZeldovich1970}
{Sunyaev}, R.~A. \& {Zel\textquotesingle dovich}, Y.~B. 1970, \apss, 7, 3

\bibitem[{{Sunyaev} \& {Zel\textquotesingle
  dovich}(1972)}]{SunyaevZeldovich1972}
{Sunyaev}, R.~A. \& {Zel\textquotesingle dovich}, Y.~B. 1972, Comments on
  Astrophysics and Space Physics, 4, 173

\bibitem[{{Sylos Labini}(2011)}]{SylosLabini2011}
{Sylos Labini}, F. 2011, Classical and Quantum Gravity, 28, 164003

\bibitem[{{Taylor} \& {Jagannathan}(2016)}]{Taylor2016}
{Taylor}, A.~R. \& {Jagannathan}, P. 2016, \mnras, 459, L36

\bibitem[{{Tegmark} {et~al.}(2004){Tegmark}, {Strauss}, {Blanton}, {Abazajian},
  {Dodelson}, {Sandvik}, {Wang}, {Weinberg}, {Zehavi}, {Bahcall}, {Hoyle},
  {Schlegel}, {Scoccimarro}, {Vogeley}, {Berlind}, {Budavari}, {Connolly},
  {Eisenstein}, {Finkbeiner}, {Frieman}, {Gunn}, {Hui}, {Jain}, {Johnston},
  {Kent}, {Lin}, {Nakajima}, {Nichol}, {Ostriker}, {Pope}, {Scranton},
  {Seljak}, {Sheth}, {Stebbins}, {Szalay}, {Szapudi}, {Xu}, {Annis},
  {Brinkmann}, {Burles}, {Castander}, {Csabai}, {Loveday}, {Doi}, {Fukugita},
  {Gillespie}, {Hennessy}, {Hogg}, {Ivezi{\'c}}, {Knapp}, {Lamb}, {Lee},
  {Lupton}, {McKay}, {Kunszt}, {Munn}, {O'Connell}, {Peoples}, {Pier},
  {Richmond}, {Rockosi}, {Schneider}, {Stoughton}, {Tucker}, {vanden Berk},
  {Yanny}, \& {York}}]{Tegmark2004}
{Tegmark}, M., {Strauss}, M.~A., {Blanton}, M.~R., {et~al.} 2004, \prd, 69,
  103501

\bibitem[{{Tempel} \& {Tamm}(2015)}]{Tempel2015}
{Tempel}, E. \& {Tamm}, A. 2015, \aap, 576, L5

\bibitem[{{Tiwari} \& {Jain}(2013)}]{Tiwari2013}
{Tiwari}, P. \& {Jain}, P. 2013, International Journal of Modern Physics D, 22,
  1350089

\bibitem[{{Tiwari} \& {Jain}(2019)}]{Tiwari2019}
{Tiwari}, P. \& {Jain}, P. 2019, \aap, 622, A113

\bibitem[{{Tully} {et~al.}(2015){Tully}, {Libeskind}, {Karachentsev},
  {Karachentseva}, {Rizzi}, \& {Shaya}}]{Tully2015}
{Tully}, R.~B., {Libeskind}, N.~I., {Karachentsev}, I.~D., {et~al.} 2015,
  \apjl, 802, L25

\bibitem[{{Turner}(1991)}]{Turner1991}
{Turner}, E.~L. 1991, \aj, 101, 5

\bibitem[{Upton \& Fingleton(1989)}]{Upton1989}
Upton, G.~J. \& Fingleton, B. 1989, Spatial Data Analysis by Example-Volume 2:
  Categorical and Directional Data (John Wiley \& Sons)

\bibitem[{{Vanden Berk} {et~al.}(2005){Vanden Berk}, {Schneider}, {Richards},
  {Hall}, {Strauss}, {Brunner}, {Fan}, {Baldry}, {York}, {Gunn}, {Nichol},
  {Meiksin}, \& {Brinkmann}}]{VandenBerk2005}
{Vanden Berk}, D.~E., {Schneider}, D.~P., {Richards}, G.~T., {et~al.} 2005,
  \aj, 129, 2047

\bibitem[{{Vielva} {et~al.}(2004){Vielva}, {Mart{\'\i}nez-Gonz{\'a}lez},
  {Barreiro}, {Sanz}, \& {Cay{\'o}n}}]{Vielva2004}
{Vielva}, P., {Mart{\'\i}nez-Gonz{\'a}lez}, E., {Barreiro}, R.~B., {Sanz},
  J.~L., \& {Cay{\'o}n}, L. 2004, \apj, 609, 22

\bibitem[{{Virtanen} {et~al.}(2019){Virtanen}, {Gommers}, {Oliphant},
  {Haberland}, {Reddy}, {Cournapeau}, {Burovski}, {Peterson}, {Weckesser},
  {Bright}, {van der Walt}, {Brett}, {Wilson}, {Jarrod Millman}, {Mayorov},
  {Nelson}, {Jones}, {Kern}, {Larson}, {Carey}, {Polat}, {Feng}, {Moore}, {Vand
  erPlas}, {Laxalde}, {Perktold}, {Cimrman}, {Henriksen}, {Quintero}, {Harris},
  {Archibald}, {Ribeiro}, {Pedregosa}, {van Mulbregt}, \&
  {Contributors}}]{Virtanen2019}
{Virtanen}, P., {Gommers}, R., {Oliphant}, T.~E., {et~al.} 2019, arXiv
  e-prints, arXiv:1907.10121

\bibitem[{Voronoi(1908)}]{Voronoi1908}
Voronoi, G. 1908, Journal für die Reine und Angewandte Mathematik, 134

\bibitem[{{Webster}(1982)}]{Webster1982}
{Webster}, A. 1982, \mnras, 199, 683

\bibitem[{Welch(1938)}]{Welch1938}
Welch, B.~L. 1938, Biometrika, 29, 350

\bibitem[{{Welker} {et~al.}(2020){Welker}, {Bland-Hawthorn}, {Van de Sande},
  {Lagos}, {Elahi}, {Obreschkow}, {Bryant}, {Pichon}, {Cortese}, {Richards},
  {Croom}, {Goodwin}, {Lawrence}, {Sweet}, {Lopez-Sanchez}, {Medling}, {Owers},
  {Dubois}, \& {Devriendt}}]{Welker2020}
{Welker}, C., {Bland-Hawthorn}, J., {Van de Sande}, J., {et~al.} 2020, \mnras,
  491, 2864

\bibitem[{{West} {et~al.}(2017){West}, {de Propris}, {Bremer}, \&
  {Phillipps}}]{West2017}
{West}, M.~J., {de Propris}, R., {Bremer}, M.~N., \& {Phillipps}, S. 2017,
  Nature Astronomy, 1, 0157

\bibitem[{{\noopsort{Weygaert}}{van de Weygaert}(1994)}]{vandeWeygaert1994}
{\noopsort{Weygaert}}{van de Weygaert}, R. 1994, \aap, 283, 361

\bibitem[{{\noopsort{Weygaert}}{van de Weygaert}(2007)}]{vandeWeygaert2007}
{\noopsort{Weygaert}}{van de Weygaert}, R. 2007, arXiv e-prints,
  arXiv:0707.2877

\bibitem[{{\noopsort{Weygaert}}{van de Weygaert} \&
  {Schaap}(2001)}]{vandeWeygaert2001}
{\noopsort{Weygaert}}{van de Weygaert}, R. \& {Schaap}, W. 2001, in Mining the
  Sky, ed. A.~J. {Banday}, S.~{Zaroubi}, \& M.~{Bartelmann}, 268

\bibitem[{{Williams} {et~al.}(1996){Williams}, {Blacker}, {Dickinson}, {Dixon},
  {Ferguson}, {Fruchter}, {Giavalisco}, {Gilliland}, {Heyer}, {Katsanis},
  {Levay}, {Lucas}, {McElroy}, {Petro}, {Postman}, {Adorf}, \&
  {Hook}}]{Williams1996}
{Williams}, R.~E., {Blacker}, B., {Dickinson}, M., {et~al.} 1996, \aj, 112,
  1335

\bibitem[{{Williger} {et~al.}(2002){Williger}, {Campusano}, {Clowes}, \&
  {Graham}}]{Williger2002}
{Williger}, G.~M., {Campusano}, L.~E., {Clowes}, R.~G., \& {Graham}, M.~J.
  2002, \apj, 578, 708

\bibitem[{{Yadav} {et~al.}(2010){Yadav}, {Bagla}, \& {Khandai}}]{Yadav2010}
{Yadav}, J.~K., {Bagla}, J.~S., \& {Khandai}, N. 2010, \mnras, 405, 2009

\bibitem[{Yamada {et~al.}(2009)Yamada, Funatsuki, Hagihara, FUJITA, TANAKA,
  TSUJI, ISHIMOTO, FUJINO, \& HAJIKA}]{Yamada2009}
Yamada, T., Funatsuki, H., Hagihara, S., {et~al.} 2009, Breeding science, 59,
  435

\bibitem[{{Yang} {et~al.}(2020){Yang}, {Hudson}, \& {Afshordi}}]{Yang2020}
{Yang}, T., {Hudson}, M.~J., \& {Afshordi}, N. 2020, arXiv e-prints,
  arXiv:2001.10943

\bibitem[{{Yershov} {et~al.}(2012){Yershov}, {Orlov}, \&
  {Raikov}}]{Yershov2012}
{Yershov}, V.~N., {Orlov}, V.~V., \& {Raikov}, A.~A. 2012, \mnras, 423, 2147

\bibitem[{{Yershov} {et~al.}(2014){Yershov}, {Orlov}, \&
  {Raikov}}]{Yershov2014}
{Yershov}, V.~N., {Orlov}, V.~V., \& {Raikov}, A.~A. 2014, \mnras, 445, 2440

\bibitem[{{Yershov} {et~al.}(2020){Yershov}, {Raikov}, {Lovyagin}, {Kuin}, \&
  {Popova}}]{Yershov2020}
{Yershov}, V.~N., {Raikov}, A.~A., {Lovyagin}, N.~Y., {Kuin}, N.~P.~M., \&
  {Popova}, E.~A. 2020, \mnras, 492, 5052

\bibitem[{{York} {et~al.}(2000){York}, {Adelman}, {Anderson}, {Anderson},
  {Annis}, {Bahcall}, {Bakken}, {Barkhouser}, {Bastian}, {Berman}, {Boroski},
  {Bracker}, {Briegel}, {Briggs}, {Brinkmann}, {Brunner}, {Burles}, {Carey},
  {Carr}, {Castander}, {Chen}, {Colestock}, {Connolly}, {Crocker}, {Csabai},
  {Czarapata}, {Davis}, {Doi}, {Dombeck}, {Eisenstein}, {Ellman}, {Elms},
  {Evans}, {Fan}, {Federwitz}, {Fiscelli}, {Friedman}, {Frieman}, {Fukugita},
  {Gillespie}, {Gunn}, {Gurbani}, {de Haas}, {Haldeman}, {Harris}, {Hayes},
  {Heckman}, {Hennessy}, {Hindsley}, {Holm}, {Holmgren}, {Huang}, {Hull},
  {Husby}, {Ichikawa}, {Ichikawa}, {Ivezi{\'c}}, {Kent}, {Kim}, {Kinney},
  {Klaene}, {Kleinman}, {Kleinman}, {Knapp}, {Korienek}, {Kron}, {Kunszt},
  {Lamb}, {Lee}, {Leger}, {Limmongkol}, {Lindenmeyer}, {Long}, {Loomis},
  {Loveday}, {Lucinio}, {Lupton}, {MacKinnon}, {Mannery}, {Mantsch}, {Margon},
  {McGehee}, {McKay}, {Meiksin}, {Merelli}, {Monet}, {Munn}, {Narayanan},
  {Nash}, {Neilsen}, {Neswold}, {Newberg}, {Nichol}, {Nicinski}, {Nonino},
  {Okada}, {Okamura}, {Ostriker}, {Owen}, {Pauls}, {Peoples}, {Peterson},
  {Petravick}, {Pier}, {Pope}, {Pordes}, {Prosapio}, {Rechenmacher}, {Quinn},
  {Richards}, {Richmond}, {Rivetta}, {Rockosi}, {Ruthmansdorfer}, {Sand ford},
  {Schlegel}, {Schneider}, {Sekiguchi}, {Sergey}, {Shimasaku}, {Siegmund},
  {Smee}, {Smith}, {Snedden}, {Stone}, {Stoughton}, {Strauss}, {Stubbs},
  {SubbaRao}, {Szalay}, {Szapudi}, {Szokoly}, {Thakar}, {Tremonti}, {Tucker},
  {Uomoto}, {Vanden Berk}, {Vogeley}, {Waddell}, {Wang}, {Watanabe},
  {Weinberg}, {Yanny}, {Yasuda}, \& {SDSS Collaboration}}]{York2000}
{York}, D.~G., {Adelman}, J., {Anderson}, John~E., J., {et~al.} 2000, \aj, 120,
  1579

\bibitem[{{Zel'Dovich}(1970)}]{ZelDovich1970}
{Zel'Dovich}, Y.~B. 1970, \aap, 500, 13

\bibitem[{{Zhang} {et~al.}(2013){Zhang}, {Yang}, {Wang}, {Wang}, {Mo}, \& {van
  den Bosch}}]{Zhang2013}
{Zhang}, Y., {Yang}, X., {Wang}, H., {et~al.} 2013, \apj, 779, 160

\bibitem[{Zonca {et~al.}(2019)Zonca, Singer, Lenz, Reinecke, Rosset, Hivon, \&
  Gorski}]{Zonca2019}
Zonca, A., Singer, L., Lenz, D., {et~al.} 2019, Journal of Open Source
  Software, 4, 1298

\end{thebibliography}
